\documentclass[a4paper,11pt]{extarticle}
\pdfoutput=1 

\usepackage{jheppub} 
\usepackage{hyperref}
\usepackage[toc,page]{appendix}
\hypersetup{
    bookmarks=true,         
    unicode=false,          
    pdftoolbar=true,        
   pdfmenubar=true,        
    pdffitwindow=false,     
    pdfstartview={FitH},    
    pdftitle={My title},    
    pdfauthor={Author},     
    pdfsubject={Subject},   
    pdfcreator={Creator},   
    pdfproducer={Producer}, 
    pdfkeywords={keyword1} {key2} {key3}, 
    pdfnewwindow=true,      
    colorlinks=true,       
    linkcolor=red,          
    citecolor=purple,        
    filecolor=magenta,      
    urlcolor=blue,           
    linktocpage=true
}
\usepackage{graphics}
\usepackage{rotating}
\usepackage{subfigure}
\usepackage{pstricks}
\usepackage{color}
\usepackage{amsfonts}
\usepackage{mathrsfs,amsmath}
\usepackage{epsfig}
\usepackage{amsmath,amssymb,amsthm,graphicx,latexsym}
\usepackage{ulem}
\renewcommand{\arraystretch}{2}

\newcommand{\be}{\begin{equation}}
\newcommand{\ee}{\end{equation}}
\newcommand{\bea}{\begin{eqnarray}}
\newcommand{\eea}{\end{eqnarray}}
\newcommand{\bml}{\begin{subequations}}
\newcommand{\eml}{\end{subequations}}
\newcommand{\bfig}{\begin{figure}}
\newcommand{\efig}{\end{figure}}

\newcommand{\bfx}{\mbox{\boldmath{$x$}}}

\newcommand{\slashed}{\hspace{-1.1ex}/}

\newcommand{\bmat}{\begin{pmatrix}}
\newcommand{\emat}{\end{pmatrix}}
\begin{document}
	\title{
	\textcolor{purple}{CMB from EFT}}
	\author[a,b]{Sayantan Choudhury,
		\footnote{\textcolor{purple}{\bf Alternative
				E-mail: sayanphysicsisi@gmail.com}. \newline \textcolor{blue}{\bf NOTE: This project is the part of the non-profit virtual international research consortium ``Quantum Structures of the Space-Time \& Matter''.}}}
				
				\affiliation[a]{Quantum Gravity and Unified Theory and Theoretical Cosmology Group, Max Planck Institute for Gravitational Physics (Albert Einstein Institute), Am M$\ddot{\text{u}}$hlenberg 1, 14476 Potsdam-Golm, Germany.}
	\affiliation[b]{
		Inter-University Centre for Astronomy and Astrophysics, Post Bag 4,
		Ganeshkhind, Pune 411007, India.
	}	
	\emailAdd{sayantan.choudhury@aei.mpg.de}

	\abstract{In this work, we study the key role of generic Effective Field Theory (EFT) framework to quantify the correlation functions in a quasi de Sitter background for an arbitrary initial choice of the quantum vacuum state. We perform the computation in unitary gauge, in which we apply the St$\ddot{\text{u}}$ckelberg trick in lowest dimensional EFT operators which are broken under time diffeomorphism. In particular, using this non-linear realization of broken time diffeomorphism and truncating the action by considering the contribution from two derivative terms in the metric, we compute the two-point and three-point correlations from scalar perturbations and two-point correlation from tensor perturbations to quantify the quantum fluctuations observed in the Cosmic Microwave Background (CMB) map. We also use equilateral limit and squeezed limit configurations for the scalar three-point correlations in Fourier space. To give future predictions from EFT setup and to check the consistency of our derived results for correlations, we use the results obtained from all classes of the canonical single-field and general single-field $P(X,\phi)$ model. This analysis helps us to fix the coefficients of the relevant operators in EFT in terms of the slow-roll parameters and effective sound speed. Finally, using CMB observations from Planck we constrain all these coefficients of EFT operators for the single-field slow-roll inflationary paradigm. }

	\keywords{Effective field theories, Cosmology of Theories beyond the SM, Quantum Field Theory of De Sitter space.}

	\maketitle
	\flushbottom
	\section{{Introduction}}
	\label{v1}
	The basic idea of Effective Field Theory (EFT) is very useful in many branches in theoretical physics including particle physics~\cite{Pich:1998xt,Burgess:2007pt}, condensed matter physics~\cite{Shankar:1996vk}, gravity~\cite{Donoghue:1995cz,Donoghue:2012zc}, cosmology~\cite{Cheung:2007st,Weinberg:2008hq,Agarwal:2012mq,Giblin:2017qjp,Ozsoy:2015rna,Burgess:2017ytm,Baumann:2014nda,Baumann:2009ds,Choudhury:2016wlj,Choudhury:2013iaa,Choudhury:2014sua,Delacretaz:2016nhw,Delacretaz:2015edn,LopezNacir:2011kk,Naskar:2017ekm,Senatore:2010wk,Senatore:2009gt,Behbahani:2011it,Cheung:2007sv,Baumann:2015nta,Assassi:2013gxa} and hydrodynamics~\cite{Dubovsky:2011sj,Crossley:2015evo}. In~a more technical ground, EFT framework is an approximated model-independent version of the underlying physical theory which is valid up to a specified cut-off scale at high energies, commonly known as UV cut-off scale ($\Lambda_{\rm UV}$), which is in usual practice fixed at the Planck scale $M_p$. EFT prescription deal with all possible relevant and irrelevant operators allowed by the underlying symmetry in the effective action and all the higher-dimensional non-renormalizable operators are accordingly suppressed by the UV cut-off scale ($\Lambda_{\rm UV}\sim M_p$). There are two possible approaches within the framework of quantum field theory (QFT) using which one can explain the origin of EFT, which are appended~below:
	\begin{enumerate}
		\item \underline{\textcolor{red}{\bf Top-down~approach:}}
		 In this case, the usual idea is to start with a UV complete fundamental QFT framework which contain all possible degrees of freedom. Furthermore, using this setup one can finally derive the EFT of relevant degrees of freedom at low energy scale $\Lambda_{s}<\Lambda_{\rm UV}\sim M_p$ by doing path integration over all irrelevant field contents~\cite{Baumann:2014nda,Choudhury:2016wlj}. To~demonstrate this idea in a more technical ground let us consider a visible sector light scalar field $\phi$ which has a very small mass $m_{\phi}<\Lambda_{\rm UV}\sim M_p$ and heavy scalar fields $\Psi_{i}\forall i=1,2,\cdots, N$ with mass $M_{\Psi_{i}}>\Lambda_{\rm UV}\sim M_p$, in~the hidden sector of the theory. the representative action of the theory is described by the following action~\cite{Baumann:2014nda,Choudhury:2016wlj}:
		\bea \label{we1} {S[\phi,\Psi_{i},g_{\mu\nu}]=\int d^4x \sqrt{-g}\left[\frac{M^2_p}{2}R+{\cal L}_{\rm vis}[\phi]+\sum^{N}_{i=1}{\cal L}^{(i)}_{\rm hid}[\Psi_{i}]+\sum^{N}_{j=1}{\cal L}^{(j)}_{\rm int}[\phi,\Psi_{j}]\right]}~,\eea
		where $g_{\mu\nu}$ is the classical background metric, ${\cal L}_{\rm vis}[\phi]$ is the Lagrangian density of the visible sector light field, $ {\cal L}^{(i)}_{\rm hid}[\Psi_{i}]\forall i=1,2,\cdots, N$ is the Lagrangian density of the hidden-sector heavy field and ${\cal L}^{(j)}_{\rm int}[\phi,\Psi_{j}]\forall j=1,2,\cdots, N$ is the Lagrangian density of the interaction between hidden-sector and visible-sector field. Furthermore, using Equation~(\ref{we1}) one can construct an EFT by performing path integration over the contributions from all hidden-sector heavy fields and all possible high-frequency 		contributions as given by:
		\bea S_{EFT}\left[\phi,g_{\mu\nu}\right]=-i\ln\left[\prod^{N}_{j=1}\int\left[{\cal D}\Psi_{j}\right]e^{iS[\phi,\Psi_{j},g_{\mu\nu}]}\right]=-i\sum^{N}_{j=1}\ln\left[\int\left[{\cal D}\Psi_{j}\right]e^{S[\phi,\Psi_{j},g_{\mu\nu}]}\right].~~~~~\eea
		Finally, one can express the EFT action in terms of the systematic series expansion of visible sector light degrees of freedom and classical gravitational background as~\cite{Baumann:2014nda,Choudhury:2016wlj}:
		\bea \label{we2}{S_{EFT}\left[\phi,g_{\mu\nu}\right]=\int d^4x \sqrt{-g}\left[\frac{M^2_p}{2}R+{\cal L}_{\rm vis}[\phi]+\sum_{\gamma}\sum^{N}_{j=1}{\cal C}^{(j)}_{\gamma}(g_{c})\frac{\widetilde{\cal O}^{(j)}_{\gamma}[\phi]}{M^{\Delta_{\gamma}-4}_{\Psi_{j}}}\right]}~,\eea
		where ${\cal C}^{(j)}_{\gamma}(g_{c})\forall \gamma,~\forall j=1,2,\cdots, N$ represent dimensionless coupling constants which depend on the parameter $g_{c}$ of the UV complete QFT. Also $\widetilde{\cal O}^{(j)}_{\gamma}[\phi]\forall \gamma,~\forall j=1,2,\cdots, N$ represent $\Delta_{\gamma}$ mass dimensional local EFT operators suppressed by the scale $M^{\Delta_{\gamma}-4}_{\Psi_{j}}$. In~this connection one of the best possible example of UV complete field theoretic setup is string theory from which one can derive an EFT setup at the string scale $\Lambda_{s}$ which is identified with $M_{\Psi_{j}}$ in Equation~(\ref{we2}).
		
		\item\underline{\textcolor{red}{\bf Bottom-up~approach:}}~~In this case, the usual idea is to start with a low-energy model-independent effective action allowed by the symmetry requirements. Using such a setup, the prime job is to find out the appropriate UV complete field theoretic setup allowed by the underlying \mbox{symmetries~\cite{Baumann:2014nda,Choudhury:2016wlj}}. This identification allows us to determine the coefficients of the EFT operators in terms of the model parameters of UV complete field theories. In~this paper we follow this approach to write down the most generic EFT framework using which we describe the theory of quantum fluctuations observed in CMB around a quasi de Sitter inflationary background solution of Einstein's equations. 
	\end{enumerate}
	
	In this paper, our prime objective is to compute the expressions for the cosmological two- and three-point correlation functions in unitary gauge using the well-known St$\ddot{\text{u}}$ckelberg trick~\cite{Ruegg:2003ps,GrosseKnetter:1992nn} along with the arbitrary choice of initial quantum vacuum state. The~working principle of the St$\ddot{\text{u}}$ckelberg trick in quasi de Sitter background is to break the time diffeomorphism symmetry to generate all the required quantum fluctuations observed in CMB. This is exactly same as applicable in the context of $SU(N)$ non-abelian gauge theory to describe the spontaneous symmetry breaking. In~the present context 
the scalar modes which are appearing from the quantum fluctuation exactly mimic the role of Goldstone mode as appearing in $SU(N)$ non-abelian gauge theory. After~breaking the time diffeomorphism in the unitary gauge, scalar Goldstone-like degrees of freedom are eaten by the metric. In~unitary gauge, to~write a most generic EFT in terms of operators which breaks time diffeomorphism symmetry, the~following contributions play significant roles in quasi de Sitter~background:
\begin{itemize}
	\item Polynomial powers of the time fluctuation of the component in the metric, $g^{00}$ such as, \mbox{$\delta g^{00}= g^{00}+1$,}
	\item Polynomial powers of the time fluctuation in the extrinsic curvature at constant time surfaces, $K_{\mu \nu}$ such as, $\delta K_{\mu\nu}=\left(K_{\mu\nu}-a^2Hh_{\mu\nu}\right)$, where $a$ is the scale factor in quasi de Sitter background.
\end{itemize} 

 Construction of EFT action using the St$\ddot{\text{u}}$ckelberg trick also allows us to characterize all the possible contributions to the model-independent simple versions of field theoretic framework based on the models of the inflationary paradigm described by single field, where the observables are constrained by the CMB observation appearing from Planck data. It is important to note that this idea of constructing EFT action using the St$\ddot{\text{u}}$ckelberg trick can also be generalized to the EFT framework guided by multiple number of scalar fields as~well.

 The main highlighting points of this paper are appended below~pointwise:
 \begin{enumerate}
 	\item We have presented all the results by restricting up to all possible contributions coming from the two derivative terms in the metric which finally give rise to a consistently truncated EFT action\footnote{In the context of EFT one can in principle consider terms containing more than two derivatives in the metric, which will give rise to appearance of many higher derivative operators, i.e.,~$\displaystyle \sum^{\infty}_{n=2} R^{n}$, $\displaystyle\sum^{\infty}_{m=1} (R_{\mu\nu}R^{\mu\nu})^{m}$, $\displaystyle\sum^{\infty}_{p=1} (R_{\mu\nu\alpha\beta}R^{\mu\nu\alpha\beta})^{p}$,  $\displaystyle\sum^{\infty}_{q=1} (R_{\mu\nu\alpha\beta}R^{\mu\nu}R^{\alpha\beta})^{q}$ and various other terms which contain the quantum fluctuation on the trace of the extrinsic curvature terms i.e.,~$\displaystyle\sum^{\infty}_{m=1}(\delta K^{\mu}_{\mu})^{m+2}$ and other possible terms which are appearing due to all possible index contraction of extrinsic curvature terms i.e.,~$\displaystyle \sum^{\infty}_{m=1}\delta K^{\mu_{1}}_{\mu_{2}}\delta K^{\mu_{2}}_{\mu_{3}}\delta K^{\mu_{3}}_{\mu_{4}}\cdots\delta K^{\mu_{m+1}}_{\mu_{m+2}}\delta K^{\mu_{m+2}}_{\mu_{1}}$ in the gravity sector of the EFT action. However, in~the present work our prime objective is to compute the expressions for cosmological correlation (two- and three-point) functions from quasi-de Sitter space. For~this reason it is sufficient enough to consider the two derivative terms in the metric as such contributions will appear in the two- and three-point correlation functions in the leading order. If~one condensed the effects of higher derivative terms in the metric it will appear at the sub-leading or sub-sub-leading-order expressions for the correlation functions, which are highly suppressed due to the validity of slow-roll  approximations, i.e.,~$\displaystyle\epsilon=-\frac{\dot{H}}{H^2}<<1$ and $\displaystyle|\eta|=\left|\epsilon-\frac{\dot{\epsilon}}{2\epsilon H}\right|<<1$ during inflation. Thus, it implies that due to very small numerical contributions one can easily neglect the terms which contain the higher-order slow-roll contributions in the two- and three-point cosmological correlation functions. In~our computations performed in this paper we have also maintained these approximations everywhere and this will give rise to the leading-order result which we have presented explicitly later. This is the main reason for which we have restricted up to two derivative terms in the metric in this paper. }. Consequently, we get consistent predictions for single-field slow-roll~\cite{Maldacena:2002vr,Ade:2015lrj,Shukla:2016bnu,Choudhury:2017cos,Choudhury:2015pqa,Choudhury:2014hja,Choudhury:2013jya,Choudhury:2013zna,Choudhury:2011jt,Choudhury:2011sq,Choudhury:2012ib,Choudhury:2011rz} and Generalized Single-Field $P(X,\phi)$ models of inflation~\cite{Chen:2006nt,Alishahiha:2004eh,Choudhury:2015yna,Choudhury:2012yh,Choudhury:2014sxa,Choudhury:2014uxa,Choudhury:2015hvr,Bhattacharjee:2014toa,Chingangbam:2004ng,Mazumdar:2001mm,Choudhury:2003vr,Choudhury:2002xu,Panda:2005sg,Baumann:2011ws}. In~earlier works various efforts are made to derive cosmological three-point correlation functions by writing a consistent EFT action in the similar theoretical framework. However, the~earlier results are not consistent with the single-field slow-roll inflation with effective sound speed $c_S=1$ as it predicts vanishing three-point correlation function for scalar fluctuations. See ref.~\cite{Cheung:2007st} for more details. The~main reason for this inconsistency was ignoring specific contributions from the fluctuation in the EFT action, which give rise to improper~truncation.  
 	
 	\item We have computed the analytical expression for the two-point and three-point correlation function for the scalar fluctuation in quasi-de Sitter inflationary background in the presence of generalized initial quantum state. Also, for the first time we have presented the result for two-point correlation function for the tensor fluctuation in this context. To~simplify our results we have also presented the results for Bunch–Davies vacuum and $\alpha,\beta$ vacuums\footnote{In QFT of quasi de Sitter space we deal with a class of non-thermal quantum states, characterized by infinite family of two real parameters $\alpha$ and $\beta$, commonly known as $\alpha,\beta$ vacuums. It is important to note that $\alpha,\beta$ quantum states are CP invariant under the ${\rm SO(1,4)}$ de Sitter isometry group. On~the other hand, we fix $\beta=0$ then we get $\alpha$ vacuum which is actually CPT-invariant under the ${\rm SO(1,4)}$ de Sitter isometry group. Furthermore, if~we fix both $\alpha=0$ and $\beta=0$ then we get the thermal Bunch–Davies vacuum state.}.
 	
 	\item We have presented the exact analytical expressions for all the coefficients of EFT operators for single-field slow-roll and Generalized Single-Field $P(X,\phi)$ models of inflation in terms of the time-dependent slow-roll parameters as well the parameters which characterize the generalized initial quantum state. To~give numerical estimates we have further presented the results for Bunch–Davies vacuum and $\alpha,\beta$ vacuums.

 \end{enumerate}

This paper is organized as follows. In~Section \ref{v2}, we discuss the overview of the EFT framework under consideration, which includes the construction of the EFT action under broken time diffeomorphism in quasi de Sitter background. In~Section \ref{v3}, we derive the expression for the two-point correlation function from EFT using scalar and tensor mode fluctuation. Furthermore, in Section \ref{v4}, we~derive the expression for the scalar three-point function from EFT using scalar mode fluctuation in equilateral and squeezed limit configurations. After~that, in Section \ref{v5}, we derive the exact analytical expressions for coefficients of EFT operators for both single-field slow-roll inflation and generalized single-field $P(X,\phi)$ models of inflation. Finally, we conclude in Section \ref{v6} with some future prospects of the present~work.

\section{{Overview of~EFT}\label{v2}}
\unskip
\subsection{Construction of the Generic EFT Action\label{v2a}}

In this section, our motivation is to construct the most generic EFT action in the background of quasi de Sitter space.  Before~going into the further technical details it is important to note that the method of implementing cosmological perturbation using a scalar field is different compared to the generic EFT framework.  However, the underlying connection can be explained by interpreting the scalar (inflaton) field as a scalar under all space-time diffeomorphisms in General Relativity:
	\be\label{xc1} {\textcolor{red}{\textbf{Space-time~diffeomorphism:}}~~ x^{\mu}\Longrightarrow  x^{\mu}+\xi^{\mu}(t,{\bf x})~~~\forall~ \mu=0,1,2,3}~.\ee  
	
	Consequently, in the cosmological perturbation the scalar field $\delta \phi$ transforms like a scalar under the operation of spatial diffeomorphisms; 
on the other hand, it transforms in non-linear fashion with respect to time diffeomorphisms. The~space and time diffeomorphic transformation rules are appended~bellow:
\begin{small}
\begin{equation}
{\begin{array}{rl}
\textcolor{red}{\bf Spatial~diffeomorphism:}~~	t&\Longrightarrow t,~x^{i}\Longrightarrow  x^{i}+\xi^{i}(t,{\bf x})~~~\forall~ i=1,2,3\longrightarrow
\delta\phi\Longrightarrow \delta\phi,\\	
\textcolor{red}{\bf Time~diffeomorphism:}~~	t&\Longrightarrow t+ \xi^{0}(t,{\bf x}),~x^{i}\Longrightarrow x^{i}~~~\forall~ i=1,2,3\longrightarrow
	\delta\phi\Longrightarrow \delta\phi +\dot{\phi}_{0}(t)\xi^{0}(t,{\bf x}).
	\end{array}}
\end{equation}
\end{small}

Here $\xi^{0}(t,{\bf x})$ and $\xi^{i}(t,{\bf x})\forall i=1,2,3$ are the diffeomorphism parameter. In~this context one can choose a specific gauge in which we set the background scalar degrees of freedom as, $\phi(t,{\bf x})=\phi_{0}(t),$
which is consistent with the requirement that the perturbation in the scalar field vanishes: \be {\textcolor{red}{\bf Unitary~gauge~fixing}~~~~\Rightarrow~~~~\delta \phi(t,{\bf x})=0}~,\ee 

In cosmological perturbation theory this is known as unitary gauge in which all degrees of freedom are preserved in the metric of quasi de Sitter space.
This phenomenon is analogous to the spontaneous symmetry breaking as appearing in the context of $SU(N)$ gauge theory where the Goldstone mode transform in a non-linear fashion  and 
destroyed by the $SU(N)$ gauge boson in unitary gauge to give a massive spin $1$ degrees of freedom after symmetry breaking. In~an alternative way one can present the framework of EFT by describing cosmological perturbation theory during inflation where time 
diffeomorphisms are realized in non-linear~fashion.

Now to construct a most general structure of the EFT action suitable for the inflationary paradigm we need to follow the step appended~below:
\begin{enumerate}
	\item One must write down the EFT operators that are functions of the metric $g_{\mu\nu}$. Here one of the possibilities is Riemann tensor.
	\item Also the EFT operators are invariant under the linearly realized time-dependent spatial diffeomorphic transformation:
	\be\label{xc} {\textcolor{red}{\bf Spatial~diffeomorphism:}~~ 	t\Longrightarrow t,~~x^{i}\Longrightarrow x^{i}+\xi^{i}(t,{\bf x})~~~\forall~ i=1,2,3}~.\ee
	For an example, one can consider an EFT operator constructed by $g^{00}$ or its polynomials without derivatives which transform like a scalar under Equation~(\ref{xc}).
	\item  Due to the reduced symmetry of the physical system many more extra contributions are allowed in the EFT action.
	\item In the EFT action one can also allow geometrical quantities in a preferred space-time slice. For example, one can consider the extrinsic curvature $K_{\mu\nu}$ of surfaces at constant time, which transform like a tensor under Equation~(\ref{xc}).
\end{enumerate}

   Consequently, the most general 
EFT action can be written in terms of all possible allowed operators by the space-time diffeomorphism as~\cite{Cheung:2007st,Naskar:2017ekm}:
\begin{small}
\begin{equation}\label{Eqv1}
{
	\begin{array}{rl}
	S&=\displaystyle\int d^{4}x \sqrt{-g}\left[\frac{M^2_p}{2}R+M^2_p \dot{H} g^{00}-M^2_p \left(3H^2+\dot{H}\right)+\sum^{\infty}_{n=2}\frac{M^4_n(t)}{n!}
	(\delta g^{00})^n~~~~~~~~\right.\\&
	\left.\displaystyle~~~~~~~-\sum^{\infty}_{q=0}\frac{\bar{M}^{3-q}_1(t)}{(q+2)!}\delta g^{00}
	\left(\delta K_{\mu}^{\mu}\right)^{q+1}-\sum^{\infty}_{m=0}\frac{\bar{M}^{2-m}_2(t)}{(m+2)!}
	\left(\delta K_{\mu}^{\mu}\right)^{m+2}-\sum^{\infty}_{m=0}\frac{\bar{M}^{2-m}_3(t)}{(m+2)!}
	\left[\delta K\right]^{m+2}+\cdots \right].
	\end{array}}
\end{equation}
\end{small}
\hspace{-3pt}where the dots stand for higher-order fluctuations in the EFT action which contains operators with more derivatives in space-time metric. 
Here we use the following sets of definitions for extrinsic curvature $K_{\mu\nu}$, unit normal $n_{\mu}$ and 
induced metric $h_{\mu\nu}$:
\bea K_{\mu \nu}&=&h^{\sigma}_{\mu}\nabla_{\sigma} n_{\nu}=\frac{\delta^{0}_{\mu}\partial_{\nu}g^{00}+\delta^{0}_{\nu}\partial_{\mu}g^{00}}{2(-g^{00})^{3/2}}
+\frac{\delta^{0}_{\mu}\delta^{0}_{\nu}g^{0\sigma}\partial_{\sigma}g^{00}}{2(-g^{00})^{5/2}}-\frac{g^{0\rho}\left(\partial_{\mu}g_{\rho\nu}+\partial_{\nu}g_{\rho\mu}-\partial_{\rho}g_{\mu\nu}\right)}{2(-g^{00})^{1/2}}\nonumber,\\
h_{\mu \nu}&=&g_{\mu \nu}+n_{\mu} n_{\nu},~~~~
n_{\mu}=\frac{\partial_{\mu}t}{\sqrt{-g^{\mu \nu}\partial_{\mu}t \partial_{\nu}t}}
=\frac{\delta_{\mu}^0}{\sqrt{-g^{00}}}.\eea	

Here $\delta K_{\mu\nu}$ represents the variation of the extrinsic curvature of constant time surfaces with respect to the unperturbed background FLRW metric in quasi de Sitter space-time:
\bea \delta g^{00}&=&g^{00}+1,~~~~
\delta K_{\mu\nu}=K_{\mu\nu}-a^2H h_{\mu\nu}
.\eea	

Additionally, we have used a shorthand notation $[\delta K]$ to define the following tensor contraction rule useful to quantify the EFT action~\cite{Naskar:2017ekm}:
\bea	\left[\delta K\right]^{m+2} &=&\delta K^{\mu_{1}}_{\mu_{2}}\delta K^{\mu_{2}}_{\mu_{3}}\delta K^{\mu_{3}}_{\mu_{4}}\cdots\delta K^{\mu_{m+1}}_{\mu_{m+2}}\delta K^{\mu_{m+2}}_{\mu_{1}}.
\eea

Before going into the further details let us first point out the few important characteristics of the EFT action which are appended~bellow:
\begin{itemize}
	\item In the EFT action the operators $M^2_p\dot{H}g^{00}$ and $M^2_p\left(3H^2+\dot{H}\right)$ are completely specified by the Hubble parameter $H(t)$ which is the solution of Friedman's equations in unperturbed~background. 
	
	\item The rest of the contributions in EFT action captures the effect of quantum fluctuations, which are characterized by the perturbation around the background FLRW solution of all UV complete theories of inflation. 
	\item  The coefficients of the operators appearing in the EFT action
are in general time-dependent. 
\end{itemize}

 Now as we are interested to compute the two- and three-point correlation function, we have restricted to the following truncated EFT action~\cite{Cheung:2007st,Naskar:2017ekm}:
\begin{small}
\begin{equation}\label{Eqv2}
{
	\begin{array}{rl}
 S&=\displaystyle\int d^{4}x \sqrt{-g}\left[\frac{M^2_p}{2}R+M^2_p \dot{H} g^{00}-M^2_p \left(3H^2+\dot{H}\right)+\frac{M^{4}_{2}(t)}{2!}\left(g^{00}+1\right)^2+\frac{M^{4}_{3}(t)}{3!}\left(g^{00}+1\right)^3~~~~~~~~\right.\\&
	\left.\displaystyle~~~~~~~~~~~~~~~~~~~~~~~~-\frac{\bar{M}^{3}_{1}(t)}{2}\left(g^{00}+1\right)\delta K^{\mu}_{\mu}-\frac{\bar{M}^{2}_{2}(t)}{2}(\delta K^{\mu}_{\mu})^2-\frac{\bar{M}^{2}_{3}(t)}{2}\delta K^{\mu}_{\nu}\delta K^{\nu}_{\mu}\right].
	\end{array}}
\end{equation}
\end{small}
\hspace{-3pt}where we have considered the terms in two derivatives in the metric\footnote{As we are dealing with EFT, in~principle one can consider operators which includes higher derivatives in the metric i.e.,~$\left(g^{00}+1\right)^2\delta K^2$, $\delta K^2 \delta K^{\nu}_{\mu}\delta K^{\mu}_{\nu}$, $\delta K^3$, $\delta K\delta N^2$ (here $\delta N=N-1$, where $N$ is the lapse function in ADM formalism. See~\mbox{ref.~\cite{Pirtskhalava:2015zwa}} for more details). But~since we have considered the terms two derivative in the metric we have truncated the EFT action in the form presented in Equation~(\ref{Eqv2}) and the form of the EFT action is exactly similar to ref.~\cite{Cheung:2007st}. In~this paper our prime objective is to concentrate only on the leading-order tree-level contributions and for this reason we have not considered any sub-leading suppressed contributions or any other contributions which are coming from the quantum loop corrections. Additionally, we have also neglected the term $\left(g^{00}+1\right)^2\delta K$ in the EFT action as this term is suppressed by the contribution $H^2\epsilon<<1$ in the decoupling limit and also the higher derivatives of the Goldstone mode $\pi$ after implementing the symmetry breaking through the St$\ddot{\text{u}}$ckelberg trick.  }. 
 
\subsection{EFT as a Theory of Goldstone~Boson}
\label{v2b}
\subsubsection{St${\ddot{\text{u}}}$ckelberg Trick I: an Example from $SU(N)$ Gauge Theory with Massive-Gauge Boson in Flat~Background}
\label{v2b1}
In the unitary gauge, the EFT action consists of graviton mode, two helicities, and scalar mode, respectively. In~this context first we apply a broken time diffeomorphic transformation on the Goldstone boson. As~a result, $SU(N)$ gauge symmetry~\cite{Cheung:2007st,Notes} is non-linearly realized in the framework of EFT. This~mechanism is commonly known as the 
{St$\ddot{\text{u}}$ckelberg trick}. Let us mention two crucial roles of the {St$\ddot{\text{u}}$ckelberg trick} in gauge~theory:
\begin{enumerate}
	\item Using this trick in $SU(N)$ gauge theory~\cite{Cheung:2007st,Notes} one can study the physical implications  from longitudinal components of a massive-gauge boson degrees of~freedom.
	
	\item  It is expected that in the weak coupling limit the contributions from the mixing terms are very small and consequently Goldstone modes decouple from the theory.
\end{enumerate}

To give a specific example of the {St$\ddot{\text{u}}$ckelberg trick} we consider $SU(N)$ gauge theory  characterized by a non-abelian gauge field $A^{a}_{\mu}$ in the background of Minkowski flat space-time. In~unitary gauge this theory is described by the following action:
\bea\label{po1} {S=\int d^4x\left[-\frac{1}{4}{\rm Tr}(F_{\mu\nu}F^{\mu\nu})-\frac{m^2}{2}{\rm Tr}(A_{\mu}A^{\mu})\right]}~,\eea
where $A_{\mu}=A^{a}_{\mu}T_{a}$ and $F^{a}_{\mu\nu}=\partial_{[\mu}A^a_{\nu]}$. Here the label $a=1,2,\cdots,N$ for $SU(N)$ gauge theory. Also $T_a$ are the generators of the non-abelian gauge group which satisfy the following properties:
\bea \left[T^a, T^b\right]&=& if^{abc}T_c,~~~~~~~~~
{\rm Tr}(T^a)=0,~~~~~~~{\rm Tr}(T^aT^b)=\frac{\delta^{ab}}{2}.\eea

Here $f^{abc}~\forall a,b,c =1,2,\cdots,N$ are the structure constants of the non-abelian $SU(N)$ gauge~theory.

It is important to mention that in~this context the $SU(N)$ gauge transformation on the non-abelian gauge field can be written as:
\bea A_{\mu}\Longrightarrow \tilde{A}_{\mu}=\frac{i}{g}UD_{\mu}U^{\dagger},~~~~~{\rm with}~~D_{\mu}=\partial_{\mu}-igA_{\mu}\eea
where $D_{\mu}$ is the covariant derivative. Here $g$ is the gauge coupling parameter for $SU(N)$ non-abelian gauge theory. Under~this gauge transformation each of the terms in the action stated in Equation~(\ref{po1}) transform as:
\bea {\rm Tr}(F_{\mu\nu}F^{\mu\nu})&\Longrightarrow&{\rm Tr}(\tilde{F}_{\mu\nu}\tilde{F}^{\mu\nu~})={\rm Tr}(F_{\mu\nu}F^{\mu\nu}),\\
\frac{m^2}{2}{\rm Tr}(A_{\mu}A^{\mu})&\Longrightarrow&\frac{m^2}{2}{\rm Tr}(\tilde{A}_{\mu}\tilde{A}^{\mu~})=\frac{m^2}{2g}{\rm Tr}[(D_{\mu}U^{\dagger})(D^{\mu}U)],\eea
where $U$ is the unitary operator in $SU(N)$ non-abelian gauge~theory.

Consequently, after doing $SU(N)$ gauge transformation action can be expressed as:
\begin{equation}\label{Eqv3}
{
	\begin{array}{rl}
	S\Longrightarrow \tilde{S}&= S+\underbrace{\int d^4x\left[\frac{m^2}{2}{\rm Tr}(A_{\mu}A^{\mu})-\frac{m^2}{2g}{\rm Tr}[(D_{\mu}U^{\dagger})(D^{\mu}U)]\right]}_{\textcolor{red}{\bf  Additional~part~which~breaks~SU(N)~gauge~symmetry}}.
	\end{array}}
\end{equation}
where $\underbrace{}$ term signifies the gauge symmetry breaking contribution in the unitary~gauge.

Furthermore, it is important to note that the $SU(N)$ gauge symmetry can be restored by defining the previously mentioned unitary operator in a following fashion:
\be U=\exp\left[iT^a\pi^a(t,{\bf x})\right],\ee
where one can identify the $\pi^a~\forall ~a=1,2,\cdots,N$~s with the Goldstone modes, which transform in a linear fashion under the action of the following gauge transformation:
\bea U\Longrightarrow \tilde{U}&=&\exp\left[iT^a\tilde{\pi}^a(t,{\bf x})\right]=\Sigma(t,{\bf x})\exp\left[iT^a\pi^a(t,{\bf x})\right]=\underbrace{\Sigma(t,{\bf x})}_{Local~operator}U.\eea

For the sake of simplicity one can rescale the Goldstone modes by absorbing the mass of the $SU(N)$ gauge field $m$ and the $SU(N)$ gauge coupling parameter $g$ by introducing the following canonical normalization as given by:
\bea {{\rm Canonical~normalization:}~~~~~~~~~\pi_c= \frac{m}{g}\pi~~}~.\eea

Consequently, the~action in terms of canonically normalized field $\pi_c$ can be written after $SU(N)$ gauge transformation as:
\begin{equation}\label{Eqv4}
{
	\begin{array}{rl}S\Longrightarrow \tilde{S}&=\displaystyle S+\int d^4x\left[\frac{m^2}{2}{\rm Tr}(A_{\mu}A^{\mu})-\underbrace{\frac{1}{2}{\rm Tr}[(\partial_{\mu}\pi_c)(\partial^{\mu}\pi_c)]}_{\textcolor{red}{\bf Kinetic~term~of~Goldstone}}\right.\\ & \left.~~~~~~~~~~~~~~~~~~~~~~~~~~~~~~~~\displaystyle\underbrace{-\frac{2g^2}{m}{\rm Tr}(A_{\mu}\partial^{\mu}\pi_c)+\frac{g^2}{2}{\rm Tr}(A_{\mu}A^{\mu}\pi^2_c)+ig{\rm Tr}(\pi_c A_{\mu}\partial^{\mu}\pi_c)}_{\textcolor{red}{\bf Mixing~terms~after~canonical~normalization}}\right].	\end{array}}
\end{equation}

It is important to note the important facts from Equation~(\ref{Eqv4}) which are appended~below:
\begin{itemize}
	\item The last two terms in Equation~(\ref{Eqv4}) are the mixing terms between the transverse component of the  $SU(N)$ gauge field, the Goldstone boson, and its kinetic term,~respectively.
	
	\item Here one can neglect all such mixing contributions at the energy scale $E_{mix}>>m$. Consequently, two sectors decouple from each other as they are weakly coupled in the energy scale $E_{mix}>>m$ and Equation~(\ref{Eqv4}) takes the following form:
	\vspace{12pt}
\begin{equation}\label{Eqv5}
	{
		\begin{array}{rl}
		S\Longrightarrow \tilde{S}&=\displaystyle S+\int d^4x\left[\frac{m^2}{2}{\rm Tr}(A_{\mu}A^{\mu})-\frac{1}{2}{\rm Tr}[(\partial_{\mu}\pi_c)(\partial^{\mu}\pi_c)]\right].
		\end{array}}
	\end{equation}
\end{itemize}

\subsubsection{St$\ddot{\text{u}}$ckelberg Trick II: Broken Time Diffeomorphism in Quasi-de Sitter~Background}
\label{v2b2}
Here one needs to perform a time diffeomorphism with a local
parameter $\xi^{0}(t,{\bf x})$, which is interpreted as a Goldstone field $\pi(t, {\bf x})$. These Goldstone modes shifts under the application of time
diffeomorphism, as~given by:
\begin{small}
\begin{equation}
{\begin{array}{rl}
	\textcolor{red}{\bf Time~diffeomorphism:}~~	t&\Longrightarrow t+ \xi^{0}(t,{\bf x}),~x^{i}\Longrightarrow x^{i}~~~\forall~ i=1,2,3\longrightarrow \pi(t, {\bf x})\rightarrow\pi(t, {\bf x})-\xi^{0}(t,{\bf x}).
	\end{array}}
\end{equation}
\end{small}

The $\pi$ is the Goldstone mode which describes the scalar perturbations around the background FLRW metric.
The effective action in the unitary gauge can be reproduced by gauge-fixing the time diffeomorphism as:
\be {\textcolor{red}{\bf Unitary~gauge~fixing}~~~~\Rightarrow~~~~\pi(t,{\bf x})=0~~~~\Rightarrow~~~~\tilde{\pi}(t,{\bf x})=-\xi^{0}(t,{\bf x})}~.\ee 

To construct the EFT action, it is  important to write down the transformation property of each operators under the application of broken time diffeomorphism, which are given~by: 
\begin{enumerate}
	\item \underline{\textcolor{red}{\bf Rule for metric:}} Under broken time diffeomorphism contravariant and covariant metric transform as:
\begin{equation}
	{
		\begin{array}{rl}
		\textcolor{red}{\bf Contravariant~metric:}~~	{g}^{00}\Longrightarrow& 
		(1+\dot{\pi})^2 {g}^{00}+2(1+\dot{\pi}){g}^{0 i}\partial_{i}\pi+{g}^{ij}\partial_{i}\pi\partial_{j}\pi,\\ {g}^{0i}\Longrightarrow& 
		(1+\dot{\pi}){g}^{0i}+{g}^{ij}\partial_j \pi,\\
		{g}^{ij}\Longrightarrow&{g}^{ij}.
		\end{array}}
	\end{equation}
\begin{equation}
	{
		\begin{array}{rl}
		\textcolor{red}{\bf Covariant~metric:}~~	g_{00}&\Longrightarrow  (1+\dot{\pi})^2 g_{00},\\
		g_{0i}&\Longrightarrow  (1+\dot{\pi}){g}_{0i}+{g}_{00}\dot{\pi}\partial_{i}\pi,\\ g_{ij}&\Longrightarrow  {g}_{ij}+{g}_{0j}\partial_{i}\pi+{g}_{i0}\partial_{j}\pi.
		\end{array}}
	\end{equation}
	\item \underline{\textcolor{red}{\bf Rule for Ricci scalar and Ricci tensor:}} Under broken time diffeomorphism Ricci scalar and the spatial component of the Ricci tensor on 3-hypersurface transform as:
\begin{equation}
		{
			\begin{array}{rl}
			\textcolor{red}{\bf Ricci~ scalar:}~~ {}^{(3)}R&\Longrightarrow\displaystyle {}^{(3)}R+\frac{4}{a^2}H(\partial^2\pi),\\
			\textcolor{red}{\bf Spatial~ Ricci~ tensor:}~~ {}^{(3)}R_{ij}&\Longrightarrow{}^{(3)}R_{ij}+H(\partial_{i}\partial_{j}\pi+\delta_{ij}\partial^2\pi).
			\end{array}}
		\end{equation}
	
	\item \underline{\textcolor{red}{\bf Rule for extrinsic curvature:}} Under broken time diffeomorphism trace and the spatial, time and mixed component of the extrinsic curvature transform as:
\begin{equation}
	{
		\begin{array}{rl}
		\textcolor{red}{\bf Trace:}~~ \delta K&\Longrightarrow\displaystyle \delta K-3\pi\dot{H}-\frac{1}{a^2}(\partial^2\pi),\\
		\textcolor{red}{\bf Spatial~ extrinsic~ curvature:}~~\delta K_{ij}&\Longrightarrow\delta K_{ij}-\pi\dot{H}h_{ij}-\partial_{i}\partial_{j}\pi\\
		\textcolor{red}{\bf Temporal~ extrinsic~ curvature:}~~\delta K^{0}_{0}&\Longrightarrow\delta K^{0}_{0},\\
		\textcolor{red}{\bf Mixed~ extrinsic~ curvature:}~~\delta K^{0}_{i}&\Longrightarrow\delta K^{0}_{i},\\
		\textcolor{red}{\bf Mixed~ extrinsic~ curvature:}~~\delta K^{i}_{0}&\Longrightarrow\delta K^{i}_{0}+2Hg^{ij}\partial_j\pi.
		\end{array}}
	\end{equation}
	\item \underline{\textcolor{red}{\bf Rule for time-dependent EFT coefficients:}} Under broken time diffeomorphism time-dependent EFT coefficients transform after canonical normalization $\pi_c=F^2(t)\pi$ as:
\begin{equation}
	{	\begin{array}{rl}
		\textcolor{red}{\bf EFT~coefficient}:~~F(t)\Longrightarrow F(t+\pi)&=\displaystyle \left[\sum^{\infty}_{n=0}\frac{\pi^{n}}{n!}\frac{d^{n}}{dt^n}\right]F(t)\\
		&=\displaystyle \left[\sum^{\infty}_{n=0}\underbrace{\frac{\pi^{n}_c}{n!F^{2n}}}_{\textcolor{red}{\bf Suppression}}\frac{d^{n}}{dt^n}\right]F(t)\approx F(t)~.\end{array}}
\end{equation}
Here $F(t)$ corresponds to all EFT coefficients mention in the EFT action.
\item \underline{\textcolor{red}{\bf Rule for Hubble parameter:}} Under broken time diffeomorphism, time-dependent EFT coefficients transform after using the following canonical normalization:
\be {\textcolor{red}{\bf Canonical~normalization:}~~~~~\pi_c=F^2(t)\pi}~,\ee as given by:
\begin{small}
\begin{equation}
{	\begin{array}{rl}
	\textcolor{red}{\bf Hubble~parameter}:~~H(t)\Longrightarrow H(t+\pi)&=\displaystyle \left[\sum^{\infty}_{n=0}\frac{\pi^{n}}{n!}\frac{d^{n}}{dt^n}\right]H(t)\\
	&=\displaystyle\left[1-\underbrace{\pi H(t) \epsilon-\frac{\pi^2H(t)}{2}\left(\dot{\epsilon}-2\epsilon^2\right)+\cdots}_{\textcolor{red}{\bf Correction~terms}}\right]H(t)~.\end{array}}
\end{equation}
\end{small}
Here $\epsilon=-\dot{H}/H^2$ is the slow-roll parameter.
\end{enumerate}

Now to construct the EFT action we need to also understand the behavior of all the operators appearing in the weak coupling regime of EFT. In~this regime one can neglect the mixing contributions between the gravity and Goldstone modes. To~demonstrate this explicitly let us start with the EFT~operator:
\be {\cal O}_{1}(t)=-\dot{H}M_{p}^2g^{00}.\ee  

Under  broken time diffeomorphism, the~operator ${\cal O}(t)$ transform as:
\be\label{zx}  {{\cal O}_{1}(t)\Longrightarrow \left[1+\frac{\pi}{\epsilon}\left(\dot{\epsilon}-2H\epsilon^2\right)+\cdots\right]\left[ (1+\dot{\pi})^2{\cal O}_{1}(t)-\dot{H}M^2_p\left(2(1+\dot{\pi})\partial_i \pi g^{0i}+g^{ij}\partial_i\pi \partial_j \pi\right)\right]}~.\ee

For further simplification the temporal component of the metric $g^{00}$ can be written as, 
$g^{00}=\bar{g}^{00}+\delta g^{00}$, where the background metric is given by, $\bar{g}^{00}=-1$ and the metric fluctuation is characterized by $\delta g^{00}$~\cite{Cheung:2007st,Naskar:2017ekm}. Using this in Equation~(\ref{zx}) and considering only the first term in Equation~(\ref{zx}) we get a kinetic term, $M_{p}^2\dot{H}\dot{\pi}^2\bar{g^{00}}$ and a mixing contribution, $M_{p}^2\dot{H}\dot{\pi}\delta g^{00}$ respectively. Furthermore, we use a canonical normalized metric fluctuation from the mixing contribution as given by:
\be {\textcolor{red}{\bf Canonical~normalization:}~~~\delta g^{00}_c=M_{p}\delta g^{00}}~,\ee
in terms of which  one can write, $M_{p}^2\dot{H}\dot{\pi}\delta g^{00}= \sqrt{\dot{H}}\dot{\pi}_c\delta g^{00}_c$. Consequently, at~above the energy scale $E_{mix}=\sqrt{\dot{H}}$, we can neglect this mixing term in the weak coupling~regime. 

One can also consider mixing contributions $M_{p}^2\dot{H}\dot{\pi}^2\delta{g^{00}}$ and $\pi M_{p}^2\ddot{H}\dot{\pi}\bar{g}^{00}$, which can be recast after canonical normalization as, 
$M_{p}^2\dot{H}\dot{\pi}^2\delta{g^{00}}=\dot{\pi}_c^2\delta{g^{00}_c}/M_{p}$ and $\pi M_{p}^2\ddot{H}\dot{\pi}\bar{g}^{00}=\ddot{H}\pi_c\dot{\pi}_c\bar{g}^{00}/\dot{H}$ with $\ddot{H}/\dot{H}<<1$. Here all higher-order terms in $\dot{\pi}$ will lead to additional Planck-suppression after canonical normalization. Consequently, we can neglect the contribution from $M_{p}^2\dot{H}\dot{\pi}\delta{g^{00}}$ term at the scale $E>E_{mix}$. 
Finally, in~the weak coupling regime one can recast Equation~(\ref{zx}) as:
\be\label{zx1}  {{\cal O}_{1}(t)\Longrightarrow {\cal O}_{1}(t)\left[\dot{\pi}^2-\frac{1}{a^2}(\partial_i\pi)^2\right]}~.\ee

   \subsubsection{The Goldstone Action from~EFT}
\label{v2b4}
   \scalebox{.95}[1.0]{Finally, in the weak coupling limit (or decoupling limit) we get the following simplified EFT action:}
   \bea S_{EFT}&=&S_{g}+S_{\pi},\eea 
   where the gravitational part and the Goldstone action is given by:
   \bea S_{g}&=& \int d^{4}x \sqrt{-g}\left[\frac{M^2_{p}}{2}R-M^2_p \left(3H^2+\dot{H}\right)\right],\\
   S_{\pi}&=&S^{(2)}_{\pi}+S^{(3)}_{\pi}+\cdots,\eea
   where the second and third-order Goldstone action can be written as:
\begin{equation}
   {	\begin{array}{rl}
   	S^{(2)}_{\pi}&=\displaystyle \int d^{4}x ~a^3\left[-M^2_{p}\dot{H}\left(\dot{\pi}^2-\frac{1}{a^2}(\partial_{i}\pi)^2\right)
   	+2M^4_2 \dot{\pi}^2\right.\\ &\left. ~~~~~~~~\displaystyle
   	+\frac{1}{2}\left(\bar{M}^2_3+3\bar{M}^2_2\right)H^2(1-\epsilon)\frac{\left(\partial_{i}\pi\right)^2}{a^2}-\left(\bar{M}^2_3+3\bar{M}^2_2\right)H^2\frac{\left(\partial_{i}\pi\right)^2}{a^2}
   	-\bar{M}^3_{1}\dot{\pi}\frac{1}{a^2}(\partial^2_{i}\pi)\right]~.\end{array}}
   \end{equation}
\begin{equation}
   {	\begin{array}{rl}
   	S^{(3)}_{\pi}&=\displaystyle \int d^{4}x ~a^3\left[\left(2M^4_2-\frac{4}{3}M^4_3\right)
   	\dot{\pi}^3-2M^4_2\dot{\pi}\frac{1}{a^2}(\partial_{i}\pi)^2
   	\right.\\ & \left.~~~~~~~~~~\displaystyle
   	-\bar{M}^2_3 \pi\dot{H}\frac{1}{a^2}\partial^2_{i}\pi-3\bar{M}^2_2\dot{H}\pi 
   	\frac{1}{a^2}(\partial^2_{i}\pi)+\frac{3}{2}\bar{M}^3_{1}\pi \dot{H}\frac{1}{a^2}(\partial_{i}\pi)^2
   	\right.\\ & \left.~~~~~~~~~~\displaystyle
   	-\frac{3}{2}\bar{M}^3_{1}\dot{H}\pi\dot{\pi}^2-\bar{M}^3_{1}\dot{\pi}\frac{1}{a^2}(\partial_{i}\pi)^2\right]~.\end{array}}
   \end{equation}
 
   Here we introduce EFT sound speed $c_{S}$ as:
   \bea\label{op1} c_{S}\equiv \frac{1}{\sqrt{1-\frac{2M^4_2}{\dot{H}M^2_p}}}.\eea 
   
   Here if we set $M_{2}=0$ or equivalently if we say that $\frac{M^4_2}{2!}(g^{00}+1)^2$ term is absent in the effective Lagrangian then Equation~(\ref{op1})
   suggests that in that case sound speed $c_{S}=1$, which is true for single-field canonical slow-roll inflation. 
  Next using Equation~(\ref{op1}) and applying integration by parts in the Goldstone part of the Lagrangian we get\footnote{Let us concentrate on the following contribution in the second- and third-order perturbed EFT action, which can be written after integration by parts as:\bea S^{2}_{\pi}&\supset& -\int d^3x~ dt~a^3~\bar{M}^3_1~\frac{\dot{\pi}}{a^2}~\left(\partial^2_i\pi\right)=\int d^3x~ dt~a^3~\frac{\bar{M}^3_1}{a^2}~\left[-\partial_{i}\left(\dot{\pi}\partial_{i}\pi\right)+\frac{1}{2}\frac{d}{dt}\left(\partial_i\pi\right)^2\right]\nonumber\\&&~~~~~~~~~~~~~~~~~~~~~~~~~~~~~~~~~~~~~~=\int d^3x~ dt~a^3~\frac{\bar{M}^3_1}{2}~\left[\frac{d}{dt}\left(\frac{\left(\partial_i\pi\right)^2}{a^2}\right)-\frac{H}{a^2}\left(\partial_i\pi\right)^2\right]\nonumber\\
  	&&~~~~~~~~~~~~~~~~~~~~~~~~~~~~~~~~~~~~~~=-\int d^3x~ dt~a^3~\frac{\bar{M}^3_1}{2}~H~\frac{\left(\partial_i\pi\right)^2}{a^2}. \\ S^{3}_{\pi}&\supset& -\int d^3x~ dt~a^3~\bar{M}^3_1~\frac{3}{2}~\dot{H}~\pi~\dot{\pi}^2=\int d^3x~ dt~a^3~\left[\frac{3}{2}\bar{M}^3_1 H \dot{\pi}^3-\frac{9}{2}H^2\bar{M}^3_1\pi\dot{\pi}^2\right]. \\ S^{3}_{\pi}&\supset& \int d^3x~ dt~a^3~\bar{M}^3_1~\frac{3}{2}~\dot{H}~\pi~\frac{\left(\partial_{i}\pi\right)^2}{a^2}=-\int d^3x~ dt~a^3~\left[\frac{3}{2}\bar{M}^3_1  \frac{H\pi}{a^2}\left(\partial_{i}\pi\right)^2+\frac{\dot{\pi}}{a^2}\frac{3}{2}\bar{M}^3_1\left(\partial_{i}\pi\right)^2\right]. \\ S^{3}_{\pi}&\supset& -3\int d^3x~ dt~a^3~\bar{M}^2_2~\dot{H}~\pi~\frac{\pi}{a^2}\left(\partial^2_{i}\pi\right)=\int d^3x~ dt~a^3~3 \bar{M}^2_2\left[\frac{H\pi}{a^2}\left(\partial_{i}\pi\right)^2+\frac{\dot{\pi}}{a^2}\left(\partial_{i}\pi\right)^2\right]. \eea}:
   \bea {S^{(2)}_{\pi}=\int d^{4}x ~a^3~\left(-\frac{M^2_p\dot{H}}{c^2_S}\right)\left[\dot{\pi}^2
   -c^2_S\left(1-\frac{\bar{M}^3_1 H}{M^2_p \dot{H}}-\left[\bar{M}^2_3 +3\bar{M}^2_2\right]\frac{H^2(1+\epsilon)}{2M^2_p\dot{H}}\right)\frac{1}{a^2}(\partial_{i}\pi)^2
   \right]}~.~~~~~~~\eea
\begin{equation}
 {	\begin{array}{rl}
 	S^{(3)}_{\pi}&= \displaystyle\int d^{4}x ~a^3\left[\left\{\left(1-\frac{1}{c^2_{S}}\right)\dot{H}M^2_p +\frac{3}{2}\bar{M}^3_1 H-\frac{4}{3}M^4_3\right\}
 	\dot{\pi}^3\right.\\ & \left.~~~~~~~~~~~~~~~~~~~~~~~~~\displaystyle
 	-\left\{\left(1-\frac{1}{c^2_{S}}\right)\dot{H}M^2_p +\frac{3}{2}\bar{M}^3_{1} H\right\}\frac{1}{a^2}\dot{\pi}(\partial_{i}\pi)^2
 	\right.\\ & \left.~~~~~~~~~~~~~~~~~~~~~~~~~~~~~~~~~~~~~~~~~~\displaystyle
 	-\frac{9}{2}\bar{M}^3_{1}H^2\pi \dot{\pi}^2+\frac{3}{2}\bar{M}^3_1 H \frac{1}{a^2} \pi \frac{d}{dt}\left(\partial_{i}\pi\right)^2\right]~.\end{array}}
 \end{equation}
 
   In the present context metric fluctuation of the spatial components are given by:
   \bea\label{eq1} g_{ij}=a^{2}(t)\left[\left(1+2\zeta(t,{\bf x})\right)\delta_{ij}+\gamma_{ij}\right]~~\forall~~~i=1,2,3,\eea
   where $a(t)$ is the scale factor in FLRW quasi de Sitter background space-time. 
   Also $\zeta(t,{\bf x})$ is known as curvature perturbation which signifies scalar fluctuation. On~the other hand, tensor fluctuations are identified with
   $\gamma_{ij}$, which is spin-2, transverse, and traceless rank 2 tensor. Here under the broken time diffeomorphism the scale factor $a(t)$ transforms 
   in the following fashion:
   \bea\label{eq2}{ a(t)\Longrightarrow a(t-\pi(t,{\bf x}))=a(t)-H\pi(t,{\bf x})a(t)+\cdots\approx a(t)\left(1-H\pi(t,{\bf x})\right)}~. \eea
   
   Furthermore, using Equations~(\ref{eq1}) and~(\ref{eq2}), we get:
   \bea a^2(t)\left(1-H\pi(t,{\bf x})\right)^2\approx a^2(t)\left(1-2H\pi(t,{\bf x})\right)=a^{2}(t)\left(1+2\zeta(t,{\bf x})\right).\eea
   
  This implies that the curvature perturbation $\zeta(t,{\bf x})$ can be written in terms of Goldstone modes $\pi(t,{\bf x})$ in the following way\footnote{Here we have considered the linear relation between the curvature perturbation ($\zeta$) and the Goldstone mode ($\pi$). In~this context one can consider the following non-linear relation to compute the three-point correlation function from the present~setup:
  	\be \zeta(t,{\bf x})=-H\pi(t,{\bf x})-\frac{(\epsilon-\eta)}{2}H^2\pi^2(t,{\bf x})+\cdots,\ee
  	where the slow-roll parameters are given by, $\epsilon=-\dot{H}/H^2$ and $\eta=\epsilon-\frac{1}{2}\frac{d\ln \epsilon}{d {\cal N}}$. Here ${\cal N}=\int H ~dt,$ represents the number of e-foldings. However, the contribution from such non-linear term is extremely small and proportional to sub-leading terms $\epsilon^2$, $\eta^2$ and $\epsilon\eta$ in the expression for the three-point function and the associated bispectrum. From~the observational perspective such contributions also not so important and can be treated as very small correction to the leading-order result computed in this paper. }:
  \be \label{eq3} {\textcolor{red}{\bf Quantum~fluctuation~in~terms~of~Goldstone~mode:}~~~~~\zeta(t,{\bf x})=-H\pi(t,{\bf x})}~.\ee
  
  Furthermore, using Equation~(\ref{eq3}), the~effective action for the Goldstone part of the Lagrangian can be recast in terms of curvature perturbation $\zeta(t,{\bf x})$ as:
  \bea {S^{(2)}_{\zeta}\approx\int d^{4}x ~a^3~\left(\frac{M^2_p\epsilon}{c^2_S}\right)\left[\dot{\zeta}^2
   -c^2_S\left(1-\frac{\bar{M}^3_1 H}{M^2_p \dot{H}}-\left[\bar{M}^2_3 +3\bar{M}^2_2\right]\frac{H^2(1+\epsilon)}{2M^2_p\dot{H}}\right)
   \frac{1}{a^2}(\partial_{i}\zeta)^2
   \right]}~.~~~~~~~\eea
\begin{equation}
{	\begin{array}{rl}
	S^{(3)}_{\zeta}&\approx \displaystyle\int d^{4}x ~\frac{a^3}{H^3}\left[-\left\{\left(1-\frac{1}{c^2_{S}}\right)\dot{H}M^2_p +\frac{3}{2}\bar{M}^3_1 H-\frac{4}{3}M^4_3\right\}
	\dot{\zeta}^3\right.\\ & \left.\displaystyle~~~~~~~~~~~~~~~~~~~~~~~~~
	+\left\{\left(1-\frac{1}{c^2_{S}}\right)\dot{H}M^2_p +\frac{3}{2}\bar{M}^3_{1} H\right\}\frac{1}{a^2}\dot{\zeta}(\partial_{i}\zeta)^2
	\right.\\ & \left.\displaystyle~~~~~~~~~~~~~~~~~~~~~~~~~~~~~~~~~~~~~~~~~~
	+\frac{9}{2}\bar{M}^3_{1}H^2\zeta \dot{\zeta}^2-\frac{3}{2}\bar{M}^3_1 H \frac{1}{a^2} \zeta \frac{d}{dt}\left(\partial_{i}\zeta\right)^2\right]~.\end{array}}
\end{equation}
 
   For further simplification we introduce a few new parameters which are appended bellow\footnote{Here we have used a few choices for the simplifications of the further computation of the two- and three-point correlation function in the EFT coefficients which are partly motivated by ref.~\cite{Senatore:2010jy}. Also it is important to note that since we are restricted our computation up to tree-level and not considering any quantum effects through loop correction, we have discussed the radiative stability or naturalness of these choices under quantum corrections. }:
   \begin{itemize}
    \item First we define an effective sound speed $\tilde{c}_{S}$, which can be expressed in terms of the usual EFT sound speed $c_{S}$ as\footnote{Here it is important to point out that in~the case when $M_2=0$ we have the EFT sound speed $c_S=1$ exactly, which is true for all canonical slow-roll models of inflation driven by a single field. But~since here the EFT coefficients are sufficiently small $\bar{M}_{i}\forall i=1,2,3(\sim {\cal O}(10^{-2}-10^{-3}))$ it is expected that $\tilde{c}_S \approx c_S$ and for the situation $c_S=1$ one can approximately fix $\tilde{c}_S \approx 1$. Thus, for the canonical slow-roll model one can easily approximate the redefined sound speed $\tilde{c}_S$ with the usual EFT sound speed $c_{S}$ without losing any generality. But~such small EFT coefficients $\bar{M}_{i}\forall i=1,2,3(\sim {\cal O}(10^{-2}-10^{-3}))$ play significant roles in the computation of the three-point function and the associated bispectrum as in the absence of these coefficients the amplitude of the bispectrum $f_{NL}$ is zero. This also implies that for the canonical slow-roll model of single-field inflation the amount of non-Gaussianity is not very large and this completely consistent with the previous finding that in that case the amplitude of the bispectrum $f_{NL}\propto \epsilon$~(where $\epsilon$ is the slow-roll parameter), at~the leading order of the computation. See ref.~\cite{Maldacena:2002vr} for details.  }:
       \bea \tilde{c}_{S}=c_{S}~\sqrt{1-\frac{\bar{M}^3_1 H}{M^2_p \dot{H}}-\left[\bar{M}^2_3 +3\bar{M}^2_2\right]\frac{H^2(1+\epsilon)}{2M^2_p\dot{H}}}.
       \eea
       Since the following approximations:
       \bea \left|\frac{\bar{M}^3_1 H}{M^2_p \dot{H}}\right|&<<&1,~~~~
       \left|\left[\bar{M}^2_3 +3\bar{M}^2_2\right]\frac{H^2(1+\epsilon)}{2M^2_p\dot{H}}\right|<<1,\eea
       are valid in the present context of discussion, one can recast the effective sound
       speed in the following simplified form as:
       \bea \tilde{c}_{S}&\approx& c_{S}~\left\{1+\frac{1}{2\epsilon H M^2_p}\left[\bar{M}^3_1+\left(\bar{M}^2_3 +3\bar{M}^2_2\right)\frac{H(1+\epsilon)}{2}\right]\right\} .
       \eea
       \item Secondly, we introduce the following connecting relationship between $M_3$ and $M_2$ given by:
       \bea \label{eq5} M^4_3 c^2_S &=&-\tilde{c}_{3} M^4_2.\eea
       When $M_2=0$ then from Equation~(\ref{op1}) we can see that the sound speed $c_{S}=1$ and Equation~(\ref{eq5}) also implies that $M_3=0$ in that~case.
       
       \item Next we define the following connecting relationship between $M_3$ and $\bar{M}_1$ given by:
       \bea \label{eq6} M^4_3 \tilde{c}_4 &=&-H\bar{M}^3_{1}\tilde{c}_{3}.\eea
       When $M_2=0$ then from Equation~(\ref{op1}) we can see that the sound speed $c_{S}=1$ (which is actually the result for single-field canonical slow-roll models of inflation) and Equations~(\ref{eq5}) and~(\ref{eq6})
       also implies the following~possibilities:
       \begin{enumerate}
        \item $M_3=0$, $\bar{M}_{1}\neq 0$ and ${\Large\frac{\tilde{c}_{3}}{\tilde{c}_{4}}\rightarrow 0}$. We will look into this possibility in detail during our computation for $c_{S}=1$ case as this will finally give rise to non-vanishing three-point function (non-Gaussianity).
        \item $M_3=0$, $\bar{M}_{1}=0$ and ${\Large\frac{\tilde{c}_{3}}{\tilde{c}_{4}}\neq 0}$. We do not consider this possibility for $c_{S}=1$ case because for this case third ($S^{(3)}_{\zeta}$) action for curvature perturbation vanishes, which will give rise to zero three-point function (non-Gaussianity).
       \end{enumerate}
        \item For further simplification one can also assume that:
        \be \bar{M}^2_3+ 3\bar{M}^2_2 = \frac{\bar{M}^3_1}{H\tilde{c}_{5}}\ee
        so that one can write:
        \be \frac{1}{\epsilon H M^2_p}\left[\bar{M}^3_1+\left(\bar{M}^2_3 
       +3\bar{M}^2_2\right)\frac{H(1+\epsilon)}{2}\right]=\frac{\bar{M}^3_1}{\epsilon H M^2_p}\left[1+
       \frac{(1+\epsilon)}{2\tilde{c}_{5}}\right].\ee
        For $c_{S}=1$ this implies the following two~possibilities:
        \begin{enumerate}
         \item $\bar{M}_{1}\neq 0$ and $\tilde{c}_{5}= -\frac{1}{2}(1+\epsilon)$. We will look into this possibility in detail during our computation for $c_{S}=1$ case as this will finally give rise to non-vanishing three-point function (non-Gaussianity).
         
         \item $\bar{M}_{1}=0$.  We do not consider this possibility for $c_{S}=1$ case because for this case third ($S^{(3)}_{\zeta}$) action for curvature perturbation vanishes, which will give rise to zero three-point function (non-Gaussianity).
         \end{enumerate}
         Consequently, the~effective sound speed can be recast as:
        \bea \tilde{c}_{S}&=&c_{S}~\sqrt{1+\frac{\Delta \bar{M}^3_1}{2\epsilon H M^2_p}}\approx c_{S}~\left\{1+\frac{\Delta \bar{M}^3_1}{4\epsilon H M^2_p}\right\}
       \eea
         where $\Delta$ is defined as, 
         $\Delta= 2+\frac{1+\epsilon}{\tilde{c}_{5}}.$
        Here $\Delta=0$ for $\tilde{c}_{5}= -\frac{1}{2}(1+\epsilon)$ when $c_{S}=1$. Consequently, we have $\tilde{c}_{S}=c_{S}=1$ in that case.
        \item For further simplification one can also assume that:
                \bea \bar{M}^2_3 \approx \bar{M}^2_2=\frac{\bar{M}^3_1}{4H\tilde{c}_{5}}.\eea
                Here $c_{S}=1$ this implies the following two~possibilities:
                \begin{enumerate}
                 \item $ \bar{M}^2_3 \approx \bar{M}^2_2\neq 0, \bar{M}_{1}\neq 0$ and $\tilde{c}_{5}= -\frac{1}{2}(1+\epsilon)$ as mentioned earlier.  We will investigate this possibility in detail during our computation for $c_{S}=1$ case as this will finally give rise to non-vanishing~non-Gaussianity.
                 
                 \item $ \bar{M}^2_3 \approx \bar{M}^2_2= 0,\bar{M}_{1}=0$. As~mentioned earlier here we do not consider this possibility for $c_{S}=1$ case because for this case second ($S^{(2)}_{\zeta}$) and third-order ($S^{(3)}_{\zeta}$) action for curvature perturbation vanishes, which will give rise to zero non-Gaussianity.
                 \end{enumerate}
\item Next we define the following connecting relationship between $M_4$ and $M_3$ given by:
       \bea \label{eq8} M^4_4 \tilde{c}_6=M^4_3 \tilde{c}_4 &=&-H\bar{M}^3_{1}\tilde{c}_{3}.\eea
       When $M_2=0$ then from Equation~(\ref{op1}) we can see that the sound speed $c_{S}=1$ and Equations~(\ref{eq5}) and~(\ref{eq8})
       also implies the following~possibilities:
       \begin{enumerate}
        \item $M_4\neq 0$, $M_3=0$, $\bar{M}_{1}\neq 0$ and ${\Large\frac{\tilde{c}_{3}}{\tilde{c}_{4}}\rightarrow 0}$.  We will look into this possibility in detail during our computation for $c_{S}=1$ case as this will finally give rise to non-vanishing three-point function (non-Gaussianity).
        \item $M_4=0$, $M_3=0$, $\bar{M}_{1}=0$ and ${\Large\frac{\tilde{c}_{3}}{\tilde{c}_{4}}\neq 0}$. We do not consider this possibility for $c_{S}=1$ case because for this case third ($S^{(3)}_{\zeta}$)  order action for curvature perturbation vanishes, which will give rise to zero three-point function (non-Gaussianity).
\end{enumerate}

   \end{itemize}
   
Furthermore, using all such new defined parameters the EFT action for the Goldstone boson can be recast~as\footnote{Here it is important to note that for~the case $c_S=1$ we have written an approximated form of the second and third-order action by assuming that $\tilde{c}_S\approx c_{S}\sim1$, which is true for all canonical slow-roll models of inflation driven by a single field. Here the EFT coefficients are sufficiently small $\bar{M}_{i}\forall i=1,2,3(\sim {\cal O}(10^{-2}-10^{-3}))$ for which it is expected that $\tilde{c}_S \approx c_S$ and for the situation $c_S=1$ one can approximately fix $\tilde{c}_S \approx 1$. }: \bea \underline{\bf \textcolor{red}{\bf For ~c_{S}=1:}}\nonumber~~~~~~~~~~~~~~~~~~~~~~~~~~~~~~~~~~~~~~~~~~~~~~~~~~~~~~~~~~~~~~~~~~~~~~~~~~~~~~~~~~~~~~~~~~~~~~~~~~~\\ { \begin{array}{rl}
    		S^{(2)}_{\zeta}&\approx\displaystyle \int d^{4}x ~a^3~M^2_p\epsilon\left[\dot{\zeta}^2
    		-
    		\frac{1}{a^2}(\partial_{i}\zeta)^2
    		\right].~~~~~~~~~~~~\end{array}}~~~~~~~~~~~~~~~~~~~~~~~~~~~~~~~~~~~~~~~~~~~~~~~~~~~~~~~~\\
    { \begin{array}{rl}
    		S^{(3)}_{\zeta}&\approx\displaystyle \int d^{4}x ~\frac{a^3}{H^3}\left[-\left\{\frac{3}{2}\bar{M}^3_1 H\right\}
    		\dot{\zeta}^3 
    		+\left\{\frac{3}{2}\bar{M}^3_{1} H\right\}\frac{1}{a^2}\dot{\zeta}(\partial_{i}\zeta)^2
    		\right.\\ & \left.~~~~~~~~~~~~~~~~~~~~~~~~~~~~~~~~~~~~~~~~~~\displaystyle
    		+\frac{9}{2}\bar{M}^3_{1}H^2\zeta \dot{\zeta}^2-\frac{3}{2}\bar{M}^3_1 H \frac{1}{a^2} \zeta \frac{d}{dt}\left(\partial_{i}\zeta\right)^2\right].~~~~~~~~~~~~\end{array}}~ ~~~~~~
 \eea\bea \underline{\bf \textcolor{red}{\bf For ~c_{S}<1:}}\nonumber~~~~~~~~~~~~~~~~~~~~~~~~~~~~~~~~~~~~~~~~~~~~~~~~~~~~~~~~~~~~~~~~~~~~~~~~~~~~~~~~~~~~~~~~~~~~~~~~~~~\\ { \begin{array}{rl}
S^{(2)}_{\zeta}&\approx\displaystyle \int d^{4}x ~a^3~\left(\frac{M^2_p\epsilon}{c^2_S}\right)\left[\dot{\zeta}^2
   -\tilde{c}^2_S
   \frac{1}{a^2}(\partial_{i}\zeta)^2
   \right].~~~~~~~~~~~~\end{array}}~~~~~~~~~~~~~~~~~~~~~~~~~~~~~~~~~~~~~~~~~~~~~~\\
 { \begin{array}{rl}
 	S^{(3)}_{\zeta}&\approx\displaystyle \int d^{4}x ~a^3~\frac{\epsilon M^2_p}{H}\left(1-\frac{1}{c^2_{S}}\right)\left[\left\{1+\frac{3\tilde{c}_{4}}{4c^2_{S}}+\frac{2\tilde{c}_{3}}{3c^2_{S}}\right\}
 	\dot{\zeta}^3
 	-\left\{1+\frac{3\tilde{c}_{4}}{4c^2_{S}}\right\}\frac{1}{a^2}\dot{\zeta}(\partial_{i}\zeta)^2
 	\right.\\ & \left.~~~~~~~~~~~~~~~~~~~~~~~~~~~~~~~~~~~~~~~~~~\displaystyle 
 	-\frac{9H\tilde{c}_{4}}{4c^2_S}\zeta \dot{\zeta}^2+\frac{3\tilde{c}_{4}}{4c^2_{S}}\frac{1}{a^2} \zeta \frac{d}{dt}\left(\partial_{i}\zeta\right)^2\right].~~~~~~~~~~~~\end{array}}~~~~~~~~~~~~~
      \eea
\section{{Two-Point Correlation Function from~EFT}\label{v3}}
\unskip
	\subsection{For Scalar~Modes	\label{v3a}}
	\unskip
		\subsubsection{Mode Equation and Solution for Scalar~Perturbation\label{v3a1}}
		
		Here we compute the two-point correlation from scalar perturbation. For~this purpose we consider the second-order perturbed action as given by\footnote{See also ref.~\cite{Maldacena:2002vr,Chen:2006nt}, where similar computations have been performed for canonical single-field slow-roll and generalized slow-roll models of inflation in the presence of {Bunch–Davies vacuum} state.}:
		 \bea S^{(2)}_{\zeta}&\approx&\int d^{4}x ~a^3~\left(\frac{M^2_p\epsilon}{c^2_S}\right)\left[\dot{\zeta}^2
		   -c^2_S\left(1-\frac{\bar{M}^3_1 H}{M^2_p \dot{H}}-\left[\bar{M}^2_3 +3\bar{M}^2_2\right]\frac{H^2(1+\epsilon)}{2M^2_p\dot{H}}\right)
		   \frac{1}{a^2}(\partial_{i}\zeta)^2
		   \right],\eea
		   which can be recast for $c_{S}=1$ and $c_{S}<1$ case as:
		    \be {\underline{\bf \textcolor{red}{\bf For ~c_{S}=1:}}~~~~~
		    S^{(2)}_{\zeta}\approx\int d^{4}x ~a^3~M^2_p\epsilon\left[\dot{\zeta}^2
		      -
		      \frac{1}{a^2}(\partial_{i}\zeta)^2
		      \right]}~,~~~~~~~~~~~~\ee
		      \be{\underline{\bf \textcolor{red}{\bf For ~c_{S}<1:}}\\~~~~~
		   S^{(2)}_{\zeta}\approx\int d^{4}x ~a^3~\left(\frac{M^2_p\epsilon}{c^2_S}\right)\left[\dot{\zeta}^2
		      -\tilde{c}^2_S
		      \frac{1}{a^2}(\partial_{i}\zeta)^2
		      \right]}~,~~~~~~~~~~~~\ee
		      where the effective sound speed $\tilde{c}_{S}$ is defined~earlier.
		      
		      Next we define Mukhanov–Sasaki variable $v(\eta,{\bf x})$ which is defined as:
		      \be  {\textcolor{red}{\bf Mukhanov–Sasaki~variable:}~~~~~~~~~~v(\eta,{\bf x})=z~\zeta(\eta,{\bf x})~M_p=-z~H~\pi(\eta,{\bf x})~M_p}~.\ee
		      
		      In general, the parameter $z$
		      is defined for the present EFT setup as,
$z =\frac{a\sqrt{2\epsilon}}{\tilde{c}_{S}}.$		      
		      Now in terms of $v(\eta,{\bf x})$ the second-order action for the curvature perturbation can be recast as:
		      \bea {S^{(2)}_{\zeta}\approx\int d^{3}x~ d\eta ~\left[v^{'2}
		      		   -\tilde{c}^2_S(\partial_{i}v)^2
		      		   \frac{1}{a^2}(\partial_{i}\zeta)^2-m^{2}_{eff}(\eta)v^2
		      		   \right]}~,~~~~~~~~~~~~\eea
		      		   where the effective mass parameter $m_{eff}(\eta)$ is defined as, 
		      		   $m^{2}_{eff}(\eta)=-\frac{1}{z}\frac{d^2z}{d\eta^2}.$
		      		   Here $\eta$ is the conformal time which can be expressed in terms of physical time $t$ as, 
		      		   $\eta=\int \frac{dt}{a(t)}.$
		      		   The conformal time described here is negative and lying within $-\infty<\eta<0$.
		      		    During inflation, the scale factor and the parameter $z$ can be expressed in terms of the conformal time $\eta$ as:
		      		   \be\begin{array}{lll}\label{kg1}
		      		    \displaystyle a(\eta) =\footnotesize\left\{\begin{array}{ll}
		      		                       \displaystyle   -\frac{1}{H\eta}~~~~ &
		      		    \mbox{\small {\bf for ~dS}}  \\ 
		      		   	\displaystyle  -\frac{1}{H\eta}\left(1+\epsilon\right)~~~~ & \mbox{\small {\bf for~ qdS}}.
		      		             \end{array}
		      		   \right.
		      		   \end{array}\ee
		      		   and 
		      		   \be\begin{array}{lll}\label{kg0}
		      		   		      		    \displaystyle z =\frac{a\sqrt{2\epsilon}}{c_{S}}=\footnotesize\left\{\begin{array}{ll}
		      		   		      		                       \displaystyle   -\frac{1}{H\eta}\frac{\sqrt{2\epsilon}}{c_{S}}~~~~ &
		      		   		      		    \mbox{\small {\bf for ~dS}}  \\ 
		      		   		      		   	\displaystyle  -\frac{1}{H\eta}\frac{\sqrt{2\epsilon}}{c_{S}}\left(1+\epsilon\right)~~~~ & \mbox{\small {\bf for~ qdS}}.
		      		   		      		             \end{array}
		      		   		      		   \right.
		      		   		      		   \end{array}\ee	
		      		   		      		   
		      		   Additionally, it is important to note that for de Sitter and quasi de Sitter case the relation between conformal time $\eta$ and physical time $t$ can be expressed 
		      		   as, 
		      		   $t=-\frac{1}{H}\ln(-H\eta).$
		      		   Within this setup inflation ends when the conformal time $\eta\sim 0$.
		      		   
		      		   Now further doing the Fourier transform:
		      		   \be v(\eta,{\bfx})=\int \frac{d^3k}{(2\pi)^3}~v_{\bf k}(\eta)~e^{i{\bf k}.{\bf x}}\ee
		      		   one can write down the equation of motion for scalar fluctuation as:
		      		   \be {\textcolor{red}{\textbf{Mukhanov–Sasaki~Eqn~for~scalar~mode:}}~~~~~~~~~~v^{''}_{\bf k}+\left(\tilde{c}^2_{S}k^2+m^{2}_{eff}(\eta)\right)v_{\bf k}=0}~.\ee
		      		   
		      		   Here it is important to note that for de Sitter and quasi de Sitter case the effective mass parameter can be expressed as:
		      		   \be\begin{array}{lll}\label{kg11}
		      		   \displaystyle m^{2}_{eff}(\eta) =\footnotesize\left\{\begin{array}{ll}
		      		   		      		                       \displaystyle   -\frac{2}{\eta^2}~~~~ &
		      		   		      		    \mbox{\small {\bf for ~dS}}  \\ 
		      		   		      		   	\displaystyle  -\frac{\left(\nu^2-\frac{1}{4}\right)}{\eta^2}~~~~ & \mbox{\small {\bf for~ qdS}}.
		      		   		      		             \end{array}
		      		   		      		   \right.
		      		   		      		   \end{array}\ee
		      		   		      		   
		      		   		      		   Here in the de Sitter and quasi de Sitter case the parameter $\nu$ can be written as:
		      \be\begin{array}{lll}\label{kg111}
		      		      		   \displaystyle \nu =\footnotesize\left\{\begin{array}{ll}
		      		      		   		      		                       \displaystyle   \frac{3}{2}~~~~ &
		      		      		   		      		    \mbox{\small {\bf for ~dS}}  \\ 
		      		      		   		      		   	\displaystyle  \frac{3}{2}+3\epsilon-\eta+\frac{s}{2}~~~~ & \mbox{\small {\bf for~ qdS}},
		      		      		   		      		             \end{array}
		      		      		   		      		   \right.
		      		      		   		      		   \end{array}\ee		   		    
		      		   		      		   where $\epsilon$, $\eta$ and $s$ are the  slow-roll parameter defined as:
		      		   		      		   \bea \label{rt1}\epsilon &=& -\frac{\dot{H}}{H^2},~~~~
		      		   		      		   \eta = 2\epsilon-\frac{\dot{\epsilon}}{2H\epsilon},~~~~
		      		   		      		   s=\frac{\dot{c_{S}}}{Hc_{S}}.\eea
		      		   		      		  
		      		   		      		   In the slow-roll regime of inflation $\epsilon<<1$ and $|\eta|<<1$ and at the end of inflation, 
		      		   		      		   the slow-roll condition breaks when any of the criteria satisfy (1) $\epsilon=1$ or $|\eta|=1$, (2) $\epsilon=1=|\eta|$.  
		      		   		      		   
		      		The general solution for $v_{\bf k}(\eta)$ thus can be written as:
		      		\be{\begin{array}{lll}\label{yu2}
		      		 \displaystyle v_{\bf k}(\eta) =\footnotesize\left\{\begin{array}{ll}
		      		                    \displaystyle   \sqrt{-\eta}\left[C_1  H^{(1)}_{\frac{3}{2}} \left(-k\tilde{c}_{S}\eta\right) 
		      		+ C_2 H^{(2)}_{\frac{3}{2}} \left(-k\tilde{c}_{S}\eta\right)\right]~~~~ &
		      		 \mbox{\small {\bf for ~dS}}  \\ 
		      			\displaystyle \sqrt{-\eta}\left[C_1  H^{(1)}_{\nu} \left(-k\tilde{c}_{S}\eta\right) 
		      		+ C_2 H^{(2)}_{\nu} \left(-k\tilde{c}_{S}\eta\right)\right]~~~~ & \mbox{\small {\bf for~ qdS}}.
		      		          \end{array}
		      		\right.
		      		\end{array}}\ee
		      		
		      		Here $C_{1}$ and $C_{2}$ are the arbitrary integration constants and the numerical values depend on the choice of the 
		      		initial vacuum. In~the present context we consider the following 
		      		choice of the vacuum for the~computation:
		      		\begin{enumerate}
		      		 \item \underline{{\bf Bunch–Davies vacuum:}} In this case, we choose, $C_{1}=1, C_{2}= 0~.$
		      		 \item \underline{{\bf ${\alpha,\beta}$ vacuum:}} In this case, we choose $C_{1}=\cosh\alpha, C_{2}= e^{i\beta}\sinh\alpha~.$ 
		      		 Here $\beta$ is a phase factor.
		      		\end{enumerate}
		      		
		      		For the most general solution as stated in Equation~(\ref{yu2}) one can consider the limiting physical 
		      		 situations, as~given by, 
		      		 I. {\bf Superhorizon regime:}~~ $k\tilde{c}_{S}\eta<<-1$,
		      		  II. {\bf Horizon crossing:}~~ $k\tilde{c}_{S}\eta= -1$,
		      		 III. {\bf Subhorizon regime:}~~ $k\tilde{c}_{S}\eta>>-1$.

Finally, considering the behavior of the mode function in the {subhorizon regime} and {superhorizon regime} one can write the expression in de Sitter and quasi de Sitter case as:
\bea\label{yu2zxzx}
 \displaystyle v_{\bf k} (\eta) &=&
 \footnotesize\left\{\begin{array}{ll}
                    \displaystyle  \frac{1}{i\eta}\frac{1}{\sqrt{2}\left(k\tilde{c}_{S}\right)^{\frac{3}{2}}}\left[C_1  e^{- ik\tilde{c}_{S}\eta}
                    \left(1+ik\tilde{c}_{S}\eta\right)e^{-i\pi}
                    - C_2 e^{ ikc_{S}\eta}
                    \left(1-ik\tilde{c}_{S}\eta\right)e^{i\pi}\right]~~~~ &
 \mbox{\small {\bf for ~dS}}  \nonumber\\ 
	\displaystyle  2^{\nu-\frac{3}{2}}\frac{1}{i\eta}\frac{1}{\sqrt{2}\left(k\tilde{c}_{S}\right)^{\frac{3}{2}}}
	(-k
					\tilde{c}_{S}\eta)^{\frac{3}{2}-\nu}\left|\frac{\Gamma(\nu)}{\Gamma\left(\frac{3}{2}\right)}\right|\left[C_1  e^{- ik\tilde{c}_{S}\eta}
	                    \left(1+ik\tilde{c}_{S}\eta\right)
e^{-\frac{i\pi}{2}\left(\nu+\frac{1}{2}\right)} 
\right.\\ \left.\displaystyle~~~~~~~~~~~~~~~~~~~~~~~~~~~~~~~~~~~~~~~- C_2 e^{ ikc_{S}\eta}
                    \left(1-ik\tilde{c}_{S}\eta\right)
e^{\frac{i\pi}{2}\left(\nu+\frac{1}{2}\right)}\right]~~~~ & \mbox{\small {\bf for~ qdS}}.
          \end{array}
\right.\nonumber\eea 
		
Furthermore, using Equation~(\ref{yu2zxzx}) one can write down the expression for the curvature perturbation $\zeta(\eta,{\bf k})=\frac{v_{\bf k}(\eta)}{z~M_p}$ as:      
\bea\label{yu2zxzxcva}
 \displaystyle \zeta(\eta,{\bf k}) &=&\footnotesize\left\{\begin{array}{ll}
                    \displaystyle  \frac{iH\tilde{c}_{S}}{2~M_p\sqrt{\epsilon}\left(k\tilde{c}_{S}\right)^{\frac{3}{2}}}\left[C_1  e^{- ik\tilde{c}_{S}\eta}
                    \left(1+ik\tilde{c}_{S}\eta\right)e^{-i\pi}
                    - C_2 e^{ ikc_{S}\eta}
                    \left(1-ik\tilde{c}_{S}\eta\right)e^{i\pi}\right]~~~~ &
 \mbox{\small {\bf for ~dS}}  \nonumber\\ 
	\displaystyle 2^{\nu-\frac{3}{2}}\frac{iH\tilde{c}_{S}}{2~M_p\sqrt{\epsilon}(1+\epsilon)
	(\tilde{c}_{S}k)^{\frac{3}{2}}}(-k\tilde{c}_{S}\eta)^{\frac{3}{2}-\nu}\left|\frac{\Gamma(\nu)}{\Gamma\left(\frac{3}{2}\right)}\right|\left[C_1  e^{- ik\tilde{c}_{S}\eta}
	                    \left(1+ik\tilde{c}_{S}\eta\right)
e^{-\frac{i\pi}{2}\left(\nu+\frac{1}{2}\right)} 
\right.\\ \left.\displaystyle~~~~~~~~~~~~~~~~~~~~~~~~~~~~~~~~~~~~~~~- C_2 e^{ ikc_{S}\eta}
                    \left(1-ik\tilde{c}_{S}\eta\right)
e^{\frac{i\pi}{2}\left(\nu+\frac{1}{2}\right)}\right]~~~~ & \mbox{\small {\bf for~ qdS}}.
          \end{array}
\right.\nonumber\eea 	

	One can further compute the two-point function for scalar fluctuation as:
\bea	{\langle \zeta(\eta,{\bf k}) \zeta(\eta,{\bf q})\rangle
=(2\pi)^3\delta^{(3)}({\bf k}+{\bf q})P_{\zeta}(k,\eta)}~~,\eea
	where $P_{\zeta}(k,\eta)$ is the power spectrum at time $\eta$ for scalar fluctuations and in the present context it is defined as:
	\bea\label{powa}
	 \displaystyle P_{\zeta}(k,\eta)&=&\frac{|v_{\bf k}(\eta)|^2}{z^2M^2_p}=\footnotesize\left\{\begin{array}{ll}
	                    \displaystyle  \frac{H^2}{4~M^2_p\epsilon \tilde{c}_{S}}\frac{1}{k^{3}}\left|C_1  e^{- ik\tilde{c}_{S}\eta}
	                    \left(1+ik\tilde{c}_{S}\eta\right)e^{-i\pi}
	                    - C_2 e^{ ikc_{S}\eta}
	                    \left(1-ik\tilde{c}_{S}\eta\right)e^{i\pi}\right|^2~~~~ &
	 \mbox{\small {\bf for ~dS}}  \\ 
		\displaystyle 2^{2\nu-3}\frac{H^2}{4~M^2_p\epsilon(1+\epsilon)^2\tilde{c}_{S} }\frac{1}{k^3}(-k\tilde{c}_{S}\eta)^{3-2\nu}\left|\frac{\Gamma(\nu)}{\Gamma\left(\frac{3}{2}\right)}\right|^2
\\ \displaystyle~\left|C_1  e^{- ik\tilde{c}_{S}\eta}
	\left(1+ik\tilde{c}_{S}\eta\right)
	e^{-\frac{i\pi}{2}\left(\nu+\frac{1}{2}\right)} - C_2 e^{ ikc_{S}\eta}
	                    \left(1-ik\tilde{c}_{S}\eta\right)
	e^{\frac{i\pi}{2}\left(\nu+\frac{1}{2}\right)}\right|^2~~~~ & \mbox{\small {\bf for~ qdS}}.
	          \end{array}
	\right.\eea 
		\subsubsection{Primordial Power Spectrum for Scalar~Perturbation}
		\label{v3a2}
	Finally, at the { horizon crossing} one can furthermore write the two-point correlation function as\footnote{See also ref.~\cite{Maldacena:2002vr,Chen:2006nt,Agarwal:2012mq}, where similar computation have been performed for canonical single-field slow-roll and generalized slow-roll models of inflation in the presence of {Bunch–Davies vacuum} state and {general initial} state.}:
	\bea	{\langle \zeta({\bf k}) \zeta({\bf q})\rangle
	=(2\pi)^3\delta^{(3)}({\bf k}+{\bf q})P_{\zeta}(k)}~~,\eea
		where $P_{\zeta}(k)$ is the power spectrum at time $\eta$ for scalar fluctuations and it is defined as: 
		\begin{small}	
\bea\label{pows}
	 P_{\zeta}(k)&=&\left[\frac{|v_{\bf k}(\eta)|^2}{z^2M^2_p}\right]_{|k\tilde{c}_{S}\eta|=1}=P_{\zeta}(k_*)\frac{1}{k^3}=\footnotesize\left\{\begin{array}{ll}
	                    \displaystyle  \frac{H^2}{4~M^2_p\epsilon \tilde{c}_{S}}\frac{1}{k^{3}}\left[|C_1|^2+|C_2|^2 -\left(C^{*}_{1}C_{2}+C_{1}C^{*}_{2}\right)\right]~~~~ &
	 \mbox{\small {\bf for ~dS}}  \\ 
		\displaystyle 2^{2\nu-3}\frac{H^2}{4~M^2_p\epsilon(1+\epsilon)^2\tilde{c}_{S} }\frac{1}{k^3}\left|\frac{\Gamma(\nu)}{\Gamma\left(\frac{3}{2}\right)}\right|^2\left[|C_1|^2+|C_2|^2 \right.\\ \left.\displaystyle ~~~-\left(C^{*}_{1}C_{2}e^{i\pi\left(\nu+\frac{1}{2}\right)}+C_{1}C^{*}_{2}e^{-i\pi\left(\nu+\frac{1}{2}\right)}\right)\right]~~~~ & \mbox{\small {\bf for~ qdS}},
	          \end{array}
	\right.\eea 
			\end{small}	
\hspace{-3pt}where $P_{\zeta}(k_*)$ is power spectrum for scalar fluctuation at the pivot scale $k=k_{*}$. 
	For simplicity one can keep $k^3/2\pi^2$ dependence outside and further define amplitude of the power spectrum $\Delta_{\zeta}(k_*)$ at the pivot scale $k=k_*$ as:
			\bea\label{powasss}
				 \displaystyle \Delta_{\zeta}(k_*)&=&\frac{k^3}{2\pi^2}P_{\zeta}(k)=\frac{1}{2\pi^2}P_{\zeta}(k_*)=\footnotesize\left\{\begin{array}{ll}
				                    \displaystyle  \frac{H^2}{8\pi^2~M^2_p\epsilon \tilde{c}_{S}}\left[|C_1|^2+|C_2|^2 -\left(C^{*}_{1}C_{2}+C_{1}C^{*}_{2}\right)\right]~~~~ &
				 \mbox{\small {\bf for ~dS}}  \\ 
					\displaystyle 2^{2\nu-3}\frac{H^2}{8\pi^2~M^2_p\epsilon(1+\epsilon)^2\tilde{c}_{S} }\left|\frac{\Gamma(\nu)}{\Gamma\left(\frac{3}{2}\right)}\right|^2\left[|C_1|^2+|C_2|^2 \right.\\ \left.\displaystyle ~~~~~~~~~~~~~~~~~~~~-\left(C^{*}_{1}C_{2}e^{i\pi\left(\nu+\frac{1}{2}\right)}+C_{1}C^{*}_{2}e^{-i\pi\left(\nu+\frac{1}{2}\right)}\right)\right]~~~~ & \mbox{\small {\bf for~ qdS}}.
				          \end{array}
				\right.\eea
				
				For {{\bf  Bunch–Davies}} and ${{{\alpha},{\beta}}}$ vacuum power spectrum can be written~as:
				\begin{itemize}
				\item \underline{{{\bf For Bunch–Davies vacuum }}:}\\
				In this case, by setting $C_1=1$ and $C_2=0$ we get the following expression for the power spectrum:
				\bea\label{powzx}
					 \displaystyle P_{\zeta}(k)&=&\footnotesize\left\{\begin{array}{ll}
					                    \displaystyle  \frac{H^2}{4~M^2_p\epsilon \tilde{c}_{S}}\frac{1}{k^{3}}~~~~ &
					 \mbox{\small {\bf for ~dS}} \\ 
						\displaystyle 2^{2\nu-3}\frac{H^2}{4~M^2_p\epsilon(1+\epsilon)^2\tilde{c}_{S} }\frac{1}{k^3}\left|\frac{\Gamma(\nu)}{\Gamma\left(\frac{3}{2}\right)}\right|^2~~~~ & \mbox{\small {\bf for~ qdS}}.
					          \end{array}
					\right.\eea 
					Also, the power spectrum $\Delta_{\zeta}(k_*)$ at the pivot scale $k=k_*$ as:
							\bea\label{xpowas}
								 \displaystyle \Delta_{\zeta}(k_*)&=&
									 \footnotesize\left\{\begin{array}{ll}
									                    \displaystyle  \frac{H^2}{8\pi^2~M^2_p\epsilon \tilde{c}_{S}}~~~~ &
									 \mbox{\small {\bf for ~dS}}  \\ 
										\displaystyle 2^{2\nu-3}\frac{H^2}{8\pi^2~M^2_p\epsilon(1+\epsilon)^2\tilde{c}_{S} }\left|\frac{\Gamma(\nu)}{\Gamma\left(\frac{3}{2}\right)}\right|^2~~~~ & \mbox{\small {\bf for~ qdS}}.
									          \end{array}
									\right.\eea
				\item \underline{{{\bf For ${\alpha,\beta}$ vacuum }}:}\\
				In this case, by setting $C_1=\cosh\alpha$ and $C_2=e^{i\beta}\sinh\alpha$ we get the following expression for the power spectrum:
				\bea\label{xpow}
					 \displaystyle P_{\zeta}(k)&=&\footnotesize
				\left\{\begin{array}{ll}
					                    \displaystyle  \frac{H^2}{4~M^2_p\epsilon \tilde{c}_{S}}\frac{1}{k^{3}}\left[\cosh2\alpha -\sinh2\alpha\cos\beta\right]~~~~ &
					 \mbox{\small {\bf for ~dS}}  \\ 
						\displaystyle 2^{2\nu-3}\frac{H^2}{4~M^2_p\epsilon(1+\epsilon)^2\tilde{c}_{S} }\frac{1}{k^3}\left|\frac{\Gamma(\nu)}{\Gamma\left(\frac{3}{2}\right)}\right|^2\\ 
						\displaystyle \left[\cosh2\alpha -\sinh2\alpha\cos\left(\pi\left(\nu+\frac{1}{2}\right)+\beta\right)\right]~~~~ & \mbox{\small {\bf for~ qdS}}.
					          \end{array}
					\right.\eea
Also, the power spectrum $\Delta_{\zeta}(k_*)$ at the pivot scale $k=k_*$ as:
					\bea\label{powas1}
									 \displaystyle \Delta_{\zeta}(k_*)&=&\footnotesize\left\{\begin{array}{ll}
									                    \displaystyle  \frac{H^2}{8\pi^2~M^2_p\epsilon \tilde{c}_{S}}\left[\cosh2\alpha -\sinh2\alpha\cos\beta\right]~~~~ &
									 \mbox{\small {\bf for ~dS}}  \\ 
										\displaystyle 2^{2\nu-3}\frac{H^2}{8\pi^2~M^2_p\epsilon(1+\epsilon)^2\tilde{c}_{S} }\left|\frac{\Gamma(\nu)}{\Gamma\left(\frac{3}{2}\right)}\right|^2\\
										\displaystyle\left[\cosh2\alpha -\sinh2\alpha\cos\left(\pi\left(\nu+\frac{1}{2}\right)+\beta\right)\right]~~~~ & \mbox{\small {\bf for~ qdS}}.
									          \end{array}
									\right.\eea 
				\end{itemize}
				
Finally, at the {horizon crossing} we get the following
expression for the spectral tilt for scalar fluctuation at the pivot scale $k=k_*$ as:
\bea\label{nbpowas}
				 \displaystyle n_{\zeta}(k_*)-1=\left[\frac{d\ln \Delta_{\zeta}(k)}{d\ln k}\right]_{|k\tilde{c}_{S}\eta|=1}= 2\eta-4\epsilon-\tilde{s},\eea	
				 where $\tilde{s}$ is defined as,
				 $\tilde{s}=\frac{\dot{\tilde{c}}_{S}}{H\tilde{c}_{S}}.$	
	\subsection{For Tensor~Modes\label{v3b}}
\unskip
		\subsubsection{Mode Equation and Solution for Tensor~Perturbation}
		\label{v3b1}
	Here we compute the two-point correlation from tensor perturbation. For~this purpose we consider the second-order perturbed action as given by\footnote{See also ref.~\cite{Maldacena:2002vr,Chen:2006nt,Agarwal:2012mq}, where similar computation have been performed for canonical single-field slow-roll and generalized slow-roll models of inflation in the presence of {Bunch–Davies vacuum} state and {general initial} state.}:
\begin{small}
			 \be \label{kl1}{S^{(2)}_{\gamma}\approx\int d^{4}x ~a^3~\frac{M^2_p}{8}\left[\left(1-\frac{\bar{M}^2_{3}}{M^2_p}\right)\dot{\gamma}_{ij}\dot{\gamma}_{ij}
			   -
			   \frac{1}{a^2}(\partial_{m}\gamma_{ij})^2
			   \right]=\int d^{3}x~d\eta ~a^2~\frac{M^2_p}{8}\left[\left(1-\frac{\bar{M}^2_{3}}{M^2_p}\right)\gamma^{'2}_{ij}
			   -
			   (\partial_{m}\gamma_{ij})^2\right]}~.\ee
			   \end{small}
			   
			   In Fourier space one can write $\gamma_{ij}(\eta,{\bf x})$ as:
			   \be \gamma_{ij}(\eta,{\bf x})=\sum_{\lambda=\times,+}\int\frac{d^3 k}{(2\pi)^{\frac{3}{2}}}\epsilon^{\lambda}_{ij}(k)~\gamma_{\lambda}(\eta,{\bf k})~e^{i{\bf k}.{\bf x}},\ee
			      where the rank-2 polarization tensor $\epsilon^{\lambda}_{ij}$ satisfies the properties, 
			      $\epsilon^{\lambda}_{ii}=k^{i}\epsilon^{\lambda}_{ij}=0,$ 
			      $\sum_{i,j}\epsilon^{\lambda}_{ij}\epsilon^{\lambda^{'}}_{ij}=2\delta_{\lambda\lambda^{'}}.$
			      Similar to scalar fluctuation here we also define a new variable $u_{\lambda}(\eta,{\bf k})$ in Fourier space as:
			      \be\begin{array}{lll}\label{kg1xcv}
			      		      		    \displaystyle u_{\lambda}(\eta,{\bf k}) =\frac{a}{\sqrt{2}}~M_p~\gamma_{\lambda}(\eta,{\bf k})=\footnotesize\left\{\begin{array}{ll}
			      		      		                       \displaystyle   -\frac{1}{\sqrt{2}H\eta}~M_p~\gamma_{\lambda}(\eta,{\bf k})~~~~ &
			      		      		    \mbox{\small {\bf for ~dS}}  \\ 
			      		      		   	\displaystyle  -\frac{1}{\sqrt{2}H\eta}\left(1+\epsilon\right)~M_p~\gamma_{\lambda}(\eta,{\bf k})~~~~ & \mbox{\small {\bf for~ qdS}}.
			      		      		             \end{array}
			      		      		   \right.
			      		      		   \end{array}\ee
			      		      		   
			      		      		   Using $u_{\lambda}(\eta,{\bf k})$ one can further 
			      		      		   write Equation~(\ref{kl1}) as:
\bea {S^{(2)}_{\gamma}\approx\int d^{3}x~d\eta ~a^2~\frac{M^2_p}{4}\left[\left(1-\frac{\bar{M}^2_{3}}{M^2_p}\right)u^{'2}_{\lambda}(\eta,{\bf k})
  -\left(k^2-\frac{a^{''}}{a}\right) (u_{\lambda}(\eta,{\bf k}))^2
			   			   \right]}~.\eea	
			   			   	
			   			   From this action one can find out the mode equation for tensor fluctuation as:
			   			   \bea {\textcolor{red}{\textbf{Mukhanov–Sasaki~Eqn~for~tensor~mode:}}~~~~~~
	u^{''}_{\lambda}(\eta,{\bf k})+\frac{\left(k^2-\frac{a^{''}}{a}\right)}{\left(1-\frac{\bar{M}^2_{3}}{M^2_p}\right)}u_{\lambda}(\eta,{\bf k})=0}~.~~~~\eea
	
	Furthermore, we introduce a new parameter $c_{T}$ defined as:
	\be c_{T}=\frac{1}{\sqrt{1-\frac{\bar{M}^2_{3}}{M^2_p}}} .\ee
	
	The general solution for the mode equation for graviton fluctuation can finally written as:	\be{\begin{array}{lll}\label{gu2}
			      		 \displaystyle u_{\lambda}(\eta,{\bf k}) =\footnotesize\left\{\begin{array}{ll}
			      		                    \displaystyle   \sqrt{-\eta}\left[D_1  H^{(1)}_{\frac{1}{2}\sqrt{1+8c^2_{T}}} \left(-kc_{T}\eta\right) 
			      		+ D_2 H^{(2)}_{\frac{1}{2}\sqrt{1+8c^2_{T}}} \left(-kc_{T}\eta\right)\right]~~~~ &
			      		 \mbox{\small {\bf for ~dS}}  \\ 
			      			\displaystyle \sqrt{-\eta}\left[D_1  H^{(1)}_{\frac{1}{2}\sqrt{1+4c^2_{T}\left(\nu^2-\frac{1}{4}\right)}} \left(-kc_{T}\eta\right) 
			      		+ D_2 H^{(2)}_{\frac{1}{2}\sqrt{1+4c^2_{T}\left(\nu^2-\frac{1}{4}\right)}} \left(-kc_{T}\eta\right)\right]~~~~ & \mbox{\small {\bf for~ qdS}}.
			      		          \end{array}
			      		\right.
			      		\end{array}}\ee
			      		\newpage
			      		Here $D_{1}$ and $D_{2}$ are the arbitrary integration constants and the numerical values depend on the choice of the 
			      		initial vacuum. In~the present context we consider the following 
			      		choice of the vacuum for the~computation:
			      		\begin{enumerate}
			      		 \item \underline{{\bf Bunch–Davies vacuum:}}~ In this case, we choose, $D_{1}=1,~~ D_{2}= 0$.
			      		 \item \underline{{\bf ${\alpha,\beta}$ vacuum:}}~ In this case, we choose $D_{1}=\cosh\alpha,~~ D_{2}= e^{i\beta}\sinh\alpha$. Here $\beta$ is a phase factor.
			      		\end{enumerate}
			      		
			      		For the most general solution as stated in Equation~(\ref{gu2}) one can consider the limiting physical 
			      		 situations, as~given by, I. {\bf Superhorizon regime:}~~ $|kc_{T}\eta|<<1$,
			      		  II. {\bf Horizon crossing:}~~ $|kc_{T}\eta|= 1$,
			      		  III. {\bf Subhorizon regime:}~~ $|kc_{T}\eta|>>1$.
			      		 
			      		 Finally, considering the behavior of the mode function in the {subhorizon regime} and {superhorizon regime} we get:
			      		 \begin{tiny}
			      		 \bea\label{gu2zxzx}
			      		  \displaystyle u_{\lambda}(\eta,{\bf k}) 
			      		  =\footnotesize\left\{\begin{array}{ll}
			      		                     \displaystyle  2^{\frac{1}{2}\sqrt{1+8c^2_{T}}-\frac{3}{2}}\frac{1}{i\eta}\frac{1}{\sqrt{2}\left(kc_{T}\right)^{\frac{3}{2}}}(-k
			      		                     			      		 					c_{T}\eta)^{\frac{3}{2}-\frac{1}{2}\sqrt{1+8c^2_{T}}}\left|\frac{\Gamma\left(\frac{1}{2}\sqrt{1+8c^2_{T}}\right)}{\Gamma\left(\frac{3}{2}\right)}\right|\\
			      		                     			      		 					\displaystyle \left[D_1  e^{- ikc_{T}\eta}
			      		                     \left(1+ikc_{T}\eta\right)e^{-\frac{i\pi}{2}\left(\frac{1}{2}\sqrt{1+8c^2_{T}}+\frac{1}{2}\right)}
			      		                     - D_2 e^{ ikc_{T}\eta}
			      		                     \left(1-ikc_{T}\eta\right)e^{\frac{i\pi}{2}\left(\frac{1}{2}\sqrt{1+8c^2_{T}}+\frac{1}{2}\right)}\right]&
			      		  \mbox{\small {\bf for ~dS}}  \\ 
			      		 	\displaystyle  2^{\frac{1}{2}\sqrt{1+4c^2_{T}\left(\nu^2-\frac{1}{4}\right)}-\frac{3}{2}}\frac{1}{i\eta}\frac{1}{\sqrt{2}\left(kc_{T}\right)^{\frac{3}{2}}}
			      		 	(-k
			      		 					c_{T}\eta)^{\frac{3}{2}-\frac{1}{2}\sqrt{1+4c^2_{T}\left(\nu^2-\frac{1}{4}\right)}}\left|\frac{\Gamma\left(\frac{1}{2}\sqrt{1+4c^2_{T}\left(\nu^2-\frac{1}{4}\right)}\right)}{\Gamma\left(\frac{3}{2}\right)}\right|\\
			      		 					\displaystyle \left[D_1  e^{- ikc_{T}\eta}
			      		 	                    \left(1+ikc_{T}\eta\right)
			      		 e^{-\frac{i\pi}{2}\left(\frac{1}{2}\sqrt{1+4c^2_{T}\left(\nu^2-\frac{1}{4}\right)}+\frac{1}{2}\right)} 
			      		 - D_2 e^{ ikc_{T}\eta}
			      		                     \left(1-ikc_{T}\eta\right)
			      		 e^{\frac{i\pi}{2}\left(\frac{1}{2}\sqrt{1+4c^2_{T}\left(\nu^2-\frac{1}{4}\right)}+\frac{1}{2}\right)}\right] & \mbox{\small {\bf for~ qdS}}.
			      		           \end{array}
			      		 \right.\eea		
			      		 \end{tiny}
			      		 
			      		 Furthermore, using Equation~(\ref{yu2zxzx}) one can write down the expression for the curvature perturbation $\zeta(\eta,{\bf k})$ as:     
			      		  \begin{tiny} 
			      		 \bea\label{yu2zxzxcv}
			      		  \displaystyle h_{\lambda}(\eta,{\bf k}) &=& \frac{u_{\lambda}(\eta,{\bf k})}{a~M_p} 
			      		   =\footnotesize\left\{\begin{array}{ll}
						      		 			      		                     \displaystyle  2^{\frac{1}{2}\sqrt{1+8c^2_{T}}-\frac{3}{2}}\frac{iH}{M_p}\frac{1}{\left(kc_{T}\right)^{\frac{3}{2}}}(-k
						      		 			      		                     			      		 					c_{T}\eta)^{\frac{3}{2}-\frac{1}{2}\sqrt{1+8c^2_{T}}}\left|\frac{\Gamma\left(\frac{1}{2}\sqrt{1+8c^2_{T}}\right)}{\Gamma\left(\frac{3}{2}\right)}\right|\\
						      		 			      		                     			      		 					\displaystyle \left[D_1  e^{- ikc_{T}\eta}
						      		 			      		                     \left(1+ikc_{T}\eta\right)e^{-\frac{i\pi}{2}\left(\frac{1}{2}\sqrt{1+8c^2_{T}}+\frac{1}{2}\right)}
						      		 			      		                     - D_2 e^{ ikc_{T}\eta}
						      		 			      		                     \left(1-ikc_{T}\eta\right)e^{\frac{i\pi}{2}\left(\frac{1}{2}\sqrt{1+8c^2_{T}}+\frac{1}{2}\right)}\right] &
						      		 			      		  \mbox{\small {\bf for ~dS}}  \\ 
						      		 			      		 	\displaystyle  2^{\frac{1}{2}\sqrt{1+4c^2_{T}\left(\nu^2-\frac{1}{4}\right)}-\frac{3}{2}}\frac{iH}{M_p \left(1+\epsilon\right)}\frac{1}{\left(kc_{T}\right)^{\frac{3}{2}}}
						      		 			      		 	(-k
						      		 			      		 					c_{T}\eta)^{\frac{3}{2}-\frac{1}{2}\sqrt{1+4c^2_{T}\left(\nu^2-\frac{1}{4}\right)}}\\
						      		 			      		 					\displaystyle \left|\frac{\Gamma\left(\frac{1}{2}\sqrt{1+4c^2_{T}\left(\nu^2-\frac{1}{4}\right)}\right)}{\Gamma\left(\frac{3}{2}\right)}\right|\left[D_1  e^{- ikc_{T}\eta}
						      		 			      		 	                    \left(1+ikc_{T}\eta\right)
						      		 			      		 e^{-\frac{i\pi}{2}\left(\frac{1}{2}\sqrt{1+4c^2_{T}\left(\nu^2-\frac{1}{4}\right)}+\frac{1}{2}\right)} 
						      		 			      		 \right.\\ \left.\displaystyle~~~~~~~~~~~~~~~~~~~~~~~~~~~~~~~~~~~~~~~- D_2 e^{ ikc_{T}\eta}
						      		 			      		                     \left(1-ikc_{T}\eta\right)
						      		 			      		 e^{\frac{i\pi}{2}\left(\frac{1}{2}\sqrt{1+4c^2_{T}\left(\nu^2-\frac{1}{4}\right)}+\frac{1}{2}\right)}\right] & \mbox{\small {\bf for~ qdS}}.
						      		 			      		           \end{array}
						      		 			      		 \right.\eea 		
						      		 			      		  \end{tiny}	
	\subsubsection{Primordial Power Spectrum for Tensor~Perturbation}
	\label{v3b2}			      		 
			      		 	One can further compute the two-point function for tensor fluctuation as:
			      		 \bea	{\langle h(\eta,{\bf k}) h(\eta,{\bf q})\rangle=\sum_{\lambda,\lambda^{'}} \langle h_{\lambda}(\eta,{\bf k})h_{\lambda^{'}}(\eta,{\bf q})\rangle
			      		 =(2\pi)^3\delta^{(3)}({\bf k}+{\bf q})P_{h}(k,\eta)}~,\eea
			      		 	where $P_{h}(k,\eta)$ is the power spectrum at time $\eta$ for tensor fluctuations and in the present context it is defined as:
			      		 	\vspace{12pt}
	  \begin{tiny} 
			      		 	\bea\label{spow}
			      		 	 \displaystyle P_{h}(k,\eta)&=&\frac{4| h_{\lambda}(\eta,{\bf k})|^2}{a^2M^2_p} 
			      		   =\footnotesize\left\{\begin{array}{ll}
						      		 			      		                     \displaystyle  2^{\sqrt{1+8c^2_{T}}-3}\frac{4H^2}{M^2_p}\frac{1}{\left(kc_{T}\right)^{3}}(-k
						      		 			      		                     			      		 					c_{T}\eta)^{3-\sqrt{1+8c^2_{T}}}\left|\frac{\Gamma\left(\frac{1}{2}\sqrt{1+8c^2_{T}}\right)}{\Gamma\left(\frac{3}{2}\right)}\right|^2\\
						      		 			      		                     			      		 					\displaystyle \left|D_1  e^{- ikc_{T}\eta}
						      		 			      		                     \left(1+ikc_{T}\eta\right)e^{-\frac{i\pi}{2}\left(\frac{1}{2}\sqrt{1+8c^2_{T}}+\frac{1}{2}\right)}
						      		 			      		                    - D_2 e^{ ikc_{T}\eta}
						      		 			      		                     \left(1-ikc_{T}\eta\right)e^{\frac{i\pi}{2}\left(\frac{1}{2}\sqrt{1+8c^2_{T}}+\frac{1}{2}\right)}\right|^2 &
						      		 			      		  \mbox{\small {\bf for ~dS}}  \\ 
						      		 			      		 	\displaystyle  2^{\sqrt{1+4c^2_{T}\left(\nu^2-\frac{1}{4}\right)}-3}\frac{4H^2}{M^2_p \left(1+\epsilon\right)^2}\frac{1}{\left(kc_{T}\right)^{3}}
						      		 			      		 	(-k
						      		 			      		 					c_{T}\eta)^{3-\sqrt{1+4c^2_{T}\left(\nu^2-\frac{1}{4}\right)}}\\
						      		 			      		 					\displaystyle \left|\frac{\Gamma\left(\frac{1}{2}\sqrt{1+4c^2_{T}\left(\nu^2-\frac{1}{4}\right)}\right)}{\Gamma\left(\frac{3}{2}\right)}\right|^2\left|D_1  e^{- ikc_{T}\eta}
						      		 			      		 	                    \left(1+ikc_{T}\eta\right)
						      		 			      		 e^{-\frac{i\pi}{2}\left(\frac{1}{2}\sqrt{1+4c^2_{T}\left(\nu^2-\frac{1}{4}\right)}+\frac{1}{2}\right)} 
						      		 			      		 \right.\\ \left.\displaystyle~~~- D_2 e^{ ikc_{T}\eta}
						      		 			      		                     \left(1-ikc_{T}\eta\right)
						      		 			      		 e^{\frac{i\pi}{2}\left(\frac{1}{2}\sqrt{1+4c^2_{T}\left(\nu^2-\frac{1}{4}\right)}+\frac{1}{2}\right)}\right|^2& \mbox{\small {\bf for~ qdS}}.
						      		 			      		           \end{array}
						      		 			      		 \right.\eea 
			      		 		  \end{tiny} 
			      		 		  
			      		 	Finally, at the  { horizon crossing} we get the following two-point correlation function for tensor perturbation as:
			      		 	\bea	{\langle h({\bf k})h({\bf q})\rangle
			      		 	=(2\pi)^3\delta^{(3)}({\bf k}+{\bf q})P_{h}(k)}~,\eea
			      		 		where $P_{h}(k)$ is known as the power spectrum at the  { horizon crossing} for tensor fluctuations and in the present context it is defined as: 	
			      		 			  \begin{tiny} 
			      		 \bea\label{pfow}
			      		 	 \displaystyle P_{h}(k)&=&P_{h}(k_*)\frac{1}{k^3}=\footnotesize\left\{\begin{array}{ll}
			      		 	                    \displaystyle  2^{\sqrt{1+8c^2_{T}}-3}\frac{4H^2}{M^2_p c^3_T}\frac{1}{k^{3}}\left|\frac{\Gamma\left(\frac{1}{2}\sqrt{1+8c^2_{T}}\right)}{\Gamma\left(\frac{3}{2}\right)}\right|^2\left[|D_1|^2+|D_2|^2 \right.\\ \left.\displaystyle ~-\left(D^{*}_{1}D_{2}e^{i\pi\left(\frac{1}{2}\sqrt{1+8c^2_{T}}+\frac{1}{2}\right)}+D_{1}D^{*}_{2}e^{-i\pi\left(\frac{1}{2}\sqrt{1+8c^2_{T}}+\frac{1}{2}\right)}\right)\right]~ &
			      		 	 \mbox{\small {\bf for ~dS}}  \\ 
			      		 		\displaystyle  2^{\sqrt{1+4c^2_{T}\left(\nu^2-\frac{1}{4}\right)}-3}\frac{4H^2}{M^2_p \left(1+\epsilon\right)^2 c^3_T}\frac{1}{k^3}\left|\frac{\Gamma\left(\frac{1}{2}\sqrt{1+4c^2_{T}\left(\nu^2-\frac{1}{4}\right)}\right)}{\Gamma\left(\frac{3}{2}\right)}\right|^2\left[|D_1|^2+|D_2|^2 \right.\\ \left.\displaystyle ~-\left(D^{*}_{1}D_{2}e^{i\pi\left(\frac{1}{2}\sqrt{1+4c^2_{T}\left(\nu^2-\frac{1}{4}\right)}+\frac{1}{2}\right)}+D_{1}D^{*}_{2}e^{-i\pi\left(\frac{1}{2}\sqrt{1+4c^2_{T}\left(\nu^2-\frac{1}{4}\right)}+\frac{1}{2}\right)}\right)\right]~ & \mbox{\small {\bf for~ qdS}}.
			      		 	          \end{array}
			      		 	\right.\eea 
			      		 		  \end{tiny} 
\hspace{-3pt}where $P_{h}(k_*)$ is power spectrum for tensor fluctuation at the pivot scale $k=k_{*}$. 
			      		 	For simplicity one can keep $k^3/2\pi^2$ dependence outside and further define amplitude of the power spectrum $\Delta_{h}(k_*)$ at the pivot scale $k=k_*$ as:	
			      		 			\bea\label{poxwas}
			      		 				 \displaystyle \Delta_{h}(k_*)&=&\frac{k^3}{2\pi^2}P_{h}(k)=\frac{1}{2\pi^2}P_{h}(k_*)\nonumber\\
			      		 				 &=&
			      		 				 			      		 \footnotesize\left\{\begin{array}{ll}
			      		 				 			      		 	                    \displaystyle  2^{\sqrt{1+8c^2_{T}}-3}\frac{2H^2}{\pi^2M^2_p c^3_T}\left|\frac{\Gamma\left(\frac{1}{2}\sqrt{1+8c^2_{T}}\right)}{\Gamma\left(\frac{3}{2}\right)}\right|^2\left[|D_1|^2+|D_2|^2 \right.\\ \left.\displaystyle ~~~~~~~~~~~~~~~~~~~~-\left(C^{*}_{1}C_{2}e^{i\pi\left(\frac{1}{2}\sqrt{1+8c^2_{T}}+\frac{1}{2}\right)}+D_{1}D^{*}_{2}e^{-i\pi\left(\frac{1}{2}\sqrt{1+8c^2_{T}}+\frac{1}{2}\right)}\right)\right]~~~~ &
			      		 				 			      		 	 \mbox{\small {\bf for ~dS}}  \\ 
			      		 				 			      		 		\displaystyle  2^{\sqrt{1+4c^2_{T}\left(\nu^2-\frac{1}{4}\right)}-3}\frac{2H^2}{\pi^2 M^2_p \left(1+\epsilon\right)^2 c^3_T}\left|\frac{\Gamma\left(\frac{1}{2}\sqrt{1+4c^2_{T}\left(\nu^2-\frac{1}{4}\right)}\right)}{\Gamma\left(\frac{3}{2}\right)}\right|^2\left[|D_1|^2+|D_2|^2 \right.\\ \left.\displaystyle ~~~~~~~~~~~~~~~~~~~~-\left(D^{*}_{1}D_{2}e^{i\pi\left(\frac{1}{2}\sqrt{1+4c^2_{T}\left(\nu^2-\frac{1}{4}\right)}+\frac{1}{2}\right)}+D_{1}D^{*}_{2}e^{-i\pi\left(\frac{1}{2}\sqrt{1+4c^2_{T}\left(\nu^2-\frac{1}{4}\right)}+\frac{1}{2}\right)}\right)\right]~~~~ & \mbox{\small {\bf for~ qdS}}.
			      		 				 			      		 	          \end{array}
			      		 				 			      		 	\right.\eea 
			      		 				 			      		 	
			      		 				For {{\bf  Bunch–Davies}} and ${{{\alpha,\beta}}}$ vacuums we~get:
			      		 				\begin{itemize}
			      		 				\item \underline{{{\bf For Bunch–Davies vacuum }}:}\\
			      		 								In this case, by setting $D_1=1$ and $D_2=0$ we get the following expression for the power spectrum:
			      		 								\vspace{12pt}

		 \bea\label{ddpow}
			      		 	 \displaystyle P_{h}(k)&=&\footnotesize\left\{\begin{array}{ll}
			      		 	                    \displaystyle  2^{\sqrt{1+8c^2_{T}}-3}\frac{4H^2}{M^2_p c^3_T}\frac{1}{k^{3}}\left|\frac{\Gamma\left(\frac{1}{2}\sqrt{1+8c^2_{T}}\right)}{\Gamma\left(\frac{3}{2}\right)}\right|^2~~~~ &
			      		 	 \mbox{\small {\bf for ~dS}}  \\ 
			      		 		\displaystyle  2^{\sqrt{1+4c^2_{T}\left(\nu^2-\frac{1}{4}\right)}-3}\frac{4H^2}{M^2_p \left(1+\epsilon\right)^2 c^3_T}\frac{1}{k^3}\left|\frac{\Gamma\left(\frac{1}{2}\sqrt{1+4c^2_{T}\left(\nu^2-\frac{1}{4}\right)}\right)}{\Gamma\left(\frac{3}{2}\right)}\right|^2~~~~ & \mbox{\small {\bf for~ qdS}}.
			      		 	          \end{array}
			      		 	\right.\eea 
Also, the power spectrum $\Delta_{\zeta}(k_*)$ at the pivot scale $k=k_*$ as:			      		 					      		 
			      		 			\bea\label{dpowas}
			      		 				 \displaystyle \Delta_{h}(k_*)&=&\footnotesize\left\{\begin{array}{ll}
			      		 				 			      		 	                    \displaystyle  2^{\sqrt{1+8c^2_{T}}-3}\frac{2H^2}{\pi^2M^2_p c^3_T}\left|\frac{\Gamma\left(\frac{1}{2}\sqrt{1+8c^2_{T}}\right)}{\Gamma\left(\frac{3}{2}\right)}\right|^2~~~~ &
			      		 				 			      		 	 \mbox{\small {\bf for ~dS}} \\ 
			      		 				 			      		 		\displaystyle  2^{\sqrt{1+4c^2_{T}\left(\nu^2-\frac{1}{4}\right)}-3}\frac{2H^2}{\pi^2 M^2_p \left(1+\epsilon\right)^2 c^3_T}\left|\frac{\Gamma\left(\frac{1}{2}\sqrt{1+4c^2_{T}\left(\nu^2-\frac{1}{4}\right)}\right)}{\Gamma\left(\frac{3}{2}\right)}\right|^2~~~~ & \mbox{\small {\bf for~ qdS}}.
			      		 				 			      		 	          \end{array}
			      		 				 			      		 	\right.\eea 
			      		 				\item \underline{{{\bf For $~{\alpha,\beta}$ vacuum }}:}\\
			      		 								In this case, by setting $D_1=\cosh\alpha$ and $D_2=e^{i\beta}\sinh\alpha$ we get the following expression for the power spectrum:
 \bea\label{vpow}
			      		 	 \displaystyle P_{h}(k)&=&\footnotesize\left\{\begin{array}{ll}
			      		 	                    \displaystyle  2^{\sqrt{1+8c^2_{T}}-3}\frac{4H^2}{M^2_p c^3_T}\frac{1}{k^{3}}\left|\frac{\Gamma\left(\frac{1}{2}\sqrt{1+8c^2_{T}}\right)}{\Gamma\left(\frac{3}{2}\right)}\right|^2
			      		 	                   \\ \displaystyle  \left[\cosh2\alpha -\sinh2\alpha\cos\left(\pi\left(\frac{1}{2}\sqrt{1+8c^2_{T}}+\frac{1}{2}\right)+\beta\right)\right]~~~~ &
			      		 	 \mbox{\small {\bf for ~dS}}  \\ 
			      		 		\displaystyle  2^{\sqrt{1+4c^2_{T}\left(\nu^2-\frac{1}{4}\right)}-3}\frac{4H^2}{M^2_p \left(1+\epsilon\right)^2 c^3_T}\frac{1}{k^3}\left|\frac{\Gamma\left(\frac{1}{2}\sqrt{1+4c^2_{T}\left(\nu^2-\frac{1}{4}\right)}\right)}{\Gamma\left(\frac{3}{2}\right)}\right|^2
			      		 		 \\ \displaystyle  \left[\cosh2\alpha -\sinh2\alpha\cos\left(\pi\left(\frac{1}{2}\sqrt{1+4c^2_{T}\left(\nu^2-\frac{1}{4}\right)}+\frac{1}{2}\right)+\beta\right)\right]~~~~ & \mbox{\small {\bf for~ qdS}}.
			      		 	          \end{array}
			      		 	\right.\eea 
Also, the power spectrum $\Delta_{\zeta}(k_*)$ at the pivot scale $k=k_*$ as:	
			      		 			\bea\label{powaus}
			      		 				 \displaystyle \Delta_{h}(k_*)&=&\footnotesize\left\{\begin{array}{ll}
		\displaystyle  2^{\sqrt{1+8c^2_{T}}-3}\frac{2H^2}{\pi^2M^2_p c^3_T}\left|\frac{\Gamma\left(\frac{1}{2}\sqrt{1+8c^2_{T}}\right)}{\Gamma\left(\frac{3}{2}\right)}\right|^2
\\ \displaystyle  \left[\cosh2\alpha -\sinh2\alpha\cos\left(\pi\left(\frac{1}{2}\sqrt{1+8c^2_{T}}+\frac{1}{2}\right)+\beta\right)\right]			      		 				 			      		 	                    ~~~~ &
			      		 				 			      		 	 \mbox{\small {\bf for ~dS}}  \\ 
			      		 				 			      		 		\displaystyle  2^{\sqrt{1+4c^2_{T}\left(\nu^2-\frac{1}{4}\right)}-3}\frac{2H^2}{\pi^2 M^2_p \left(1+\epsilon\right)^2 c^3_T}\left|\frac{\Gamma\left(\frac{1}{2}\sqrt{1+4c^2_{T}\left(\nu^2-\frac{1}{4}\right)}\right)}{\Gamma\left(\frac{3}{2}\right)}\right|^2
\\ \displaystyle  \left[\cosh2\alpha -\sinh2\alpha\cos\left(\pi\left(\frac{1}{2}\sqrt{1+4c^2_{T}\left(\nu^2-\frac{1}{4}\right)}+\frac{1}{2}\right)+\beta\right)\right]			      		 				 			      		 		~~~~ & \mbox{\small {\bf for~ qdS}}.
			      		 				 			      		 	          \end{array}
			      		 				 			      		 	\right.\eea 			      		 									
			      		 				\end{itemize}	  	
			      		 				
Now let us consider a special case for tensor fluctuation where $c_{T}=1$ and it implies the following two~possibilities:
\begin{enumerate}
\item $\bar{M}_{3}=0$. But~for this case as we have assumed earlier $\bar{M}^2_3\approx \bar{M}^2_3=\bar{M}^3_1 /4H\tilde{c}_{5}$, then $\bar{M}_{1}=0$ which is not our matter of interest in this work as this leads to zero three-point function for scalar fluctuation. But~if we assume that $\bar{M}^2_3\neq \bar{M}^2_3$ but $\bar{M}^2_3=\bar{M}^3_1 /4H\tilde{c}_{5}$ then by setting $\bar{M}_{3}=0$ one can get $\bar{M}_{1}\neq 0$, which is necessarily required for non-vanishing three-point function for scalar~fluctuation. 

\item $\bar{M}_{3}<<M_p$. In~this case if we assume  $\bar{M}^2_3\approx \bar{M}^2_3=\bar{M}^3_1 /4H\tilde{c}_{5}$, then $\bar{M}^3_1 /4H\tilde{c}_{5}M^2_p<<1$ and $\bar{M}_2<<M_p$. This is perfectly ok of generating non-vanishing three-point function for scalar fluctuation. 
\end{enumerate}	

If we set $c_{T}=1$ then for {{\bf  Bunch–Davies}} and ${{\bf \alpha,\beta}}$ vacuum power spectrum can be recast into the following simplified~form:
			      		 				\begin{itemize}
			      		 				\item \underline{{{\bf For Bunch–Davies vacuum }}:}\\
	In this case, by setting $D_1=1$ and $D_2=0$ we get the following expression for the power spectrum:
		 \bea\label{powww}
			      		 	 \displaystyle P_{h}(k)&=&\footnotesize\left\{\begin{array}{ll}
			      		 	                    \displaystyle  \frac{4H^2}{M^2_p }\frac{1}{k^{3}}~~~~ &
			      		 	 \mbox{\small {\bf for ~dS}} \\ 
			      		 		\displaystyle  2^{2\nu-3}\frac{4H^2}{M^2_p \left(1+\epsilon\right)^2 }\frac{1}{k^3}\left|\frac{\Gamma\left(\nu\right)}{\Gamma\left(\frac{3}{2}\right)}\right|^2~~~~ & \mbox{\small {\bf for~ qdS}}.
			      		 	          \end{array}
			      		 	\right.\eea 
Also, the power spectrum $\Delta_{\zeta}(k_*)$ at the pivot scale $k=k_*$ as:			      		 			      		 
			      		 			\bea\label{powkas}
			      		 				 \displaystyle \Delta_{h}(k_*)&=&\footnotesize\left\{\begin{array}{ll}
			      		 				 			      		 	                    \displaystyle \frac{2H^2}{\pi^2M^2_p }~~~~ &
			      		 				 			      		 	 \mbox{\small {\bf for ~dS}}  \\ 
			      		 				 			      		 		\displaystyle  2^{\nu-3}\frac{2H^2}{\pi^2 M^2_p \left(1+\epsilon\right)^2 }\left|\frac{\Gamma\left(\nu\right)}{\Gamma\left(\frac{3}{2}\right)}\right|^2~~~~ & \mbox{\small {\bf for~ qdS}}.
			      		 				 			      		 	          \end{array}
			      		 				 			      		 	\right.\eea 
			      		 				\item \underline{{{\bf For $~{\alpha,\beta}$ vacuum }}:}\\
In this case, by setting $D_1=\cosh\alpha$ and $D_2=e^{i\beta}\sinh\alpha$ we get the following expression for the power spectrum:
 \bea\label{powl}
			      		 	 \displaystyle P_{h}(k)&=&\footnotesize\left\{\begin{array}{ll}
			      		 	                    \displaystyle  \frac{4H^2}{M^2_p }\frac{1}{k^{3}} \left[\cosh2\alpha -\sinh2\alpha\cos\beta\right]~~~~ &
			      		 	 \mbox{\small {\bf for ~dS}}  \\ 
			      		 		\displaystyle  2^{2\nu-3}\frac{4H^2}{M^2_p \left(1+\epsilon\right)^2 }\frac{1}{k^3}\left|\frac{\Gamma\left(\nu\right)}{\Gamma\left(\frac{3}{2}\right)}\right|^2\left[\cosh2\alpha -\sinh2\alpha\cos\left(\pi\left(\nu+\frac{1}{2}\right)+\beta\right)\right]~~~~ & \mbox{\small {\bf for~ qdS}}.
			      		 	          \end{array}
			      		 	\right.\eea 
Also, the power spectrum $\Delta_{\zeta}(k_*)$ at the pivot scale $k=k_*$ as:		      		 	
			      		 			\bea\label{powasp}
			      		 				 \displaystyle \Delta_{h}(k_*)&=&\footnotesize\left\{\begin{array}{ll}
\displaystyle  \frac{2H^2}{\pi^2M^2_p }
 \left[\cosh2\alpha -\sinh2\alpha\cos\beta\right]			      		 				 			      		 	                    ~~~~ &
			      		 				 			      		 	 \mbox{\small {\bf for ~dS}}  \\ 
			      		 				 			      		 		\displaystyle  2^{2\nu-3}\frac{2H^2}{\pi^2 M^2_p \left(1+\epsilon\right)^2 }\left|\frac{\Gamma\left(\nu\right)}{\Gamma\left(\frac{3}{2}\right)}\right|^2
\\ \displaystyle  \left[\cosh2\alpha -\sinh2\alpha\cos\left(\pi\left(\nu+\frac{1}{2}\right)+\beta\right)\right]			      		 				 			      		 		~~~~ & \mbox{\small {\bf for~ qdS}}.
			      		 				 			      		 	          \end{array}
			      		 				 			      		 	\right.\eea 			      		 									
			      		 				\end{itemize}	  			      		 					      		   
	\section{{Scalar Three-Point Correlation Function from~EFT}\label{v4} }
\unskip
	\subsection{Basic~Setup}
	\label{v4c}  
	Here we compute the three-point correlation function for perturbations from scalar modes. For~this purpose we consider the third-order perturbed action
	for the scalar modes as given by\footnote{Here it is important to note that the \textcolor{red}{\bf red} colored terms are the new contribution in the EFT action considered in this paper, which are not present in ref.~\cite{Cheung:2007st}. From~the EFT action itself it is clear that for effective sound speed $c_S=1$ three-point correlation function and the associated bispectrum vanishes if we do not contribution these \textcolor{red}{\bf red} colored terms. This is obviously true if we fix $c_S=1$ in the result obtained in ref.~\cite{Cheung:2007st}. On~the other hand, if we consider these \textcolor{red}{\bf red} colored terms then the result is consistent with ref.~\cite{Maldacena:2002vr} with $c_S=1$ and with ref.~\cite{Chen:2006nt} with $c_S\neq 1$. This implies that $c_S=1$ is not fully radiatively stable in single-field slow-roll inflation. However, if~we include the effects produced by quantum correction through loop effects, then a small deviation in the effective sound speed $1-c_S\sim \epsilon\left(H/M_p\right)^2$ can be produced. See ref.~\cite{Cheung:2007st} where this fact is clearly pointed. But~for inflation we know that in the inflationary regime the slow-roll parameter $\epsilon<1$ and the scale of inflation is $H/M_p<<1$, which imply this deviation is also very small and not very interesting for our purpose studied in this paper.  Also see ref.~\cite{Agarwal:2012mq} for more details.}:
	\vspace{12pt}

	\be{\begin{array}{rl} S^{(3)}_{\zeta}&\approx \displaystyle \int d^{4}x ~\frac{a^3}{H^3}\left[-\left\{\left(1-\frac{1}{c^2_{S}}\right)\dot{H}M^2_p \textcolor{red}{+\frac{3}{2}\bar{M}^3_1 H}-\frac{4}{3}M^4_3\right\}
	   \dot{\zeta}^3\right.\\ & \left.~~~~~~~~~~~~~~~~~~~~~~~~~\displaystyle
	   +\left\{\left(1-\frac{1}{c^2_{S}}\right)\dot{H}M^2_p \textcolor{red}{+\frac{3}{2}\bar{M}^3_{1} H}\right\}\frac{1}{a^2}\dot{\zeta}(\partial_{i}\zeta)^2
	   \right.\\ & \left.~~~~~~~~~~~~~~~~~~~~~~~~~~~~~~~~~~~~~~~~~~\displaystyle
	   \textcolor{red}{+\frac{9}{2}\bar{M}^3_{1}H^2\zeta \dot{\zeta}^2}\textcolor{red}{-\frac{3}{2}\bar{M}^3_1 H \frac{1}{a^2} \zeta \frac{d}{dt}\left(\partial_{i}\zeta\right)^2}\right],\end{array}}\ee
	which can be recast for $c_S = 1$ and $c_S <1$ case as:
		\be{\begin{array}{rl} \underline{\textcolor{red}{\bf For ~c_{S}=1:}}&\\
				S^{(3)}_{\zeta}&\approx \displaystyle \int d^{4}x ~\frac{a^3}{H^3}\left[-\left\{\textcolor{red}{\frac{3}{2}\bar{M}^3_1 H}\right\}
				\dot{\zeta}^3 
				+\left\{\textcolor{red}{\frac{3}{2}\bar{M}^3_{1} H}\right\}\frac{1}{a^2}\dot{\zeta}(\partial_{i}\zeta)^2
				\right.\\ & \left.~~~~~~~~~~~~~~~~~~~~~~~~~~~~~~~~~~~~~~~~~~\displaystyle
				\textcolor{red}{+\frac{9}{2}\bar{M}^3_{1}H^2\zeta \dot{\zeta}^2}\textcolor{red}{-\frac{3}{2}\bar{M}^3_1 H \frac{1}{a^2} \zeta \frac{d}{dt}\left(\partial_{i}\zeta\right)^2}\right],\end{array}}\ee
		\be{\begin{array}{rl} \underline{\textcolor{red}{\bf\bf For ~c_{S}<1:}}&\\
				S^{(3)}_{\zeta}&\approx \displaystyle\int d^{4}x ~a^3~\frac{\epsilon M^2_p}{H}\left(1-\frac{1}{c^2_{S}}\right)\left[\left\{1+\frac{3\tilde{c}_{4}}{4c^2_{S}}	\textcolor{red}{+\frac{2\tilde{c}_{3}}{3c^2_{S}}}\right\}
				\dot{\zeta}^3
				-\left\{1	\textcolor{red}{+\frac{3\tilde{c}_{4}}{4c^2_{S}}}\right\}\frac{1}{a^2}\dot{\zeta}(\partial_{i}\zeta)^2
				\right.\\ & \left.~~~~~~~~~~~~~~~~~~~~~~~~~~~~~~~~~~~~~~~~~~\displaystyle
				\textcolor{red}{-\frac{9H\tilde{c}_{4}}{4c^2_S}\zeta \dot{\zeta}^2}\textcolor{red}{+\frac{3\tilde{c}_{4}}{4c^2_{S}}\frac{1}{a^2} \zeta \frac{d}{dt}\left(\partial_{i}\zeta\right)^2}\right],\end{array}}\ee   
				
	   To extract further information from third-order action, first one needs to start with the Fourier transform of the curvature perturbation $\zeta(\eta,{\bf x})$ defined as:
	   \bea \zeta(\eta,{\bf x})&=&\int \frac{d^{3}k}{(2\pi)^3}\zeta_{\bf k}(\eta)\exp(i{\bf k}.{\bf x}),\eea
	   where $\zeta_{\bf k}(\eta)$ is the time-dependent part of the curvature fluctuation after Fourier transform and can be expressed 
	   in terms of the normalized time-dependent scalar mode function $v_{\bf{k}}(\eta)$ as:
\begin{equation}\label{sdfc}
	  { \hat{\zeta}(\eta,{\bf k}) = \frac{v_{{\bf k}}(\eta)}{zM_{p}} = \frac{\zeta\left(\eta , \bf{k}\right) a \left(\bf{k}\right) 
	   + \zeta^* \left(\eta ,-\bf{k}\right) a^\dagger \left(-\bf{k}\right)}{zM_{p}}}
	   \end{equation}
	   where $z$ is explicitly defined earlier
	   and $a({\bf k}), a^{\dagger}({\bf k})$ are the creation and annihilation operator satisfies
	   the following commutation relations:
	   \bea \left[a({\bf k}),a^{\dagger}(-{\bf k}^{'})\right]&=&(2\pi)^3\delta^{3}({\bf k}+{\bf k}^{'}),~~
	   \left[a({\bf k}),a({\bf k}^{'})\right]=0,~~ 
	   \left[a^{\dagger}({\bf k}),a^{\dagger}({\bf k}^{'})\right]=0.\eea
	  
	   \subsection{Computation of Scalar Three-Point Function in Interaction~Picture}
	   \label{v4a}
	   Presently our prime objective is to compute the three-point function of the curvature fluctuation in momentum space from $S^{2}_{\zeta}$ with respect to the arbitrary choice of vacuum, 
	   which leads to important result in the context of primordial cosmology.
	   Furthermore, using the interaction picture the three-point function of the curvature fluctuation in momentum space can be expressed as:
\begin{equation}\label{opioz}
	   {\langle\zeta({\bf k_{1}})\zeta({\bf k_{2}})\zeta({\bf k_{3}})\rangle = -i \int\limits_{\eta_{i}=-\infty}^{\eta_f=0} {d\eta}~ a(\eta)~ \langle0|\left[\hat{\zeta}(\eta_f,{\bf k_{1}})\hat{\zeta}(\eta_f,{\bf k_{2}})\hat{\zeta}(\eta_f,{\bf k_{3}}), H_{int} \left(\eta\right)\right]|0\rangle}~,
	   \end{equation}
	   where $a(\eta)$ is the scale factor defined in the earlier section in terms of Hubble parameter $H$ and conformal time scale $\eta$. Here $|0\rangle$ represents any arbitrary vacuum state and for discussion we will only derive the results for {Bunch–Davies vacuum} and {$\alpha,\beta$~vacuum}. In~the interaction picture the Hamiltonian can written as\footnote{See also ref.~\cite{Maldacena:2002vr,Chen:2006nt}, where similar computations have been performed for canonical single-field slow-roll and generalized slow-roll models of inflation in the presence of {Bunch–Davies vacuum} state and {general initial} state.}:
	   \be\label{hamil}{\begin{array}{rl}
	   H_{int}(\eta) &= \displaystyle - \frac{1}{H^3}\left[-\left\{\left(1-\frac{1}{c^2_{S}}\right)\dot{H}M^2_p +\frac{3}{2}\bar{M}^3_1 H-\frac{4}{3}M^4_3\right\}
	   	   \dot{\zeta}^3\right.\\ & \left.~~~~~~~~~~~~~~~~~~~~~~~~~\displaystyle
	   	   +\left\{\left(1-\frac{1}{c^2_{S}}\right)\dot{H}M^2_p +\frac{3}{2}\bar{M}^3_{1} H\right\}\frac{1}{a^2}\dot{\zeta}(\partial_{i}\zeta)^2
	   	   \right.\\ & \left.~~~~~~~~~~~~~~~~~~~~~~~~~~~~~~~~~~~~~~~~~~\displaystyle
	   	   +\frac{9}{2}\bar{M}^3_{1}H^2\zeta \dot{\zeta}^2-\frac{3}{2}\bar{M}^3_1 H \frac{1}{a^2} \zeta \frac{d}{dt}\left(\partial_{i}\zeta\right)^2\right].\end{array}}
	   \ee
	   which gives the primary information to compute the explicit expression for the three-point function in the present context.
	   After substituting the Hamiltonian interaction, we finally get the following expression for the three-point function for the scalar fluctuation:
\begin{small}
	 \bea\label{opio}
	 \langle\zeta({\bf k_{1}})\zeta({\bf k_{2}})\zeta({\bf k_{3}})\rangle &=& -i \int\limits_{\eta_{i}=-\infty}^{\eta_f=0} {d\eta}~ a(\eta)~\int\frac{d^3x}{H^3}\int\int\int\frac{d^3k_{4}}{(2\pi)^3}\frac{d^3k_{5}}{(2\pi)^3}\frac{d^3k_{6}}{(2\pi)^3}~e^{i({\bf k_{4}}+{\bf k_{5}}+{\bf k_{6}}).{\bf x}}\nonumber\\
	 &&~~~~~~~
	 \left\{\alpha_{1}\langle0|\left[\hat{\zeta}(\eta_f,{\bf k_{1}})\hat{\zeta}(\eta_f,{\bf k_{2}})\hat{\zeta}(\eta_f,{\bf k_{3}}), \hat{\zeta}^{'}(\eta,{\bf k_{4}})\hat{\zeta}^{'}(\eta,{\bf k_{5}})\hat{\zeta}^{'}(\eta,{\bf k_{6}})\right]|0\rangle\right.\nonumber\\
	 &&\left.~~-\alpha_{2}({\bf k_{5}}.{\bf k_{6}})\langle0|\left[\hat{\zeta}(\eta_f,{\bf k_{1}})\hat{\zeta}(\eta_f,{\bf k_{2}})\hat{\zeta}(\eta_f,{\bf k_{3}}), \hat{\zeta}^{'}(\eta,{\bf k_{4}})\hat{\zeta}(\eta,{\bf k_{5}})\hat{\zeta}(\eta,{\bf k_{6}})\right]|0\rangle\right.\\
	 &&\left.~~+\alpha_{3}~a(\eta)~\langle0|\left[\hat{\zeta}(\eta_f,{\bf k_{1}})\hat{\zeta}(\eta_f,{\bf k_{2}})\hat{\zeta}(\eta_f,{\bf k_{3}}), \hat{\zeta}(\eta,{\bf k_{4}})\hat{\zeta}^{'}(\eta,{\bf k_{5}})\hat{\zeta}^{'}(\eta,{\bf k_{6}})\right]|0\rangle\right.\nonumber\\
	 &&\left.~~
	 -\alpha_{4}({\bf k_{5}}.{\bf k_{6}})\langle0|\left[\hat{\zeta}(\eta_f,{\bf k_{1}})\hat{\zeta}(\eta_f,{\bf k_{2}})\hat{\zeta}(\eta_f,{\bf k_{3}}), \hat{\zeta}(\eta,{\bf k_{4}})\hat{\zeta}^{'}(\eta,{\bf k_{5}})\hat{\zeta}(\eta,{\bf k_{6}})\right]|0\rangle	   \right.\nonumber\\
	 &&\left.~~	    -\alpha_{5}({\bf k_{5}}.{\bf k_{6}})\langle0|\left[\hat{\zeta}(\eta_f,{\bf k_{1}})\hat{\zeta}(\eta_f,{\bf k_{2}})\hat{\zeta}(\eta_f,{\bf k_{3}}), \hat{\zeta}(\eta,{\bf k_{4}})\hat{\zeta}(\eta,{\bf k_{5}})\hat{\zeta}^{'}(\eta,{\bf k_{6}})\right]|0\rangle	 \right\},\nonumber~~~~~~~
	 \eea
	 \end{small}
\hspace{-3pt}where the coefficients $\alpha_{j}\forall j=1,2,3,4,5$ are defined as\footnote{Here it is clearly observed that for canonical single-field slow-roll model, which is described by $c_S=1$ we have $M_3=0$ and other EFT coefficients are sufficiently small, $\bar{M}_{i}\forall i=1,2,3(\sim{\cal O}()10^{-3}-10^{-2})$. This directly implies that the contribution in the three-point function and in the associated bispectrum is very small and consistent with the previous result as obtained in ref.~\cite{Maldacena:2002vr}. Additionally, it is important to mention that in~momentum space the bispectrum contains additional terms in the presence of any arbitrary choice of the quantum vacuum initial state. Also, if~we compare with ref.~\cite{Cheung:2007st}.}:
\bea \alpha_{1}&=&\left\{\left(1-\frac{1}{c^2_{S}}\right)\dot{H}M^2_p +\frac{3}{2}\bar{M}^3_1 H-\frac{4}{3}M^4_3\right\},\\
\alpha_{2}&=&-\left\{\left(1-\frac{1}{c^2_{S}}\right)\dot{H}M^2_p +\frac{3}{2}\bar{M}^3_{1} H\right\},\\
\alpha_{3}&=&-\frac{9}{2}\bar{M}^3_{1}H^2,\\
\alpha_{4}&=&\frac{3}{2}\bar{M}^3_1 H,\\
\alpha_{5}&=&\frac{3}{2}\bar{M}^3_1 H.\eea

Now let us evaluate the coefficients of $\alpha_{1},\alpha_{2},\alpha_{3},\alpha_{4}$ in the present context using Wick's~theorem:
\begin{enumerate}
	\item \underline{\textcolor{red}{\bf Coefficient of $\alpha_{1}$}}:
	\bea \langle0|\left[\hat{\zeta}(\eta_f,{\bf k_{1}})\hat{\zeta}(\eta_f,{\bf k_{2}})\hat{\zeta}(\eta_f,{\bf k_{3}}), \hat{\zeta}^{'}(\eta,{\bf k_{4}})\hat{\zeta}^{'}(\eta,{\bf k_{5}})\hat{\zeta}^{'}(\eta,{\bf k_{6}})\right]|0\rangle\nonumber\\
	=\langle 0|a({\bf k_{1}})a({\bf k_{2}})a({\bf k_{3}})a^{\dagger}({\bf -k_{4}})a^{\dagger}({\bf -k_{5}})a^{\dagger}({\bf -k_{6}})|0\rangle\nonumber\\
	\bar{v}(\eta_{f},{\bf k_{1}})\bar{v}(\eta_{f},{\bf k_{2}})\bar{v}(\eta_{f},{\bf k_{3}})\bar{v}^{'*}(\eta_{f},-{\bf k_{4}})\bar{v}^{'*}(\eta_{f},-{\bf k_{5}})
	\bar{v}^{'*}(\eta_{f},-{\bf k_{6}})
	\\+\langle 0|a({\bf k_{4}})a({\bf k_{5}})a({\bf k_{6}})a^{\dagger}({\bf -k_{1}})a^{\dagger}({\bf -k_{2}})a^{\dagger}({\bf -k_{3}})|0\rangle\nonumber\\
	\bar{v}^{*}(\eta_{f},-{\bf k_{1}})\bar{v}^{*}(\eta_{f},-{\bf k_{2}})\bar{v}^{*}(\eta_{f},-{\bf k_{3}})\bar{v}^{'}(\eta,{\bf k_{4}})\bar{v}^{'}(\eta,{\bf k_{5}})
	\bar{v}^{'}(\eta,{\bf k_{6}}).\nonumber\eea
	
	\item \underline{\textcolor{red}{\bf Coefficient of $\alpha_{2}$}}:
	\bea ({\bf k_{5}}.{\bf k_{6}})\langle0|\left[\hat{\zeta}(\eta_f,{\bf k_{1}})\hat{\zeta}(\eta_f,{\bf k_{2}})\hat{\zeta}(\eta_f,{\bf k_{3}}), \hat{\zeta}^{'}(\eta,{\bf k_{4}})\hat{\zeta}(\eta,{\bf k_{5}})\hat{\zeta}(\eta,{\bf k_{6}})\right]|0\rangle\nonumber\\
	=~~~~~~~~~~~~~~~~~~~~~~~~~~~~~~({\bf k_{5}}.{\bf k_{6}})\langle 0|a({\bf k_{1}})a({\bf k_{2}})a({\bf k_{3}})a^{\dagger}({\bf -k_{4}})a^{\dagger}({\bf -k_{5}})a^{\dagger}({\bf -k_{6}})|0\rangle\nonumber\\
	\bar{v}(\eta_{f},{\bf k_{1}})\bar{v}(\eta_{f},{\bf k_{2}})\bar{v}(\eta_{f},{\bf k_{3}})\bar{v}^{'*}(\eta_{f},-{\bf k_{4}})\bar{v}^{*}(\eta_{f},-{\bf k_{5}})
	\bar{v}^{*}(\eta_{f},-{\bf k_{6}})
	\\+({\bf k_{5}}.{\bf k_{6}})\langle 0|a({\bf k_{4}})a({\bf k_{5}})a({\bf k_{6}})a^{\dagger}({\bf -k_{1}})a^{\dagger}({\bf -k_{2}})a^{\dagger}({\bf -k_{3}})|0\rangle\nonumber\\
	\bar{v}^{*}(\eta_{f},-{\bf k_{1}})\bar{v}^{*}(\eta_{f},-{\bf k_{2}})\bar{v}^{*}(\eta_{f},-{\bf k_{3}})\bar{v}^{'}(\eta_{f},{\bf k_{4}})\bar{v}(\eta_{f},{\bf k_{5}})
	\bar{v}(\eta_{f},{\bf k_{6}}).\nonumber\eea
	
	\item \underline{\textcolor{red}{\bf Coefficient of $\alpha_{3}$}}:
	\bea \langle0|\left[\hat{\zeta}(\eta_f,{\bf k_{1}})\hat{\zeta}(\eta_f,{\bf k_{2}})\hat{\zeta}(\eta_f,{\bf k_{3}}), \hat{\zeta}(\eta,{\bf k_{4}})\hat{\zeta}^{'}(\eta,{\bf k_{5}})\hat{\zeta}^{'}(\eta,{\bf k_{6}})\right]|0\rangle\nonumber\\
	=~~~~~~~~~~~~~~~~~~~~~~~~~~~~~~\langle 0|a({\bf k_{1}})a({\bf k_{2}})a({\bf k_{3}})a^{\dagger}({\bf -k_{4}})a^{\dagger}({\bf -k_{5}})a^{\dagger}({\bf -k_{6}})|0\rangle\nonumber\\
	\bar{v}(\eta_{f},{\bf k_{1}})\bar{v}(\eta_{f},{\bf k_{2}})\bar{v}(\eta_{f},{\bf k_{3}})\bar{v}^{*}(\eta_{f},-{\bf k_{4}})\bar{v}^{'*}(\eta_{f},-{\bf k_{5}})
	\bar{v}^{'*}(\eta_{f},-{\bf k_{6}})
	\\+\langle 0|a({\bf k_{4}})a({\bf k_{5}})a({\bf k_{6}})a^{\dagger}({\bf -k_{1}})a^{\dagger}({\bf -k_{2}})a^{\dagger}({\bf -k_{3}})|0\rangle\nonumber\\
	\bar{v}^{*}(\eta_{f},-{\bf k_{1}})\bar{v}^{*}(\eta_{f},-{\bf k_{2}})\bar{v}^{*}(\eta_{f},-{\bf k_{3}})\bar{v}(\eta_{f},{\bf k_{4}})\bar{v}^{'}(\eta_{f},{\bf k_{5}})
	\bar{v}^{'}(\eta_{f},{\bf k_{6}}).\nonumber\eea
	
	\item \underline{\textcolor{red}{\bf Coefficient of $\alpha_{4}$}}:\\
	\bea ({\bf k_{5}}.{\bf k_{6}})\langle0|\left[\hat{\zeta}(\eta_f,{\bf k_{1}})\hat{\zeta}(\eta_f,{\bf k_{2}})\hat{\zeta}(\eta_f,{\bf k_{3}}), \hat{\zeta}(\eta,{\bf k_{4}})\hat{\zeta}^{'}(\eta,{\bf k_{5}})\hat{\zeta}(\eta,{\bf k_{6}})\right]|0\rangle	\nonumber\\
	=~~~~~~~~~~~~~~~~~~~~~~~~~~~~~~({\bf k_{5}}.{\bf k_{6}})\langle 0|a({\bf k_{1}})a({\bf k_{2}})a({\bf k_{3}})a^{\dagger}({\bf -k_{4}})a^{\dagger}({\bf -k_{5}})a^{\dagger}({\bf -k_{6}})|0\rangle\nonumber\\
	\bar{v}(\eta_{f},{\bf k_{1}})\bar{v}(\eta_{f},{\bf k_{2}})\bar{v}(\eta_{f},{\bf k_{3}})\bar{v}^{*}(\eta_{f},-{\bf k_{4}})\bar{v}^{'*}(\eta_{f},-{\bf k_{5}})
	\bar{v}^{*}(\eta_{f},-{\bf k_{6}})
	\\+({\bf k_{5}}.{\bf k_{6}})\langle 0|a({\bf k_{4}})a({\bf k_{5}})a({\bf k_{6}})a^{\dagger}({\bf -k_{1}})a^{\dagger}({\bf -k_{2}})a^{\dagger}({\bf -k_{3}})|0\rangle\nonumber\\
	\bar{v}^{*}(\eta_{f},-{\bf k_{1}})\bar{v}^{*}(\eta_{f},-{\bf k_{2}})\bar{v}^{*}(\eta_{f},-{\bf k_{3}})\bar{v}(\eta_{f},{\bf k_{4}})\bar{v}^{'}(\eta_{f},{\bf k_{5}})
	\bar{v}(\eta_{f},{\bf k_{6}}).\nonumber\eea
	\item \underline{\textcolor{red}{\bf Coefficient of $\alpha_{5}$}}:\\  	  	  	  
	\bea ({\bf k_{5}}.{\bf k_{6}})\langle0|\left[\hat{\zeta}(\eta_f,{\bf k_{1}})\hat{\zeta}(\eta_f,{\bf k_{2}})\hat{\zeta}(\eta_f,{\bf k_{3}}), \hat{\zeta}(\eta,{\bf k_{4}})\hat{\zeta}(\eta,{\bf k_{5}})\hat{\zeta}^{'}(\eta,{\bf k_{6}})\right]|0\rangle\nonumber\\
	=~~~~~~~~~~~~~~~~~~~~~~~~~~~~~~\langle 0|a({\bf k_{1}})a({\bf k_{2}})a({\bf k_{3}})a^{\dagger}({\bf -k_{4}})a^{\dagger}({\bf -k_{5}})a^{\dagger}({\bf -k_{6}})|0\rangle\nonumber\\
	\bar{v}(\eta_{f},{\bf k_{1}})\bar{v}(\eta_{f},{\bf k_{2}})\bar{v}(\eta_{f},{\bf k_{3}})\bar{v}^{*}(\eta_{f},-{\bf k_{4}})\bar{v}^{*}(\eta_{f},-{\bf k_{5}})
	\bar{v}^{'*}(\eta_{f},-{\bf k_{6}})
	\\+\langle 0|a({\bf k_{4}})a({\bf k_{5}})a({\bf k_{6}})a^{\dagger}({\bf -k_{1}})a^{\dagger}({\bf -k_{2}})a^{\dagger}({\bf -k_{3}})|0\rangle\nonumber\\
	\bar{v}^{*}(\eta_{f},-{\bf k_{1}})\bar{v}^{*}(\eta_{f},-{\bf k_{2}})\bar{v}^{*}(\eta_{f},-{\bf k_{3}})\bar{v}(\eta_{f},{\bf k_{4}})\bar{v}(\eta_{f},{\bf k_{5}})
	\bar{v}^{'}(\eta_{f},{\bf k_{6}}).\nonumber\eea
	
\end{enumerate} 
where we define $\bar{v}$ as:
\be \bar{v}(\eta,{\bf k})=\frac{v(\eta,{\bf k})}{zM_p}.\ee

  Furthermore, we also use the following result in simplify the coefficients of $\alpha_{1},\alpha_{2},\alpha_{3},\alpha_{4}$:
   	  \bea&& \langle 0|a({\bf k_{1}})a({\bf k_{2}})a({\bf k_{3}})a^{\dagger}({\bf -k_{4}})a^{\dagger}({\bf -k_{5}})a^{\dagger}({\bf -k_{6}})|0\rangle \nonumber\\
   	  &=&\langle 0|a({\bf k_{4}})a({\bf k_{5}})a({\bf k_{6}})a^{\dagger}({\bf -k_{1}})a^{\dagger}({\bf -k_{2}})a^{\dagger}({\bf -k_{3}})|0\rangle\nonumber\\ 
   	  &=& (2\pi)^9 \left\{\delta^{(3)}({\bf k_{4}}+{\bf k_{1}})\left[\delta^{(3)}({\bf k_{5}}+{\bf k_{2}})\delta^{(3)}({\bf k_{6}}+{\bf k_{3}})+\delta^{(3)}({\bf k_{5}}+{\bf k_{3}})\delta^{(3)}({\bf k_{6}}+{\bf k_{2}})\right]\right.\nonumber\\ &&\left.+\delta^{(3)}({\bf k_{4}}+{\bf k_{2}})\left[\delta^{(3)}({\bf k_{5}}+{\bf k_{1}})\delta^{(3)}({\bf k_{6}}+{\bf k_{3}})+\delta^{(3)}({\bf k_{5}}+{\bf k_{3}})\delta^{(3)}({\bf k_{6}}+{\bf k_{1}})\right]\right.\nonumber\\ &&\left.+\delta^{(3)}({\bf k_{4}}+{\bf k_{3}})\left[\delta^{(3)}({\bf k_{5}}+{\bf k_{1}})\delta^{(3)}({\bf k_{6}}+{\bf k_{3}})+\delta^{(3)}({\bf k_{5}}+{\bf k_{2}})\delta^{(3)}({\bf k_{6}}+{\bf k_{1}})\right]\right\}. \eea
   	  
   	  \scalebox{.95}[1.0]{Finally, one can write the following expression for the three-point function for the scalar fluctuation}\footnote{See also ref.~\cite{Maldacena:2002vr,Chen:2006nt,Agarwal:2012mq}, where similar computations have been performed for canonical single-field slow-roll and generalized slow-roll models of inflation in the presence of {Bunch–Davies vacuum} and {general initial} state.}:
\bea\label{op}
		   	   {\langle\zeta({\bf k_{1}})\zeta({\bf k_{2}})\zeta({\bf k_{3}})\rangle = 
		   	   (2\pi)^3\delta^{(3)}({\bf k_{1}}+{\bf k_{2}}+{\bf k_{3}})B_{EFT}(k_{1},k_{2},k_{3})}~.
		   	   \eea	
		   	   where $B_{EFT}(k_{1},k_{2},k_{3})$ is the bispectrum for scalar fluctuation. 
		   	   In the present computation, one can further write down the expression for the 
		   	   bispectrum as:
		   	   \bea  {B_{EFT}(k_{1},k_{2},k_{3})=\sum^{5}_{j=1}\alpha_{j}\Theta_{j}(k_{1},k_{2},k_{3})}~,\eea
		   	   where $\Theta_{j}(k_{1},k_{2},k_{3})\forall j=1,2,3,4,5$ is defined in the next subsections. Here it is important to note that we have derived the expression for the three-point function and the associated bispectrum for effective sound speed $c_S=1$ and $c_S<1$ with a choice of general quantum vacuum~state.

		   	   \subsubsection{Function $\Theta_{1}(k_{1},k_{2},k_{3})$}
		   	   \label{v4a1}
		   	   Here we can write the function $\Theta_{1}(k_{1},k_{2},k_{3})$ as:
		   	   \bea \Theta_{1}(k_{1},k_{2},k_{3})&=&6i\int^{\eta_{f}=0}_{\eta_{i}=-\infty}d\eta~\frac{a(\eta)}{H^3}\left[\bar{v}(\eta_{f},{\bf k_{1}})\bar{v}(\eta_{f},{\bf k_{2}})\bar{v}(\eta_{f},{\bf k_{3}})\bar{v}^{'*}(\eta,{\bf k_{1}})\bar{v}^{'*}(\eta,{\bf k_{2}})\bar{v}^{'*}(\eta,{\bf k_{3}})\right.\nonumber\\
		   	    && \left.~~~~~~~~~~~~~~~~~+\bar{v}^{*}(\eta_{f},{\bf k_{1}})\bar{v}^{*}(\eta_{f},{\bf k_{2}})\bar{v}^{*}(\eta_{f},{\bf k_{3}})\bar{v}^{'}(\eta,-{\bf k_{1}})\bar{v}^{'}(\eta,-{\bf k_{2}})\bar{v}^{'}(\eta,-{\bf k_{3}})\right].~~~~~~~~~~~~\eea
		   	    
		   	  Furthermore, using the integrals from the Appendix we finally get the following simplified expression for the three-point function for the scalar fluctuations\footnote{Here it is important to point out that in de Sitter space if we consider the Bunch–Davies vacuum state then here only the term with $1/K^3$ will appear explicitly in the expression for the three-point function and in the associated bispectrum. On~the other hand, if we consider all other non-trivial quantum vacuum states in our computation, then the rest of the contribution will explicitly appear. From~the perspective of observation, this is obviously important information as for the non-trivial quantum vacuum state we get additional contribution in the bispectrum which may enhance the amplitude of the non-Gaussianity in squeezed limiting configuration. Additionally, it is important to mention that in quasi de Sitter case we get extra contributions $1/\tilde{c}^{6\nu-9}_S$ and $1/(1+\epsilon)^5$. Also, the factor $1/(k_1 k_2 k_3)$ will be replaced by  $1/(k_1 k_2 k_3)^{2(\nu-1)}$. Consequently, in~quasi de Sitter case this contribution in the bispectrum can be recast as:
		   	  \begin{scriptsize}
		   	 \be \begin{array}{rl} \Theta_{1}(k_{1},k_{2},k_{3})&=\displaystyle  \frac{3H^2}{16\epsilon^3M^6_p \tilde{c}^{6\nu-9}_S (1+\epsilon)^5}\frac{1}{(k_{1}k_{2}k_{3})^{2(\nu-1)}}\left[\left\{\frac{1}{K^3}\left[\left(C_{1}-C_{2}\right)^3\left(C^{*3}_{1}+C^{*3}_{2}\right) +\left(C^{*}_{1}-C^{*}_{2}\right)^3\left(C^{3}_{1}+C^{3}_{2}\right)\right]\right.\right.\\ & \left.\left.~~~+\left[\left(C_{1}-C_{2}\right)^3C^{*}_{1}C^{*}_{2}\left(C^{*}_{1}-C^{*}_{2}\right) +\left(C^{*}_{1}-C^{*}_{2}\right)^3C_{1}C_{2}\left(C_{1}-C_{2}\right)\right]\displaystyle \sum^{3}_{i=1}\frac{1}{(2k_{i}-K)^3}\right\}\right],\end{array}~~~~~~~~\ee\end{scriptsize} }:	  	    	  
		   	    	    \begin{footnotesize}
		   	   \be {\begin{array}{rl} \Theta_{1}(k_{1},k_{2},k_{3})&=\displaystyle  \frac{3H^2}{16\epsilon^3M^6_p}\frac{1}{k_{1}k_{2}k_{3}}\left[\left\{\frac{1}{K^3}\left[\left(C_{1}-C_{2}\right)^3\left(C^{*3}_{1}+C^{*3}_{2}\right) +\left(C^{*}_{1}-C^{*}_{2}\right)^3\left(C^{3}_{1}+C^{3}_{2}\right)\right]\right.\right.\\ & \left.\left.~~~+\left[\left(C_{1}-C_{2}\right)^3C^{*}_{1}C^{*}_{2}\left(C^{*}_{1}-C^{*}_{2}\right) +\left(C^{*}_{1}-C^{*}_{2}\right)^3C_{1}C_{2}\left(C_{1}-C_{2}\right)\right]\displaystyle \sum^{3}_{i=1}\frac{1}{(2k_{i}-K)^3}\right\}\right],\end{array}}~~~~~~~~\ee
		   	   \end{footnotesize}

		   	   Finally, for {Bunch–Davies} and {$\alpha,\beta$} vacuum we get the following contribution in the three-point function for scalar~fluctuations:   
\begin{itemize}
\item \underline{\bf {For Bunch–Davies vacuum}:} \\After setting $C_{1}=1$ and $C_{2}=0$ we get:
\bea \Theta_{1}(k_{1},k_{2},k_{3})= \frac{6H^2}{16\epsilon^3M^6_p}\frac{1}{k_{1}k_{2}k_{3}}\frac{1}{K^3},~~~~~~~~\eea
\item \underline{\bf {For $~{\alpha,\beta}$ vacuum}:}\\After setting $C_{1}=\cosh\alpha$ and $C_{2}=e^{i\beta}~\sinh\alpha$ we get:
\begin{small}
\be \begin{array}{rl} \Theta_{1}(k_{1},k_{2},k_{3})&=\displaystyle  \frac{3H^2}{16\epsilon^3M^6_p}\frac{1}{k_{1}k_{2}k_{3}}\left\{\frac{1}{K^3}\left[\left(\cosh\alpha-e^{i\beta}\sinh\alpha\right)^3\left(\cosh^3\alpha+e^{-3i\beta}\sinh^3\alpha\right)\right.\right.\\ &\left.\left.\displaystyle+\left(\cosh\alpha-e^{-i\beta}\sinh\alpha\right)^3\left(\cosh^3\alpha+e^{3i\beta}\sinh^3\alpha\right)\right]\right.\\ & \left.\displaystyle+\frac{1}{2}\left[\left(\cosh\alpha-e^{i\beta}\sinh\alpha\right)^3e^{-i\beta}\sinh2\alpha \left(\cosh\alpha-e^{-i\beta}\sinh\alpha\right)\right.\right.\\ & \left.\left.\displaystyle+\left(\cosh\alpha-e^{-i\beta}\sinh\alpha\right)^3e^{i\beta}\sinh2\alpha\left(\cosh\alpha-e^{i\beta}\sinh\alpha\right)\right]\displaystyle\sum^{3}_{i=1}\frac{1}{(2k_{i}-K)^3}\right\}.\end{array}~~~~~~~~\ee
\end{small}
\end{itemize}
		   	     \subsubsection{Function $\Theta_{2}(k_{1},k_{2},k_{3})$}
		\label{v4a2}   	
		Here we can write the function $\Theta_{2}(k_{1},k_{2},k_{3})$ as: 
		\begin{small}    
 \bea \Theta_{2}(k_{1},k_{2},k_{3})&=&i\int^{\eta_{f}=0}_{\eta_{i}=-\infty}d\eta~\frac{a(\eta)}{H^3}\left\{2({\bf k_{2}}.{\bf k_{3}})\left[\bar{v}(\eta_{f},{\bf k_{1}})\bar{v}(\eta_{f},{\bf k_{2}})\bar{v}(\eta_{f},{\bf k_{3}})\bar{v}^{'*}(\eta,{\bf k_{1}})\bar{v}^{*}(\eta,{\bf k_{2}})\bar{v}^{*}(\eta,{\bf k_{3}})\right.\right.\nonumber\\
 && \left.\left.~~~~~~~~~~~~~~~~~+\bar{v}^{*}(\eta_{f},-{\bf k_{1}})\bar{v}^{*}(\eta_{f},-{\bf k_{2}})\bar{v}^{*}(\eta_{f},-{\bf k_{3}})\bar{v}^{'}(\eta,-{\bf k_{1}})\bar{v}(\eta,-{\bf k_{2}})\bar{v}(\eta,-{\bf k_{3}})\right]\right.\nonumber\\ &&\left.~~~~~~~~~~~+2({\bf k_{3}}.{\bf k_{1}})\left[\bar{v}(\eta_{f},{\bf k_{1}})\bar{v}(\eta_{f},{\bf k_{2}})\bar{v}(\eta_{f},{\bf k_{3}})\bar{v}^{'*}(\eta,{\bf k_{2}})\bar{v}^{*}(\eta,{\bf k_{1}})\bar{v}^{*}(\eta,{\bf k_{3}})\right.\right.\nonumber\\
 && \left.\left.~~~~~~~~~~~~~~~~~+\bar{v}^{*}(\eta_{f},-{\bf k_{1}})\bar{v}^{*}(\eta_{f},-{\bf k_{2}})\bar{v}^{*}(\eta_{f},-{\bf k_{3}})\bar{v}^{'}(\eta,-{\bf k_{2}})\bar{v}(\eta,-{\bf k_{1}})\bar{v}(\eta,-{\bf k_{3}})\right]
 \right.\nonumber\\ &&\left.~~~~~~~~~~~+2({\bf k_{1}}.{\bf k_{2}})\left[\bar{v}(\eta_{f},{\bf k_{1}})\bar{v}(\eta_{f},{\bf k_{2}})\bar{v}(\eta_{f},{\bf k_{3}})\bar{v}^{'*}(\eta,{\bf k_{3}})\bar{v}^{*}(\eta,{\bf k_{1}})\bar{v}^{*}(\eta,{\bf k_{2}})\right.\right.\nonumber\\
 && \left.\left.~~~~~~~~~~~~~~~~~+\bar{v}^{*}(\eta_{f},-{\bf k_{1}})\bar{v}^{*}(\eta_{f},-{\bf k_{2}})\bar{v}^{*}(\eta_{f},-{\bf k_{3}})\bar{v}^{'}(\eta,-{\bf k_{3}})\bar{v}(\eta,-{\bf k_{1}})\bar{v}(\eta,-{\bf k_{2}})		   	   		   	   \right]
 \right\}.\eea		
 \end{small}   	  
  
Using the results derived in Appendix we finally get the following simplified expression for the three-point function for the scalar fluctuations\footnote{Here it is important to point out that in de Sitter space if we consider the Bunch–Davies vacuum state then here only the term with $1/K^3$ will appear explicitly in the expression for the three-point function and in the associated bispectrum. On~the other hand, if we consider all other non-trivial quantum vacuum states in our computation, then the rest of the contribution will explicitly appear. From~the perspective of observation, this is obviously important information as for the non-trivial quantum vacuum state we get additional contribution in the bispectrum which may enhance the amplitude of the non-Gaussianity in squeezed limiting configuration. Additionally, it is important to mention that in quasi de Sitter case we get extra contributions $1/\tilde{c}^{6\nu-7}_S$ and $1/(1+\epsilon)^5$. Also, the factor $1/(k_1 k_2 k_3)^3$ will be replaced by  $1/(k_1 k_2 k_3)^{2\nu}$. Consequently, in~quasi de Sitter case this contribution in the bispectrum can be recast as:
\be \begin{array}{rl}  \Theta_{2}(k_{1},k_{2},k_{3})&=\displaystyle \frac{H^2}{32\epsilon^3M^6_p \tilde{c}^{6\nu-7}_{S}(1+\epsilon)^5}\frac{1}{(k_{1}k_{2}k_{3})^{2\nu}}\left[k^2_{1}({\bf k_{2}}.{\bf k_{3}})G_{1}(k_{1},k_{2},k_{3})\right.\\ & \left.~~~~~~~~~~~~~~~~~~~~~~~~~\displaystyle +k^2_{2}({\bf k_{1}}.{\bf k_{3}})G_{2}(k_{1},k_{2},k_{3})+k^2_{3}({\bf k_{1}}.{\bf k_{2}})G_{3}(k_{1},k_{2},k_{3})\right],\end{array}~~~~~~~~\ee}:
\vspace{12pt}
		   	   \be {\begin{array}{rl}  \Theta_{2}(k_{1},k_{2},k_{3})&=\displaystyle \frac{H^2}{32\epsilon^3M^6_p \tilde{c}^2_{S}}\frac{1}{(k_{1}k_{2}k_{3})^3}\left[k^2_{1}({\bf k_{2}}.{\bf k_{3}})G_{1}(k_{1},k_{2},k_{3})\right.\\ & \left.~~~~~~~~~~~~~~~~~~~~~~~~~\displaystyle +k^2_{2}({\bf k_{1}}.{\bf k_{3}})G_{2}(k_{1},k_{2},k_{3})+k^2_{3}({\bf k_{1}}.{\bf k_{2}})G_{3}(k_{1},k_{2},k_{3})\right],\end{array}}~~~~~~~~\ee
		   	   where the momentum dependent functions $G_{1}(k_{1},k_{2},k_{3})$, $G_{2}(k_{1},k_{2},k_{3})$ and $G_{3}(k_{1},k_{2},k_{3})$ are defined~as:
		   	   \begin{small}
\bea		   
G_{1}(k_{1},k_{2},k_{3})&=&\frac{1}{K^3}\left[K^2+2k_{2}k_{3}+K(K-k_{1})\right]\left[(C_{1}-C_{2})^3(C^{*3}_{1}+C^{*3}_{2})+(C_{1}-C_{2})^3(C^{*3}_{1}+C^{*3}_{2})\right]\nonumber\\&&
+\left\{\frac{1}{(2k_{1}-K)^3}\left[K^2+2k_{2}k_{3}+K(K-5k_{1})-2(K-k_{1})k_{1}+4k^2_1\right]\nonumber\right.\\
&&\left.
+\frac{1}{(2k_{2}-K)^3}\left[K^2-4k_{2}k_{3}+K(k_{3}-5k_{2})+6k^2_2\right]\nonumber\right.\\
&&\left.
+\frac{1}{(2k_{1}-K)^3}\left[K^2-4k_{2}k_{3}+K(k_{2}-5k_{3})+6k^2_3\right]\right\}\nonumber\\
&&~~~~~~~~~~~\left[(C_{1}-C_{2})^3 C^{*}_{1}C^{*}_{2}(C^{*}_{1}+C^{*}_{2})+(C^{*}_{1}-C^{*}_{2})^3 C_{1}C_{2}(C_{1}+C_{2})\right].
\\		   
G_{2}(k_{1},k_{2},k_{3})&=&\frac{1}{K^3}\left[K^2+2k_{1}k_{3}+K(K-k_{2})\right]\left[(C_{1}-C_{2})^3(C^{*3}_{1}+C^{*3}_{2})+(C_{1}-C_{2})^3(C^{*3}_{1}+C^{*3}_{2})\right]\nonumber\\&&
+\left\{\frac{1}{(2k_{2}-K)^3}\left[K^2+2k_{1}k_{3}+K(K-5k_{2})-2(K-k_{2})k_{2}+4k^2_2\right]\nonumber\right.\\
&&\left.
+\frac{1}{(2k_{1}-K)^3}\left[K^2-4k_{1}k_{3}+K(k_{3}-5k_{1})+6k^2_1\right]\nonumber\right.\\
&&\left.
+\frac{1}{(2k_{3}-K)^3}\left[K^2-4k_{1}k_{3}+K(k_{1}-5k_{3})+6k^2_3\right]\right\}\nonumber\\
&&~~~~~~~~~~~\left[(C_{1}-C_{2})^3 C^{*}_{1}C^{*}_{2}(C^{*}_{1}+C^{*}_{2})+(C^{*}_{1}-C^{*}_{2})^3 C_{1}C_{2}(C_{1}+C_{2})\right].
\\		   
G_{3}(k_{1},k_{2},k_{3})&=&\frac{1}{K^3}\left[K^2+2k_{1}k_{2}+K(K-k_{3})\right]\left[(C_{1}-C_{2})^3(C^{*3}_{1}+C^{*3}_{2})+(C_{1}-C_{2})^3(C^{*3}_{1}+C^{*3}_{2})\right]\nonumber\\&&
+\left\{\frac{1}{(2k_{3}-K)^3}\left[K^2+2k_{1}k_{2}+K(K-5k_{3})-2(K-k_{3})k_{3}+4k^2_3\right]\nonumber\right.\\
&&\left.
+\frac{1}{(2k_{2}-K)^3}\left[K^2-4k_{1}k_{2}+K(k_{1}-5k_{2})+6k^2_2\right]\nonumber\right.\\
&&\left.
+\frac{1}{(2k_{1}-K)^3}\left[K^2-4k_{1}k_{2}+K(k_{2}-5k_{1})+6k^2_1\right]\right\}\nonumber\\
&&~~~~~~~~~~~\left[(C_{1}-C_{2})^3 C^{*}_{1}C^{*}_{2}(C^{*}_{1}+C^{*}_{2})+(C^{*}_{1}-C^{*}_{2})^3 C_{1}C_{2}(C_{1}+C_{2})\right].
\eea
	   	   \end{small}
	   	   
Here  
$\sum^{3}_{i=1}{\bf k_{i}}={\bf k_{1}}+{\bf k_{2}}+{\bf k_{3}}=0.$
Consequently, one can write:
\bea {\bf k_{1}}.{\bf k_{2}}&=&\frac{1}{2}\left(k^2_3 -k^2_2 -k^2_1\right),~~
{\bf k_{1}}.{\bf k_{3}}=\frac{1}{2}\left(k^2_2 -k^2_1 -k^2_3\right),~~
{\bf k_{2}}.{\bf k_{3}}=\frac{1}{2}\left(k^2_1 -k^2_2 -k^2_3\right),
\eea
and using these results one can further recast the three-point function for the scalar fluctuation as:
\be {\begin{array}{rl} \Theta_{2}(k_{1},k_{2},k_{3})&= \displaystyle\frac{H^2}{64\epsilon^3M^6_p \tilde{c}^2_{S}}\frac{1}{(k_{1}k_{2}k_{3})^3}\left[k^2_{1}\left(k^2_1 -k^2_2 -k^2_3\right)G_{1}(k_{1},k_{2},k_{3})\right. \\ &\displaystyle \left.~+k^2_{2}\left(k^2_2 -k^2_1 -k^2_3\right)G_{2}(k_{1},k_{2},k_{3})+k^2_{3}\left(k^2_3 -k^2_2 -k^2_1\right)G_{3}(k_{1},k_{2},k_{3})\right].\end{array}}\ee

 Finally, for {Bunch–Davies} and {$\alpha,\beta$} vacuum we get the following contribution in the three-point function for scalar~fluctuations:   
\begin{itemize}
\item \underline{\bf {For Bunch–Davies vacuum}:} \\After setting $C_{1}=1$ and $C_{2}=0$ we get:
\bea		   
G_{1}(k_{1},k_{2},k_{3})&=&\frac{2}{K^3}\left[K^2+2k_{2}k_{3}+K(K-k_{1})\right].
\\		   
G_{2}(k_{1},k_{2},k_{3})&=&\frac{2}{K^3}\left[K^2+2k_{1}k_{3}+K(K-k_{2})\right].
\\		   
G_{3}(k_{1},k_{2},k_{3})&=&\frac{2}{K^3}\left[K^2+2k_{1}k_{2}+K(K-k_{3})\right].
\eea
Consequently, we get:
\bea \Theta_{2}(k_{1},k_{2},k_{3})&=& \frac{H^2}{32\epsilon^3M^6_p \tilde{c}^2_{S}}\frac{1}{(k_{1}k_{2}k_{3})^3}\frac{1}{K^3}\left[k^2_{1}\left(k^2_1 -k^2_2 -k^2_3\right)\left[K^2+2k_{2}k_{3}+K(K-k_{1})\right]\right.\nonumber \\ && \left.~~~~~~~~~~~~~~~~+k^2_{2}\left(k^2_2 -k^2_1 -k^2_3\right)\left[K^2+2k_{1}k_{3}+K(K-k_{2})\right]\right.\nonumber \\ && \left.~~~~~~~~~~~~~~~~+k^2_{3}\left(k^2_3 -k^2_2 -k^2_1\right)\left[K^2+2k_{1}k_{2}+K(K-k_{3})\right]\right].\eea
\item \underline{\bf {For $~{\alpha,\beta}$ vacuum}:}\\After setting $C_{1}=\cosh\alpha$ and $C_{2}=e^{i\beta}~\sinh\alpha$ we get:
\bea		   
G_{1}(k_{1},k_{2},k_{3})&=&\frac{1}{K^3}\left[K^2+2k_{2}k_{3}+K(K-k_{1})\right]J_{1}(\alpha,\beta)\nonumber\\&&
+\left\{\frac{1}{(2k_{1}-K)^3}\left[K^2+2k_{2}k_{3}+K(K-5k_{1})-2(K-k_{1})k_{1}+4k^2_1\right]\nonumber\right.\\
&&\left.
+\frac{1}{(2k_{2}-K)^3}\left[K^2-4k_{2}k_{3}+K(k_{3}-5k_{2})+6k^2_2\right]\nonumber\right.\\
&&\left.
+\frac{1}{(2k_{1}-K)^3}\left[K^2-4k_{2}k_{3}+K(k_{2}-5k_{3})+6k^2_3\right]\right\}J_{2}(\alpha,\beta).
\\		   
G_{2}(k_{1},k_{2},k_{3})&=&\frac{1}{K^3}\left[K^2+2k_{1}k_{3}+K(K-k_{2})\right]J_{1}(\alpha,\beta)\nonumber\\&&
+\left\{\frac{1}{(2k_{2}-K)^3}\left[K^2+2k_{1}k_{3}+K(K-5k_{2})-2(K-k_{2})k_{2}+4k^2_2\right]\nonumber\right.\\
&&\left.
+\frac{1}{(2k_{1}-K)^3}\left[K^2-4k_{1}k_{3}+K(k_{3}-5k_{1})+6k^2_1\right]\nonumber\right.\\
&&\left.
+\frac{1}{(2k_{3}-K)^3}\left[K^2-4k_{1}k_{3}+K(k_{1}-5k_{3})+6k^2_3\right]\right\}J_{2}(\alpha,\beta).
\\		   
G_{3}(k_{1},k_{2},k_{3})&=&\frac{1}{K^3}\left[K^2+2k_{1}k_{2}+K(K-k_{3})\right]J_{1}(\alpha,\beta)\nonumber\\&&
+\left\{\frac{1}{(2k_{3}-K)^3}\left[K^2+2k_{1}k_{2}+K(K-5k_{3})-2(K-k_{3})k_{3}+4k^2_3\right]\nonumber\right.\\
&&\left.
+\frac{1}{(2k_{2}-K)^3}\left[K^2-4k_{1}k_{2}+K(k_{1}-5k_{2})+6k^2_2\right]\nonumber\right.\\
&&\left.
+\frac{1}{(2k_{1}-K)^3}\left[K^2-4k_{1}k_{2}+K(k_{2}-5k_{1})+6k^2_1\right]\right\}J_{2}(\alpha,\beta).
\eea
where $J_{1}(\alpha,\beta)$ and $J_{2}(\alpha,\beta)$ are defined as:
\bea J_{1}(\alpha,\beta)&=&\left[\left(\cosh\alpha-e^{i\beta}\sinh\alpha\right)^3\left(\cosh^3\alpha+e^{-3i\beta}\sinh^3\alpha\right)\right.\nonumber\\ &&\left.+\left(\cosh\alpha-e^{-i\beta}\sinh\alpha\right)^3\left(\cosh^3\alpha+e^{3i\beta}\sinh^3\alpha\right)\right],\eea\bea
J_{2}(\alpha,\beta)&=&\frac{1}{2}\left[\left(\cosh\alpha-e^{i\beta}\sinh\alpha\right)^3e^{-i\beta}\sinh2\alpha \left(\cosh\alpha-e^{-i\beta}\sinh\alpha\right)\right.\nonumber\\ && \left.+\left(\cosh\alpha-e^{-i\beta}\sinh\alpha\right)^3e^{i\beta}\sinh2\alpha\left(\cosh\alpha-e^{i\beta}\sinh\alpha\right)\right].
\eea
Consequently the three-point function for the scalar fluctuation can also be written after substituting all the momentum dependent functions $G_{1}(k_{1},k_{2},k_{3})$, $G_{2}(k_{1},k_{2},k_{3})$ and $G_{3}(k_{1},k_{2},k_{3})$ for $\alpha,\beta$ vacuum.

\end{itemize}

		   	   \subsubsection{Function $\Theta_{3}(k_{1},k_{2},k_{3})$}
	\label{v4a3}
		Here we can write the function $\Theta_{3}(k_{1},k_{2},k_{3})$ as: 	 
			   	   \begin{footnotesize}  	   
\bea \Theta_{3}(k_{1},k_{2},k_{3})&=&2i\int^{\eta_{f}=0}_{\eta_{i}=-\infty}d\eta~\frac{a^2(\eta)}{H^3}\left\{\left[\bar{v}(\eta_{f},{\bf k_{1}})\bar{v}(\eta_{f},{\bf k_{2}})\bar{v}(\eta_{f},{\bf k_{3}})\bar{v}^{*}(\eta,{\bf k_{1}})\bar{v}^{'*}(\eta,{\bf k_{2}})\bar{v}^{'*}(\eta,{\bf k_{3}})\right.\right.\nonumber\\
&& \left.\left.~~~~~~~~~~~~~~~~~+\bar{v}^{*}(\eta_{f},-{\bf k_{1}})\bar{v}^{*}(\eta_{f},-{\bf k_{2}})\bar{v}^{*}(\eta_{f},-{\bf k_{3}})\bar{v}(\eta,-{\bf k_{1}})\bar{v}^{'}(\eta,-{\bf k_{2}})\bar{v}^{'}(\eta,-{\bf k_{3}})\right]\right.\nonumber\\ &&\left.~~~~~~~~~~~+\left[\bar{v}(\eta_{f},{\bf k_{1}})\bar{v}(\eta_{f},{\bf k_{2}})\bar{v}(\eta_{f},{\bf k_{3}})\bar{v}^{*}(\eta,{\bf k_{2}})\bar{v}^{'*}(\eta,{\bf k_{1}})\bar{v}^{'*}(\eta,{\bf k_{3}})\right.\right.\nonumber\\
&& \left.\left.~~~~~~~~~~~~~~~~~+\bar{v}^{*}(\eta_{f},-{\bf k_{1}})\bar{v}^{*}(\eta_{f},-{\bf k_{2}})\bar{v}^{*}(\eta_{f},-{\bf k_{3}})\bar{v}(\eta,-{\bf k_{2}})\bar{v}^{'}(\eta,-{\bf k_{1}})\bar{v}^{'}(\eta,-{\bf k_{3}})\right]
\right.\nonumber\\ &&\left.~~~~~~~~~~~+\left[\bar{v}(\eta_{f},{\bf k_{1}})\bar{v}(\eta_{f},{\bf k_{2}})\bar{v}(\eta_{f},{\bf k_{3}})\bar{v}^{*}(\eta,{\bf k_{3}})\bar{v}^{'*}(\eta,{\bf k_{1}})\bar{v}^{'*}(\eta,{\bf k_{2}})\right.\right.\nonumber\\
&& \left.\left.~~~~~~~~~~~~~~~~~+\bar{v}^{*}(\eta_{f},-{\bf k_{1}})\bar{v}^{*}(\eta_{f},-{\bf k_{2}})\bar{v}^{*}(\eta_{f},-{\bf k_{3}})\bar{v}(\eta,-{\bf k_{3}})\bar{v}^{'}(\eta,-{\bf k_{1}})\bar{v}^{'}(\eta,-{\bf k_{2}})		   	   		   	   \right]
\right\}.\eea		   
	   	   \end{footnotesize}	 
	   	     
 Using the results obtained in the Appendix we finally get the following simplified expression for the three-point function for the scalar fluctuations\footnote{Here it is important to point out that in de Sitter space if we consider the Bunch–Davies vacuum state then here only the term with $1/K^2$ will appear explicitly in the expression for the three-point function and in the associated bispectrum. On~the other hand, if we consider all other non-trivial quantum vacuum states in our computation, then the rest of the contribution will explicitly appear. From~the perspective of observation, this is obviously important information as for the non-trivial quantum vacuum state we get additional contribution in the bispectrum which may enhance the amplitude of the non-Gaussianity in squeezed limiting configuration. Additionally, it is important to mention that in quasi de Sitter case we get extra contributions $1/\tilde{c}^{6\nu-9}_S$ and $1/(1+\epsilon)^3$. Also, the factor $1/(k_1 k_2 k_3)^3$ will be replaced by  $1/(k_1 k_2 k_3)^{2\nu}$. Consequently, in~quasi de Sitter case this contribution in the bispectrum can be recast as:
   \be \begin{array}{rl} \Theta_{3}(k_{1},k_{2},k_{3})&=\displaystyle \frac{H}{32\epsilon^3M^6_p \tilde{c}^{6\nu-9}_S (1+\epsilon)^3}\frac{1}{(k_{1}k_{2}k_{3})^{2\nu}}\left[(k_{2}k_{3})^2M_{1}(k_{1},k_{2},k_{3})\right. \\ & \left.~~~~~~~~~~~~~~~~~~~~~~~~~\displaystyle+(k_{1}k_{3})^2M_{2}(k_{1},k_{2},k_{3})+(k_{1}k_{2})^2M_{3}(k_{1},k_{2},k_{3})\right],\end{array}~~~~~~~~\ee}:
			   	   \be {\begin{array}{rl} \Theta_{3}(k_{1},k_{2},k_{3})&=\displaystyle \frac{H}{32\epsilon^3M^6_p }\frac{1}{(k_{1}k_{2}k_{3})^3}\left[(k_{2}k_{3})^2M_{1}(k_{1},k_{2},k_{3})\right. \\ & \left.~~~~~~~~~~~~~~~~~~~~~~~~~\displaystyle+(k_{1}k_{3})^2M_{2}(k_{1},k_{2},k_{3})+(k_{1}k_{2})^2M_{3}(k_{1},k_{2},k_{3})\right],\end{array}}~~~~~~~~\ee
			   	   where the momentum dependent functions $M_{1}(k_{1},k_{2},k_{3})$, $M_{2}(k_{1},k_{2},k_{3})$ and $M_{3}(k_{1},k_{2},k_{3})$ are defined~as:
	\bea		   
	M_{1}(k_{1},k_{2},k_{3})&=&\frac{1}{K^2}(K+k_{1})\left[(C_{1}-C_{2})^3(C^{*3}_{1}+C^{*3}_{2})+(C_{1}-C_{2})^3(C^{*3}_{1}+C^{*3}_{2})\right]\nonumber\\&&
	+\left\{\frac{(K-3k_1)}{(2k_{1}-K)^2}
	+\frac{(K+k_1-2k_2)}{(2k_{2}-K)^2}
	+\frac{(K+k_1-2k_3)}{(2k_{3}-K)^2}\right\}\nonumber\\
	&&~~~\left[(C_{1}-C_{2})^3 C^{*}_{1}C^{*}_{2}(C^{*}_{1}+C^{*}_{2})+(C^{*}_{1}-C^{*}_{2})^3 C_{1}C_{2}(C_{1}+C_{2})\right].~~~~~~
	\eea	
	
	\bea	   
	M_{2}(k_{1},k_{2},k_{3})&=&\frac{1}{K^2}(K+k_{2})\left[(C_{1}-C_{2})^3(C^{*3}_{1}+C^{*3}_{2})+(C_{1}-C_{2})^3(C^{*3}_{1}+C^{*3}_{2})\right]\nonumber\\&&
		+\left\{\frac{(K-3k_2)}{(2k_{2}-K)^2}
		+\frac{(K+k_2-2k_1)}{(2k_{1}-K)^2}
		+\frac{(K+k_2-2k_3)}{(2k_{3}-K)^2}\right\}\nonumber\\
		&&~~~\left[(C_{1}-C_{2})^3 C^{*}_{1}C^{*}_{2}(C^{*}_{1}+C^{*}_{2})+(C^{*}_{1}-C^{*}_{2})^3 C_{1}C_{2}(C_{1}+C_{2})\right].~~~~~~
	\\		   
	M_{3}(k_{1},k_{2},k_{3})&=&\frac{1}{K^2}(K+k_{3})\left[(C_{1}-C_{2})^3(C^{*3}_{1}+C^{*3}_{2})+(C_{1}-C_{2})^3(C^{*3}_{1}+C^{*3}_{2})\right]\nonumber\\&&
		+\left\{\frac{(K-3k_3)}{(2k_{3}-K)^2}
		+\frac{(K+k_3-2k_2)}{(2k_{2}-K)^2}
		+\frac{(K+k_3-2k_1)}{(2k_{1}-K)^2}\right\}\nonumber\\
		&&~~~\left[(C_{1}-C_{2})^3 C^{*}_{1}C^{*}_{2}(C^{*}_{1}+C^{*}_{2})+(C^{*}_{1}-C^{*}_{2})^3 C_{1}C_{2}(C_{1}+C_{2})\right].~~~~~~
	\eea
	
	 Finally, for {Bunch–Davies} and {$\alpha,\beta$} vacuum we get the following contribution in the three-point function for scalar~fluctuations:   
	\begin{itemize}
	\item \underline{\bf {For Bunch–Davies vacuum}:} \\After setting $C_{1}=1$ and $C_{2}=0$ we get:
	\bea		   
	M_{1}(k_{1},k_{2},k_{3})&=&\frac{2}{K^2}(K+k_1).
	\\		   
	M_{2}(k_{1},k_{2},k_{3})&=&\frac{2}{K^2}(K+k_2).\\
	M_{3}(k_{1},k_{2},k_{3})&=&\frac{2}{K^2}(K+k_3).
	\eea
	
	Consequently, we get the following contribution:
	\begin{footnotesize}	
	\bea \Theta_{3}(k_{1},k_{2},k_{3})&=& \frac{H}{16\epsilon^3M^6_p }\frac{1}{(k_{1}k_{2}k_{3})^3}\frac{1}{K^2}\left[(k_{2}k_{3})^2(K+k_1)]+(k_{1}k_{3})^2(K+k_2)+(k_{1}k_{2})^2(K+k_3)\right],\eea
	\end{footnotesize}	
	\item \underline{\bf {For $~{\alpha,\beta}$ vacuum}:}\\After setting $C_{1}=\cosh\alpha$ and $C_{2}=e^{i\beta}~\sinh\alpha$ we get:
	\begin{footnotesize}	
	\bea		   
		M_{1}(k_{1},k_{2},k_{3})&=&\frac{(K+k_{1})J_{1}(\alpha,\beta)}{K^2}
		+\left\{\frac{(K-3k_1)}{(2k_{1}-K)^2}
		+\frac{(K+k_1-2k_2)}{(2k_{2}-K)^2}
		+\frac{(K+k_1-2k_3)}{(2k_{3}-K)^2}\right\}J_{2}(\alpha,\beta).~~~~~~~~~
	\\	   
		M_{2}(k_{1},k_{2},k_{3})&=&\frac{(K+k_{2})J_{1}(\alpha,\beta)}{K^2}
			+\left\{\frac{(K-3k_2)}{(2k_{2}-K)^2}
			+\frac{(K+k_2-2k_1)}{(2k_{1}-K)^2}
			+\frac{(K+k_2-2k_3)}{(2k_{3}-K)^2}\right\}J_{2}(\alpha,\beta).~~~~~~~~~
	\\		   
		M_{3}(k_{1},k_{2},k_{3})&=&\frac{(K+k_{3})J_{1}(\alpha,\beta)}{K^2}
			+\left\{\frac{(K-3k_3)}{(2k_{3}-K)^2}
			+\frac{(K+k_3-2k_2)}{(2k_{2}-K)^2}
			+\frac{(K+k_3-2k_1)}{(2k_{1}-K)^2}\right\}J_{2}(\alpha,\beta).~~~~~~~~~
		\eea
		\end{footnotesize}	
\hspace{-3pt}where $J_{1}(\alpha,\beta)$ and $J_{2}(\alpha,\beta)$ are defined earlier. Consequently the three-point function for the scalar fluctuation can also be written after substituting all the momentum dependent functions $M_{1}(k_{1},k_{2},k_{3})$, $M_{2}(k_{1},k_{2},k_{3})$ and $M_{3}(k_{1},k_{2},k_{3})$ for $\alpha,\beta$ vacuum.
	
	\end{itemize}
		   	    \subsubsection{Function $\Theta_{4}(k_{1},k_{2},k_{3})$}
	\label{v4a4}
		Here we can write the function $\Theta_{4}(k_{1},k_{2},k_{3})$ as: 
		\begin{footnotesize}		   	    
	\bea \Theta_{4}(k_{1},k_{2},k_{3})&=&i\int^{\eta_{f}=0}_{\eta_{i}=-\infty}d\eta~\frac{a(\eta)}{H^3}\left\{({\bf k_{2}}.{\bf k_{3}})X_{1}(k_1,k_2,k_3)+({\bf k_{1}}.{\bf k_{3}})X_{2}(k_1,k_2,k_3) 
	+({\bf k_{1}}.{\bf k_{2}})X_{3}(k_1,k_2,k_3)\right\},~~~~~~~\eea	
	\end{footnotesize}	
\hspace{-3pt}where the momentum dependent functions $X_{1}(k_1,k_2,k_3)$, $X_{2}(k_1,k_2,k_3)$ and $X_{3}(k_1,k_2,k_3)$ can be expressed in terms of the various combinations of the scalar mode functions as:   
\begin{small}
	\bea X_{1}(k_1,k_2,k_3)&=&\bar{v}(\eta_{f},{\bf k_{1}})\bar{v}(\eta_{f},{\bf k_{2}})\bar{v}(\eta_{f},{\bf k_{3}})\bar{v}^{*}(\eta,{\bf k_{1}})\bar{v}^{'*}(\eta,{\bf k_{2}})\bar{v}^{*}(\eta,{\bf k_{3}})\nonumber\\
	&& ~~~~~~~~+\bar{v}^{*}(\eta_{f},-{\bf k_{1}})\bar{v}^{*}(\eta_{f},-{\bf k_{2}})\bar{v}^{*}(\eta_{f},-{\bf k_{3}})\bar{v}(\eta,-{\bf k_{1}})\bar{v}^{'}(\eta,-{\bf k_{2}})\bar{v}(\eta,-{\bf k_{3}})\nonumber\\ &&~~~~~~~~~~~+\bar{v}(\eta_{f},{\bf k_{1}})\bar{v}(\eta_{f},{\bf k_{2}})\bar{v}(\eta_{f},{\bf k_{3}})\bar{v}^{*}(\eta,{\bf k_{1}})\bar{v}^{'*}(\eta,{\bf k_{3}})\bar{v}^{*}(\eta,{\bf k_{2}})\nonumber\\
	&& ~~~~~~~+\bar{v}^{*}(\eta_{f},-{\bf k_{1}})\bar{v}^{*}(\eta_{f},-{\bf k_{2}})\bar{v}^{*}(\eta_{f},-{\bf k_{3}})\bar{v}(\eta,-{\bf k_{1}})\bar{v}^{'}(\eta,-{\bf k_{3}})\bar{v}(\eta,-{\bf k_{2}}),\\ X_{2}(k_1,k_2,k_3)&=&\bar{v}(\eta_{f},{\bf k_{1}})\bar{v}(\eta_{f},{\bf k_{2}})\bar{v}(\eta_{f},{\bf k_{3}})\bar{v}^{*}(\eta,{\bf k_{2}})\bar{v}^{'*}(\eta,{\bf k_{1}})\bar{v}^{*}(\eta,{\bf k_{3}})\nonumber\\
	&& ~~~~~~~~~~~~~~~~~+\bar{v}^{*}(\eta_{f},-{\bf k_{1}})\bar{v}^{*}(\eta_{f},-{\bf k_{2}})\bar{v}^{*}(\eta_{f},-{\bf k_{3}})\bar{v}(\eta,-{\bf k_{2}})\bar{v}^{'}(\eta,-{\bf k_{1}})\bar{v}(\eta,-{\bf k_{3}})
	\nonumber\\&& ~~~~~~~~~~~~~~~~~   
	+\bar{v}(\eta_{f},{\bf k_{1}})\bar{v}(\eta_{f},{\bf k_{2}})\bar{v}(\eta_{f},{\bf k_{3}})\bar{v}^{*}(\eta,{\bf k_{2}})\bar{v}^{'*}(\eta,{\bf k_{3}})\bar{v}^{*}(\eta,{\bf k_{1}})\nonumber\\
	&& ~~~~~~~~~~~~~~~~~+\bar{v}^{*}(\eta_{f},-{\bf k_{1}})\bar{v}^{*}(\eta_{f},-{\bf k_{2}})\bar{v}^{*}(\eta_{f},-{\bf k_{3}})\bar{v}(\eta,-{\bf k_{2}})\bar{v}^{'}(\eta,-{\bf k_{3}})\bar{v}(\eta,-{\bf k_{1}}),\\ X_{3}(k_1,k_2,k_3)&=&\bar{v}(\eta_{f},{\bf k_{1}})\bar{v}(\eta_{f},{\bf k_{2}})\bar{v}(\eta_{f},{\bf k_{3}})\bar{v}^{*}(\eta,{\bf k_{3}})\bar{v}^{'*}(\eta,{\bf k_{1}})\bar{v}^{*}(\eta,{\bf k_{2}})\nonumber\\
	&& ~~~~~~~~~~~~~~~~~+\bar{v}^{*}(\eta_{f},-{\bf k_{1}})\bar{v}^{*}(\eta_{f},-{\bf k_{2}})\bar{v}^{*}(\eta_{f},-{\bf k_{3}})\bar{v}(\eta,-{\bf k_{3}})\bar{v}^{'}(\eta,-{\bf k_{1}})\bar{v}(\eta,-{\bf k_{2}})\nonumber\\ &&~~~~~~~~~~~+({\bf k_{1}}.{\bf k_{2}})\bar{v}(\eta_{f},{\bf k_{1}})\bar{v}(\eta_{f},{\bf k_{2}})\bar{v}(\eta_{f},{\bf k_{3}})\bar{v}^{*}(\eta,{\bf k_{3}})\bar{v}^{'*}(\eta,{\bf k_{2}})\bar{v}^{*}(\eta,{\bf k_{1}})\nonumber\\
	&& ~~~~~~~~~~~~~~~~~+\bar{v}^{*}(\eta_{f},-{\bf k_{1}})\bar{v}^{*}(\eta_{f},-{\bf k_{2}})\bar{v}^{*}(\eta_{f},-{\bf k_{3}})\bar{v}(\eta,-{\bf k_{3}})\bar{v}^{'}(\eta,-{\bf k_{2}})\bar{v}^{'}(\eta,-{\bf k_{1}}),~~~~~~~~~\eea		
	\end{small}	

 Using the results obtained in the Appendix we finally get the following simplified expression for the three-point function for the scalar fluctuations\footnote{Here it is important to point out that in de Sitter space if we consider the Bunch–Davies vacuum state then here only the term with $1/K^3$ will appear explicitly in the expression for the three-point function and in the associated bispectrum. On~the other hand, if we consider all other non-trivial quantum vacuum states in our computation, then the rest of the contribution will explicitly appear. From~the perspective of observation, this is obviously important information as for the non-trivial quantum vacuum state we get additional contribution in the bispectrum which may enhance the amplitude of the non-Gaussianity in squeezed limiting configuration. Additionally, it is important to mention that in quasi de Sitter case we get an extra contribution  $1/(1+\epsilon)^5$. Also the factor $1/(k_1 k_2 k_3)^3$ will be replaced by  $1/(k_1 k_2 k_3)^{2\nu}$ and $1/\tilde{c}^{2}_S$ is replaced by $1/\tilde{c}^{6\nu-7}_S$. Consequently, in~quasi de Sitter case this contribution in the bispectrum can be recast as:\begin{scriptsize}\be\begin{array}{rl} \Theta_{4}(k_{1},k_{2},k_{3})&=\displaystyle -\frac{H^2}{64\tilde{c}^{6\nu-7}_{S}\epsilon^3M^6_p (1+\epsilon)^5}\frac{1}{(k_{1}k_{2}k_{3})^{2\nu}}\left[k^2_2({\bf k_2}.{\bf k_3}){\cal F}_{1}(k_1,k_2,k_3)+k^2_3({\bf k_2}.{\bf k_3}){\cal F}_{2}(k_1,k_2,k_3)\right.\\ & \left.~~~\displaystyle+k^2_1({\bf k_1}.{\bf k_3}){\cal F}_{3}(k_1,k_2,k_3)+k^2_3({\bf k_1}.{\bf k_3}){\cal F}_{4}(k_1,k_2,k_3)+k^2_1({\bf k_1}.{\bf k_2}){\cal F}_{5}(k_1,k_2,k_3)+k^2_2({\bf k_1}.{\bf k_2}){\cal F}_{6}(k_1,k_2,k_3)\right],\end{array}\ee\end{scriptsize}}:
		   	  \begin{small}\be\label{azokl5}{\begin{array}{rl} \Theta_{4}(k_{1},k_{2},k_{3})&=\displaystyle -\frac{H^2}{64\tilde{c}^2_{S}\epsilon^3M^6_p}\frac{1}{(k_{1}k_{2}k_{3})^3}\left[k^2_2({\bf k_2}.{\bf k_3}){\cal F}_{1}(k_1,k_2,k_3)+k^2_3({\bf k_2}.{\bf k_3}){\cal F}_{2}(k_1,k_2,k_3)\right.\\ & \left.~~~~~~~~~~~~~~\displaystyle+k^2_1({\bf k_1}.{\bf k_3}){\cal F}_{3}(k_1,k_2,k_3)+k^2_3({\bf k_1}.{\bf k_3}){\cal F}_{4}(k_1,k_2,k_3)\right.\\ & \left.~~~~~~~~~~~~~~\displaystyle+k^2_1({\bf k_1}.{\bf k_2}){\cal F}_{5}(k_1,k_2,k_3)+k^2_2({\bf k_1}.{\bf k_2}){\cal F}_{6}(k_1,k_2,k_3)\right],\end{array}}~~~~~~~~~~~~\ee  \end{small} 	   
		   	   where the momentum dependent functions ${\cal F}_{i}(k_1,k_2,k_3)\forall i=1,2,\cdots,6$ are defined as:
\begin{small}
\bea		   
{\cal F}_{1}(k_1,k_2,k_3)&=&\frac{1}{K^3}\left[K^2+2k_{1}k_{3}+K(K-k_{2})\right]\left[(C_{1}-C_{2})^3(C^{*3}_{1}+C^{*3}_{2})+(C_{1}-C_{2})^3(C^{*3}_{1}+C^{*3}_{2})\right]\nonumber\\&&
+\left\{\frac{1}{(2k_{1}-K)^3}\left[K^2-4k_{1}k_{3}+K(k_3-5k_{1})+6k^2_1\right]\nonumber\right.\\
&&\left.
+\frac{1}{(2k_{2}-K)^3}\left[(K-2k_2)(K-2k_2+k_1)+(K+2k_1-2k_2)k_3\right]\nonumber\right.\\
&&\left.
+\frac{1}{(2k_{3}-K)^3}\left[K^2-4k_{1}k_{3}+K(k_{1}-5k_{3})+6k^2_3\right]\right\}\nonumber\\
&&~~~~~~~~~~~\left[(C_{1}-C_{2})^3 C^{*}_{1}C^{*}_{2}(C^{*}_{1}+C^{*}_{2})+(C^{*}_{1}-C^{*}_{2})^3 C_{1}C_{2}(C_{1}+C_{2})\right].
\eea
\end{small}
\begin{small}
\bea	   
{\cal F}_{2}(k_1,k_2,k_3)&=&\frac{1}{K^3}\left[K^2+2k_{1}k_{2}+K(K-k_{3})\right]\left[(C_{1}-C_{2})^3(C^{*3}_{1}+C^{*3}_{2})+(C_{1}-C_{2})^3(C^{*3}_{1}+C^{*3}_{2})\right]\nonumber\\&&
+\left\{\frac{1}{(2k_{1}-K)^3}\left[K^2-4k_{1}k_{2}+K(k_2-5k_{1})+6k^2_1\right]\nonumber\right.\\
&&\left.
+\frac{1}{(2k_{3}-K)^3}\left[(K-2k_3)(K-2k_3+k_1)+(K+2k_1-2k_3)k_2\right]\nonumber\right.\\
&&\left.
+\frac{1}{(2k_{2}-K)^3}\left[K^2-4k_{1}k_{2}+K(k_{1}-5k_{2})+6k^2_2\right]\right\}\nonumber\\
&&~~~~~~~~~~~\left[(C_{1}-C_{2})^3 C^{*}_{1}C^{*}_{2}(C^{*}_{1}+C^{*}_{2})+(C^{*}_{1}-C^{*}_{2})^3 C_{1}C_{2}(C_{1}+C_{2})\right].
\\	   
{\cal F}_{3}(k_1,k_2,k_3)&=&\frac{1}{K^3}\left[K^2+2k_{2}k_{3}+K(K-k_{1})\right]\left[(C_{1}-C_{2})^3(C^{*3}_{1}+C^{*3}_{2})+(C_{1}-C_{2})^3(C^{*3}_{1}+C^{*3}_{2})\right]\nonumber\\&&
+\left\{\frac{1}{(2k_{2}-K)^3}\left[K^2-4k_{2}k_{3}+K(k_3-5k_{2})+6k^2_2\right]\nonumber\right.\\
&&\left.
+\frac{1}{(2k_{1}-K)^3}\left[(K-2k_1)(K-2k_1+k_2)+(K+2k_2-2k_1)k_2\right]\nonumber\right.\\
&&\left.
+\frac{1}{(2k_{3}-K)^3}\left[K^2-4k_{2}k_{3}+K(k_{2}-5k_{3})+6k^2_3\right]\right\}\nonumber\\
&&~~~~~~~~~~~\left[(C_{1}-C_{2})^3 C^{*}_{1}C^{*}_{2}(C^{*}_{1}+C^{*}_{2})+(C^{*}_{1}-C^{*}_{2})^3 C_{1}C_{2}(C_{1}+C_{2})\right].
\\	   
{\cal F}_{4}(k_1,k_2,k_3)&=&\frac{1}{K^3}\left[K^2+2k_{1}k_{2}+K(K-k_{3})\right]\left[(C_{1}-C_{2})^3(C^{*3}_{1}+C^{*3}_{2})+(C_{1}-C_{2})^3(C^{*3}_{1}+C^{*3}_{2})\right]\nonumber\\&&
+\left\{\frac{1}{(2k_{2}-K)^3}\left[K^2-4k_{1}k_{2}+K(k_1-5k_{2})+6k^2_2\right]\nonumber\right.\\
&&\left.
+\frac{1}{(2k_{3}-K)^3}\left[(K-2k_3)(K-2k_3+k_2)+(K+2k_2-2k_3)k_2\right]\nonumber\right.\\
&&\left.
+\frac{1}{(2k_{1}-K)^3}\left[K^2-4k_{1}k_{2}+K(k_{2}-5k_{1})+6k^2_1\right]\right\}\nonumber\\
&&~~~~~~~~~~~\left[(C_{1}-C_{2})^3 C^{*}_{1}C^{*}_{2}(C^{*}_{1}+C^{*}_{2})+(C^{*}_{1}-C^{*}_{2})^3 C_{1}C_{2}(C_{1}+C_{2})\right].
\\		   
{\cal F}_{5}(k_1,k_2,k_3)&=&\frac{1}{K^3}\left[K^2+2k_{2}k_{3}+K(K-k_{1})\right]\left[(C_{1}-C_{2})^3(C^{*3}_{1}+C^{*3}_{2})+(C_{1}-C_{2})^3(C^{*3}_{1}+C^{*3}_{2})\right]\nonumber\\&&
+\left\{\frac{1}{(2k_{3}-K)^3}\left[K^2-4k_{2}k_{3}+K(k_2-5k_{3})+6k^2_3\right]\nonumber\right.\\
&&\left.
+\frac{1}{(2k_{1}-K)^3}\left[(K-2k_1)(K-2k_1+k_2)+(K+2k_2-2k_1)k_3\right]\nonumber\right.\\
&&\left.
+\frac{1}{(2k_{2}-K)^3}\left[K^2-4k_{2}k_{3}+K(k_{3}-5k_{2})+6k^2_2\right]\right\}\nonumber\\
&&~~~~~~~~~~~\left[(C_{1}-C_{2})^3 C^{*}_{1}C^{*}_{2}(C^{*}_{1}+C^{*}_{2})+(C^{*}_{1}-C^{*}_{2})^3 C_{1}C_{2}(C_{1}+C_{2})\right].\\
 {\cal F}_{6}(k_1,k_2,k_3)&=&\frac{1}{K^3}\left[K^2+2k_{1}k_{3}+K(K-k_{2})\right]\left[(C_{1}-C_{2})^3(C^{*3}_{1}+C^{*3}_{2})+(C_{1}-C_{2})^3(C^{*3}_{1}+C^{*3}_{2})\right]\nonumber\\&&
+\left\{\frac{1}{(2k_{3}-K)^3}\left[K^2-4k_{1}k_{3}+K(k_1-5k_{3})+6k^2_3\right]\nonumber\right.\\
&&\left.
+\frac{1}{(2k_{2}-K)^3}\left[(K-2k_2)(K-2k_2+k_3)+(K+2k_3-2k_2)k_1\right]\nonumber\right.\\
&&\left.
+\frac{1}{(2k_{1}-K)^3}\left[K^2-4k_{1}k_{3}+K(k_{3}-5k_{1})+6k^2_1\right]\right\}\nonumber\\
&&~~~~~~~~~~~\left[(C_{1}-C_{2})^3 C^{*}_{1}C^{*}_{2}(C^{*}_{1}+C^{*}_{2})+(C^{*}_{1}-C^{*}_{2})^3 C_{1}C_{2}(C_{1}+C_{2})\right].
\eea
\end{small}

Furthermore, after simplification one can recast the three-point function for the scalar fluctuation as:
\begin{footnotesize}
\be\label{xzokl5}{\begin{array}{rl} \Theta_{4}(k_{1},k_{2},k_{3})&=\displaystyle-\frac{H^2}{128\tilde{c}^2_{S}\epsilon^3M^6_p}\frac{1}{(k_{1}k_{2}k_{3})^3}\left[k^2_2\left(k^2_1 -k^2_2 -k^2_3\right){\cal F}_{1}(k_1,k_2,k_3)\displaystyle+k^2_3\left(k^2_1 -k^2_2 -k^2_3\right){\cal F}_{2}(k_1,k_2,k_3)\right.\\ & \left.~~~~~~~~~~~~~~~~~\displaystyle+k^2_1\left(k^2_2 -k^2_1 -k^2_3\right){\cal F}_{3}(k_1,k_2,k_3)\displaystyle+k^2_3\left(k^2_2 -k^2_1 -k^2_3\right){\cal F}_{4}(k_1,k_2,k_3)\right.\\ & \left.~~~~~~~~~~~~~~~~~\displaystyle+k^2_1\left(k^2_3 -k^2_2 -k^2_1\right){\cal F}_{5}(k_1,k_2,k_3)\displaystyle+k^2_2\left(k^2_3 -k^2_2 -k^2_1\right){\cal F}_{6}(k_1,k_2,k_3)\right],\end{array}}~~~~~~~~~~~~\ee
\end{footnotesize}

 Finally, for {Bunch–Davies} and {$\alpha,\beta$} vacuum we get the following contribution in the three-point function for scalar~fluctuations:   
\begin{itemize}
\item \underline{\bf {For Bunch–Davies vacuum}:} \\After setting $C_{1}=1$ and $C_{2}=0$ we get:
\bea		   
{\cal F}_{1}(k_1,k_2,k_3)&=&\frac{1}{K^3}\left[K^2+2k_{1}k_{3}+K(K-k_{2})\right].
\\		   
{\cal F}_{2}(k_1,k_2,k_3)&=&\frac{1}{K^3}\left[K^2+2k_{1}k_{2}+K(K-k_{3})\right].
\\		   
{\cal F}_{3}(k_1,k_2,k_3)&=&\frac{1}{K^3}\left[K^2+2k_{2}k_{3}+K(K-k_{1})\right].
\\		   
{\cal F}_{4}(k_1,k_2,k_3)&=&\frac{1}{K^3}\left[K^2+2k_{1}k_{2}+K(K-k_{3})\right].
\\		   
{\cal F}_{5}(k_1,k_2,k_3)&=&\frac{1}{K^3}\left[K^2+2k_{2}k_{3}+K(K-k_{1})\right].
\\	   
{\cal F}_{6}(k_1,k_2,k_3)&=&\frac{1}{K^3}\left[K^2+2k_{1}k_{3}+K(K-k_{2})\right].
\eea
Consequently, the three-point function for the scalar fluctuation can be expressed as:
\bea\label{okl5xx} \Theta_{4}(k_{1},k_{2},k_{3})&=&-\frac{H^2}{128\tilde{c}^2_{S}\epsilon^3M^6_p}\frac{1}{(k_{1}k_{2}k_{3})^3}\left[k^2_2\left(k^2_1 -k^2_2 -k^2_3\right)\left[K^2+2k_{1}k_{3}+K(K-k_{2})\right]\right.\nonumber\\ && \left.~~~~~~~~~~~~~~~~~~~~~~~~~~~~~~~+k^2_3\left(k^2_1 -k^2_2 -k^2_3\right)\left[K^2+2k_{1}k_{2}+K(K-k_{3})\right]\right.\nonumber\\ && \left.~~~~~~~~~~~~~~~~~~~~~~~~~~~~~~~+k^2_1\left(k^2_2 -k^2_1 -k^2_3\right)\left[K^2+2k_{2}k_{3}+K(K-k_{1})\right]\right.\nonumber\\ && \left.~~~~~~~~~~~~~~~~~~~~~~~~~~~~~~~+k^2_3\left(k^2_2 -k^2_1 -k^2_3\right)\left[K^2+2k_{1}k_{2}+K(K-k_{3})\right]\right.\nonumber\\ && \left.~~~~~~~~~~~~~~~~~~~~~~~~~~~~~~~+k^2_1\left(k^2_3 -k^2_2 -k^2_1\right)\left[K^2+2k_{2}k_{3}+K(K-k_{1})\right]\right.\nonumber\\ && \left.~~~~~~~~~~~~~~~~~~~~~~~~~~~~~~~+k^2_2\left(k^2_3 -k^2_2 -k^2_1\right)\left[K^2+2k_{1}k_{3}+K(K-k_{2})\right]\right],~~~~~~~~~~~~\eea 
\item \underline{\bf {For $~{\alpha,\beta}$ vacuum}:}\\After setting $C_{1}=\cosh\alpha$ and $C_{2}=e^{i\beta}~\sinh\alpha$ we get:
\bea		   
{\cal F}_{1}(k_1,k_2,k_3)&=&\frac{1}{K^3}\left[K^2+2k_{1}k_{3}+K(K-k_{2})\right]J_{1}(\alpha,\beta)\nonumber\\&&
+\left\{\frac{1}{(2k_{1}-K)^3}\left[K^2-4k_{1}k_{3}+K(k_3-5k_{1})+6k^2_1\right]\nonumber\right.\\
&&\left.
+\frac{1}{(2k_{2}-K)^3}\left[(K-2k_2)(K-2k_2+k_1)+(K+2k_1-2k_2)k_3\right]\nonumber\right.\\
&&\left.
+\frac{1}{(2k_{3}-K)^3}\left[K^2-4k_{1}k_{3}+K(k_{1}-5k_{3})+6k^2_3\right]\right\}J_{2}(\alpha,\beta).~~~~~~~~~~~
\eea\bea		   
{\cal F}_{2}(k_1,k_2,k_3)&=&\frac{1}{K^3}\left[K^2+2k_{1}k_{2}+K(K-k_{3})\right]J_{1}(\alpha,\beta)\nonumber\\&&
+\left\{\frac{1}{(2k_{1}-K)^3}\left[K^2-4k_{1}k_{2}+K(k_2-5k_{1})+6k^2_1\right]\nonumber\right.\\
&&\left.
+\frac{1}{(2k_{3}-K)^3}\left[(K-2k_3)(K-2k_3+k_1)+(K+2k_1-2k_3)k_2\right]\nonumber\right.\\
&&\left.
+\frac{1}{(2k_{2}-K)^3}\left[K^2-4k_{1}k_{2}+K(k_{1}-5k_{2})+6k^2_2\right]\right\}J_{2}(\alpha,\beta).~~~~~~~~~~~
\eea\bea		   
{\cal F}_{3}(k_1,k_2,k_3)&=&\frac{1}{K^3}\left[K^2+2k_{2}k_{3}+K(K-k_{1})\right]J_{1}(\alpha,\beta)\nonumber\\&&
+\left\{\frac{1}{(2k_{2}-K)^3}\left[K^2-4k_{2}k_{3}+K(k_3-5k_{2})+6k^2_2\right]\nonumber\right.\\
&&\left.
+\frac{1}{(2k_{1}-K)^3}\left[(K-2k_1)(K-2k_1+k_2)+(K+2k_2-2k_1)k_2\right]\nonumber\right.\\
&&\left.
+\frac{1}{(2k_{3}-K)^3}\left[K^2-4k_{2}k_{3}+K(k_{2}-5k_{3})+6k^2_3\right]\right\}J_{2}(\alpha,\beta).~~~~~~~~~~~
\eea\bea		   
{\cal F}_{4}(k_1,k_2,k_3)&=&\frac{1}{K^3}\left[K^2+2k_{1}k_{2}+K(K-k_{3})\right]J_{1}(\alpha,\beta)\nonumber\\&&
+\left\{\frac{1}{(2k_{2}-K)^3}\left[K^2-4k_{1}k_{2}+K(k_1-5k_{2})+6k^2_2\right]\nonumber\right.\\
&&\left.
+\frac{1}{(2k_{3}-K)^3}\left[(K-2k_3)(K-2k_3+k_2)+(K+2k_2-2k_3)k_2\right]\nonumber\right.\\
&&\left.
+\frac{1}{(2k_{1}-K)^3}\left[K^2-4k_{1}k_{2}+K(k_{2}-5k_{1})+6k^2_1\right]\right\}J_{2}(\alpha,\beta).~~~~~~~~~~~
\eea\bea		   
{\cal F}_{5}(k_1,k_2,k_3)&=&\frac{1}{K^3}\left[K^2+2k_{2}k_{3}+K(K-k_{1})\right]J_{1}(\alpha,\beta)\nonumber\\&&
+\left\{\frac{1}{(2k_{3}-K)^3}\left[K^2-4k_{2}k_{3}+K(k_2-5k_{3})+6k^2_3\right]\nonumber\right.\\
&&\left.
+\frac{1}{(2k_{1}-K)^3}\left[(K-2k_1)(K-2k_1+k_2)+(K+2k_2-2k_1)k_3\right]\nonumber\right.\\
&&\left.
+\frac{1}{(2k_{2}-K)^3}\left[K^2-4k_{2}k_{3}+K(k_{3}-5k_{2})+6k^2_2\right]\right\}J_{2}(\alpha,\beta).~~~~~~~~~~~
\eea\bea		   
{\cal F}_{6}(k_1,k_2,k_3)&=&\frac{1}{K^3}\left[K^2+2k_{1}k_{3}+K(K-k_{2})\right]J_{1}(\alpha,\beta)\nonumber\\&&
+\left\{\frac{1}{(2k_{3}-K)^3}\left[K^2-4k_{1}k_{3}+K(k_1-5k_{3})+6k^2_3\right]\nonumber\right.\\
&&\left.
+\frac{1}{(2k_{2}-K)^3}\left[(K-2k_2)(K-2k_2+k_3)+(K+2k_3-2k_2)k_1\right]\nonumber\right.\\
&&\left.
+\frac{1}{(2k_{1}-K)^3}\left[K^2-4k_{1}k_{3}+K(k_{3}-5k_{1})+6k^2_1\right]\right\}J_{2}(\alpha,\beta).~~~~~~~~~~~
\eea
where $J_{1}(\alpha,\beta)$ and $J_{2}(\alpha,\beta)$ are defined earlier. Consequently the three-point function for the scalar fluctuation can also be written after substituting all the momentum dependent functions ${\cal F}_{i}(k_{1},k_{2},k_{3})\forall i=1,2,\cdots,6$ for $\alpha,\beta$ vacuum.

\end{itemize}		

	  \subsubsection{Function $\Theta_{5}(k_{1},k_{2},k_{3})$}
	  \label{v4a5}
\bea \Theta_{5}(k_{1},k_{2},k_{3})&=&i\int^{\eta_{f}=0}_{\eta_{i}=-\infty}d\eta~\frac{a(\eta)}{H^3}\left\{({\bf k_{2}}.{\bf k_{3}})Y_{1}(k_1,k_2,k_3)+({\bf k_{1}}.{\bf k_{3}})Y_{2}(k_1,k_2,k_3)\right.\nonumber\\&&\left.~~~~~~~~~~~~~~~~~~~~~~~+({\bf k_{1}}.{\bf k_{2}})Y_{3}(k_1,k_2,k_3)\right\},\eea	
where the momentum dependent functions $Y_{1}(k_1,k_2,k_3)$, $Y_{2}(k_1,k_2,k_3)$ and $Y_{3}(k_1,k_2,k_3)$ can be expressed in terms of the various combinations of the scalar mode functions as:
\begin{small}
\bea Y_{1}(k_1,k_2,k_3)&=&\bar{v}(\eta_{f},{\bf k_{1}})\bar{v}(\eta_{f},{\bf k_{2}})\bar{v}(\eta_{f},{\bf k_{3}})\bar{v}^{*}(\eta,{\bf k_{1}})\bar{v}^{*}(\eta,{\bf k_{2}})\bar{v}^{*'}(\eta,{\bf k_{3}})\nonumber\\
&& ~~~~~~~~+\bar{v}^{*}(\eta_{f},-{\bf k_{1}})\bar{v}^{*}(\eta_{f},-{\bf k_{2}})\bar{v}^{*}(\eta_{f},-{\bf k_{3}})\bar{v}(\eta,-{\bf k_{1}})\bar{v}(\eta,-{\bf k_{2}})\bar{v}^{'}(\eta,-{\bf k_{3}})\nonumber\\ &&~~~~~~~~~~~+\bar{v}(\eta_{f},{\bf k_{1}})\bar{v}(\eta_{f},{\bf k_{2}})\bar{v}(\eta_{f},{\bf k_{3}})\bar{v}^{*}(\eta,{\bf k_{1}})\bar{v}^{*}(\eta,{\bf k_{3}})\bar{v}^{*'}(\eta,{\bf k_{2}})\nonumber\\
&& ~~~~~~~+\bar{v}^{*}(\eta_{f},-{\bf k_{1}})\bar{v}^{*}(\eta_{f},-{\bf k_{2}})\bar{v}^{*}(\eta_{f},-{\bf k_{3}})\bar{v}(\eta,-{\bf k_{1}})\bar{v}(\eta,-{\bf k_{3}})\bar{v}^{'}(\eta,-{\bf k_{2}}),\\
 Y_{2}(k_1,k_2,k_3)&=&\bar{v}(\eta_{f},{\bf k_{1}})\bar{v}(\eta_{f},{\bf k_{2}})\bar{v}(\eta_{f},{\bf k_{3}})\bar{v}^{*}(\eta,{\bf k_{2}})\bar{v}^{*}(\eta,{\bf k_{1}})\bar{v}^{*'}(\eta,{\bf k_{3}})\nonumber\\
&& ~~~~~~~~~~~~~~~~~+\bar{v}^{*}(\eta_{f},-{\bf k_{1}})\bar{v}^{*}(\eta_{f},-{\bf k_{2}})\bar{v}^{*}(\eta_{f},-{\bf k_{3}})\bar{v}(\eta,-{\bf k_{2}})\bar{v}(\eta,-{\bf k_{1}})\bar{v}^{'}(\eta,-{\bf k_{3}})
\nonumber\\&& ~~~~~~~~~~~~~~~~~   
+\bar{v}(\eta_{f},{\bf k_{1}})\bar{v}(\eta_{f},{\bf k_{2}})\bar{v}(\eta_{f},{\bf k_{3}})\bar{v}^{*}(\eta,{\bf k_{2}})\bar{v}^{*}(\eta,{\bf k_{3}})\bar{v}^{*'}(\eta,{\bf k_{1}})\nonumber\\
&& ~~~~~~~~~~~~~~~~~+\bar{v}^{*}(\eta_{f},-{\bf k_{1}})\bar{v}^{*}(\eta_{f},-{\bf k_{2}})\bar{v}^{*}(\eta_{f},-{\bf k_{3}})\bar{v}(\eta,-{\bf k_{2}})\bar{v}(\eta,-{\bf k_{3}})\bar{v}^{'}(\eta,-{\bf k_{1}}),\\ 
Y_{3}(k_1,k_2,k_3)&=&\bar{v}(\eta_{f},{\bf k_{1}})\bar{v}(\eta_{f},{\bf k_{2}})\bar{v}(\eta_{f},{\bf k_{3}})\bar{v}^{*}(\eta,{\bf k_{3}})\bar{v}^{*}(\eta,{\bf k_{1}})\bar{v}^{'*}(\eta,{\bf k_{2}})\nonumber\\
&& ~~~~~~~~~~~~~~~~~+\bar{v}^{*}(\eta_{f},-{\bf k_{1}})\bar{v}^{*}(\eta_{f},-{\bf k_{2}})\bar{v}^{*}(\eta_{f},-{\bf k_{3}})\bar{v}(\eta,-{\bf k_{3}})\bar{v}(\eta,-{\bf k_{1}})\bar{v}^{'}(\eta,-{\bf k_{2}})\nonumber\\ &&~~~~~~~~~~~+\bar{v}(\eta_{f},{\bf k_{1}})\bar{v}(\eta_{f},{\bf k_{2}})\bar{v}(\eta_{f},{\bf k_{3}})\bar{v}^{*}(\eta,{\bf k_{3}})\bar{v}^{*}(\eta,{\bf k_{2}})\bar{v}^{*'}(\eta,{\bf k_{1}})\nonumber\\
&& ~~~~~~~~~~~~~~~~~+\bar{v}^{*}(\eta_{f},-{\bf k_{1}})\bar{v}^{*}(\eta_{f},-{\bf k_{2}})\bar{v}^{*}(\eta_{f},-{\bf k_{3}})\bar{v}(\eta,-{\bf k_{3}})\bar{v}^{'}(\eta,-{\bf k_{2}})\bar{v}^{'}(\eta,-{\bf k_{1}}),~~~~~~~~\eea	
\end{small}	
				   	   		   	   	  	  
Here we get the following contribution in the three-point function for scalar fluctuations\footnote{Here it is important to point out that in de Sitter space if we consider the Bunch–Davies vacuum state then here only the term with $1/K^3$ will appear explicitly in the expression for the three-point function and in the associated bispectrum. On~the other hand, if we consider all other non-trivial quantum vacuum states in our computation, then the rest of the contribution will explicitly appear. From~the perspective of observation, this is obviously important information as for the non-trivial quantum vacuum state we get additional contribution in the bispectrum which may enhance the amplitude of the non-Gaussianity in squeezed limiting configuration. Additionally, it is important to mention that in quasi de Sitter case we get an extra contribution  $1/(1+\epsilon)^5$. Also the factor $1/(k_1 k_2 k_3)^3$ will be replaced by  $1/(k_1 k_2 k_3)^{2\nu}$ and $1/\tilde{c}^{2}_S$ is replaced by $1/\tilde{c}^{6\nu-7}_S$. Consequently, in~quasi de Sitter case this contribution in the bispectrum can be recast as:
\begin{scriptsize}\be\begin{array}{rl} \Theta_{5}(k_{1},k_{2},k_{3})&=\Theta_{4}(k_{1},k_{2},k_{3})=\displaystyle -\frac{H^2}{64\tilde{c}^{6\nu-7}_{S}\epsilon^3M^6_p (1+\epsilon)^5}\frac{1}{(k_{1}k_{2}k_{3})^{2\nu}}\left[k^2_2({\bf k_2}.{\bf k_3}){\cal F}_{1}(k_1,k_2,k_3)+k^2_3({\bf k_2}.{\bf k_3}){\cal F}_{2}(k_1,k_2,k_3)\right.\\ & \left.~~~\displaystyle+k^2_1({\bf k_1}.{\bf k_3}){\cal F}_{3}(k_1,k_2,k_3)+k^2_3({\bf k_1}.{\bf k_3}){\cal F}_{4}(k_1,k_2,k_3)+k^2_1({\bf k_1}.{\bf k_2}){\cal F}_{5}(k_1,k_2,k_3)+k^2_2({\bf k_1}.{\bf k_2}){\cal F}_{6}(k_1,k_2,k_3)\right],\end{array}~~~~~~~~~~~~\ee	\end{scriptsize}}:
	\bea\label{zokl5} \Theta_{5}(k_{1},k_{2},k_{3})=\Theta_{4}(k_{1},k_{2},k_{3}),~~~~~~~~~~~~\eea	
	 where $\Theta_{4}(k_{1},k_{2},k_{3})$ 
	is defined earlier. Here the result is exactly same as derived for the coefficient $\alpha_4$.
 
\subsection{Limiting Configurations of Scalar~Bispectrum}
\label{v4b}
To analyze the features of the bispectrum computed from the present setup here we further consider the following two~configurations:
\subsubsection{Equilateral Limit~Configuration}
\label{v4b1}
Equilateral limit configuration is characterized by the condition, 
$k_{1}=k_{2}=k_{3}=k,$
where $k_{i}=|{\bf k_{i}}|\forall i=1,2,3$. Consequently,  we have,
$K=3k.$

For this case, the~bispectrum can be written as:
\be {B_{EFT}(k,k,k)=\sum^{5}_{j=1}\alpha_{j}\Theta_{j}(k,k,k)}~,\ee
where $\alpha_{j}\forall j=1,2,\cdots,5$ are defined earlier and $\Theta_{j}(k,k,k)\forall j=1,2,\cdots,5$ are given by:
\bea \Theta_{1}(k,k,k)&=& \frac{3H^2}{16\epsilon^3M^6_p}\frac{1}{k^{6}}\left[\frac{1}{27}U_1-3U_2\right],\\ \Theta_{2}(k,k,k)&=&  -\frac{3H^2}{64\epsilon^3M^6_p \tilde{c}^2_{S}}\frac{1}{k^{6}}\left[\frac{17}{27}U_1-3U_2\right],\\  \Theta_{3}(k,k,k)&=& \frac{3H}{32\epsilon^3M^6_p }\frac{1}{k^6}\left[\frac{10}{9}U_1-\frac{22}{49}U_2\right],\\  \Theta_{4}(k,k,k)&=& \frac{3H^2}{64\tilde{c}^2_{S}\epsilon^3M^6_p}\frac{1}{k^6}\left[\frac{17}{27}U_1-3U_2\right]=\Theta_{5}(k,k,k),\eea
 where $U_{1}$ and $U_{2}$ are defined as:
\bea U_1&=&\left[\left(C_{1}-C_{2}\right)^3\left(C^{*3}_{1}+C^{*3}_{2}\right)+\left(C^{*}_{1}-C^{*}_{2}\right)^3\left(C^{3}_{1}+C^{3}_{2}\right)\right],\nonumber\\ 
U_2&=&\left[\left(C_{1}-C_{2}\right)^3C^{*}_{1}C^{*}_{2}\left(C^{*}_{1}-C^{*}_{2}\right)+\left(C^{*}_{1}-C^{*}_{2}\right)^3C_{1}C_{2}\left(C_{1}-C_{2}\right)\right].\eea

Furthermore, substituting the explicit expressions for $\alpha_{j}\forall j=1,2,\cdots,5$ and $\Theta_{j}(k,k,k)\forall j=1,2,\cdots,5$ we get the following expression for the bispectrum 
for scalar fluctuations:
\bea {B_{EFT}(k,k,k)=\frac{3H^2}{16\epsilon^3\tilde{c}_{S}M^6_p}\frac{1}{k^6}\sum^{2}_{p=1}f_{p}U_{p}}~,\eea
where $f_{p}\forall p=1,2$ are defined as:
\bea f_{1}&=&\frac{\tilde{c}_{s}}{27}\left\{\left(1-\frac{1}{c^2_{S}}\right)\dot{H}M^2_p +\frac{3}{2}\bar{M}^3_1 H-\frac{4}{3}M^4_3\right\} +\frac{17}{108\tilde{c}_{s}}\left\{\left(1-\frac{1}{c^2_{S}}\right)\dot{H}M^2_p +\frac{3}{2}\bar{M}^3_{1} H\right\}\nonumber\\
&&~~~~~~~~~~~~~~~~~~~~~~~~~~~~~~~~~~~~~~~~~~~~~~~~~~~~~~~~~~-\frac{5}{2}\bar{M}^3_{1}H\tilde{c}_{s}+\frac{17}{36\tilde{c}_{s}}\bar{M}^3_1 H,\\
f_{2}&=&-3\tilde{c}_{s}\left\{\left(1-\frac{1}{c^2_{S}}\right)\dot{H}M^2_p +\frac{3}{2}\bar{M}^3_1 H-\frac{4}{3}M^4_3\right\} -\frac{3}{4\tilde{c}_{s}}\left\{\left(1-\frac{1}{c^2_{S}}\right)\dot{H}M^2_p +\frac{3}{2}\bar{M}^3_{1} H\right\}\nonumber\\
&&~~~~~~~~~~~~~~~~~~~~~~~~~~~~~~~~~~~~~~~~~~~~~~~~~~~~~~~~~~+\frac{99}{98}\bar{M}^3_{1}H\tilde{c}_{s}-\frac{9}{4\tilde{c}_{s}}\bar{M}^3_1 H.\eea
\begin{itemize}
\item \underline{\bf {For Bunch–Davies vacuum}:} \\After setting $C_{1}=1$ and $C_{2}=0$ we get:
\bea U_{1}=2,~~~
U_{2}=0.\eea
Consequently, we get the following expression for the bispectrum 
for scalar fluctuations:
\bea B_{EFT}(k,k,k)&=&\frac{H^2}{4\epsilon\tilde{c}_{S}M^2_p}\frac{1}{M^4_p\epsilon^2}\frac{1}{k^6}\left[\frac{\tilde{c}_{s}}{18}\left\{\left(1-\frac{1}{c^2_{S}}\right)\dot{H}M^2_p +\frac{3}{2}\bar{M}^3_1 H-\frac{4}{3}M^4_3\right\}\nonumber\right.\\
&&\left.~~ +\frac{17}{72\tilde{c}_{s}}\left\{\left(1-\frac{1}{c^2_{S}}\right)\dot{H}M^2_p +\frac{3}{2}\bar{M}^3_{1} H\right\}-\frac{15}{4}\bar{M}^3_{1}H\tilde{c}_{s}+\frac{17}{24\tilde{c}_{s}}\bar{M}^3_1 H\right].~~~~~~~~~~\eea
For $\tilde{c}_{S}=1=c_{S}$ case we know that $M_2=0$ and $M_3=0$ which we have already shown earlier. As~a result the bispectrum for scalar fluctuation can be expressed in the following simplified form:
\bea B_{EFT}(k,k,k)&=&-\frac{H^2}{4\epsilon M^2_p}\frac{1}{M^4_p\epsilon^2}\frac{1}{k^6}\frac{125}{48}\bar{M}^3_1 H.\eea
For $\tilde{c}_{S}<1$ and $c_{S}<1$ case one can also recast the bispectrum 
for scalar fluctuations in the following simplified form:
\bea B_{EFT}(k,k,k)&=&\frac{H^2}{4\epsilon\tilde{c}_{S}M^2_p}\frac{1}{M^4_p\epsilon^2}\frac{1}{k^6}\bar{M}^3_1 H\left[\frac{\tilde{c}_{S}}{18}\left\{\frac{3}{2}+\frac{4}{3}\frac{\tilde{c}_{3}}{\tilde{c}_{4}}+\frac{2c^2_S}{\tilde{c}_{4}}\right\}\nonumber\right.\\
&&\left.~~~~~~~~~~~ +\frac{17}{72\tilde{c}_{S}}\left\{\frac{2c^2_S}{\tilde{c}_{4}}+\frac{3}{2}\right\} 
-\frac{15}{4}\tilde{c}_{S}+\frac{17}{24\tilde{c}_{S}}\right].\eea
\item \underline{\bf {For $~{\alpha,\beta}$ vacuum}:}\\After setting $C_{1}=\cosh\alpha$ and $C_{2}=e^{i\beta}~\sinh\alpha$ we get:
\bea U_{1}=J_{1}(\alpha,\beta),~~~
U_{2}=J_{2}(\alpha,\beta),\eea
where $J_{1}(\alpha,\beta)$ and $J_{2}(\alpha,\beta)$ are defined~earlier.

Consequently, we get the following expression for the bispectrum 
for scalar fluctuations:
\begin{small}
\bea B_{EFT}(k,k,k)&=&\frac{H^2}{4\epsilon\tilde{c}_{S}M^2_p}\frac{1}{M^4_p\epsilon^2}\frac{1}{k^6}\left[\left(\frac{\tilde{c}_{s}}{36}\left\{\left(1-\frac{1}{c^2_{S}}\right)\dot{H}M^2_p +\frac{3}{2}\bar{M}^3_1 H-\frac{4}{3}M^4_3\right\}\nonumber\right.\right.\\
&&\left.\left.~~~~ +\frac{17}{144\tilde{c}_{s}}\left\{\left(1-\frac{1}{c^2_{S}}\right)\dot{H}M^2_p +\frac{3}{2}\bar{M}^3_{1} H\right\}-\frac{15}{8}\bar{M}^3_{1}H\tilde{c}_{s}+\frac{17}{48\tilde{c}_{s}}\bar{M}^3_1 H\right)J_{1}(\alpha,\beta)\nonumber\right.\\
&&\left.~~~~~~~+\left(-\frac{9}{4}\tilde{c}_{s}\left\{\left(1-\frac{1}{c^2_{S}}\right)\dot{H}M^2_p +\frac{3}{2}\bar{M}^3_1 H-\frac{4}{3}M^4_3\right\}\nonumber\right.\right.\\
&&\left.\left.~~~ -\frac{9}{16\tilde{c}_{s}}\left\{\left(1-\frac{1}{c^2_{S}}\right)\dot{H}M^2_p +\frac{3}{2}\bar{M}^3_{1} H\right\}+\frac{297}{392}\bar{M}^3_{1}H\tilde{c}_{s}-\frac{27}{16\tilde{c}_{s}}\bar{M}^3_1 H\right)J_{2}(\alpha,\beta)\right].~~~~~~~~~\eea
\end{small}
For $\tilde{c}_{S}=1=c_{S}$ case we know that $M_2=0$ and $M_3=0$ which we have already shown earlier. As~a result the bispectrum for scalar fluctuation can be expressed in the following simplified form:
\bea B(k,k,k)&=&-\frac{H^2}{4\epsilon M^2_p}\frac{1}{M^4_p\epsilon^2}\frac{1}{k^6}\bar{M}^3_{1}H\left[\frac{125}{96}J_{1}(\alpha,\beta)+\frac{8073}{1568}J_{2}(\alpha,\beta)\right].\eea
For $\tilde{c}_{S}<1$ and $c_{S}<1$ case one can also recast the bispectrum 
for scalar fluctuations in the following simplified form:
\begin{small}
\bea B_{EFT}(k,k,k)&=&\frac{H^2}{4\epsilon\tilde{c}_{S}M^2_p}\frac{1}{M^4_p\epsilon^2}\frac{1}{k^6}\bar{M}^3_1 H\left[\left(\frac{\tilde{c}_{S}}{36}\left\{\frac{3}{2}+\frac{4}{3}\frac{\tilde{c}_{3}}{\tilde{c}_{4}}+\frac{2c^2_S}{\tilde{c}_{4}}\right\} 
 +\frac{17}{144\tilde{c}_{S}}\left\{\frac{2c^2_S}{\tilde{c}_{4}}+\frac{3}{2}\right\}\nonumber\right.\right. \\
&&\left.\left.~~~-\frac{15}{8}\tilde{c}_{S}+\frac{17}{48\tilde{c}_{S}}\right)J_{1}(\alpha,\beta)+\left(-\frac{9}{4}\tilde{c}_{S}\left\{\frac{3}{2}+\frac{4}{3}\frac{\tilde{c}_{3}}{\tilde{c}_{4}}+\frac{2c^2_S}{\tilde{c}_{4}}\right\}
 -\frac{9}{16\tilde{c}_{S}}\left\{\frac{2c^2_S}{\tilde{c}_{4}}+\frac{3}{2}\right\}\nonumber\right.\right. \\
&&\left.\left.~~~~~~~~~~~~~~~~~~~~~~~~~~~~~~~~~~~~~~~~~~~~+\frac{297}{392}\tilde{c}_{S}-\frac{27}{16\tilde{c}_{S}}\right)J_{2}(\alpha,\beta)\right].\eea
\end{small}
\end{itemize}		

\subsubsection{Squeezed Limit~Configuration}
\label{v4b2}
 Squeezed limit configuration is characterized by the condition,
$k_{1}\approx k_{2}(=k_{L})>>k_{3}(=k_{S}),$
where $k_{i}=|{\bf k_{i}}|\forall i=1,2,3$. Also $k_{L}$ and $k_{S}$ characterize long and short mode momentum, respectively. Consequently,  we have,
$K=2k_{L}+k_{S}.$
For this case, the~bispectrum can be written as:
\be {B_{EFT}(k_{L},k_{L},k_{S})=\sum^{5}_{j=1}\alpha_{j}\Theta_{j}(k_{L},k_{L},k_{S})}~,\ee
where $\alpha_{j}\forall j=1,2,\cdots,5$ are defined earlier and $\Theta_{j}(k_{L},k_{L},k_{S})\forall j=1,2,\cdots,5$ are given by:
\bea \Theta_{1}(k_{L},k_{L},k_{S})&\approx& \frac{3H^2}{128\epsilon^3M^6_p}\frac{1}{k^{5}_L k_S}\left[U_1-16U_2\left(\frac{k_L}{k_S}\right)^3\right],\\ \Theta_{2}(k_{L},k_{L},k_{S})&=&  -\frac{H^2}{64\epsilon^3M^6_p \tilde{c}^2_{S}}\frac{1}{k^{5}_Lk_S}\left\{
\frac{3}{4}\left[U_1-3U_2\right]+\frac{3}{4}\left[U_1-U_2 \left(1+\frac{8}{3}\left(\frac{k_L}{k_S}\right)^2\right)\right]\right.\\&& \left.~~~~~~~~~~~~~~~~~~+\frac{5}{4}\left(2-\left(\frac{k_S}{k_L}\right)^2\right)\left[U_1-U_2 \left(1-\frac{8}{5}\left(\frac{k_L}{k_S}\right)^3\right)\right]\right\},\nonumber\\  \Theta_{3}(k_{L},k_{L},k_{S})&=& \frac{H}{64\epsilon^3M^6_p }\frac{1}{k^{5}_Lk_S}\left\{3\left[U_1-U_2\right]+\left(\frac{k_L}{k_S}\right)^2\left[U_1-\left(1+8\left(\frac{k_L}{k_S}\right)\right)U_2\right]\right\},~~~~~~~~\\  \Theta_{4}(k_{L},k_{L},k_{S})&=& -\frac{H^2}{64\tilde{c}^2_{S}\epsilon^3M^6_p}\left\{
\frac{3}{4}\left(\frac{1}{k^5_Lk_S}+\frac{k_S}{k^7_L}\right)\left[U_1-U_2\left(1+\frac{4}{3}\left(\frac{k_L}{k_S}\right)^2\right)\right]\right.\nonumber\\&& \left.\nonumber~~~~~~~~~~~~~~~~~~+\frac{5}{4}\left(\frac{1}{k^5_Lk_S}+\frac{k_S}{k^7_L}\right)\left[U_1-U_2 \left(1-\frac{16}{5}\left(\frac{k_L}{k_S}\right)^2\right)\right]\right.\\&& \left.~~~~+\frac{3}{2k^3_Lk^3_S}\left(2-\left(\frac{k_S}{k_L}\right)^2\right)\left[U_1-U_2 \left(1+\frac{8}{3}\left(\frac{k_L}{k_S}\right)+\frac{4}{3}\left(\frac{k_L}{k_S}\right)^2\right)\right]\right\}\nonumber\\&=&\Theta_{5}(k_{L},k_{L},k_{S}),\eea where $U_{1}$ and $U_{2}$ are defined~earlier.
 
Furthermore, substituting the explicit expressions for $\alpha_{j}\forall j=1,2,\cdots,5$ and $\Theta_{j}(k_{L},k_{L},k_{S})\forall j=1,2,\cdots,5$ we get the following expression for the bispectrum 
for scalar fluctuations:
\bea B_{EFT}(k_{L},k_{L},k_{S})&=&\frac{H^2}{64\epsilon^3\tilde{c}_{S}M^6_p}\sum^{2}_{p=1}g_{p}(k_L,k_S)U_{p},\eea
where $g_{p}(k_L,k_S)\forall p=1,2$ are defined as:
\begin{footnotesize}
\bea g_{1}(k_L,k_S)&=&\frac{3\tilde{c}_{S}}{2k^5_Lk_S}\left\{\left(1-\frac{1}{c^2_{S}}\right)\dot{H}M^2_p +\frac{3}{2}\bar{M}^3_1 H-\frac{4}{3}M^4_3\right\} \nonumber\\
&&~~~~~~~~~+\frac{1}{\tilde{c}_{S}k^5_Lk_S}\left\{\left(1-\frac{1}{c^2_{S}}\right)\dot{H}M^2_p +\frac{3}{2}\bar{M}^3_{1} H\right\}\left(2-\frac{5}{4}\left(\frac{k_S}{k_L}\right)^2\right)\\
&&~~~-\frac{9}{2}\bar{M}^3_{1}H\frac{\tilde{c}_{S}}{k^5_Lk_S}\left(3+\left(\frac{k_L}{k_S}\right)^2\right)+\frac{3}{\tilde{c}_{S}}\bar{M}^3_1 H\left\{2\left(\frac{1}{k^5_Lk_S}+\frac{k_S}{k^7_L}\right)+\frac{3}{2k^3_Lk^3_S}\left(2-\left(\frac{k_S}{k_L}\right)^2\right)\right\},\nonumber\\
g_{2}(k_L,k_S)&=&\frac{24\tilde{c}_{S}}{k^5_Lk_S}\left\{\left(1-\frac{1}{c^2_{S}}\right)\dot{H}M^2_p +\frac{3}{2}\bar{M}^3_1 H-\frac{4}{3}M^4_3\right\}\left(\frac{k_L}{k_S}\right)^2 \nonumber\\
&&~+\frac{1}{\tilde{c}_{S}k^5_Lk_S}\left\{\left(1-\frac{1}{c^2_{S}}\right)\dot{H}M^2_p +\frac{3}{2}\bar{M}^3_{1} H\right\}\left\{3+2\left(\frac{k_S}{k_L}\right)^2\nonumber\right.\\
&&\left.~~~+\frac{5}{4}\left(2-\frac{5}{4}\left(\frac{k_S}{k_L}\right)^2\right)\left(1-\frac{8}{5}\left(\frac{k_S}{k_L}\right)^3\right)\right\}-\frac{9}{2}\bar{M}^3_{1}H\frac{\tilde{c}_{S}}{k^5_Lk_S}\left\{3+\left(\frac{k_L}{k_S}\right)^2\left(1+8\left(\frac{k_L}{k_S}\right)\right)\right\}\nonumber\\
&&~~~~~~~~~+\frac{3}{\tilde{c}_{S}}\bar{M}^3_1 H\left\{2\left(\frac{1}{k^5_Lk_S}+\frac{k_S}{k^7_L}\right)+\frac{3}{2k^3_Lk^3_S}\left(2-\left(\frac{k_S}{k_L}\right)^2\right)\right\}.~~~~~~\eea\end{footnotesize}
\begin{itemize}
\item \underline{\bf {For Bunch–Davies vacuum}:} \\After setting $C_{1}=1$ and $C_{2}=0$ we get:
\bea U_{1}=2,~~~
U_{2}=0.\eea
Consequently, we get the following expression for the bispectrum 
for scalar fluctuations:
\bea B_{EFT}(k_{L},k_{L},k_{S})&=&\frac{H^2}{4\epsilon\tilde{c}_{S}M^2_p}\frac{1}{8M^4_p\epsilon^2}\left[\frac{3\tilde{c}_{S}}{2k^5_Lk_S}\left\{\left(1-\frac{1}{c^2_{S}}\right)\dot{H}M^2_p +\frac{3}{2}\bar{M}^3_1 H-\frac{4}{3}M^4_3\right\} \nonumber\right.\\
&&\left.~~~~~~~~~+\frac{1}{\tilde{c}_{S}k^5_Lk_S}\left\{\left(1-\frac{1}{c^2_{S}}\right)\dot{H}M^2_p +\frac{3}{2}\bar{M}^3_{1} H\right\}\left(2-\frac{5}{4}\left(\frac{k_S}{k_L}\right)^2\right)\nonumber\right.\\
&&\left.~~~~~~~~~-\frac{9}{2}\bar{M}^3_{1}H\frac{\tilde{c}_{S}}{k^5_Lk_S}\left(3+\left(\frac{k_L}{k_S}\right)^2\right)\nonumber\right.\\
&&\left.~~~~~~~~~+\frac{3}{\tilde{c}_{S}}\bar{M}^3_1 H\left\{2\left(\frac{1}{k^5_Lk_S}+\frac{k_S}{k^7_L}\right)+\frac{3}{2k^3_Lk^3_S}\left(2-\left(\frac{k_S}{k_L}\right)^2\right)\right\}\right].~~~~~~~~~~~~~\eea
For $\tilde{c}_{S}=1=c_{S}$ case we know that $M_2=0$ and $M_3=0$ which we have already shown earlier. As~a result the bispectrum for scalar fluctuation can be expressed in the following simplified~form:
\begin{footnotesize}
\bea B_{EFT}(k_{L},k_{L},k_{S})&=&\frac{H^2}{4\epsilon M^2_p}\frac{1}{8M^4_p\epsilon^2}\bar{M}^3_1 H \left[\frac{9}{4k^5_Lk_S} 
+\frac{3}{2k^5_Lk_S} \left(2-\frac{5}{4}\left(\frac{k_S}{k_L}\right)^2\right)\right.\\
&&\left.~~-\frac{9}{2}\frac{1}{k^5_Lk_S}\left(3+\left(\frac{k_L}{k_S}\right)^2\right)+3\left\{2\left(\frac{1}{k^5_Lk_S}+\frac{k_S}{k^7_L}\right)+\frac{3}{2k^3_Lk^3_S}\left(2-\left(\frac{k_S}{k_L}\right)^2\right)\right\}\right].\nonumber\eea
\end{footnotesize}

For $\tilde{c}_{S}<1$ and $c_{S}<1$ case one can also recast the bispectrum 
for scalar fluctuations in the following simplified form:
\bea B_{EFT}(k_{L},k_{L},k_{S})&=&\frac{H^2}{4\epsilon\tilde{c}_{S}M^2_p}\frac{1}{8M^4_p\epsilon^2}\bar{M}^3_1 H\left[\frac{3\tilde{c}_{S}}{2k^5_Lk_S}\left\{\frac{3}{2}+\frac{4}{3}\frac{\tilde{c}_{3}}{\tilde{c}_{4}}+\frac{2c^2_S}{\tilde{c}_{4}}\right\}\nonumber\right.\\
&&\left.~~~~~~~~~+\frac{1}{\tilde{c}_{S}k^5_Lk_S}\left\{\frac{2c^2_S}{\tilde{c}_{4}}+\frac{3}{2}\right\}\left(2-\frac{5}{4}\left(\frac{k_S}{k_L}\right)^2\right)\nonumber\right.\\
&&\left.~~~~~~~~~-\frac{9}{2}\frac{\tilde{c}_{S}}{k^5_Lk_S}\left(3+\left(\frac{k_L}{k_S}\right)^2\right)\nonumber\right.\\
&&\left.~~~~~~~~~+\frac{3}{\tilde{c}_{S}}\left\{2\left(\frac{1}{k^5_Lk_S}+\frac{k_S}{k^7_L}\right)+\frac{3}{2k^3_Lk^3_S}\left(2-\left(\frac{k_S}{k_L}\right)^2\right)\right\}\right].~~~~~~~~~~~~~\eea
\item \underline{\bf {For $~{\alpha,\beta}$ vacuum}:}\\After setting $C_{1}=\cosh\alpha$ and $C_{2}=e^{i\beta}~\sinh\alpha$ we get:
\bea U_{1}=J_{1}(\alpha,\beta),~~~
U_{2}=J_{2}(\alpha,\beta),\eea
where $J_{1}(\alpha,\beta)$ and $J_{2}(\alpha,\beta)$ are defined~earlier.

Consequently, we get the following expression for the bispectrum 
for scalar fluctuations:
\bea B_{EFT}(k_{L},k_{L},k_{S})&=&\frac{H^2}{4\epsilon\tilde{c}_{S}M^2_p}\frac{1}{16M^4_p\epsilon^2}\left[g_{1}(k_L,k_S)J_{1}(\alpha,\beta)+g_{2}(k_L,k_S)J_{2}(\alpha,\beta)\right].~~~~~~\eea
For $\tilde{c}_{S}=1=c_{S}$ case we know that $M_2=0$ and $M_3=0$ which we have already shown earlier. As~a result the factors $g_{1}(k_L,k_S)$ and $g_{2}(k_L,k_S)$ appearing in the expression for bispectrum for scalar fluctuation can be expressed in the following simplified form:
\bea g_{1}(k_L,k_S)&=&\frac{9}{4k^5_Lk_S}\bar{M}^3_1 H+\frac{3}{2k^5_Lk_S}\bar{M}^3_{1} H\left(2-\frac{5}{4}\left(\frac{k_S}{k_L}\right)^2\right)\nonumber\\
&&~~~~~~~~~-\frac{9}{2}\bar{M}^3_{1}H\frac{1}{k^5_Lk_S}\left(3+\left(\frac{k_L}{k_S}\right)^2\right)\nonumber\\
&&~~~~~~~~~+3\bar{M}^3_1 H\left\{2\left(\frac{1}{k^5_Lk_S}+\frac{k_S}{k^7_L}\right)+\frac{3}{2k^3_Lk^3_S}\left(2-\left(\frac{k_S}{k_L}\right)^2\right)\right\},~~~~~~~~~~~\eea\bea
g_{2}(k_L,k_S)&=&\frac{36}{k^5_Lk_S}\bar{M}^3_1 H\left(\frac{k_L}{k_S}\right)^2 
+\frac{3}{2k^5_Lk_S}\bar{M}^3_{1} H\left\{3+2\left(\frac{k_S}{k_L}\right)^2\nonumber\right.\\
&&\left.~~~~~+\frac{5}{4}\left(2-\frac{5}{4}\left(\frac{k_S}{k_L}\right)^2\right)\left(1-\frac{8}{5}\left(\frac{k_S}{k_L}\right)^3\right)\right\}\nonumber\\
&&~~~~~~~~~-\frac{9}{2}\bar{M}^3_{1}H\frac{1}{k^5_Lk_S}\left\{3+\left(\frac{k_L}{k_S}\right)^2\left(1+8\left(\frac{k_L}{k_S}\right)\right)\right\}\nonumber\\
&&~~~~~~~~~+3\bar{M}^3_1 H\left\{2\left(\frac{1}{k^5_Lk_S}+\frac{k_S}{k^7_L}\right)++\frac{3}{2k^3_Lk^3_S}\left(2-\left(\frac{k_S}{k_L}\right)^2\right)\right\}.~~~~~~~~~~\eea
For $\tilde{c}_{S}<1$ and $c_{S}<1$ case one can also recast the factors $g_{1}(k_L,k_S)$ and $g_{2}(k_L,k_S)$ as appearing in the expression for bispectrum
for scalar fluctuations in the following simplified form:
\begin{small}
\bea g_{1}(k_L,k_S)&=&\frac{3\tilde{c}_{S}}{2k^5_Lk_S}\left\{\frac{3}{2}+\frac{4}{3}\frac{\tilde{c}_{3}}{\tilde{c}_{4}}+\frac{2c^2_S}{\tilde{c}_{4}}\right\} \nonumber\\
&&~~~~~~~~~+\frac{1}{\tilde{c}_{S}k^5_Lk_S}\left\{\frac{2c^2_S}{\tilde{c}_{4}}+\frac{3}{2}\right\}\left(2-\frac{5}{4}\left(\frac{k_S}{k_L}\right)^2\right)\nonumber\\
&&~~~~~~~~~-\frac{9}{2}\bar{M}^3_{1}H\frac{\tilde{c}_{S}}{k^5_Lk_S}\left(3+\left(\frac{k_L}{k_S}\right)^2\right)\nonumber\\
&&~~~~~~~~~+\frac{3}{\tilde{c}_{S}}\bar{M}^3_1 H\left\{2\left(\frac{1}{k^5_Lk_S}+\frac{k_S}{k^7_L}\right)+\frac{3}{2k^3_Lk^3_S}\left(2-\left(\frac{k_S}{k_L}\right)^2\right)\right\},~~~~~~~~~~~\\
g_{2}(k_L,k_S)&=&\frac{24\tilde{c}_{S}}{k^5_Lk_S}\left\{\frac{3}{2}+\frac{4}{3}\frac{\tilde{c}_{3}}{\tilde{c}_{4}}+\frac{2c^2_S}{\tilde{c}_{4}}\right\}\left(\frac{k_L}{k_S}\right)^2 \nonumber\\
&&~+\frac{1}{\tilde{c}_{S}k^5_Lk_S}\left\{\frac{2c^2_S}{\tilde{c}_{4}}+\frac{3}{2}\right\}\left\{3+2\left(\frac{k_S}{k_L}\right)^2 
+\frac{5}{4}\left(2-\frac{5}{4}\left(\frac{k_S}{k_L}\right)^2\right)\left(1-\frac{8}{5}\left(\frac{k_S}{k_L}\right)^3\right)\right\}\nonumber\\
&&~~~~~~~~~-\frac{9}{2}\bar{M}^3_{1}H\frac{\tilde{c}_{S}}{k^5_Lk_S}\left\{3+\left(\frac{k_L}{k_S}\right)^2\left(1+8\left(\frac{k_L}{k_S}\right)\right)\right\}\nonumber\\
&&~~~~~~~~~+\frac{3}{\tilde{c}_{S}}\bar{M}^3_1 H\left\{2\left(\frac{1}{k^5_Lk_S}+\frac{k_S}{k^7_L}\right)++\frac{3}{2k^3_Lk^3_S}\left(2-\left(\frac{k_S}{k_L}\right)^2\right)\right\}.~~~~~~~~~~\eea
\end{small}
\end{itemize}

\section{{Determination of EFT Coefficients and Future~Predictions }}
\label{v5}
In this section, we compute the exact analytical expression for the EFT coefficients for two specific cases---1. Canonical single-field slow-roll inflation and 2. General single-field $P(X,\phi)$ models of inflation, where $ X=-\frac{1}{2}g^{\mu\nu}\partial_{\mu}\phi\partial_{\nu}\phi$ is the kinetic term. To~determine the EFT coefficients for the canonical single-field slow-roll model or from general single-field $P(X,\phi)$ model of inflation we will follow the following~strategy:
	\begin{enumerate}
	\item First, we will start with the general expression for the three-point function and the bispectrum for scalar perturbations with an arbitrary choice of quantum vacuum. Then we take the {Bunch–Davies} and {$\alpha,\beta$} vacuum to match with the standard results of scalar three-point function.
	\item Next we take the {equilateral limit} and {squeezed limit} configuration of the bispectrum obtained from the single-field slow-roll model and general single-field $P(X,\phi)$ model.
	\item Furthermore, we equate the {equilateral limit} and {squeezed limit} configuration of the bispectrum computed from the EFT of inflation with the single-field slow-roll or from the general single-field $P(X,\phi)$ model.
	\item Finally, for~sound speed $c_{S}=1$ and $c_{S}<1$ we get the analytical expressions for all the EFT coefficients for canonical single-field slow-roll models or from generalized single-field $P(X,\phi)$ models of inflation.
\end{enumerate}
\subsection{For Canonical Single-Field Slow-Roll~Inflation}
\label{v5b}	
Here our prime objective is to derive the EFT coefficients by computing the most general expression for the three-point function for scalar fluctuations from the canonical single-field slow-roll model of inflation for arbitrary vacuum. Then we give specific example for {Bunch–Davies} and {$\alpha,\beta$} vacuum for~completeness.
\subsubsection{Basic~Setup}
\label{v5b1}
Let us start with the action for single scalar field (inflaton) which has canonical kinetic term as given by:
\be\label{ouw1} {\textcolor{red}{\bf Canonical~model:}~~S=\int d^4x\sqrt{-g}\left[\frac{M^2_p}{2}R+X-V(\phi)\right]}~,\ee
where $V(\phi)$ is the potential which satisfies the slow-roll condition for~inflation.

It is important to mention here that perturbations to the homogeneous situation discussed above are introduced in the ADM
formalism where the metric takes the form~\cite{Maldacena:2002vr}:
\be\label{gr1} {\textcolor{red}{\bf ADM~metric:}~~ds^{2}= -N^{2}dt^{2} + g_{ij} \left(dx^{i} + N^{i} dt\right)\left(dx^{j} + N^{j} dt \right)}~,\ee 
where $g_{ij}$ is the metric on the spatial three-surface characterized by $t$, lapse $N$ and shift $N_i$. Here we choose synchronous gauge to maintain diffeomorphism invariance of the theory where the gauge-fixing conditions are given by:
\bea\label{io1} {\textcolor{red}{\bf Synchronous~ gauge:}~~~~ N=1,~~~N^i=0}~,\eea
and the perturbed metric is given by:
\bea {g_{ij}=a^2(t)\left[(1+2\zeta(t,{\bf x}))\delta_{ij}+\gamma_{ij}\right],~~~~
\gamma_{ii}=0}~,\eea
where $\zeta(t,{\bf x})$ and $\gamma_{ij}$ are defined earlier. Here it is important to note that the~structure of $g_{ij}$ is exactly same that we have mentioned in the case of EFT framework discussed in this paper. Please note that in the context of ADM formalism one can treat the scalar field $\phi$, induced metric $g_{ij}$ as the dynamical variables. On~the other hand, $N$ and $N^i$ mimics the role of Lagrange multipliers in ADM formalism. Consequently, one needs to impose the equations of motion of $N, N^i$ as additional constraints
in the synchronous gauge where the gauge condition as stated in Equation~(\ref{io1}) holds good perfectly. More precisely, in~this context the equations
of motion of $N$ and $N^i$ correspond to time and spatial reparametrization~invariance.

Furthermore, using the ADM metric as stated in Equation~(\ref{gr1}), the~action for the single scalar field Equation~(\ref{ouw1}) can be recast as~\cite{Maldacena:2002vr}:
\bea\label{bnv} {S =\int d^4x\sqrt{-g}\left[\frac{M^2_p}{2}N~{}^{(3)}R-NV+\frac{1}{2N}\left(E_{ij}E^{ij}-E^2\right)+\frac{1}{2N}\left(\dot{\phi}-N^i\partial_i\phi\right)^2-Ng^{ij}\partial_{i}\phi\partial_{j}\phi\right]}~,~~~~~~\eea
where ${}^{(3)}R$ is the Ricci scalar curvature of the spatial slice. Also, here 
$E_{ij} $ and $E$ is defined as~\cite{Maldacena:2002vr}:
\bea E_{ij}:&=& \frac{1}{2}\left(\dot{g}_{ij}-\nabla_{i}N_j-\nabla_{j}N_i\right)=NK_{ij},\\
E:&=& E^{i}_{i}=g_{ij}E^{ij}=g_{ij}g^{im}g^{jn}E_{mn}=Ng_{ij}g^{im}g^{jn}K_{mn}.\eea

Here the covariant derivative $\nabla_{i}$, is taken with respect to the 3-metric $g_{ij}$. Also, in this context the extrinsic curvature $K_{ij}$ is defined as~\cite{Maldacena:2002vr}:
\bea K_{ij}&=&\frac{1}{N}E_{ij}=\frac{1}{2N}\left(\dot{g}_{ij}-\nabla_{i}N_j-\nabla_{j}N_i\right).\eea

Additionally, we choose the following two gauges:
\bea {\underline{\textcolor{red}{\bf Gauge~I}:}~~~~~~\delta\phi(t,{\bf x})= 0,~~~
\zeta(t,{\bf x})\neq 0,~~~
\partial_{i}\gamma_{ij}=0,~~~
\gamma_{ii}=0}~.\\ {\underline{\textcolor{red}{\bf Gauge~II}:}~~~~~~\delta\phi(t,{\bf x})\neq 0,~~~
\zeta(t,{\bf x})=0,~~~
\partial_{i}\gamma_{ij}=0,~~~
\gamma_{ii}=0}~.\eea

For our present computations, we will work in \textcolor{red}{\bf Gauge I} as this is exactly same as the unitary gauge that we have used in the context of EFT framework. Also, the tensor perturbation $\gamma_{ij}$ is exactly same for the unitary gauge that we have used for EFT~setup. 
\subsubsection{Scalar Three-Point Function for Single-Field Slow-Roll~inflation}
\label{v5b2}
Before computing the three-point function for scalar mode fluctuation here it is important to note that the two-point function for single-field slow-roll inflation is exactly same with the results obtained for EFT of inflation with sound speed $c_{S}=1$ and $\tilde{c}_{S}=1$, which can be obtained by setting the EFT coefficients, $M_2=0$, $M_3=0$, $\bar{M}_1\neq 0$,  $M_4\neq 0$, $\bar{M}_2\neq 0$,  $\bar{M}_3\neq 0$, $\tilde{c}_{5}=-\frac{1}{2}(1+\epsilon)$\footnote{In the case of single-field slow-roll inflation amplitude of power spectrum and spectral tilt for scalar fluctuation can be written at the horizon crossing $|k\eta|=1$ as:
\bea \underline{{\bf For~Bunch–Davies~vacuum:}}~~~~ 
\Delta_{\zeta}(k_*)&=& \left\{\begin{array}{ll}
									                    \displaystyle  \frac{V(\phi_*)}{24\pi^2~M^4_p\epsilon_{V}}~~~~ &
									 \mbox{\small {\bf for ~dS}}  \nonumber\\ 
										\displaystyle 2^{6\epsilon_{V}-2\eta_{V}}\frac{V(\phi_*)}{24\pi^2~M^4_p\epsilon_V(1+\epsilon_V)^2 }\left|\frac{\Gamma(\frac{3}{2}+4\epsilon_V-\eta_V)}{\Gamma\left(\frac{3}{2}\right)}\right|^2~~~~ & \mbox{\small {\bf for~ qdS}}.
									          \end{array}
									\right.\nonumber\eea
									
	\bea \underline{{\bf For~\alpha,\beta~vacuum:}}~~~~ 
										 \displaystyle \Delta_{\zeta}(k_*)&=&\left\{\begin{array}{ll}
										                    \displaystyle  \frac{V(\phi_*)}{24\pi^2~M^4_p\epsilon_{V}}\left[\cosh2\alpha -\sinh2\alpha\cos\beta\right]~~~~ &
										 \mbox{\small {\bf for ~dS}}  \nonumber\\ 
											\displaystyle 2^{6\epsilon_{V}-2\eta_{V}}\frac{V(\phi_*)}{24\pi^2~M^4_p\epsilon_V(1+\epsilon_V)^2 }\left|\frac{\Gamma(\frac{3}{2}+4\epsilon_V-\eta_V)}{\Gamma\left(\frac{3}{2}\right)}\right|^2\\
											\displaystyle\left[\cosh2\alpha -\sinh2\alpha\cos\left(\pi\left(2+4\epsilon_V-\eta_V\right)+\beta\right)\right]~~~~ & \mbox{\small {\bf for~ qdS}}.
										          \end{array}
										\right.\nonumber\eea 
										
								and \be n_{\zeta}(k_*)-1=2\eta_V-6\epsilon_V.\ee							 }. Using three-point function we can able to fix all of these coefficients.

 We here now proceed to calculate the three-point function for the scalar fluctuation $\zeta(t,{\bf x})$ in the interacting picture with arbitrary vacuum. Then we cite results for {Bunch–Davies} and {$\alpha,\beta$} vacuum.
For single-field slow-roll inflation, the third-order term in the action Equation~(\ref{bnv}) is given by~\cite{Maldacena:2002vr}:
\be {S^{(3)}_{\zeta}=\int d^4x~\left[ a^3 \epsilon^2\tilde{\zeta}\dot{\tilde{\zeta}}^2+ a \epsilon^2\tilde{\zeta}(\partial \tilde{\zeta})^2-2 a^3 \epsilon\dot{\tilde{\zeta}}\partial_{i}\tilde{\zeta}\partial_{i}( \epsilon\partial^{-2}\dot{\tilde{\zeta}})\right]}~,\ee
which is derived from Equation~(\ref{bnv}) and here after neglecting all the contribution from the terms which are sub-leading in the
slow-roll parameters. Additionally, here we use the following field redefinition:
\be \zeta=\tilde{\zeta}+\left\{\epsilon-\frac{\eta}{2}\right\}\tilde{\zeta}^2,\ee
where $\epsilon$, $\eta$, $\delta$ and $s$ are slow-roll parameters which are defined in the context of single-field slow-roll inflation as:
\bea \label{g1}\epsilon&\sim& \frac{1}{2M^2_p}\frac{\dot{\phi}^2}{H^2},~~
\eta\sim \epsilon-\delta,~~
\delta=\frac{\ddot{\phi}}{H\dot{\phi}},~~
 s=0.\eea 
 
Here one can also express the slow-roll parameters $\epsilon$ and $\eta$ in terms of the slowly varying potential $V(\phi)$ as, 
$ \epsilon \sim \epsilon_{V},$
     $\eta\sim \eta_{V}-\epsilon_{V},$
     $\delta\sim2\epsilon_{V}-\eta_{V}.$
     where the new slow-roll parameter $\epsilon_{V}$ and $\eta_{V}$ are defined as,
     $\epsilon_{V}=\frac{M^2_p}{2}\left(\frac{V^{'}(\phi)}{V(\phi)}\right)^2,~~~~
     \eta_{V}=M^2_p\left(\frac{V^{''}(\phi)}{V(\phi)}\right).$
Here $'$ represents $d/d\phi$.

Now it is important to note that in~the present context of discussion we are interested in the three-point function for the scalar fluctuation field $\zeta$, not for the redefined scalar field fluctuation $\tilde{\zeta}$ and for this reason one can write down the exact connection between the three-point function for the scalar function field $\zeta$ and redefined scalar fluctuation field $\tilde{\zeta}$ in position space as:
\bea \langle \zeta({\bf x_1})\zeta({\bf x_2})\zeta({\bf x_3})\rangle &=&\langle \tilde{\zeta}({\bf x_1})\tilde{\zeta}({\bf x_2})\tilde{\zeta}({\bf x_3})\rangle+\left(2\epsilon-\eta\right)\left[\langle \zeta({\bf x_1})\zeta({\bf x_2})\rangle\langle \zeta({\bf x_1})\zeta({\bf x_3})\rangle\right.\\&& \left.~~~~~~+\langle \zeta({\bf x_2})\zeta({\bf x_1})\rangle\langle \zeta({\bf x_2})\zeta({\bf x_3})\rangle+\langle \zeta({\bf x_3})\zeta({\bf x_1})\rangle\langle \zeta({\bf x_3})\zeta({\bf x_2})\rangle\right]\nonumber.\eea

After taking the Fourier transform of the scalar function field $\zeta$ and redefined scalar fluctuation field $\tilde{\zeta}$ one can express connection between three-point function in momentum space and this is also our main point of interest.

The interaction Hamiltonian for the redefined scalar fluctuation $\tilde{\zeta}$ can be
expressed as:
\be H_{int}=\int d^3x~\left[ a~\epsilon^2\tilde{\zeta}{\tilde{\zeta}}^{'2}+  a~\epsilon^2\tilde{\zeta}(\partial \tilde{\zeta})^2-2 a \epsilon\tilde{\zeta}^{'}\partial_{i}\tilde{\zeta}\partial_{i}( \epsilon\partial^{-2}\tilde{\zeta}^{'})\right].\ee

Furthermore, following the in-in formalism in interaction picture the expression for the three-point function for the redefined scalar fluctuation $\tilde{\zeta}$ and then transforming the final result in terms of the scalar fluctuation $\zeta$ in momentum one can write the following expression:
\be{\begin{array}{rl}\langle \zeta({\bf k_1})\zeta({\bf k_2})\zeta({\bf k_3})\rangle&=\displaystyle -i\int^{\eta_f=0}_{\eta_i=-\infty}d\eta~a(\eta)~\langle 0|\left[\zeta(\eta_f,{\bf k_1})\zeta(\eta_f,{\bf k_2})\zeta(\eta_f,{\bf k_3}),H_{int}(\eta)\right]|0\rangle\\
&=\displaystyle(2\pi)^3\delta^{(3)}({\bf k_1}+{\bf k_2}+{\bf k_3})B_{SFSR}(k_1,k_2,k_3)~,\end{array}}\ee
where $B_{SFSR}(k_1,k_2,k_3)$ represents the bispectrum of scalar fluctuation $\zeta$,
which is computed from single-field slow-roll inflation. Here the final expression for the bispectrum of scalar fluctuation for arbitrary vacuum is given by:
\be{\begin{array}{rl} B_{SFSR}(k_1,k_2,k_3)&=\displaystyle\frac{H^4}{32\epsilon^2M^4_p}\frac{1}{(k_1k_2k_3)^3}\left[2(2\epsilon-\eta)\left(|C_1|^2+|C_2|^2\right)^2\sum^{3}_{i=1}k^3_i\right.\\ &\left.\displaystyle+\epsilon\left(|C_1|^2-|C_2|^2\right)^2\left(-\sum^{3}_{i=1}k^3_i+\sum^{3}_{i,j=1,i\neq j}k_i k^2_j+\frac{8}{K}\sum^{3}_{i,j=1,i> j}k^2_i k^2_j\right)\right.\\ &\left.\displaystyle+\epsilon\left(C^{*}_1C_2+C_1C^{*}_2\right)^2\left(-\sum^{3}_{i=1}k^3_i+\sum^{3}_{i,j=1,i\neq j}k_i k^2_j\right.\right.\\ &\left.\left.~~~~~~~~~~~~~~~~~~~~~~~~~~~~~~~\displaystyle+8\sum^{3}_{i,j=1,i> j}k^2_i k^2_j\sum^{3}_{m=1}\frac{1}{K-2k_m}\right)\right]~.\end{array}}\ee

For {Bunch–Davies} and {$\alpha,\beta$} vacuum we get the following simplified expression for the bispectrum for scalar~fluctuation:
\begin{itemize}
\item \underline{{\bf For Bunch–Davies vacuum}:}\\
After setting $C_{1}=1$ and $C_{2}=0$ we get~\cite{Maldacena:2002vr}:
\bea B_{SFSR}(k_1,k_2,k_3)&=&\frac{H^4}{32\epsilon^2M^4_p}\frac{1}{(k_1k_2k_3)^3}\left[2(2\epsilon-\eta)\sum^{3}_{i=1}k^3_i\right.\nonumber\\ &&\left.+\epsilon\left(-\sum^{3}_{i=1}k^3_i+\sum^{3}_{i,j=1,i\neq j}k_i k^2_j+\frac{8}{K}\sum^{3}_{i,j=1,i> j}k^2_i k^2_j\right)\right].~~~~~~~~\eea 
\item \underline{{\bf For $~{\alpha,\beta}$ vacuum}:}\\
After setting $C_{1}=\cosh\alpha$ and $C_{2}=e^{i\beta}\sinh\alpha$ we get~\cite{Shukla:2016bnu}:
\begin{footnotesize}
\bea B_{SFSR}(k_1,k_2,k_3)&=&\frac{H^4}{32\epsilon^2M^4_p}\frac{1}{(k_1k_2k_3)^3}\left[2(2\epsilon-\eta)\cosh^2 2\alpha\sum^{3}_{i=1}k^3_i\right.\nonumber\\ &&\left.+\epsilon\left(-\sum^{3}_{i=1}k^3_i+\sum^{3}_{i,j=1,i\neq j}k_i k^2_j+\frac{8}{K}\sum^{3}_{i,j=1,i> j}k^2_i k^2_j\right)\right.\\ &&\left.+\epsilon\sinh^2 2\alpha\cos^2\beta\left(-\sum^{3}_{i=1}k^3_i+\sum^{3}_{i,j=1,i\neq j}k_i k^2_j+8\sum^{3}_{i,j=1,i> j}k^2_i k^2_j\sum^{3}_{m=1}\frac{1}{K-2k_m}\right)\right].\nonumber\eea
\end{footnotesize}
\end{itemize}

Furthermore, we consider {equilateral limit} and {squeezed limit} in which we finally~get:
\begin{enumerate}
\item \underline{{Equilateral limit configuration}:}\\
Here the bispectrum for scalar perturbations in the presence of arbitrary quantum vacuum can be expressed as:
\be {\begin{array}{rl}B_{SFSR}(k,k,k)&=\displaystyle\frac{H^4}{32\epsilon^2M^4_p}\frac{1}{k^6}\left[6(2\epsilon-\eta)\left(|C_1|^2+|C_2|^2\right)^2 
+11\epsilon\left(|C_1|^2-|C_2|^2\right)^2\right.\\ &\left.~~~~~~~~~~~~~~~~~~~~~~~~~~~~~~~~~~\displaystyle+27\epsilon\left(C^{*}_1C_2+C_1C^{*}_2\right)^2 
\right]~.\end{array} }\ee
Now for {Bunch–Davies} and {$\alpha,\beta$} vacuum we get the following simplified expression for the bispectrum for scalar~fluctuation:
\begin{itemize}
\item \underline{{\bf For Bunch–Davies vacuum}:}\\
After setting $C_{1}=1$ and $C_{2}=0$ we get:
\bea{ B_{SFSR}(k,k,k)=\frac{H^4}{32\epsilon^2M^4_p}\frac{1}{k^6}\left[23\epsilon-6\eta 
\right]}~.~~~~~~~~\eea
\item \underline{{\bf For $~{\alpha,\beta}$ vacuum}:}\\
After setting $C_{1}=\cosh\alpha$ and $C_{2}=e^{i\beta}\sinh\alpha$ we get:
\bea {B_{SFSR}(k,k,k)=\frac{H^4}{32\epsilon^2M^4_p}\frac{1}{k^6}\left[6(2\epsilon-\eta)\cosh^2 2\alpha 
+11\epsilon 
+27\epsilon\sinh^2 2\alpha\cos^2\beta 
\right]}~.~~~~~~~~\eea 
\end{itemize} 
\item \underline{{Squeezed limit configuration}:}\\
Here the bispectrum for scalar perturbations in the presence of arbitrary quantum vacuum can be expressed as:
\bea {B_{SFSR}(k_L,k_L,k_S)=\frac{H^4}{32\epsilon^2M^4_p}\frac{1}{k^3_Lk^3_S}\sum^{3}_{j=-1}a_{j}\left(\frac{k_S}{k_L}\right)^{j}}~,~~~~~~~~\eea
where the expansion coefficients $a_{j}\forall j=-1,\cdots,3$ for arbitrary vacuum are defined as:
\bea a_{-1}&=&16\epsilon\left(C^{*}_1C_2+C_1C^{*}_2\right)^2,\nonumber\\
a_{0}&=&4(2\epsilon-\eta)\left(|C_1|^2+|C_2|^2\right)^2+4\epsilon\left(|C_1|^2-|C_2|^2\right)^2 +4\epsilon\left(C^{*}_1C_2+C_1C^{*}_2\right)^2,\nonumber\\
a_{1}&=&34\epsilon\left(C^{*}_1C_2+C_1C^{*}_2\right)^2,~~
a_{2}=10\epsilon\left(|C_1|^2-|C_2|^2\right)^2+10\epsilon\left(C^{*}_1C_2+C_1C^{*}_2\right)^2,\nonumber\\
a_{3}&=&2(2\epsilon-\eta)\left(|C_1|^2+|C_2|^2\right)^2-5\epsilon\left(|C_1|^2-|C_2|^2\right)^2 -\epsilon\left(C^{*}_1C_2+C_1C^{*}_2\right)^2.\nonumber\eea
Now for {Bunch–Davies} and {$\alpha,\beta$} vacuum we get the following simplified expression for the bispectrum for scalar~fluctuation:
\begin{itemize}
\item \underline{{\bf For Bunch–Davies vacuum}:}\\
After setting $C_{1}=1$ and $C_{2}=0$, we get the following expression for the expansion coefficients $a_{j}\forall j=-1,\cdots,3$:
\bea a_{-1}&=&0,~~
a_{0}=4(3\epsilon-\eta),~~
a_{1}=0,~~
a_{2}=10\epsilon,~~
a_{3}=-(\epsilon+2\eta).\eea
Consequently, the bispectrum can be recast as:
\bea {B_{SFSR}(k_L,k_L,k_S)=\frac{H^4}{32\epsilon^2M^4_p}\frac{1}{k^3_Lk^3_S}\left[4(3\epsilon-\eta)+10\epsilon\left(\frac{k_S}{k_L}\right)^2-(\epsilon+2\eta)\left(\frac{k_S}{k_L}\right)^3\right]}~.~~~~~~~~~~~~~ \eea
\item \underline{{\bf For $~{\alpha,\beta}$ vacuum}:}\\
After setting $C_{1}=\cosh\alpha$ and $C_{2}=e^{i\beta}\sinh\alpha$, we get the following expression for the expansion coefficients $a_{j}\forall j=-1,\cdots,3$:
\bea a_{-1}&=&16\epsilon\sinh^2 2\alpha\cos^2\beta,~~
a_{0}=4(2\epsilon-\eta)\cosh^2 2\alpha+4\epsilon +4\epsilon\sinh^2 2\alpha\cos^2\beta,\nonumber\\
a_{1}&=&34\epsilon\sinh^2 2\alpha\cos^2\beta,~~
a_{2}=10\epsilon+10\epsilon\sinh^2 2\alpha\cos^2\beta,\nonumber\\
a_{3}&=&2(2\epsilon-\eta)\cosh^2 2\alpha-5\epsilon-\epsilon\sinh^2 2\alpha\cos^2\beta.\nonumber\eea
Consequently, the bispectrum can be recast as:
\begin{footnotesize}
\bea{\begin{array}{rl} B_{SFSR}(k_L,k_L,k_S)&=\displaystyle\frac{H^4}{32\epsilon^2M^4_p}\frac{1}{k^3_Lk^3_S}\left[16\epsilon\sinh^2 2\alpha\cos^2\beta\left(\frac{k_S}{k_L}\right)^{-1}\right.\displaystyle\\ &\left.\displaystyle+\left(4(2\epsilon-\eta)\cosh^2 2\alpha+4\epsilon +4\epsilon\sinh^2 2\alpha\cos^2\beta\right)\right.\\ &\left.~~~~
+34\epsilon\sinh^2 2\alpha\cos^2\beta\left(\frac{k_S}{k_L}\right)\displaystyle+\left(10\epsilon+10\epsilon\sinh^2 2\alpha\cos^2\beta\right)\left(\frac{k_S}{k_L}\right)^2\right.\displaystyle\\ &\left.~~~~~~~~\displaystyle+\left(2(2\epsilon-\eta)\cosh^2 2\alpha-5\epsilon-\epsilon\sinh^2 2\alpha\cos^2\beta\right)\left(\frac{k_S}{k_L}\right)^3\right].\end{array}}~~~~~~~\eea
\end{footnotesize}
\end{itemize}
\end{enumerate}

\subsubsection{Expression for EFT Coefficients for Single-Field Slow-Roll~Inflation}
\label{v5b3}
Here our prime objective is to derive the analytical expressions for EFT coefficients for single-field slow-roll inflation. To~serve this purpose we start with a claim that the three-point function and the associated bispectrum for the scalar fluctuations computed from single-field slow-roll inflation is exactly same as that we have computed from EFT setup for consistent UV completion. Here we use the {equilateral limit} and {squeezed limit} configurations to extract the analytical expression for the EFT coefficients. In~the two limiting cases the results are~as follows:
\begin{enumerate}
\item \underline{{Equilateral limit configuration}:}\\
For this case with arbitrary vacuum one can write:
\bea B_{EFT}(k,k,k)&=&B_{SFSR}(k,k,k),\eea
which implies that:
\begin{small}
\be {\begin{array}{rl}
 \bar{M}_1&=\left\{\frac{HM^2_p\epsilon\left[6(\eta-2\epsilon)\left(|C_1|^2+|C_2|^2\right)^2 
-11\epsilon\left(|C_1|^2-|C_2|^2\right)^2 
-27\epsilon\left(C^{*}_1C_2+C_1C^{*}_2\right)^2 
\right]}{\left[\frac{125}{12}U_1+\frac{8073}{196}U_2\right]}\right\}^{\frac{1}{3}},\\
\bar{M}_2&\approx \bar{M}_3=\sqrt{\frac{\bar{M}^3_1}{4H\tilde{c}_{5}}}=\left\{\frac{2M^2_p\epsilon\left[6(2\epsilon-\eta)\left(|C_1|^2+|C_2|^2\right)^2 
+11\epsilon\left(|C_1|^2-|C_2|^2\right)^2 
+27\epsilon\left(C^{*}_1C_2+C_1C^{*}_2\right)^2 
\right]}{\left(1+\epsilon\right)\left[\frac{125}{3}U_1+\frac{8073}{49}U_2\right]}\right\}^{\frac{1}{2}},\\
\tilde{c}_{5}&=-\frac{1}{2}\left(1+\epsilon\right),~~~
M_2=0,~~~~
M_3=0,\\
M_4&=\left(-\frac{\tilde{c}_{3}}{\tilde{c}_{6}}H\bar{M}^3_1\right)^{\frac{1}{4}}=\left\{\frac{\tilde{c}_{3}H^2M^2_p\epsilon\left[6(2\epsilon-\eta)\left(|C_1|^2+|C_2|^2\right)^2 
+11\epsilon\left(|C_1|^2-|C_2|^2\right)^2 
+27\epsilon\left(C^{*}_1C_2+C_1C^{*}_2\right)^2 
\right]}{\tilde{c}_{6}\left[\frac{125}{12}U_1+\frac{8073}{196}U_2\right]}\right\}^{\frac{1}{4}}.\end{array}}\ee
\end{small}
\hspace{-3pt}where for arbitrary vacuum $U_1$ and $U_2$ are defined as:
\bea U_1&=&\left[\left(C_{1}-C_{2}\right)^3\left(C^{*3}_{1}+C^{*3}_{2}\right)+\left(C^{*}_{1}-C^{*}_{2}\right)^3\left(C^{3}_{1}+C^{3}_{2}\right)\right],\\ 
U_2&=&\left[\left(C_{1}-C_{2}\right)^3C^{*}_{1}C^{*}_{2}\left(C^{*}_{1}-C^{*}_{2}\right)+\left(C^{*}_{1}-C^{*}_{2}\right)^3C_{1}C_{2}\left(C_{1}-C_{2}\right)\right].\eea
To constraint all these coefficients of EFT operators using CMB observations from Planck TT+low P data we use~\cite{Ade:2015lrj}:
\bea \label{xc1x} \epsilon&<&0.012 ~~(95\%~{\rm CL}),~
\eta=-0.0080^{+0.0088}_{-0.0146} ~~(68\%~{\rm CL}),~
c_S= 1~~(95\%~{\rm CL}),\nonumber\\
H=H_{inf}&\leq &1.09\times 10^{-4}~M_p~\sqrt{\epsilon~c_S},\eea
where $M_p=2.43\times 10^{18}~{\rm GeV}$ is the reduced Planck mass.
Now for {Bunch–Davies} and {$\alpha,\beta$} vacuum we get the following simplified expression for the bispectrum for scalar~fluctuation:
\begin{itemize}
\item \underline{{\bf For Bunch–Davies vacuum}:}\\
After setting $C_{1}=1$ and $C_{2}=0$ we get:
\begin{footnotesize}
\be{\begin{array}{rl}
 \bar{M}_1&=\left\{\frac{6}{125}HM^2_p\epsilon\left[6\eta 
-23\epsilon 
\right]\right\}^{\frac{1}{3}},~~
\bar{M}_2\approx\bar{M}_3=\sqrt{\frac{\bar{M}^3_1}{4H\tilde{c}_{5}}} 
=\left\{\frac{3}{125\left(1+\epsilon\right)}M^2_p\epsilon\left[23\epsilon-6\eta 
\right]\right\}^{\frac{1}{2}},~~~~~~~~~~~\\
\tilde{c}_{5}&=-\frac{1}{2}\left(1+\epsilon\right),~~~
M_2=0,~~
M_3=0,~~
M_4=\left(-\frac{\tilde{c}_{3}}{\tilde{c}_{6}}H\bar{M}^3_1\right)^{\frac{1}{4}} 
=\left\{\frac{6\tilde{c}_{3}}{125\tilde{c}_{6}}H^2M^2_p\epsilon\left[23\epsilon-6\eta 
\right]\right\}^{\frac{1}{4}}.\end{array}}\ee
\end{footnotesize}

Furthermore, using the constraint stated in Equation~(\ref{xc1x}) we finally get the following constraints on the coefficients of EFT operators:
\begin{footnotesize}
\bea &&1.23\times 10^{-3}~M_p<|\bar{M}_1|<1.41\times 10^{-3}~M_p,~~8.79\times 10^{-3}~M_p<|\bar{M}_2|\approx|\bar{M}_3|<1.08\times 10^{-2}~M_p,\nonumber\\
      &&M_2=0,~~~M_3=0,~~ 
      3.86\times 10^{-4}~M_p<M_4\times \left(-\tilde{c}_6/\tilde{c}_3\right)^{1/4}<4.29\times 10^{-4}~M_p.~~~~~~~~   \eea
      \end{footnotesize}
\item \underline{{\bf For $~{\alpha,\beta}$ vacuum}:}\\
After setting $C_{1}=\cosh\alpha$ and $C_{2}=e^{i\beta}\sinh\alpha$ we get:
\be{\begin{array}{rl}
 \bar{M}_1&=\left\{\frac{HM^2_p\epsilon\left[6(\eta-2\epsilon)\cosh^22\alpha 
-11\epsilon 
-27\epsilon\sinh^22\alpha\cos^2\beta 
\right]}{\left[\frac{125}{12}J_1(\alpha,\beta)+\frac{8073}{196}J_2(\alpha,\beta)\right]}\right\}^{\frac{1}{3}},~~~~~~~~~~~\\
\bar{M}_2&\approx \bar{M}_3=\sqrt{\frac{\bar{M}^3_1}{4H\tilde{c}_{5}}}=\left\{\frac{2M^2_p\epsilon\left[6(2\epsilon-\eta)\cosh^22\alpha 
+11\epsilon 
+27\epsilon\sinh^22\alpha\cos^2\beta 
\right]}{\left(1+\epsilon\right)\left[\frac{125}{3}J_1(\alpha,\beta)+\frac{8073}{49}J_2(\alpha,\beta)\right]}\right\}^{\frac{1}{2}},~~~~~~~~~~~\\
\tilde{c}_{5}&=\displaystyle-\frac{1}{2}\left(1+\epsilon\right),~~~
M_2=0,~~~~
M_3=0,\\
M_4&=\left(-\frac{\tilde{c}_{3}}{\tilde{c}_{6}}H\bar{M}^3_1\right)^{\frac{1}{4}}=\left\{\frac{\tilde{c}_{3}H^2M^2_p\epsilon\left[6(2\epsilon-\eta)\cosh^22\alpha
+11\epsilon 
+27\epsilon\sinh^22\alpha\cos^2\beta 
\right]}{\tilde{c}_{6}\left[\frac{125}{12}J_1(\alpha,\beta)+\frac{8073}{196}J_2(\alpha,\beta)\right]}\right\}^{\frac{1}{4}}.\end{array}}\ee

Furthermore, using the constraint stated in Equation~(\ref{xc1x}) we finally get the following constraints on the coefficients of EFT operators for a given value of the parameters $\alpha$ and $\beta$ (say for $\alpha=0.1$ and $\beta=0.1$):
\begin{footnotesize}
\bea &&9.1\times 10^{-4}~M_p<|\bar{M}_1|<1.1\times 10^{-3}~M_p,~~1.11\times 10^{-2}~M_p<|\bar{M}_2|\approx|\bar{M}_3|<1.5\times 10^{-2}~M_p,\nonumber\\
&&M_2=0,~~~M_3=0,~~  
3.06\times 10^{-4}~M_p<M_4\times \left(-\tilde{c}_6/\tilde{c}_3\right)^{1/4}<3.54\times 10^{-4}~M_p. ~~~~~~~~   \eea
\end{footnotesize}
\end{itemize} 
\item \underline{{Squeezed limit configuration}:}\\
For this case with arbitrary vacuum one can write:
\bea B_{EFT}(k_L,k_L,k_S)&=&B_{SFSR}(k_L,k_L,k_S),\eea
which implies that:
\begin{footnotesize}
\be{\begin{array}{rl} \bar{M}_1 &=\left\{2HM^2_p\epsilon\frac{\sum^{3}_{j=-1}a_{j}\left(\frac{k_S}{k_L}\right)^{j}}{\sum^{3}_{m=-1}b_{m}\left(\frac{k_S}{k_L}\right)^{m}}\right\}^{\frac{1}{3}},~
\bar{M}_2\approx \bar{M}_3=\sqrt{\frac{\bar{M}^3_1}{4H\tilde{c}_{5}}} 
 =\left\{-\frac{M^2_p\epsilon}{\left(1+\epsilon\right)}\frac{\sum^{3}_{j=-1}a_{j}\left(\frac{k_S}{k_L}\right)^{j}}{\sum^{3}_{m=-1}b_{m}\left(\frac{k_S}{k_L}\right)^{m}}\right\}^{\frac{1}{2}},~~~~~~~~~~~\\
 \tilde{c}_{5}&=-\frac{1}{2}\left(1+\epsilon\right),~~~
 M_2=0,~~~~
 M_3=0,~
 M_4=\left(-\frac{\tilde{c}_{3}}{\tilde{c}_{6}}H\bar{M}^3_1\right)^{\frac{1}{4}} 
 =\left\{\frac{2H^2M^2_p \epsilon~\tilde{c}_{3}}{\tilde{c}_{6}}\frac{\sum^{3}_{j=-1}a_{j}\left(\frac{k_S}{k_L}\right)^{j}}{\sum^{3}_{m=-1}b_{m}\left(\frac{k_S}{k_L}\right)^{m}}\right\}^{\frac{1}{4}}.\end{array}}\ee
 \end{footnotesize}
\hspace{-3pt}where the expansion coefficients $a_j\forall j=-1,\cdots,3$ are defined earlier and here the coefficients $b_{m}\forall m=-1,\cdots,3$ for arbitrary vacuum are defined as:
\bea {b_{-1}=-36U_2,~~
b_{0}=\frac{9}{2}\left(U_1+9U_2\right),~~
b_{1}=0,~~
b_{2}=\left(\frac{27}{4}U_{1}+\frac{17}{2}U_{2}\right),~~
b_{3}=0}~,\eea
where $U_1$ and $U_2$ are already defined~earlier.

Now for {Bunch–Davies} and {$\alpha,\beta$} vacuum we get the following simplified expression for the bispectrum for scalar~fluctuation:
\begin{itemize}
\item \underline{{\bf For Bunch–Davies vacuum}:}\\
After setting $C_{1}=1$ and $C_{2}=0$, we get $U_1=2$ and $U_2=0$. Consequently, the expansion coefficients can be recast as:
\bea {a_{-1}=0,~~
a_{0}=4(3\epsilon-\eta),~~
a_{1}=0,~~
a_{2}=10\epsilon,~~
a_{3}=-(\epsilon+2\eta)}~,\eea
and 
\bea{b_{-1}=0,~~
b_{0}=9,~~
b_{1}=0,~~
b_{2}=\frac{27}{2},~~
b_{3}=0}~,~~~~~~~~~~~~~\eea
Finally, the EFT coefficients for scalar fluctuation can be written as:
\be{\begin{array}{rl} \bar{M}_1 &=\left\{\frac{2HM^2_p\epsilon\left[4(3\epsilon-\eta)+10\epsilon\left(\frac{k_S}{k_L}\right)^2-(\epsilon+2\eta)\left(\frac{k_S}{k_L}\right)^3\right]}{\left[18-\frac{27}{2}\left(\frac{k_S}{k_L}\right)^2\right]}\right\}^{\frac{1}{3}},\\
\bar{M}_2&\approx \bar{M}_3=\sqrt{\frac{\bar{M}^3_1}{4H\tilde{c}_{5}}}=\left\{\frac{M^2_p\epsilon\left[4(\eta-3\epsilon)-10\epsilon\left(\frac{k_S}{k_L}\right)^2+(\epsilon+2\eta)\left(\frac{k_S}{k_L}\right)^3\right]}{\left(1+\epsilon\right)\left[18-\frac{27}{2}\left(\frac{k_S}{k_L}\right)^2\right]}\right\}^{\frac{1}{2}},~~~~~~~~~~~\\
 \tilde{c}_{5}&=-\frac{1}{2}\left(1+\epsilon\right),~~~
 M_2=0,~~~~
 M_3=0,\\
 M_4&=\left(-\frac{\tilde{c}_{3}}{\tilde{c}_{6}}H\bar{M}^3_1\right)^{\frac{1}{4}} =\left\{\frac{2H^2M^2_p \epsilon~\tilde{c}_{3}}{\tilde{c}_{6}}\frac{\left[4(\eta-3\epsilon)-10\epsilon\left(\frac{k_S}{k_L}\right)^2+(\epsilon+2\eta)\left(\frac{k_S}{k_L}\right)^3\right]}{\left[18-\frac{27}{2}\left(\frac{k_S}{k_L}\right)^2\right]}\right\}^{\frac{1}{4}}.\end{array}}\ee
Furthermore, using the constraint stated in Equation~(\ref{xc1x}) we finally get the following constraints on the coefficients of EFT operators for a given value of the parameter $k_S/k_L$ (say for $k_S/k_L=0.1$):
\begin{footnotesize}
\bea &&1.22\times 10^{-3}~M_p<|\bar{M}_1|<1.56\times 10^{-3}~M_p,~~
8.67\times 10^{-3}~M_p<|\bar{M}_2|\approx|\bar{M}_3|<1.25\times 10^{-2}~M_p,\nonumber\\
&&M_2=0,~~~M_3=0,~~ 
3.75\times 10^{-4}~M_p<M_4\times \left(-\tilde{c}_6/\tilde{c}_3\right)^{1/4}<4.51\times 10^{-4}~M_p.~~~~~~~~   \eea
\end{footnotesize}
 
\item \underline{{\bf For $~{\alpha,\beta}$ vacuum}:}\\
After setting $C_{1}=\cosh\alpha$ and $C_{2}=e^{i\beta}\sinh\alpha$, we get $U_1=J_{1}(\alpha,\beta)$ and $U_2=J_{2}(\alpha,\beta)$. Consequently, the expansion coefficients can be recast as:
\bea a_{-1}&=&16\epsilon\sinh^2 2\alpha\cos^2\beta,~~
a_{0}=4(2\epsilon-\eta)\cosh^2 2\alpha+4\epsilon +4\epsilon\sinh^2 2\alpha\cos^2\beta,\nonumber\\
a_{1}&=&34\epsilon\sinh^2 2\alpha\cos^2\beta,~~
a_{2}=10\epsilon+10\epsilon\sinh^2 2\alpha\cos^2\beta,\nonumber\\
a_{3}&=&2(2\epsilon-\eta)\cosh^2 2\alpha-5\epsilon-\epsilon\sinh^2 2\alpha\cos^2\beta.\eea
and 
\bea b_{-1}&=&-36J_{2}(\alpha,\beta),~
b_{0}=\frac{9}{2}\left(J_{1}(\alpha,\beta)+9J_{2}(\alpha,\beta)\right),~\nonumber\\
b_{2}&=&\left(\frac{27}{4}J_{1}(\alpha,\beta)+\frac{17}{2}J_{2}(\alpha,\beta)\right),~~
b_{3}=0=b_{1}.\eea
Finally, the EFT coefficients for scalar fluctuation can be written as:
\begin{footnotesize}
\bea {\begin{array}{rl} \bar{M}_1 &=\left\{2HM^2_p\epsilon\left[-36J_{2}(\alpha,\beta)\left(\frac{k_S}{k_L}\right)^{-1}+9\left(J_{1}(\alpha,\beta)+\frac{J_{2}(\alpha,\beta)}{2}\right)\right.\right.\\& \left.\left.-\frac{3}{4}\left(9J_{1}(\alpha,\beta)-7J_{2}(\alpha,\beta)\right)\left(\frac{k_S}{k_L}\right)^2\right]^{-1}\right.\\& \left.\left[16\epsilon\sinh^2 2\alpha\cos^2\beta\left(\frac{k_S}{k_L}\right)^{-1}+\left(4(2\epsilon-\eta)\cosh^2 2\alpha+4\epsilon +4\epsilon\sinh^2 2\alpha\cos^2\beta\right)\right.\right.\\& \left.\left.+34\epsilon\sinh^2 2\alpha\cos^2\beta\left(\frac{k_S}{k_L}\right)+\left(10\epsilon+10\epsilon\sinh^2 2\alpha\cos^2\beta\right)\left(\frac{k_S}{k_L}\right)^2\right.\right.\\& \left.\left.+\left(2(2\epsilon-\eta)\cosh^2 2\alpha-5\epsilon-\epsilon\sinh^2 2\alpha\cos^2\beta\right)\left(\frac{k_S}{k_L}\right)^3\right]\right\}^{\frac{1}{3}},\\
\bar{M}_2&\approx \bar{M}_3=\sqrt{\frac{\bar{M}^3_1}{4H\tilde{c}_{5}}}=\left\{\frac{M^2_p\epsilon}{(1+\epsilon)}\left[36J_{2}(\alpha,\beta)\left(\frac{k_S}{k_L}\right)^{-1}-9\left(J_{1}(\alpha,\beta)+\frac{J_{2}(\alpha,\beta)}{2}\right)\right.\right.\\& \left.\left.+\frac{3}{4}\left(9J_{1}(\alpha,\beta)-7J_{2}(\alpha,\beta)\right)\left(\frac{k_S}{k_L}\right)^2\right]^{-1}\right.\\& \left.\left[16\epsilon\sinh^2 2\alpha\cos^2\beta\left(\frac{k_S}{k_L}\right)^{-1}+\left(4(2\epsilon-\eta)\cosh^2 2\alpha+4\epsilon +4\epsilon\sinh^2 2\alpha\cos^2\beta\right)\right.\right.\\& \left.\left.+34\epsilon\sinh^2 2\alpha\cos^2\beta\left(\frac{k_S}{k_L}\right)+\left(10\epsilon+10\epsilon\sinh^2 2\alpha\cos^2\beta\right)\left(\frac{k_S}{k_L}\right)^2\right.\right.\\& \left.\left.+\left(2(2\epsilon-\eta)\cosh^2 2\alpha-5\epsilon-\epsilon\sinh^2 2\alpha\cos^2\beta\right)\left(\frac{k_S}{k_L}\right)^3\right]\right\}^{\frac{1}{2}},\\
 \tilde{c}_{5}&=-\frac{1}{2}\left(1+\epsilon\right),~~~
 M_2=0,~~~
 M_3=0,\\
 M_4&=\left(-\frac{\tilde{c}_{3}}{\tilde{c}_{6}}H\bar{M}^3_1\right)^{\frac{1}{4}}=\left\{\frac{2H^2M^2_p \epsilon~\tilde{c}_{3}}{\tilde{c}_{6}}\left[36J_{2}(\alpha,\beta)\left(\frac{k_S}{k_L}\right)^{-1}-9\left(J_{1}(\alpha,\beta)+\frac{J_{2}(\alpha,\beta)}{2}\right)\right.\right.\\& \left.\left.+\frac{3}{4}\left(9J_{1}(\alpha,\beta)-7J_{2}(\alpha,\beta)\right)\left(\frac{k_S}{k_L}\right)^2\right]^{-1}\right.\\& \left.\left[16\epsilon\sinh^2 2\alpha\cos^2\beta\left(\frac{k_S}{k_L}\right)^{-1}+\left(4(2\epsilon-\eta)\cosh^2 2\alpha+4\epsilon +4\epsilon\sinh^2 2\alpha\cos^2\beta\right)\right.\right.\\& \left.\left.+34\epsilon\sinh^2 2\alpha\cos^2\beta\left(\frac{k_S}{k_L}\right)+\left(10\epsilon+10\epsilon\sinh^2 2\alpha\cos^2\beta\right)\left(\frac{k_S}{k_L}\right)^2\right.\right.\\& \left.\left.+\left(2(2\epsilon-\eta)\cosh^2 2\alpha-5\epsilon-\epsilon\sinh^2 2\alpha\cos^2\beta\right)\left(\frac{k_S}{k_L}\right)^3\right]\right\}^{\frac{1}{4}}.\end{array}}~~~~~~~\eea
 \end{footnotesize}
 
Furthermore, using the constraint stated in Equation~(\ref{xc1x}) we finally get the following constraints on the coefficients of EFT operators for a given value of the parameters $\alpha$, $\beta$ and $k_S/k_L$ (say for $\alpha=0.1$, $\beta=0.1$ and $k_S/k_L=0.1$):
\begin{footnotesize}
\bea &&6.05\times 10^{-4}~M_p<|\bar{M}_1|<7.15\times 10^{-4}~M_p,~~
3.03\times 10^{-3}~M_p<|\bar{M}_2|\approx|\bar{M}_3|<3.89\times 10^{-3}~M_p,\nonumber\\
&&M_2=0,~~~M_3=0,~~  
2.22\times 10^{-3}~M_p<M_4\times \left(-\tilde{c}_6/\tilde{c}_3\right)^{1/4}<2.51\times 10^{-3}~M_p. ~~~~~~~~   \eea
\end{footnotesize}
\end{itemize}
\end{enumerate}
	\subsection{For General Single-Field $P(X,\phi)$ Inflation}
	\label{v5a}
Here our prime objective is to derive the EFT coefficients by computing the most general expression for the three-point function for scalar fluctuations from the general single-field $P(X,\phi)$ model of inflation for arbitrary vacuum. Then we give specific example for {Bunch–Davies} and {$\alpha,\beta$} vacuum for~completeness.
\subsubsection{Basic~Setup}
\label{v5a1}
Let us start with the action for single scalar field (inflaton) which is described by the general function $P(X,\phi)$, contains non-canonical kinetic term in general and it is minimally coupled to the gravity~\cite{Shukla:2016bnu,Chen:2006nt}:
\be\label{st1} {S=\int d^4x\sqrt{-g}\left[\frac{M^2_p}{2}R+P(X,\phi)\right]}~.\ee

In the case of general structure of $P(X,\phi)$ the pressure $p$, the~energy density $\rho$ and effective speed of sound parameter $c_{S}$ can be written as~\cite{Chen:2006nt}:
\bea p&=&P(X,\phi),~~~~
\rho=2XP_{,X}(X,\phi)-P(X,\phi),~~~~c_{S}=\sqrt{\frac{P_{,X}(X,\phi)}{P_{,X}(X,\phi)+2XP_{,XX}(X,\phi)}}.\eea

In the case of general $P(X,\phi)$ theory the slow-roll parameters can be expressed as~\cite{Chen:2006nt}:
\bea \epsilon&=&\frac{XP_{X}(X,\phi)}{H^2M^2_p},~~~
\eta=\epsilon-\frac{\delta}{P_{,X}(X,\phi)}\left[P_{,X}(X,\phi)+X P_{,XX}(X,\phi)\right],\nonumber\\
s&=&\frac{2X\delta}{P_{,X}(X,\phi)}\frac{\left[XP^2_{,XX}(X,\phi)-P_{,X}(X,\phi) P_{,XX}(X,\phi)-XP_{,X}(X,\phi) P_{,XXX}(X,\phi)\right]}{\left[P_{,X}(X,\phi)+X P_{,XX}(X,\phi)\right]}.~~~~\eea

In the case of \textcolor{red}{\bf single-field slow-roll inflation} we have:
\be P(X,\phi)=X-V(\phi),\ee
where $V(\phi)$ is the single-field slowly varying potential.
For this case if we compute the effective sound speed then it turns out to be $c_S=1$, which is consistent with our result obtained in the previous section. Also, if we compute the expressions for the slow-roll parameters $\epsilon,\eta,\delta$ and $s$ the results also perfectly match the results obtained in Equation~(\ref{g1}). 

Similarly, in case \textcolor{red}{\bf DBI inflationary model} one can identify the function $P(X,\phi)$ as~\cite{Alishahiha:2004eh}:
\be P(X,\phi)=-\frac{1}{f(\phi)}\sqrt{1-2Xf(\phi)}+\frac{1}{f(\phi)}-V(\phi),\ee
where the inflaton $\phi$ is identified to be the position of a D3 barne which is moving in warped throat geometry and $f(\phi)$ characterize the warp factor\footnote{For AdS-like throat geometry, $f(\phi)\approx \frac{\lambda}{\phi^4},$ where $\lambda$ is the parameter in string theory which depends on the flux number.}. For~ the effective potential $V(\phi)$ one can consider following mathematical structures of the potentials in the UV and IR regime~\cite{Alishahiha:2004eh}:
\begin{itemize}
\item \underline{\textcolor{red}{\bf UV regime}:}~
In this case, the inflaton moves from the UV regime of the warped geometric space to
the IR regime under the influence of the effective potential, 
$V(\phi)\simeq\frac{1}{2}m^2\phi^2,$ 
where the inflaton mass satisfies the constraint $m>>M_p\sqrt{\lambda}$. In~this specific situation the inflaton starts rolling very far away from the origin of the effective potential and then rolls down in a relativistic fashion to the
minimum of potential situated at the origin.
\item \underline{\textcolor{red}{\bf IR regime}:}~
In this case, the inflaton started moving from the IR regime of the warped space geometry to
the UV regime under the influence of the effective potential,
$V(\phi)\simeq V_0-\frac{1}{2}m^2\phi^2,$
where the inflaton mass is comparable to the scale of inflation, as~given by, $m\approx H$. In~this specific situation, the~inflaton starts rolling down near the origin of the effective potential and rolls down in a relativistic fashion away from it.
\end{itemize}

In the case of the DBI model the pressure $p$ and the energy density $\rho$ can be written as~\cite{Alishahiha:2004eh}:
\begin{small}
\bea p&=&\frac{1}{f(\phi)}(1-c_S)-V(\phi),~~~
\rho=\frac{1}{f(\phi)}\left(\frac{1}{c_S}-1\right)+V(\phi),~~~c_{S}=\sqrt{1-2Xf(\phi)}=\sqrt{1-\dot{\phi}^2f(\phi)},~~~~~~~~~~\eea
\end{small}
where $X=\dot{\phi}^2/2$. In~this context the slow-roll parameter~\cite{Alishahiha:2004eh}: \bea \epsilon&=&\frac{3\dot{\phi}^2}{2\left[c_SV(\phi)+\frac{1}{f(\phi)}(1-c_{S})\right]}\approx \frac{3}{2\left[1+c_S f(\phi)V(\phi)\right]}.\eea is not small and as a result the effective sound speed is very small, $c_{S}<<1$. Consequently, the inflaton speed during inflation is given by the expression, 
$\dot{\phi}=\pm \frac{1}{\sqrt{f(\phi)}}.$
Additionally, it is important to note that in the context of DBI inflation the other slow-roll parameters $\eta$ and $s$ can be computed as:
\begin{small}
\bea \eta &\approx&\frac{\left[3\sqrt{1+c_S f(\phi)V(\phi)}+\frac{\sqrt{3f(\phi)c_S}}{2}M_p\dot{\phi}c_S\left\{f(\phi)V^{'}(\phi)+V(\phi)f^{'}(\phi)-\frac{1}{2c^2_S}\left(2\ddot{\phi}f(\phi)+\dot{\phi}^2f^{'}(\phi)\right)\right\}\right]}{\left[1+c_S f(\phi)V(\phi)\right]^{\frac{3}{2}}},~~~~~~~~~\nonumber\\
    s&=&-\frac{\sqrt{3f(\phi)c_S}}{2c^2_S}M_p\dot{\phi}\left[2\ddot{\phi}f(\phi)+\dot{\phi}^2f^{'}(\phi)\right].\eea
    \end{small}
    
    In the slow-roll regime to validate slow-roll approximation along with $c_{S}<<1$ we need to satisfy the constraint condition for DBI inflation, 
    $2c_{S}f(\phi)V(\phi)>>1.$
\subsubsection{Scalar Three-Point Function for General Single-Field $P(X,\phi)$ Inflation}
\label{v5a2}
Before computing the three-point function for scalar mode fluctuation here it is important to note that the two-point function for general single-field $P(X,\phi)$ inflation is exactly same with the results obtained for EFT of inflation with sound speed $c_{S}<<1$ and $\tilde{c}_{S}<<1$, which can be obtained by setting the EFT coefficients, $M_2\neq0$, $M_3\neq 0$, $\bar{M}_1\neq 0$,  $M_4\neq 0$, $\bar{M}_2\neq 0$,  $\bar{M}_3\neq 0$, $\tilde{c}_{5}\neq-\frac{1}{2}(1+\epsilon)$\footnote{In the case of general single-field $P(X,\phi)$ inflation amplitude of power spectrum and spectral tilt for scalar fluctuation can be written at the horizon crossing $|k\tilde{c}_{S}\eta|=1$ as:
\begin{footnotesize}\bea \underline{{\textbf{For~Bunch–Davies~vacuum:}}}~~~~ 
\Delta_{\zeta}(k_*)&=& \left\{\begin{array}{ll}
									                    \displaystyle  \frac{2X_*P_{,X}(X_*,\phi_*)-P(X_*,\phi_*)}{24\pi^2~M^4_p\tilde{c}_S\epsilon}~~~~ &
									 \mbox{\small {\bf for ~dS}}  \nonumber\\ 
										\displaystyle 2^{3\epsilon-\eta+\frac{s}{2}}\frac{2X_*P_{,X}(X_*,\phi_*)-P(X_*,\phi_*)}{24\pi^2~M^4_p\epsilon(1+\epsilon)^2 }\left|\frac{\Gamma(\frac{3}{2}+3\epsilon-\eta+\frac{s}{2})}{\Gamma\left(\frac{3}{2}\right)}\right|^2~~~~ & \mbox{\small {\bf for~ qdS}}.
									          \end{array}
									\right.\nonumber\eea\end{footnotesize}
									
	\bea \underline{{\bf For~\alpha,\beta~vacuum:}}~~~~ 
										 \displaystyle \Delta_{\zeta}(k_*)&=&\left\{\begin{array}{ll}
										                    \displaystyle  \frac{2X_*P_{,X}(X_*,\phi_*)-P(X_*,\phi_*)}{24\pi^2~M^4_p\tilde{c}_S\epsilon}\left[\cosh2\alpha -\sinh2\alpha\cos\beta\right]~~~~ &
										 \mbox{\small {\bf for ~dS}}  \nonumber\\ 
											\displaystyle 2^{6\epsilon-2\eta+s}\frac{2X_*P_{,X}(X_*,\phi_*)-P(X_*,\phi_*)}{24\pi^2~M^4_p\epsilon(1+\epsilon)^2 }\left|\frac{\Gamma(\frac{3}{2}+3\epsilon-\eta+\frac{s}{2})}{\Gamma\left(\frac{3}{2}\right)}\right|^2\\
											\displaystyle\left[\cosh2\alpha -\sinh2\alpha\cos\left(\pi\left(2+3\epsilon-\eta+\frac{s}{2}\right)+\beta\right)\right]~~~~ & \mbox{\small {\bf for~ qdS}}.
										          \end{array}
										\right.\nonumber\eea 
										
								and \be n_{\zeta}(k_*)-1=2\eta-6\epsilon-s.\ee							 }. Using three-point function we can able to fix all of these~coefficients.
								
								Now here before going into the details of the computation for three-point function, just using the knowledge of the two-point function, we can easily identify the exact analytical expression for the EFT coefficient $M_2$. For~this, we need to identify the effective sound speed computed from general single-field $P(X,\phi)$ inflation with the result obtained for the proposed EFT setup. Consequently, we get:
	\bea 
										 \displaystyle M_2=\left(-\frac{XP_{,XX}(X,\phi)}{P_{,X}(X,\phi)}\dot{H}M^2_p\right)^{\frac{1}{4}}=\left\{\begin{array}{ll}
										                    \displaystyle  0~~~~ &
										 \mbox{\small {\bf for ~single-field slow-roll }}  \\ 
											\displaystyle\left[\left(\frac{\dot{\phi}^2f(\phi)}{\dot{\phi}^2f(\phi)-1}\right)\frac{\dot{H}M^2_p}{2}\right]^{\frac{1}{4}} ~~~~ & \mbox{\small {\bf for~ DBI}}.
										          \end{array}
										\right.\eea 
										
 We here now proceed to calculate the three-point function for the scalar fluctuation $\zeta(t,{\bf x})$ in the interacting picture with arbitrary vacuum in the case of general single-field $P(X,\phi)$ inflation. Then we cite results for {Bunch–Davies} and {$\alpha,\beta$} vacuum and give a specific example for the DBI model of~inflation.
 
 Here we introduce two new parameters~\cite{Chen:2006nt}:
 \bea \Sigma_{1}(X,\phi) &=& XP_{,X}(X,\phi)+2X^2P_{,XX}(X,\phi)=\frac{\epsilon H^2M^2_p}{c^2_S},\\
 \Sigma_{2}(X,\phi) &=& X^2P_{,XX}(X,\phi)+\frac{2}{3}X^3P_{,XXX}(X,\phi).
 \eea
  which will appear in the expression for three-point function for the scalar fluctuation. For~\textcolor{red}{\bf single-field slow-roll inflation} and \textcolor{red}{\bf DBI inflation} we get the following expressions for these parameters~\cite{Chen:2006nt}:
  	\bea 
  										 \displaystyle \Sigma_{1}(X,\phi)&=&\left\{\begin{array}{ll}
  										                    \displaystyle  X=\epsilon H^2 M^2_p ~~~~ &
  										 \mbox{\small {\bf for ~single-field slow-roll }}  \\ 
  											\displaystyle \frac{X}{\left(1-2Xf(\phi)\right)^{\frac{3}{2}}}=\frac{\epsilon H^2 M^2_p }{c^2_S} ~~~~ & \mbox{\small {\bf for~ DBI}}.
  										          \end{array}
  										\right.\\
  											 \displaystyle \Sigma_{2}(X,\phi)&=&\left\{\begin{array}{ll}
  											                    \displaystyle 0~~~~~~~~~~~~~~~~~~~~~~~~~~ ~~~~~~~~~~~~~  &
  											 \mbox{\small {\bf for ~single-field slow-roll }}  \\ 
  												\displaystyle \frac{X^2 f(\phi)}{\left(1-2Xf(\phi)\right)^{\frac{5}{2}}}~~~~ & \mbox{\small {\bf for~ DBI}}.
  											          \end{array}
  											\right.\eea 
  												
  																For general single-field $P(X,\phi)$ inflation the third-order term in the action Equation~(\ref{st1}) is given by~\cite{Chen:2006nt}:
\bea {\begin{array}{rl}S^{(3)}_{\zeta}&=\displaystyle\int d^4x~\left[-a^3\left\{\Sigma_{1}(X,\phi)\left(1-\frac{1}{c^2_S}\right)+2\Sigma_{2}(X,\phi)\right\}\frac{\dot{\tilde{\zeta}}^3}{H^3}+ \frac{a^3 \epsilon \left(\epsilon-3+3c^2_S\right)}{c^4_S}\tilde{\zeta}\dot{\tilde{\zeta}}^2\right.\\ & \left.~~~~~~~~~~~~~~~~~~~~~~~~~~~~\displaystyle+ \frac{a \epsilon \left(\epsilon-2s+1-c^2_S\right)}{c^2_S}\tilde{\zeta}(\partial \tilde{\zeta})^2-2 a^3 \epsilon\dot{\tilde{\zeta}}\partial_{i}\tilde{\zeta}\partial_{i}\left( \frac{\epsilon}{c^2_S}\partial^{-2}\dot{\tilde{\zeta}}\right)\right],\end{array}}~~~~~~\eea
which is derived from Equation~(\ref{bnv}) and here after neglecting all the contribution from the terms which are sub-leading in the
slow-roll parameters. Additionally, here we use the following field~redefinition:
\be \zeta=\tilde{\zeta}+\frac{1}{c^2_S}\left\{\epsilon-\frac{\eta}{2}\right\}\tilde{\zeta}^2,\ee
where $\epsilon$, $\eta$, $\delta$ and $s$ are already defined earlier for general single-field $P(X,\phi)$ inflation.

Now it is important to note that in~the present context of discussion we are interested in the three-point function for the scalar fluctuation field $\zeta$, not for the redefined scalar field fluctuation $\tilde{\zeta}$ and for this reason one can write down the exact connection between the three-point function for the scalar function field $\zeta$ and redefined scalar fluctuation field $\tilde{\zeta}$ in position space as:
\bea \langle \zeta({\bf x_1})\zeta({\bf x_2})\zeta({\bf x_3})\rangle &=&\langle \tilde{\zeta}({\bf x_1})\tilde{\zeta}({\bf x_2})\tilde{\zeta}({\bf x_3})+\frac{\left(2\epsilon-\eta\right)}{c^2_S}\left[\langle \zeta({\bf x_1})\zeta({\bf x_2})\rangle\langle \zeta({\bf x_1})\zeta({\bf x_3})\rangle\right.\nonumber\\&& \left.~~~~+\langle \zeta({\bf x_2})\zeta({\bf x_1})\rangle\langle \zeta({\bf x_2})\zeta({\bf x_3})\rangle+\langle \zeta({\bf x_3})\zeta({\bf x_1})\rangle\langle \zeta({\bf x_3})\zeta({\bf x_2})\rangle\right].\eea

After taking the Fourier transform of the scalar fluctuation field $\zeta$ and redefined scalar fluctuation field $\tilde{\zeta}$ one can express connection between three-point function in momentum space and this is also our main point of interest.

The interaction Hamiltonian for the redefined scalar fluctuation $\tilde{\zeta}$ can be
expressed as:
\bea H_{int}&=&\int d^3x~ \left[-\left\{\Sigma_{1}(X,\phi)\left(1-\frac{1}{c^2_S}\right)+2\Sigma_{2}(X,\phi)\right\}\frac{{\tilde{\zeta}}^{'3}}{H^3}+\frac{a~\epsilon \left(\epsilon-3+3c^2_S\right)}{c^4_S}\tilde{\zeta}{\tilde{\zeta}}^{'2}\right.\nonumber\\ && \left.~~~~~~~~~~~~~~~~~~~~~~~~~~~~+ \frac{a~\epsilon \left(\epsilon-2s+1-c^2_S\right)}{c^2_S} \tilde{\zeta}(\partial \tilde{\zeta})^2-2 a\tilde{\zeta}^{'}\partial_{i}\tilde{\zeta}\partial_{i}\left(\frac{\epsilon}{c^2_S}\partial^{-2}\tilde{\zeta}^{'}\right)\right].~~~~~~\eea

Furthermore, following the in-in formalism in interaction picture the expression for the three-point function for the redefined scalar fluctuation $\tilde{\zeta}$ and then transforming the final result in terms of the scalar fluctuation $\zeta$ in momentum one can write the following expression:
\bea {\begin{array}{rl}\langle \zeta({\bf k_1})\zeta({\bf k_2})\zeta({\bf k_3})\rangle&=\displaystyle-i\int^{\eta_f=0}_{\eta_i=-\infty}d\eta~a(\eta)~\langle 0|\left[\zeta(\eta_f,{\bf k_1})\zeta(\eta_f,{\bf k_2})\zeta(\eta_f,{\bf k_3}),H_{int}(\eta)\right]|0\rangle\\
&=(2\pi)^3\delta^{(3)}({\bf k_1}+{\bf k_2}+{\bf k_3})B_{GSF}(k_1,k_2,k_3),\end{array}}\eea
where $B_{GSF}(k_1,k_2,k_3)$ represents the bispectrum of scalar fluctuation $\zeta$,
which is computed from general single-field $P(X,\phi)$ inflation. Here the final expression for the bispectrum of scalar fluctuation for arbitrary vacuum is given by:
\vspace{12pt}

\begin{small}
\bea {\begin{array}{rl}B_{GSF}(k_1,k_2,k_3)&=\displaystyle\frac{H^4}{32\epsilon^2M^4_p}\frac{1}{(k_1k_2k_3)^3}\left[\frac{3}{2}\left(\frac{1}{c^2_S}-1-\frac{2\Sigma_{2}(X,\phi)}{\Sigma_{1}(X,\phi)}\right)\left(|C_1|^2+|C_2|^2\right)^2\frac{(k_1k_2k_3)^2}{K^3}\right.\\ &\left.\displaystyle
+\left(\frac{1}{c^2_S}-1\right)\left(|C_1|^2+|C_2|^2\right)^2\left(\sum^{3}_{i=1}k^3_i+\frac{4}{K^2}\sum^{3}_{i,j=1,i\neq j}k^2_i k^3_j-\frac{8}{K}\sum^{3}_{i,j=1,i> j}k^2_i k^2_j\right)\right.\\ &\left.\displaystyle
+\frac{s}{c^2_S}\left(|C_1|^2+|C_2|^2\right)^2\left(-2\sum^{3}_{i=1}k^3_i+\frac{4}{K^2}\sum^{3}_{i,j=1,i\neq j}k^2_i k^3_j-\frac{8}{K}\sum^{3}_{i,j=1,i> j}k^2_i k^2_j\right)
\right.\\ &\left.\displaystyle+2(2\epsilon-\eta)\left(|C_1|^2+|C_2|^2\right)^2\sum^{3}_{i=1}k^3_i\right.\\ &\left.\displaystyle+\epsilon\left(|C_1|^2-|C_2|^2\right)^2\left(-\sum^{3}_{i=1}k^3_i+\sum^{3}_{i,j=1,i\neq j}k_i k^2_j+\frac{8}{K}\sum^{3}_{i,j=1,i> j}k^2_i k^2_j\right)\right.\\ &\left.+\epsilon\left(C^{*}_1C_2+C_1C^{*}_2\right)^2\left(-\sum^{3}_{i=1}k^3_i+\sum^{3}_{i,j=1,i\neq j}k_i k^2_j\right.\right.\\ &\left.\left.~~~~~~~~~~~~~~~~~~~~~~~~~~~~~~~\displaystyle+8\sum^{3}_{i,j=1,i> j}k^2_i k^2_j\sum^{3}_{m=1}\frac{1}{K-2k_m}\right)+{\cal O}(\epsilon)\right],\end{array}}~~~~~~~~\eea
\end{small}
\hspace{-3pt}where ${\cal O}(\epsilon)$ characterizes the sub-leading corrections in the three-point function for the scalar fluctuation computed from general single-field $P(X,\phi)$ inflation.

Furthermore, we consider a very specific class of models, where the following constraint condition\footnote{Strictly speaking, the DBI model is one of the exceptions where this condition is not applicable. On~the other hand, in~the case of single-field slow-roll inflation this condition is applicable. But~in that case one can set $c_{S}=1$ and get back all the results derived in the earlier section. Additionally, it is important to mention that here we consider those models also where $c_{S}<<1$ along with this given constraint. } $P_{,X\phi}(X,\phi)=0$ perfectly holds good. In~this case one can write down the following simplified expression for the bispectrum of scalar fluctuation for arbitrary vacuum as:
\begin{small}
\bea {\begin{array}{rl} B_{GSF}(k_1,k_2,k_3)&=\displaystyle\frac{H^4}{32\epsilon^2M^4_p}\frac{1}{(k_1k_2k_3)^3}\left[\left(\frac{1}{c^2_S}-1-\frac{s\epsilon}{3\epsilon_{X}}\right)\left(|C_1|^2+|C_2|^2\right)^2\frac{(k_1k_2k_3)^2}{K^3}\right.\\ &\left.\displaystyle
+\left(\frac{1}{c^2_S}-1\right)\left(|C_1|^2+|C_2|^2\right)^2\left(\sum^{3}_{i=1}k^3_i+\frac{4}{K^2}\sum^{3}_{i,j=1,i\neq j}k^2_i k^3_j-\frac{8}{K}\sum^{3}_{i,j=1,i> j}k^2_i k^2_j\right)\right.\\ &\left.\displaystyle
+2(2\epsilon-\eta)\left(|C_1|^2+|C_2|^2\right)^2\sum^{3}_{i=1}k^3_i\right.\\ &\left.\displaystyle+\epsilon\left(|C_1|^2-|C_2|^2\right)^2\left(-\sum^{3}_{i=1}k^3_i+\sum^{3}_{i,j=1,i\neq j}k_i k^2_j+\frac{8}{K}\sum^{3}_{i,j=1,i> j}k^2_i k^2_j\right)\right.\\ &\left.\displaystyle+\epsilon\left(C^{*}_1C_2+C_1C^{*}_2\right)^2\left(-\sum^{3}_{i=1}k^3_i+\sum^{3}_{i,j=1,i\neq j}k_i k^2_j 
+8\sum^{3}_{i,j=1,i> j}k^2_i k^2_j\sum^{3}_{m=1}\frac{1}{K-2k_m}\right)\right],\end{array}}~~~~~~~~\eea
\end{small}
\hspace{-3pt}where the new parameter $\epsilon_{X}$ is defined as\footnote{For single-field slow-roll inflation the newly introduced parameter $\epsilon_{X}$ is computed as:
\be \epsilon_{X}=\epsilon(\eta-\epsilon)\approx \epsilon_{V}\left(\eta_{V}-2\epsilon_{V}\right).\ee}:
\be \epsilon_{X}=-\frac{\dot{X}H_{,X}}{H^2}.\ee

For {Bunch–Davies} and {$\alpha,\beta$} vacuum we get the following simplified expression for the bispectrum for scalar~fluctuation:
\begin{itemize}
\item \underline{{\bf For Bunch–Davies vacuum}:}\\
After setting $C_{1}=1$ and $C_{2}=0$ we get~\cite{Chen:2006nt}:
\bea {\begin{array}{rl} B_{GSF}(k_1,k_2,k_3)&=\displaystyle\frac{H^4}{32\epsilon^2M^4_p}\frac{1}{(_1k_2k_3)^3}\left[\frac{3}{2}\left(\frac{1}{c^2_S}-1-\frac{2\Sigma_{2}(X,\phi)}{\Sigma_{1}(X,\phi)}\right)\frac{(k_1k_2k_3)^2}{K^3}\right.\\ &\displaystyle\left.
+\left(\frac{1}{c^2_S}-1\right)\left(\sum^{3}_{i=1}k^3_i+\frac{4}{K^2}\sum^{3}_{i,j=1,i\neq j}k^2_i k^3_j-\frac{8}{K}\sum^{3}_{i,j=1,i> j}k^2_i k^2_j\right)\right.\\ &\displaystyle\left.
+\frac{s}{c^2_S}\left(-2\sum^{3}_{i=1}k^3_i+\frac{4}{K^2}\sum^{3}_{i,j=1,i\neq j}k^2_i k^3_j-\frac{8}{K}\sum^{3}_{i,j=1,i> j}k^2_i k^2_j\right)
\right.\\ &\displaystyle\left.+2(2\epsilon-\eta)\sum^{3}_{i=1}k^3_i\right.\\ &\displaystyle\left.+\epsilon\left(-\sum^{3}_{i=1}k^3_i+\sum^{3}_{i,j=1,i\neq j}k_i k^2_j+\frac{8}{K}\sum^{3}_{i,j=1,i> j}k^2_i  k^2_j\right)\right].\end{array}}~~~~~~~~\eea 
Furthermore, for restricted classes of the general single-field $P(X,\phi)$ model, which satisfies the constraint $P_{,X\phi}(X,\phi)=0$, one can further write down the following expression for the bispectrum:
\bea  {\begin{array}{rl}B_{GSF}(k_1,k_2,k_3)&=\displaystyle\frac{H^4}{32\epsilon^2M^4_p}\frac{1}{(k_1k_2k_3)^3}\left[\left(\frac{1}{c^2_S}-1-\frac{s\epsilon}{3\epsilon_{X}}\right)\frac{(k_1k_2k_3)^2}{K^3}\right.\\ &\displaystyle\left.
+\left(\frac{1}{c^2_S}-1\right)\left(\sum^{3}_{i=1}k^3_i+\frac{4}{K^2}\sum^{3}_{i,j=1,i\neq j}k^2_i k^3_j-\frac{8}{K}\sum^{3}_{i,j=1,i> j}k^2_i k^2_j\right)\right.\\ &\displaystyle\left.
+2(2\epsilon-\eta)\sum^{3}_{i=1}k^3_i\right.\\ &\displaystyle\left.+\epsilon\left(-\sum^{3}_{i=1}k^3_i+\sum^{3}_{i,j=1,i\neq j}k_i k^2_j+\frac{8}{K}\sum^{3}_{i,j=1,i> j}k^2_i k^2_j\right)\right],\end{array}}~~~~~~~~\eea
\item \underline{{\bf For $~{\alpha,\beta}$ vacuum}:}\\
After setting $C_{1}=\cosh\alpha$ and $C_{2}=e^{i\beta}\sinh\alpha$ we get~\cite{Shukla:2016bnu}:
\begin{small}
\bea {\begin{array}{rl} B_{GSF}(k_1,k_2,k_3)&=\displaystyle\frac{H^4}{32\epsilon^2M^4_p}\frac{1}{(k_1k_2k_3)^3}\left[\frac{3}{2}\left(\frac{1}{c^2_S}-1-\frac{2\Sigma_{2}(X,\phi)}{\Sigma_{1}(X,\phi)}\right)\cosh^2 2\alpha\frac{(k_1k_2k_3)^2}{K^3}\right.\\ &\displaystyle\left.
+\left(\frac{1}{c^2_S}-1\right)\cosh^2 2\alpha\left(\sum^{3}_{i=1}k^3_i+\frac{4}{K^2}\sum^{3}_{i,j=1,i\neq j}k^2_i k^3_j-\frac{8}{K}\sum^{3}_{i,j=1,i> j}k^2_i k^2_j\right)\right.\\ &\displaystyle\left.
+\frac{s}{c^2_S}\cosh^2 2\alpha\left(-2\sum^{3}_{i=1}k^3_i+\frac{4}{K^2}\sum^{3}_{i,j=1,i\neq j}k^2_i k^3_j-\frac{8}{K}\sum^{3}_{i,j=1,i> j}k^2_i k^2_j\right)
\right.\\ &\displaystyle\left.+2(2\epsilon-\eta)\cosh^2 2\alpha\sum^{3}_{i=1}k^3_i+\epsilon\left(-\sum^{3}_{i=1}k^3_i+\sum^{3}_{i,j=1,i\neq j}k_i k^2_j+\frac{8}{K}\sum^{3}_{i,j=1,i> j}k^2_i k^2_j\right)\right.\\ &\displaystyle\left.+\epsilon\sinh^2 2\alpha\cos^2\beta\left(-\sum^{3}_{i=1}k^3_i+\sum^{3}_{i,j=1,i\neq j}k_i k^2_j\displaystyle+8\sum^{3}_{i,j=1,i> j}k^2_i k^2_j\sum^{3}_{m=1}\frac{1}{K-2k_m}\right)\right].\end{array}}~~~~~~~~\eea
\end{small}
Furthermore, for restricted classes of the general single-field $P(X,\phi)$ model, which satisfies the constraint $P_{,X\phi}(X,\phi)=0$, one can further write down the following expression for the bispectrum:
\begin{small}
\bea  {\begin{array}{rl} B_{GSF}(k_1,k_2,k_3)&=\displaystyle\frac{H^4}{32\epsilon^2M^4_p}\frac{1}{(k_1k_2k_3)^3}\left[\left(\frac{1}{c^2_S}-1-\frac{s\epsilon}{3\epsilon_{X}}\right)\cosh^2 2\alpha\frac{(k_1k_2k_3)^2}{K^3}\right.\\ &\displaystyle\left.
+\left(\frac{1}{c^2_S}-1\right)\cosh^2 2\alpha\left(\sum^{3}_{i=1}k^3_i+\frac{4}{K^2}\sum^{3}_{i,j=1,i\neq j}k^2_i k^3_j-\frac{8}{K}\sum^{3}_{i,j=1,i> j}k^2_i k^2_j\right)\right.\\ &\displaystyle\left.
+2(2\epsilon-\eta)\cosh^2 2\alpha\sum^{3}_{i=1}k^3_i+\epsilon\left(-\sum^{3}_{i=1}k^3_i+\sum^{3}_{i,j=1,i\neq j}k_i k^2_j+\frac{8}{K}\sum^{3}_{i,j=1,i> j}k^2_i k^2_j\right)\right.\\ &\displaystyle\left.+\epsilon\sinh^2 2\alpha\cos^2\beta\left(-\sum^{3}_{i=1}k^3_i+\sum^{3}_{i,j=1,i\neq j}k_i k^2_j+8\sum^{3}_{i,j=1,i> j}k^2_i k^2_j\sum^{3}_{m=1}\frac{1}{K-2k_m}\right)\right],\end{array}}~~~~~~~~\eea
\end{small}
\end{itemize}

Furthermore, we consider {equilateral limit} and {squeezed limit} in which we finally~get:
\begin{enumerate}
\item \underline{{Equilateral limit configuration}:}\\
Here the bispectrum for scalar perturbations in the presence of arbitrary quantum vacuum can be expressed as:
\begin{small}
\bea  {\begin{array}{rl} B_{GSF}(k,k,k)&=\displaystyle\frac{H^4}{32\epsilon^2M^4_p}\frac{1}{k^6}\left[\frac{1}{18}\left(\frac{1}{c^2_S}-1-\frac{2\Sigma_{2}(X,\phi)}{\Sigma_{1}(X,\phi)}\right)\left(|C_1|^2+|C_2|^2\right)^2\right.\\ &\displaystyle\left.
-\frac{7}{3}\left(\frac{1}{c^2_S}-1\right)\left(|C_1|^2+|C_2|^2\right)^2 
-\frac{34}{3}\frac{s}{c^2_S}\left(|C_1|^2+|C_2|^2\right)^2
\right.\\ &\displaystyle\left.+6(2\epsilon-\eta)\left(|C_1|^2+|C_2|^2\right)^2 
+11\epsilon\left(|C_1|^2-|C_2|^2\right)^2+27\epsilon\left(C^{*}_1C_2+C_1C^{*}_2\right)^2 
\right].\end{array}}~~~~~~~~\eea
\end{small}

Furthermore, for restricted classes of the general single-field $P(X,\phi)$ model, which satisfies the constraint $P_{,X\phi}(X,\phi)=0$, one can further write down the following expression for the bispectrum:
\begin{footnotesize}
\be  {\begin{array}{rl} B_{GSF}(k_1,k_2,k_3)&=\displaystyle\frac{H^4}{32\epsilon^2M^4_p}\frac{1}{k^6}\left[\frac{1}{27}\left(\frac{1}{c^2_S}-1-\frac{s\epsilon}{3\epsilon_{X}}\right)\left(|C_1|^2+|C_2|^2\right)^2
-\frac{7}{3}\left(\frac{1}{c^2_S}-1\right)\left(|C_1|^2+|C_2|^2\right)^2\right.\\ &\left.\displaystyle
+6(2\epsilon-\eta)\left(|C_1|^2+|C_2|^2\right)^2 
+11\epsilon\left(|C_1|^2-|C_2|^2\right)^2+27\epsilon\left(C^{*}_1C_2+C_1C^{*}_2\right)^2 
\right],\end{array}}~~~~~~~~\ee
\end{footnotesize}

Now for {Bunch–Davies} and {$\alpha,\beta$} vacuum we get the following simplified expression for the bispectrum for scalar~fluctuation:
\begin{itemize}
\item \underline{{\bf For Bunch–Davies vacuum}:}\\
After setting $C_{1}=1$ and $C_{2}=0$ we get:
\begin{footnotesize}
\be \begin{array}{rl} B_{GSF}(k,k,k)&=\displaystyle\frac{H^4}{32\epsilon^2M^4_p}\frac{1}{k^6}\left[\frac{1}{18}\left(\frac{1}{c^2_S}-1-\frac{2\Sigma_{2}(X,\phi)}{\Sigma_{1}(X,\phi)}\right)-\frac{7}{3}\left(\frac{1}{c^2_S}-1\right) 
-\frac{34}{3}\frac{s}{c^2_S}
+23\epsilon-6\eta 
\right].\end{array}~~~~~~~~\ee
\end{footnotesize}

Furthermore, for restricted classes of the general single-field $P(X,\phi)$ model, which satisfies the constraint $P_{,X\phi}(X,\phi)=0$, we get:
\begin{small}
\bea  \begin{array}{rl} B_{GSF}(k_1,k_2,k_3)&=\displaystyle\frac{H^4}{32\epsilon^2M^4_p}\frac{1}{k^6}\left[\frac{1}{27}\left(\frac{1}{c^2_S}-1-\frac{s\epsilon}{3\epsilon_{X}}\right)
-\frac{7}{3}\left(\frac{1}{c^2_S}-1\right) 
+23\epsilon-6\eta 
\right],\end{array}~~~~~~~~\eea
\end{small}
\item \underline{{\bf For $~{\alpha,\beta}$ vacuum}:}\\
After setting $C_{1}=\cosh\alpha$ and $C_{2}=e^{i\beta}\sinh\alpha$ we get:
\bea  \begin{array}{rl} B_{GSF}(k,k,k)&=\displaystyle\frac{H^4}{32\epsilon^2M^4_p}\frac{1}{k^6}\left[\frac{1}{18}\left(\frac{1}{c^2_S}-1-\frac{2\Sigma_{2}(X,\phi)}{\Sigma_{1}(X,\phi)}\right)\cosh^2 2\alpha\right.\\ &\displaystyle\left.
-\frac{7}{3}\left(\frac{1}{c^2_S}-1\right)\cosh^2 2\alpha 
-\frac{34}{3}\frac{s}{c^2_S}\cosh^2 2\alpha
\right.\\ &\displaystyle\left.+6(2\epsilon-\eta)\cosh^2 2\alpha 
+11\epsilon 
+27\epsilon\sinh^2 2\alpha\cos^2\beta 
\right].\end{array}~~~~~~~~\eea

Furthermore, for restricted classes of the general single-field $P(X,\phi)$ model, which satisfies the constraint $P_{,X\phi}(X,\phi)=0$, we get: 
\begin{small}
\bea \begin{array}{rl} B_{GSF}(k,k,k)&=\displaystyle\frac{H^4}{32\epsilon^2M^4_p}\frac{1}{k^6}\left[\frac{1}{27}\left(\frac{1}{c^2_S}-1-\frac{s\epsilon}{3\epsilon_{X}}\right)\cosh^2 2\alpha 
-\frac{7}{3}\left(\frac{1}{c^2_S}-1\right)\cosh^2 2\alpha 
\right.\\ &\left.~~~~~~\displaystyle+6(2\epsilon-\eta)\cosh^2 2\alpha 
+11\epsilon 
+27\epsilon\sinh^2 2\alpha\cos^2\beta 
\right].\end{array}~~~~~~~~\eea
\end{small}
\end{itemize} 
\item \underline{{Squeezed limit configuration}:}\\
Here the bispectrum for scalar perturbations in the presence of arbitrary quantum vacuum can be expressed as:
\bea {B_{GSF}(k_L,k_L,k_S)=\frac{H^4}{32\epsilon^2M^4_p}\frac{1}{k^3_Lk^3_S}\sum^{3}_{j=-1}t_{j}\left(\frac{k_S}{k_L}\right)^{j}}~,~~~~~~~~\eea
where the expansion coefficients $t_{j}\forall j=-1,\cdots,3$ for arbitrary vacuum are defined as:
\bea t_{-1}&=&16\epsilon\left(C^{*}_1C_2+C_1C^{*}_2\right)^2,\nonumber\\
t_{0}&=&\left(4(2\epsilon-\eta)-\frac{6s}{c^2_S}\right)\left(|C_1|^2+|C_2|^2\right)^2+4\epsilon\left(|C_1|^2-|C_2|^2\right)^2 +4\epsilon\left(C^{*}_1C_2+C_1C^{*}_2\right)^2,~~~~~~~~~~\nonumber\\
t_{1}&=&34\epsilon\left(C^{*}_1C_2+C_1C^{*}_2\right)^2,\\
t_{2}&=&\left\{\frac{3}{16}\left(\frac{1}{c^2_S}-1-\frac{2\Sigma_{2}(X,\phi)}{\Sigma_{1}(X,\phi)}\right)-6\left(\frac{1}{c^2_S}-1\right)-\frac{6s}{c^2_S}\right\}\left(|C_1|^2+|C_2|^2\right)^2\nonumber\\
&&~~~~~~~~~~~~~~~~~+10\epsilon\left(|C_1|^2-|C_2|^2\right)^2+10\epsilon\left(C^{*}_1C_2+C_1C^{*}_2\right)^2\nonumber,\\
t_{3}&=&\left\{2(2\epsilon-\eta)-\frac{9}{32}\left(\frac{1}{c^2_S}-1-\frac{2\Sigma_{2}(X,\phi)}{\Sigma_{1}(X,\phi)}\right)+5\left(\frac{1}{c^2_S}-1\right)+\frac{2s}{c^2_S}\right\}\left(|C_1|^2+|C_2|^2\right)^2\nonumber\\
&&~~~~~~~~~~~~~~~~~~~~~~~~-5\epsilon\left(|C_1|^2-|C_2|^2\right)^2 -\epsilon\left(C^{*}_1C_2+C_1C^{*}_2\right)^2 \nonumber.\eea
Furthermore, for restricted classes of the general single-field $P(X,\phi)$ model, which satisfies the constraint $P_{,X\phi}(X,\phi)=0$, we get: 
\bea t_{-1}&=&16\epsilon\left(C^{*}_1C_2+C_1C^{*}_2\right)^2,\nonumber\\
t_{0}&=&4(2\epsilon-\eta)\left(|C_1|^2+|C_2|^2\right)^2+4\epsilon\left(|C_1|^2-|C_2|^2\right)^2 +4\epsilon\left(C^{*}_1C_2+C_1C^{*}_2\right)^2,~~~~~~~~~~\nonumber\\
t_{1}&=&34\epsilon\left(C^{*}_1C_2+C_1C^{*}_2\right)^2,\\
t_{2}&=&\left\{\frac{1}{8}\left(\frac{1}{c^2_S}-1-\frac{s\epsilon}{3\epsilon_{X}}\right)-6\left(\frac{1}{c^2_S}-1\right)\right\}\left(|C_1|^2+|C_2|^2\right)^2\nonumber\\
&&~~~~~~~~~~~~~~~~~+10\epsilon\left(|C_1|^2-|C_2|^2\right)^2+10\epsilon\left(C^{*}_1C_2+C_1C^{*}_2\right)^2,\nonumber\\
t_{3}&=&\left\{2(2\epsilon-\eta)-\frac{3}{16}\left(\frac{1}{c^2_S}-1-\frac{s\epsilon}{3\epsilon_{X}}\right)+5\left(\frac{1}{c^2_S}-1\right)\right\}\left(|C_1|^2+|C_2|^2\right)^2\nonumber\\
&&~~~~~~~~~~~~~~~~~~~~~~~~-5\epsilon\left(|C_1|^2-|C_2|^2\right)^2 -\epsilon\left(C^{*}_1C_2+C_1C^{*}_2\right)^2.\nonumber\eea
Now for {Bunch–Davies} and {$\alpha,\beta$} vacuum we get the following simplified expression for the bispectrum for scalar~fluctuation:
\begin{itemize}
\item \underline{{\bf For Bunch–Davies vacuum}:}\\
After setting $C_{1}=1$ and $C_{2}=0$, we get the following expression for the expansion coefficients $t_{j}\forall j=-1,\cdots,3$:
\bea t_{-1}&=&0,\nonumber\\
t_{0}&=&4(3\epsilon-\eta)-\frac{6s}{c^2_S},\nonumber\\
t_{1}&=&0,\\
t_{2}&=&10\epsilon+\left\{\frac{3}{16}\left(\frac{1}{c^2_S}-1-\frac{2\Sigma_{2}(X,\phi)}{\Sigma_{1}(X,\phi)}\right)-6\left(\frac{1}{c^2_S}-1\right)-\frac{6s}{c^2_S}\right\},\nonumber\\
t_{3}&=&-(\epsilon+2\eta)+\left\{5\left(\frac{1}{c^2_S}-1\right)+\frac{2s}{c^2_S}-\frac{9}{32}\left(\frac{1}{c^2_S}-1-\frac{2\Sigma_{2}(X,\phi)}{\Sigma_{1}(X,\phi)}\right)\right\}.\nonumber\eea
Consequently, the bispectrum can be recast as:
\begin{footnotesize}
\bea \begin{array}{rl}B_{GSF}(k_L,k_L,k_S)&=\displaystyle\frac{H^4}{32\epsilon^2M^4_p}\frac{1}{k^3_Lk^3_S}\left[\left\{4(3\epsilon-\eta)-\frac{6s}{c^2_S}\right\}\right.\\
&\displaystyle \left.~~~+\left(10\epsilon+\left\{\frac{3}{16}\left(\frac{1}{c^2_S}-1-\frac{2\Sigma_{2}(X,\phi)}{\Sigma_{1}(X,\phi)}\right)-6\left(\frac{1}{c^2_S}-1\right)-\frac{6s}{c^2_S}\right\}\right)\left(\frac{k_S}{k_L}\right)^2\right.\\
&\displaystyle \left.~~~+\left(-(\epsilon+2\eta)+\left\{5\left(\frac{1}{c^2_S}-1\right)+\frac{2s}{c^2_S}\right.\right.\right.\\&\displaystyle\left.\left.\left.-\frac{9}{32}\left(\frac{1}{c^2_S}-1-\frac{2\Sigma_{2}(X,\phi)}{\Sigma_{1}(X,\phi)}\right)\right\}\right)\left(\frac{k_S}{k_L}\right)^3\right].\end{array}~~~~~~~ \eea
\end{footnotesize}

Furthermore, for restricted classes of the general single-field $P(X,\phi)$ model, which satisfies the constraint $P_{,X\phi}(X,\phi)=0$, we get: 
\vspace{12pt}

\bea t_{-1}&=&0,\nonumber\\
t_{0}&=&4(3\epsilon-\eta),~~~~~~~~~~\nonumber\\
t_{1}&=&0,\\
t_{2}&=&\left\{\frac{1}{8}\left(\frac{1}{c^2_S}-1-\frac{s\epsilon}{3\epsilon_{X}}\right)-6\left(\frac{1}{c^2_S}-1\right)\right\}+10\epsilon,\nonumber\\
t_{3}&=&\left\{5\left(\frac{1}{c^2_S}-1\right)-\frac{3}{16}\left(\frac{1}{c^2_S}-1-\frac{s\epsilon}{3\epsilon_{X}}\right)\right\}-(\epsilon+2\eta)\nonumber\eea
for such case bispectrum is given by:
\begin{small}
\bea \begin{array}{rl} B_{GSF}(k_L,k_L,k_S)&=\displaystyle\frac{H^4}{32\epsilon^2M^4_p}\frac{1}{k^3_Lk^3_S}\left[4(3\epsilon-\eta)\right.\\
&\displaystyle \left.~+\left(\left\{\frac{1}{8}\left(\frac{1}{c^2_S}-1-\frac{s\epsilon}{3\epsilon_{X}}\right)-6\left(\frac{1}{c^2_S}-1\right)\right\}+10\epsilon\right)\left(\frac{k_S}{k_L}\right)^2\right.\\
&\displaystyle \left.+\left(-(\epsilon+2\eta)+\left\{5\left(\frac{1}{c^2_S}-1\right)-\frac{3}{16}\left(\frac{1}{c^2_S}-1-\frac{s\epsilon}{3\epsilon_{X}}\right)\right\}\right)\left(\frac{k_S}{k_L}\right)^3\right].\end{array}~~~~~~~ \eea
\end{small}
\item \underline{{\bf For $~{\alpha,\beta}$ vacuum}:}\\
After setting $C_{1}=\cosh\alpha$ and $C_{2}=e^{i\beta}\sinh\alpha$, we get the following expression for the expansion coefficients $a_{j}\forall j=-1,\cdots,3$:
\bea t_{-1}&=&16\epsilon\sinh^2 2\alpha\cos^2\beta,\nonumber\\
t_{0}&=&\left(4(2\epsilon-\eta)-\frac{6s}{c^2_S}\right)\cosh^2 2\alpha+4\epsilon +4\epsilon\sinh^2 2\alpha\cos^2\beta,\nonumber\\
t_{1}&=&34\epsilon\sinh^2 2\alpha\cos^2\beta,\\
t_{2}&=&\left\{\frac{3}{16}\left(\frac{1}{c^2_S}-1-\frac{2\Sigma_{2}(X,\phi)}{\Sigma_{1}(X,\phi)}\right)-6\left(\frac{1}{c^2_S}-1\right)-\frac{6s}{c^2_S}\right\}\cosh^2 2\alpha\nonumber\\ 
&&~~~~~~~~~~~~~~~~~~~~~~~~+10\epsilon+10\epsilon\sinh^2 2\alpha\cos^2\beta,\nonumber\\
t_{3}&=&\left\{2(2\epsilon-\eta)-\frac{9}{32}\left(\frac{1}{c^2_S}-1-\frac{2\Sigma_{2}(X,\phi)}{\Sigma_{1}(X,\phi)}\right)+5\left(\frac{1}{c^2_S}-1\right)+\frac{2s}{c^2_S}\right\}\cosh^2 2\alpha\nonumber\\
&& ~~~~~~~~~~~~~~~~~~~~~~~~~~~~~-5\epsilon-\epsilon\sinh^2 2\alpha\cos^2\beta.\nonumber\eea
Consequently, the bispectrum can be recast as:
\begin{small}
\bea \begin{array}{rl} B_{GSF}(k_L,k_L,k_S)&={\displaystyle \frac{H^4}{32\epsilon^2M^4_p}\frac{1}{k^3_Lk^3_S}}\left[16\epsilon\sinh^2 2\alpha\cos^2\beta\left(\frac{k_S}{k_L}\right)^{-1}\right.\\ &\left.+\left(\left(4(2\epsilon-\eta)-\frac{6s}{c^2_S}\right)\cosh^2 2\alpha+4\epsilon +4\epsilon\sinh^2 2\alpha\cos^2\beta\right)\right.\\ &\left.~~~~~~~~~~~~~~
+34\epsilon\sinh^2 2\alpha\cos^2\beta\left(\frac{k_S}{k_L}\right)\right.\\ &\left.~~~~+\left(\left\{\frac{3}{16}\left(\frac{1}{c^2_S}-1-\frac{2\Sigma_{2}(X,\phi)}{\Sigma_{1}(X,\phi)}\right)-6\left(\frac{1}{c^2_S}-1\right)-\frac{6s}{c^2_S}\right\}\cosh^2 2\alpha\right.\right.\\& \left.\left.~~~~~~~~~~+10\epsilon+10\epsilon\sinh^2 2\alpha\cos^2\beta\right)\left(\frac{k_S}{k_L}\right)^2\right.\\ &\left.+\left(\left\{2(2\epsilon-\eta)-\frac{9}{32}\left(\frac{1}{c^2_S}-1-\frac{2\Sigma_{2}(X,\phi)}{\Sigma_{1}(X,\phi)}\right)+5\left(\frac{1}{c^2_S}-1\right)+\frac{2s}{c^2_S}\right\}\cosh^2 2\alpha\right.\right.\\& \left.\left.~~~~~~~~~~-5\epsilon-\epsilon\sinh^2 2\alpha\cos^2\beta\right)\left(\frac{k_S}{k_L}\right)^3\right].\end{array}~~~~~~\eea
\end{small}
Furthermore, for restricted classes of the general single-field $P(X,\phi)$ model, which satisfies the constraint $P_{,X\phi}(X,\phi)=0$, we get: 
\bea t_{-1}&=&16\epsilon\sinh^2 2\alpha\cos^2\beta,\nonumber\\
t_{0}&=&4(2\epsilon-\eta)\cosh^2 2\alpha+4\epsilon +4\epsilon\sinh^2 2\alpha\cos^2\beta,\nonumber\\
t_{1}&=&34\epsilon\sinh^2 2\alpha\cos^2\beta,\\
t_{2}&=&\left\{\frac{1}{8}\left(\frac{1}{c^2_S}-1-\frac{s\epsilon}{3\epsilon_{X}}\right)-6\left(\frac{1}{c^2_S}-1\right)\right\}\cosh^2 2\alpha\nonumber\\ 
&&~~~~~~~~~~~~~~~~~~~~~~~~+10\epsilon+10\epsilon\sinh^2 2\alpha\cos^2\beta,\nonumber\\
t_{3}&=&\left\{2(2\epsilon-\eta)-\frac{3}{16}\left(\frac{1}{c^2_S}-1-\frac{s\epsilon}{3\epsilon_{X}}\right)+5\left(\frac{1}{c^2_S}-1\right)\right\}\cosh^2 2\alpha\nonumber\\
&& ~~~~~~~~~~~~~~~~~~~~~~~~~~~~~-5\epsilon-\epsilon\sinh^2 2\alpha\cos^2\beta.\nonumber\eea
Consequently, the bispectrum can be recast as:
\bea \begin{array}{rl} B_{GSF}(k_L,k_L,k_S)&={\displaystyle \frac{H^4}{32\epsilon^2M^4_p}\frac{1}{k^3_Lk^3_S}}\left[16\epsilon\sinh^2 2\alpha\cos^2\beta\left(\frac{k_S}{k_L}\right)^{-1}\right.\\ &\left.+\left(4(2\epsilon-\eta)\cosh^2 2\alpha+4\epsilon +4\epsilon\sinh^2 2\alpha\cos^2\beta\right)\right.\\ &\left.~~~~~~~~~~~~~~
+34\epsilon\sinh^2 2\alpha\cos^2\beta\left(\frac{k_S}{k_L}\right)\right.\\ &\left.~~~~+\left(\left\{\frac{1}{8}\left(\frac{1}{c^2_S}-1-\frac{s\epsilon}{3\epsilon_{X}}\right)-6\left(\frac{1}{c^2_S}-1\right)\right\}\cosh^2 2\alpha\right.\right.\\& \left.\left.~~~~~~~~~~+10\epsilon+10\epsilon\sinh^2 2\alpha\cos^2\beta\right)\left(\frac{k_S}{k_L}\right)^2\right.\\ &\left.+\left(\left\{2(2\epsilon-\eta)-\frac{3}{16}\left(\frac{1}{c^2_S}-1-\frac{s\epsilon}{3\epsilon_{X}}\right)+5\left(\frac{1}{c^2_S}-1\right)\right\}\cosh^2 2\alpha\right.\right.\\& \left.\left.~~~~~~~~~~-5\epsilon-\epsilon\sinh^2 2\alpha\cos^2\beta\right)\left(\frac{k_S}{k_L}\right)^3\right].\end{array}~~~~~~~~~\eea
\end{itemize}
\end{enumerate}

\subsubsection{Expression for EFT Coefficients for General Single-Field $P(X,\phi)$ Inflation}
\label{v5a3}
Here our prime objective is to derive the analytical expressions for EFT coefficients for general single-field $P(X,\phi)$ inflation. To~serve this purpose here we start with a claim that the three-point function and the associated bispectrum for the scalar fluctuations computed from general single-field $P(X,\phi)$ inflation is exactly same as that we have computed from EFT setup. Here we use the {equilateral limit} and {squeezed limit} configurations to extract the analytical expression for the EFT coefficients. In~the two limiting cases  the results are~as follows:
\begin{enumerate}
\item \underline{{Equilateral limit configuration}:}\\
For this case with arbitrary vacuum one can write:
\bea B_{EFT}(k,k,k)&=&B_{GSF}(k,k,k),\eea
which implies that\footnote{Here we also get another solution:
\bea \bar{M}_1&=&\left\{\frac{A}{2BH}\left[-1- \sqrt{1+\frac{4BC}{A^2}}\right]\right\}^{\frac{1}{3}},\eea
which is redundant in the present context as this solution is not consistent with the 
$c_{S}=1$ and $\tilde{c}_{S}=1$ limit result as computed in the earlier section for single-field slow-roll inflation.}:
\begin{small}
\bea {\begin{array}{rl}\bar{M}_1&=\left\{\frac{\hat{A}}{2\hat{B}H}\left[-1+ \sqrt{1+\frac{4\hat{B}\hat{C}}{\hat{A}^2}}\right]\right\}^{\frac{1}{3}},~~~ 
\bar{M}_2\approx \bar{M}_3=\sqrt{\frac{\bar{M}^3_1}{4H\tilde{c}_{5}}}=\sqrt{\frac{\hat{A}}{8\hat{B}H^2\tilde{c}_{5}}\left[-1+ \sqrt{1+\frac{4\hat{B}\hat{C}}{\hat{A}^2}}\right]},\\
M_2&=\left(-\frac{XP_{,XX}(X,\phi)}{P_{,X}(X,\phi)}\dot{H}M^2_p\right)^{\frac{1}{4}},\\M_3&=\left\{-\frac{\hat{A}}{2\hat{B}}\frac{\tilde{c}_{3}}{\tilde{c}_{4}}\left[-1+ \sqrt{1+\frac{4\hat{B}\hat{C}}{\hat{A}^2}}\right]\right\}^{\frac{1}{4}},\\
M_4&=\left(-\frac{\tilde{c}_{3}}{\tilde{c}_{6}}H\bar{M}^3_1\right)^{\frac{1}{4}}=\left(-\frac{\hat{A}}{2\hat{B}}\frac{\tilde{c}_{3}}{\tilde{c}_{6}}\left[-1+ \sqrt{1+\frac{4\hat{B}\hat{C}}{\hat{A}^2}}\right]\right)^{\frac{1}{4}}.\end{array}}~~~~~~~~\eea
\end{small}

\hspace{-3pt}where the factors $\hat{A}$, $\hat{B}$ and $\hat{C}$ are defined as:
\begin{small}
\bea \hat{A}&=&\left(\frac{3}{2}+\frac{4}{3}\frac{\tilde{c}_{3}}{\tilde{c}_{4}}+\frac{2c^2_S}{\tilde{c}_{4}}\right)\left[\frac{U_1}{27}-3U_2\right]-\frac{5}{2}U_1+\frac{99}{98}U_2\nonumber\\
&&-\frac{\Delta}{4}\left[\frac{1}{18}\left(\frac{1}{c^2_S}-1-\frac{2\Sigma_{2}(X,\phi)}{\Sigma_{1}(X,\phi)}\right)\left(|C_1|^2+|C_2|^2\right)^2\right.\nonumber\\ &&\left.
-\frac{7}{3}\left(\frac{1}{c^2_S}-1\right)\left(|C_1|^2+|C_2|^2\right)^2 
-\frac{34}{3}\frac{s}{c^2_S}\left(|C_1|^2+|C_2|^2\right)^2
\right.\nonumber\\ &&\left.+6(2\epsilon-\eta)\left(|C_1|^2+|C_2|^2\right)^2 
+11\epsilon\left(|C_1|^2-|C_2|^2\right)^2 
+27\epsilon\left(C^{*}_1C_2+C_1C^{*}_2\right)^2 
\right],~~~~~~\nonumber\\
 \hat{B}&=&\left\{\left(\frac{3}{2}+\frac{4}{3}\frac{\tilde{c}_{3}}{\tilde{c}_{4}}+\frac{2c^2_S}{\tilde{c}_{4}}\right)\left[\frac{U_1}{27}-3U_2\right]-\frac{5}{2}U_1+\frac{99}{98}U_2\right\}\frac{\Delta c^2_S}{2\epsilon H^2 M^2_p},\\
 \hat{C}&=&\frac{H^2M^2_p\epsilon c^2_S}{2}\left[\frac{1}{18}\left(\frac{1}{c^2_S}-1-\frac{2\Sigma_{2}(X,\phi)}{\Sigma_{1}(X,\phi)}\right)\left(|C_1|^2+|C_2|^2\right)^2\right.\nonumber\\ &&\left.
-\frac{7}{3}\left(\frac{1}{c^2_S}-1\right)\left(|C_1|^2+|C_2|^2\right)^2 
-\frac{34}{3}\frac{s}{c^2_S}\left(|C_1|^2+|C_2|^2\right)^2
\right.\nonumber\\ &&\left.+6(2\epsilon-\eta)\left(|C_1|^2+|C_2|^2\right)^2 
+11\epsilon\left(|C_1|^2-|C_2|^2\right)^2 
+27\epsilon\left(C^{*}_1C_2+C_1C^{*}_2\right)^2 
\right].~~~~~~\nonumber\eea
\end{small}
Furthermore, for restricted classes of the general single-field $P(X,\phi)$ model, which satisfies the constraint $P_{,X\phi}(X,\phi)=0$, we get the following expression for the factors $\hat{A}$, $\hat{B}$ and $\hat{C}$ as:
\begin{small}
\bea \hat{A}&=&\left(\frac{3}{2}+\frac{4}{3}\frac{\tilde{c}_{3}}{\tilde{c}_{4}}+\frac{2c^2_S}{\tilde{c}_{4}}\right)\left[\frac{U_1}{27}-3U_2\right]-\frac{5}{2}U_1+\frac{99}{98}U_2\nonumber\\
&&-\frac{\Delta}{4}\left[\frac{1}{27}\left(\frac{1}{c^2_S}-1-\frac{s\epsilon}{3\epsilon_{X}}\right)\left(|C_1|^2+|C_2|^2\right)^2
-\frac{7}{3}\left(\frac{1}{c^2_S}-1\right)\left(|C_1|^2+|C_2|^2\right)^2 
\right.\nonumber\\ &&\left.+6(2\epsilon-\eta)\left(|C_1|^2+|C_2|^2\right)^2 
+11\epsilon\left(|C_1|^2-|C_2|^2\right)^2 
+27\epsilon\left(C^{*}_1C_2+C_1C^{*}_2\right)^2 
\right],~~~~~~\nonumber\\
 \hat{B}&=&\left\{\left(\frac{3}{2}+\frac{4}{3}\frac{\tilde{c}_{3}}{\tilde{c}_{4}}+\frac{2c^2_S}{\tilde{c}_{4}}\right)\left[\frac{U_1}{27}-3U_2\right]-\frac{5}{2}U_1+\frac{99}{98}U_2\right\}\frac{\Delta c^2_S}{2\epsilon H^2 M^2_p},\\
 \hat{C}&=&\frac{H^2M^2_p\epsilon c^2_S}{2}\left[\frac{1}{27}\left(\frac{1}{c^2_S}-1-\frac{s\epsilon}{3\epsilon_{X}}\right)\left(|C_1|^2+|C_2|^2\right)^2
-\frac{7}{3}\left(\frac{1}{c^2_S}-1\right)\left(|C_1|^2+|C_2|^2\right)^2 
\right.\nonumber\\ &&\left.+6(2\epsilon-\eta)\left(|C_1|^2+|C_2|^2\right)^2 
+11\epsilon\left(|C_1|^2-|C_2|^2\right)^2 
+27\epsilon\left(C^{*}_1C_2+C_1C^{*}_2\right)^2 
\right].~~~~~~\nonumber\eea
\end{small}

where for arbitrary vacuum $U_1$ and $U_2$ are defined as:
\bea U_1&=&\left[\left(C_{1}-C_{2}\right)^3\left(C^{*3}_{1}+C^{*3}_{2}\right)+\left(C^{*}_{1}-C^{*}_{2}\right)^3\left(C^{3}_{1}+C^{3}_{2}\right)\right],\\ 
U_2&=&\left[\left(C_{1}-C_{2}\right)^3C^{*}_{1}C^{*}_{2}\left(C^{*}_{1}-C^{*}_{2}\right)+\left(C^{*}_{1}-C^{*}_{2}\right)^3C_{1}C_{2}\left(C_{1}-C_{2}\right)\right].\eea
If we take $c_{S}=1$ and $\tilde{c}_{S}=1$ then we get then we get back all the results obtained for single-field slow-roll inflation in the previous~section.

Now for {Bunch–Davies} and {$\alpha,\beta$} vacuum we get the following simplified expression for the bispectrum for scalar~fluctuation:
\begin{itemize}
\item \underline{{\bf For Bunch–Davies vacuum}:}\\
After setting $C_{1}=1$ and $C_{2}=0$ we get the following expression for the factors $\hat{A}$, $\hat{B}$ and $\hat{C}$~as:
\bea \hat{A}&=&\frac{2}{27}\left(\frac{3}{2}+\frac{4}{3}\frac{\tilde{c}_{3}}{\tilde{c}_{4}}+\frac{2c^2_S}{\tilde{c}_{4}}\right)-5-\frac{\Delta}{4}\left[\frac{1}{18}\left(\frac{1}{c^2_S}-1-\frac{2\Sigma_{2}(X,\phi)}{\Sigma_{1}(X,\phi)}\right)\right.\nonumber\\ &&\left.
-\frac{7}{3}\left(\frac{1}{c^2_S}-1\right) 
-\frac{34}{3}\frac{s}{c^2_S}
+(23\epsilon-6\eta) 
\right],~~~~~~\nonumber\\
 \hat{B}&=&\left\{\frac{2}{27}\left(\frac{3}{2}+\frac{4}{3}\frac{\tilde{c}_{3}}{\tilde{c}_{4}}+\frac{2c^2_S}{\tilde{c}_{4}}\right)-5\right\}\frac{\Delta c^2_S}{2\epsilon H^2 M^2_p},\\
 \hat{C}&=&\frac{H^2M^2_p\epsilon c^2_S}{2}\left[\frac{1}{18}\left(\frac{1}{c^2_S}-1-\frac{2\Sigma_{2}(X,\phi)}{\Sigma_{1}(X,\phi)}\right)
-\frac{7}{3}\left(\frac{1}{c^2_S}-1\right) 
-\frac{34}{3}\frac{s}{c^2_S}
+(23\epsilon-6\eta) 
\right].~~~~~~~~~~\nonumber\eea
Furthermore, for restricted classes of the general single-field $P(X,\phi)$ model, which satisfies the constraint $P_{,X\phi}(X,\phi)=0$, we get the following expression for the factors $A$, $B$ and $C$~as:
\vspace{12pt}
\begin{small}
\bea \hat{A}&=&\frac{2}{27}\left(\frac{3}{2}+\frac{4}{3}\frac{\tilde{c}_{3}}{\tilde{c}_{4}}+\frac{2c^2_S}{\tilde{c}_{4}}\right)-5 
-\frac{\Delta}{4}\left[\frac{1}{27}\left(\frac{1}{c^2_S}-1-\frac{s\epsilon}{3\epsilon_{X}}\right)
-\frac{7}{3}\left(\frac{1}{c^2_S}-1\right) 
+(23\epsilon-6\eta) 
\right],~~~~~~~~~~\nonumber\\
 \hat{B}&=&\left\{\frac{2}{27}\left(\frac{3}{2}+\frac{4}{3}\frac{\tilde{c}_{3}}{\tilde{c}_{4}}+\frac{2c^2_S}{\tilde{c}_{4}}\right)-5\right\}\frac{\Delta c^2_S}{2\epsilon H^2 M^2_p},\\
 \hat{C}&=&\frac{H^2M^2_p\epsilon c^2_S}{2}\left[\frac{1}{27}\left(\frac{1}{c^2_S}-1-\frac{s\epsilon}{3\epsilon_{X}}\right)
-\frac{7}{3}\left(\frac{1}{c^2_S}-1\right) 
+(23\epsilon-6\eta) 
\right].~~~~~~\nonumber\eea
\end{small}
\item \underline{{\bf For $~{\alpha,\beta}$ vacuum}:}\\
After setting $C_{1}=\cosh\alpha$ and $C_{2}=e^{i\beta}\sinh\alpha$ we get the following expression for the factors $\hat{A}$, $\hat{B}$ and $\hat{C}$ as:
\begin{small}
\bea \hat{A}&=&\left(\frac{3}{2}+\frac{4}{3}\frac{\tilde{c}_{3}}{\tilde{c}_{4}}+\frac{2c^2_S}{\tilde{c}_{4}}\right)\left[\frac{J_1(\alpha,\beta)}{27}-3J_2(\alpha,\beta)\right]-\frac{5}{2}J_1(\alpha,\beta)+\frac{99}{98}J_2(\alpha,\beta)\nonumber\\
&&-\frac{\Delta}{4}\left[\left\{\frac{1}{18}\left(\frac{1}{c^2_S}-1-\frac{2\Sigma_{2}(X,\phi)}{\Sigma_{1}(X,\phi)}\right)-\frac{7}{3}\left(\frac{1}{c^2_S}-1\right)-\frac{34}{3}\frac{s}{c^2_S}\right\}\cosh^22\alpha
\right.\nonumber\\ &&\left.+6(2\epsilon-\eta)\cosh^22\alpha 
+11\epsilon 
+27\epsilon\sinh^2\alpha\cos^2\beta 
\right],~~~~~~\nonumber\\
 \hat{B}&=&\left\{\left(\frac{3}{2}+\frac{4}{3}\frac{\tilde{c}_{3}}{\tilde{c}_{4}}+\frac{2c^2_S}{\tilde{c}_{4}}\right)\left[\frac{J_1(\alpha,\beta)}{27}-3J_2(\alpha,\beta)\right]-\frac{5}{2}J_1(\alpha,\beta)+\frac{99}{98}J_2(\alpha,\beta)\right\}\frac{\Delta c^2_S}{2\epsilon H^2 M^2_p},~~~~~~~~~~\\
 \hat{C}&=&\frac{H^2M^2_p\epsilon c^2_S}{2}\left[\left\{\frac{1}{18}\left(\frac{1}{c^2_S}-1-\frac{2\Sigma_{2}(X,\phi)}{\Sigma_{1}(X,\phi)}\right)-\frac{7}{3}\left(\frac{1}{c^2_S}-1\right)-\frac{34}{3}\frac{s}{c^2_S}\right\}\cosh^22\alpha
 \right.\nonumber\\ &&\left.+6(2\epsilon-\eta)\cosh^22\alpha 
 +11\epsilon 
 +27\epsilon\sinh^2\alpha\cos^2\beta
\right].~~~~~~\nonumber\eea
\end{small}

Furthermore, for restricted classes of the general single-field $P(X,\phi)$ model, which satisfies the constraint $P_{,X\phi}(X,\phi)=0$, we get the following expression for the factors $\hat{A}$, $\hat{B}$ and $\hat{C}$~as:
\begin{small}
\bea \hat{A}&=&\left(\frac{3}{2}+\frac{4}{3}\frac{\tilde{c}_{3}}{\tilde{c}_{4}}+\frac{2c^2_S}{\tilde{c}_{4}}\right)\left[\frac{J_1(\alpha,\beta)}{27}-3J_2(\alpha,\beta)\right]-\frac{5}{2}J_1(\alpha,\beta)+\frac{99}{98}J_2(\alpha,\beta)\nonumber\\
&&-\frac{\Delta}{4}\left[\left\{\frac{1}{27}\left(\frac{1}{c^2_S}-1-\frac{s\epsilon}{3\epsilon_{X}}\right)
-\frac{7}{3}\left(\frac{1}{c^2_S}-1\right) 
+6(2\epsilon-\eta)\right\}\cosh^22\alpha\right.\nonumber\\ &&\left.
~~~~~~~~~~~~~+11\epsilon 
+27\epsilon\sinh^2\alpha\cos^2\beta 
\right],~~~~~~~~~~~~\nonumber\\
 \hat{B}&=&\left\{\left(\frac{3}{2}+\frac{4}{3}\frac{\tilde{c}_{3}}{\tilde{c}_{4}}+\frac{2c^2_S}{\tilde{c}_{4}}\right)\left[\frac{J_1(\alpha,\beta)}{27}-3J_2(\alpha,\beta)\right] 
 -\frac{5}{2}J_1(\alpha,\beta)+\frac{99}{98}J_2(\alpha,\beta)\right\}\frac{\Delta c^2_S}{2\epsilon H^2 M^2_p},~~~~~~~~~~\\
 \hat{C}&=&\frac{H^2M^2_p\epsilon c^2_S}{2}\left[\left\{\frac{1}{27}\left(\frac{1}{c^2_S}-1-\frac{s\epsilon}{3\epsilon_{X}}\right)
 -\frac{7}{3}\left(\frac{1}{c^2_S}-1\right) 
 +6(2\epsilon-\eta)\right\}\cosh^22\alpha\right.\nonumber\\ &&\left.
 ~~~~~~~~~~~~~+11\epsilon 
 +27\epsilon\sinh^2\alpha\cos^2\beta 
 \right].~~~~~~\nonumber\eea
 \end{small}
\end{itemize} 
\item \underline{{Squeezed limit configuration}:}\\
For this case with arbitrary vacuum one can write:
\bea B_{EFT}(k_L,k_L,k_S)&=&B_{GSF}(k_L,k_L,k_S),\eea
which implies that:
\begin{small}
\bea {\begin{array}{rl}\bar{M}_1&=\left\{\frac{\hat{A}}{2\hat{B}H}\left[-1+ \sqrt{1+\frac{4\hat{B}\hat{C}}{\hat{A}^2}}\right]\right\}^{\frac{1}{3}},~~~
\bar{M}_2\approx \bar{M}_3=\sqrt{\frac{\bar{M}^3_1}{4H\tilde{c}_{5}}}=\sqrt{\frac{\hat{A}}{8\hat{B}H^2\tilde{c}_{5}}\left[-1+ \sqrt{1+\frac{4\hat{B}\hat{C}}{\hat{A}^2}}\right]},\\
M_2&=\left(-\frac{XP_{,XX}(X,\phi)}{P_{,X}(X,\phi)}\dot{H}M^2_p\right)^{\frac{1}{4}},~~~M_3=\left\{-\frac{\hat{A}}{2\hat{B}}\frac{\tilde{c}_{3}}{\tilde{c}_{4}}\left[-1+ \sqrt{1+\frac{4\hat{B}\hat{C}}{\hat{A}^2}}\right]\right\}^{\frac{1}{4}},\\
M_4&=\left(-\frac{\tilde{c}_{3}}{\tilde{c}_{6}}H\bar{M}^3_1\right)^{\frac{1}{4}}=\left(-\frac{\hat{A}}{2\hat{B}}\frac{\tilde{c}_{3}}{\tilde{c}_{6}}\left[-1+ \sqrt{1+\frac{4\hat{B}\hat{C}}{\hat{A}^2}}\right]\right)^{\frac{1}{4}}.\end{array}}~~~~~~~~~~~~\eea
\end{small}
\hspace{-3pt}where the factors $\hat{A}$, $\hat{B}$ and $\hat{C}$ are defined as\footnote{Here we also get another solution:
	\bea \bar{M}_1&=&\left\{\frac{\hat{A}}{2\hat{B}H}\left[-1- \sqrt{1+\frac{4\hat{B}\hat{C}}{\hat{A}^2}}\right]\right\}^{\frac{1}{3}},\eea
	which is redundant in the present context as this solution is not consistent with the 
	$c_{S}=1$ and $\tilde{c}_{S}=1$ limit result as computed in the earlier section for general single-field $P(X,\phi)$ inflation.}:
\bea \hat{A}&=&\hat{P}_{1}+\hat{P}_{2}-\sum^{3}_{j=-1}t_{j}\left(\frac{k_S}{k_L}\right)^j,~~~~
\hat{B}=\frac{\hat{P}_{1}\Delta}{2\epsilon H^2 M^2_p},~~~~
 \hat{C}= 2\epsilon H^2 M^2_p\sum^{3}_{j=-1}t_{j}\left(\frac{k_S}{k_L}\right)^j,~~~~~~\eea
 where the expansion coefficients $t_{j}\forall j=-1,\cdots,3$ are defined earlier for the general $P(X,\phi)$ model and also for restricted classes of the model where $P_{,X\phi}(X,\phi)=0$ constraint is satisfied.
 
Here the factors $\hat{P}_{1}$ and $\hat{P}_{2}$ are defined as:
\bea \hat{P}_{1}&=&\sum^{3}_{m=-1}\hat{e}_{m}\left(\frac{k_S}{k_L}\right)^m,~~~~~~~
 \hat{P}_{2}=\sum^{3}_{m=-1}\hat{h}_{m}\left(\frac{k_S}{k_L}\right)^m,\eea
where the expansion coefficients $\hat{e}_{m}\forall m=-1,\cdots,3$ and $\hat{h}_{m}\forall m=-1,\cdots,3$ for arbitrary vacuum are defined as:
\bea \hat{e}_{-1}&=&-36U_2,~~
\hat{e}_{0}=\left[-\frac{9}{2}U_1+\left(24\left\{\frac{3}{2}+\frac{4}{3}\frac{\tilde{c}_{3}}{\tilde{c}_{4}}+\frac{2c^2_S}{\tilde{c}_{4}}\right\}-\frac{9}{2}\right)U_2\right],\nonumber\\
\hat{e}_{1}&=&0,~~
\hat{e}_{2}=\left[-\frac{27}{2}U_2+\left(\frac{3}{2}\left\{\frac{3}{2}+\frac{4}{3}\frac{\tilde{c}_{3}}{\tilde{c}_{4}}+\frac{2c^2_S}{\tilde{c}_{4}}\right\}-\frac{27}{2}\right)U_1\right],~~
\hat{e}_{3}=0,\eea
and 
\bea \hat{h}_{-1}&=&0,~~~
\hat{h}_{0}=\frac{9}{c^2_S}\left(U_1+U_2\right),~~~
\hat{h}_{1}=0,\nonumber\\
\hat{h}_{2}&=&\left[\left(15+2\left\{\frac{3}{2}+\frac{2c^2_S}{\tilde{c}_{4}}\right\}\right)U_{1}+\left(\frac{45}{2}+3\left\{\frac{3}{2}+\frac{2c^2_S}{\tilde{c}_{4}}\right\}\right)U_{2}\right],~~~
\hat{h}_{3}=0,\eea
where $U_1$ and $U_2$ are already defined~earlier.

Now for {Bunch–Davies} and {$\alpha,\beta$} vacuum we get the following simplified expression for the bispectrum for scalar~fluctuation:
\begin{itemize}
\item \underline{{\bf For Bunch–Davies vacuum}:}\\
After setting $C_{1}=1$ and $C_{2}=0$, we get $U_1=2$ and $U_2=0$. Consequently, the expansion coefficients can be recast as:
\bea \hat{e}_{-1}&=&0,~~
\hat{e}_{0}=-9,~~
\hat{e}_{1}=0,~~
\hat{e}_{2}=-27,~~
\hat{e}_{3}=0,\eea
and 
\bea \hat{h}_{-1}&=&0,~~
\hat{h}_{0}=\frac{18}{c^2_S},~~
\hat{h}_{1}=0,~~
\hat{h}_{2}=\left(30+4\left\{\frac{3}{2}+\frac{2c^2_S}{\tilde{c}_{4}}\right\}\right),~~
\hat{h}_{3}=0,\eea

\item \underline{{\bf For $~{\alpha,\beta}$ vacuum}:}\\
After setting $C_{1}=\cosh\alpha$ and $C_{2}=e^{i\beta}\sinh\alpha$, we get $U_1=J_{1}(\alpha,\beta)$ and $U_2=J_{2}(\alpha,\beta)$. Consequently, the expansion coefficients can be recast as:
\bea \hat{e}_{-1}&=&-36J_{2}(\alpha,\beta),~~
\hat{e}_{0}=\left[-\frac{9}{2}J_{1}(\alpha,\beta)+\left(24\left\{\frac{3}{2}+\frac{4}{3}\frac{\tilde{c}_{3}}{\tilde{c}_{4}}+\frac{2c^2_S}{\tilde{c}_{4}}\right\}-\frac{9}{2}\right)J_{2}(\alpha,\beta)\right],\nonumber\\
\hat{e}_{1}&=&0,~~
\hat{e}_{2}=\left[-\frac{27}{2}J_{2}(\alpha,\beta)+\left(\frac{3}{2}\left\{\frac{3}{2}+\frac{4}{3}\frac{\tilde{c}_{3}}{\tilde{c}_{4}}+\frac{2c^2_S}{\tilde{c}_{4}}\right\}-\frac{27}{2}\right)J_{1}(\alpha,\beta)\right],~~
\hat{e}_{3}=0,~~~~~~~~~~\eea
and 
\begin{small}
\bea \hat{h}_{-1}&=&0,~~
\hat{h}_{0}=\frac{9}{c^2_S}\left(U_1+J_{2}(\alpha,\beta)U_2\right),~~
\hat{h}_{1}=0,\nonumber\\
\hat{h}_{2}&=&\left[\left(15+2\left\{\frac{3}{2}+\frac{2c^2_S}{\tilde{c}_{4}}\right\}\right)J_{1}(\alpha,\beta)+\left(\frac{45}{2}+3\left\{\frac{3}{2}+\frac{2c^2_S}{\tilde{c}_{4}}\right\}\right)J_{2}(\alpha,\beta)\right],~~
\hat{h}_{3}=0.~~~~~~~~~~~\eea
\end{small}
\end{itemize}
\end{enumerate}

\section{{Conclusions}}
\label{v6}
To summarize, in~this paper, we have addressed the following~issues:
\begin{itemize}
	\item  We have derived the analytical expressions for the two-point correlation function for scalar and tensor fluctuations and three-point correlation function for scalar fluctuations from EFT framework in quasi de Sitter background in a model-independent way. For~this computation, we use an arbitrary quantum state as the initial choice of vacuum. Such a choice finally gives rise to the most general expressions for the two-point and three-point correlation functions for primordial fluctuation in EFT. Furthermore, we have simplified our results by considering the Bunch–Davies vacuum and $\alpha,\beta$ vacuum~states.
	
	\item During our computation, we have truncated the EFT action by considering the all possible two derivative terms in the metric. This allows us to derive correct expressions for the two-point and three-point correlation functions for EFT which are consistent with both the single-field slow-roll model and generalized non-canonical $P(X,\phi)$ single-field models minimally coupled with~gravity\footnote{This is really an important outcome as the earlier derived results for the three-point function for EFT in quasi de Sitter background was not consistent with the known result for the single-field slow-roll model, where effective sound speed is fixed at $\tilde{c}_{S}=1$. }. 
	
	\item  Furthermore, we have derived the analytical expressions for the coefficients of all relevant EFT operators for the single-field slow-roll model and generalized non-canonical $P(X,\phi)$ single-field models. We~have derived the results in terms of slow-roll parameters, effective sound speed parameter, and the constants which are fixed by the choice of arbitrary initial vacuum state. Next, we have simplified our results also presented the results  by considering Bunch–Davies vacuum and $\alpha,\beta$ vacuum~state.

	\item  Finally, using the CMB observation from Planck we constrain all these EFT coefficients for various single-field slow-roll models and generalized non-canonical $P(X,\phi)$ models of inflation.
\end{itemize}

The future directions of this paper are appended below~pointwise:
\begin{itemize}
	
	\item One can further carry forward this work to compute four-point scalar correlation function from EFT framework using an arbitrary initial choice of the quantum vacuum state. The~present work can also be extended for the computation of the three-point correlation from tensor fluctuation, and other three-point cross correlations between scalar and tensor mode fluctuation in the context of EFT with arbitrary initial~vacuum.   
	
	\item In the present EFT framework we have not considered the effects of any additional heavy fields ($m>>H$) in the effective action. One can redo the analysis with such additional effects in the EFT framework to study the quantum entanglement, cosmological decoherence and Bell's inequality violation in the context of primordial cosmology. Once can also further generalize this computation for any arbitrary spin fields which are consistent with the unitarity~bound.
	
	\item The analyticity property of response functions and scattering amplitudes in QFT implies significant connection between
	observables in IR regime and the underlying dynamics valid in the short-distance scale. Such analytic property is directly connected to the causality and unitarity of the QFT under consideration. Following this idea one can also study the analyticity property in the present version of EFT or including the effective of massive fields ($m>>H$) in the effective~action.
	
	\item There are other open issues as well which one can study within the framework of~EFT:
	\begin{enumerate}
		\item The role of out-of-time-ordered correlations from open quantum system~\cite{Shandera:2017qkg,Sieberer:2015svu,Avinash:2017asn}.
		
		\item EFT framework in a quantum dissipative system and its application to cosmology~\cite{Das:2014jia,Amin:2015ftc,Amin:2017wvc}.
		
		\item Thermalization, quantum critical quench and its application to the phenomena of reheating in early universe cosmology~\cite{Carrilho:2016con}. 
	\end{enumerate}
	
\end{itemize}


\acknowledgments{S.C. would like to thank Quantum Gravity and Unified Theory and Theoretical Cosmology Group, Max Planck Institute for
Gravitational Physics, Albert Einstein Institute for providing the Post-Doctoral Research Fellowship. S.C. thank Sandip. P. Trivedi for collaboration in the initial stage of the work and for various useful discussions and suggestions. S.C. take this opportunity to thank sincerely to  Shiraz Minwalla, \mbox{Sudhakar Panda}, Varun Sahni, Gautam Mandal and Jean-Luc Lehners for their constant support
and inspiration. S.C. thank the organizers of Summer School on
Cosmology 2018, ICTP, Trieste, 15th Marcel Grossman Meeting, Rome, The European Einstein Toolkit meeting 2018, Centra,
Instituto Superior Tecnico, Lisbon and The Universe as a Quantum Lab, APC, Paris, Nordic String Meeting 2019, AEI, Potsdam,
Tensor networks: from simulations to holography, DESY Zeuthen and AEI, Potsdam, XXXI Workshop Beyond the Standard
Model, Physikzentrum Bad Honnef, Workshop in String Theory and Cosmology, NISER, Bhubansewar, Indian String Meet 2016, Advanced String School 2017 and ST$^{4}$ 2017 for providing the local
hospitality during the work.
S.C. also thank IOP, CMI, SINP, 
IACS, IISER, Pune, IISER, Mohali and ISI, Kolkata for
providing the academic visit during the work. Last but not the least, we would all like to acknowledge our
debt to the people of India for their generous and steady support for research in natural sciences, especially
for theoretical high energy physics, string theory and~cosmology.}


\appendix
\section{Brief Overview on  Schwinger-Keldysh (In-In) Formalism}\label{q2}

   To compute the any n-point correlation function in quasi de Sitter space we use Schwinger-Keldysh (In-In) formalism. In~this framework the expectation value of a product of operators ${\cal O}(t)$ at time $t$ can be written as:
   \bea \label{ccc}{\langle{\cal O}(t)\rangle =\left\langle\left(T~\exp\left[-i\int^{t}_{-\infty}H_{int}(t^{'})dt^{'}\right]\right)^{\dagger}{\cal O}(t)\left(T~\exp\left[-i\int^{t}_{-\infty}H_{int}(t^{''})dt^{''}\right]\right)\right\rangle}~,\eea
   where it is important to note that all the fields appearing in the right-hand side belong to the Heisenberg picture. Here correlation function is computed with respect to the initial quantum vacuum state $|in\rangle$, which in general can be any arbitrary vacuum state. In~cosmological literature concept of Bunch–Davies and $\alpha,\beta$ vacuum are commonly used. To~mention the mathematical structure of the quantum vacuum state $|in\rangle$ we first consider an arbitrary state $|\Omega(t)\rangle$, which can be expanded in terms of the eigen basis state $|m\rangle$ of the free Hamiltonian as:
   \bea |\Omega(t)\rangle=\sum |m\rangle\langle m|\Omega(t)\rangle.\eea
   
   Furthermore, the time evolved quantum state from time $t=t_{1}$ to $t=t_{2}$ can be written as:
   \begin{small}
   \bea |\Omega(t_2)\rangle&=& T~\exp\left[-i\int^{t_2}_{t_1}H_{int}(t^{'})dt^{'}\right]|\Omega(t_1)\rangle= \underbrace{|0\rangle\langle 0|\Omega\rangle}_{\bf Free~part}+\underbrace{\sum^{\infty}_{m=1}\exp\left[iE_{m}(t_2-t_1)\right]|m\rangle\langle m|\Omega(t_1)\rangle}_{\bf Interacting~part}.~~~~~~~\eea
    \end{small}
    
   It is clearly observed that we have expressed any arbitrary quantum state in terms of the free part and the interacting part of the theory. 
   Furthermore, for further computation we set $t_2=-\infty(1-i\epsilon)$ which clearly projects all excited quantum states. Using this we have the following connecting relation between the interacting vacuum and the free vacuum state, as~given by:
      \begin{small}
   \bea |\Omega(-\infty(1-i\epsilon))\rangle\equiv|0\rangle\langle 0|\Omega\rangle. \eea
      \end{small}
      
   Finally, at~any arbitrary time the interacting vacuum can be written as: 
      \begin{small}
   \bea |in\rangle &=&T~\exp\left[-i\int^{t}_{-\infty(1-i\epsilon)}H_{int}(t^{'})dt^{'}\right]|\Omega(-\infty(1-i\epsilon))\rangle=T~\exp\left[-i\int^{t}_{-\infty(1-i\epsilon)}H_{int}(t^{'})dt^{'}\right]|0\rangle\langle 0|\Omega\rangle.~~~~~~~~~\eea
      \end{small}
      
   For our computations, initially we have written the expression which is valid for any arbitrary choice of quantum vacuum state. But~for simplicity further we consider two specific choices of vacuum state---{Bunch–Davies vacuum} and {$\alpha,\beta$ vacuum}, which are commonly used in cosmological physics. Now~in this context the total Hamiltonian of the theory can be written in terms of the free and interacting part as,
   $H=H_0+H_{int},$ where interaction Hamiltonian is described by $H_{int}$ and the free field Hamiltonian is described by $H_0$.
   
   In the context of cosmological perturbation theory one can follow the same formalism where one usually starts with the Einstein–Hilbert gravity action with the any matter content in the effective action. For~this purpose one uses the well-known ADM formalism to derive an action which contains only dynamical degrees of freedom. From this action one needs to perform the following~steps:
   \begin{itemize}
   \item First one needs to construct the canonically conjugate momenta and the Hamiltonian for the~system.
   
   \item \scalebox{.95}[1.0]{Then we need to separate out the quadratic part from the higher-order contributions in the Hamiltonian.} 
   \end{itemize}
   
   Now in this context let us consider a part of the effective action which contains the third-order contribution and all other higher-order contribution in cosmological perturbation theory, represented by $L_{int}$. In~this case the usual expression for the interaction Hamiltonian is given by, 
   $H_{int}=-L_{int}.$
   Furthermore, to make a direct connection to the in-out formalism in QFT used in the computation of S-matrix, one can further insert complete sets of states labeled by $\alpha$ and $\beta$ in Equation~(\ref{ccc}) and finally get:
      \begin{footnotesize}
    \bea \label{ccc1}\langle {\cal O}(t)\rangle &=&\int d\alpha\int d\beta \nonumber\\
    &&~\langle 0|\left(T~\exp\left[-i\int^{t}_{-\infty}H_{int}(t^{'})dt^{'}\right]\right)^{\dagger}|\alpha\rangle\overbrace{\langle \alpha| {\cal O}(t)|\beta\rangle}^{={\cal O}_{\alpha\beta}(t)}\langle \beta|\left(T~\exp\left[-i\int^{t}_{-\infty}H_{int}(t^{''})dt^{''}\right]\right)|0\rangle,
\eea
       \end{footnotesize}
       
    Here the in-in quantum correlation is interpreted as the product of the vacuum transition amplitudes and 
    in the matrix element $\langle \alpha| {\cal O}(t)|\beta\rangle\equiv {\cal O}_{\alpha\beta}(t),$ where one needs to sum over all possible quantum out states. Furthermore, to compute the quantum correlations using Schwinger-Keldysh (In-In) formalism one needs to consider the following~steps:
    \begin{itemize}
    \item First of all one needs to define the time integration in the time evolution operator $U(t)$ to go over a contour in the complex plane i.e.,
    \be U(t)=T~\exp\left[-i\int^{t}_{-\infty}H_{int}(t^{'})dt^{'}\right]\Rightarrow T~\exp\left[-i\int^{t}_{-\infty(1+i\epsilon)}H_{int}(t^{'})dt^{'}\right],\ee
    where we have redefined the time interval by including small imaginary contribution as given by $t\rightarrow t(1\pm i\epsilon)$. With~this specific choice, one can finally write the following expression for the n-point correlation function:
    \begin{small}
    \be \label{ccc2}{\langle {\cal O}(t)\rangle =\left\langle\left(T~\exp\left[-i\int^{t}_{-\infty(1-i\epsilon)}H_{int}(t^{'})dt^{'}\right]\right)^{\dagger}{\cal O}(t)\left(T~\exp\left[-i\int^{t}_{-\infty(1+i\epsilon)}H_{int}(t^{''})dt^{''}\right]\right)\right\rangle}~.~~~~~~~~~~\ee
    \end{small}
    
     Here it is important to note that complex conjugation of the time evolution operator $U(t)$ signifies the fact that the time-ordered contour does not at all coincide with the time-backward contour. 
    \item Next we analytically continue the expression for the interaction Hamiltonian as appearing in the time evolution operator $U(t)$ i.e.,~ 
    $H_{ini}(t)\rightarrow H_{int}(t(1\pm i\epsilon)).$
   \item Next we consider the following Dyson Swinger series:
   \bea T~\exp\left[-i\int^{t}_{-\infty(1+i\epsilon)}H_{int}(t^{'})dt^{'}\right]=1+\sum^{\infty}_{N=1}\frac{(-i)^{N}}{N!}\prod^{N}_{i=1}\int^{t_{i}}_{-\infty(1+i\epsilon)}dt_{i}~H_{int}(t_{i}),\eea
   using which finally we get the following simplified expression for the n-point correlation function:
   \bea \label{ccc3}{\langle {\cal O}(t)\rangle =\sum^{\infty}_{N=0}\frac{(-i)^{N}}{N!}\prod^{N}_{i=1}\int^{t_{i}}_{-\infty(1+i\epsilon)}dt_{i}~\langle 0|\left[H_{int}(t_{i}),{\cal O}(t)\right]|0\rangle=\sum^{\infty}_{n=0}\langle {\cal O}(t)\rangle^{(n)}}~.~~~~~~~~~~\eea
   where $|0\rangle$ is the initial quantum vacuum state under consideration. Here expanding in the powers of interacting Hamiltonian $H_{int}(t)$ we finally~get:
   \begin{enumerate}
   \item \underline{Zeroth order term $\langle {\cal O}(t)\rangle^{(0)}$ in Dyson Swinger series:}\\
   Here the zeroth order term in Dyson Swinger series can be expressed as:
   \bea \langle {\cal O}(t)\rangle^{(0)}&=&\langle 0|{\cal O}(t)|0\rangle.\eea
   \item \underline{First order term $\langle {\cal O}(t)\rangle^{(1)}$ in Dyson Swinger series:}\\
   Here the first order term  in Dyson Swinger series can be expressed as:
      \bea \langle {\cal O}(t)\rangle^{(1)}&=&2{\rm Re}\left[-i\int^{t}_{-\infty(1+i\epsilon)}dt^{'}\langle 0|{\cal O}(t)H_{int}(t^{'})|0\rangle\right].\eea
   \item \underline{Second-order term $\langle {\cal O}(t)\rangle^{(2)}$ in Dyson Swinger series:}\\
    Here the second-order term  in Dyson Swinger series can be expressed as:
         \bea \langle {\cal O}(t)\rangle^{(2)}&=&-2{\rm Re}\left[\int^{t_1}_{-\infty(1+i\epsilon)}dt_1\int^{t_2}_{-\infty(1+i\epsilon)}dt_2\langle 0|{\cal O}(t)H_{int}(t_1)H_{int}(t_2)|0\rangle\right]\nonumber\\
         &&~~~~~~~~+\int^{t_1}_{-\infty(1+i\epsilon)}dt_1\int^{t_2}_{-\infty(1+i\epsilon)}dt_2\langle 0|H_{int}(t_1){\cal O}(t)H_{int}(t_2)|0\rangle.\eea
         Following this trick one can easily write down the expression for any n-point correlation function of the given operator ${\cal O}(t)$.
   \end{enumerate}
    
    \end{itemize}
    \section{{Choice of Initial Quantum Vacuum~State}}\label{q3}
    In general, one can consider an arbitrary initial quantum vacuum state which is specified by the two sets of constants $(C_1,C_2)$ and $(D_1,D_2)$ as appearing in solution of the scalar and tensor mode fluctuation. In~general in this context a quantum state is described by this two number as $|C_1,C_2\rangle$ and $|D_1,D_2\rangle$ and defined as, 
    $C({\bf k})|C_1,C_2\rangle=0~\forall~{\bf k},  
    D({\bf k})|D_1,D_2\rangle=0~\forall~{\bf k},$
    where $C({\bf k})$ and $D({\bf k})$ are the annihilation operators for scalar and tensor mode fluctuations as appearing in cosmological perturbation~theory. 
    
    In general, ground one can write down the most general state $|C_1,C_2\rangle$ in terms of the well-known Bunch–Davies vacuum state as:
    \begin{small}
    \bea  |C_1,C_2\rangle&=&\prod_{\bf k}\frac{1}{\sqrt{|C_1|}}\exp\left[\frac{C^*_2}{2C^*_1}~C^{\dagger}({\bf k})C^{\dagger}(-{\bf k})\right]|0\rangle\nonumber\\
                                &=&\frac{1}{{\cal N}_{C}}~\exp\left[\frac{C^*_2}{2C^*_1}\sum_{{\bf k}}~C^{\dagger}({\bf k})C^{\dagger}(-{\bf k})\right]|0\rangle=\frac{1}{{\cal N}_{C}}~\exp\left[\frac{C^*_2}{2C^*_1}\int\frac{d^3k}{(2\pi)^3}~C^{\dagger}({\bf k})C^{\dagger}(-{\bf k})\right]|0\rangle,~~~~~~~~\eea      
                                \end{small}                        
        where ${\cal N}_{C}=\sqrt{|C_{1}|}$ are the overall normalization constant for scalar and tensor mode fluctuations. For~the tensor modes the calculation is~similar.
        
         Here it is important to mention that the quantum vacuum state 
                                    $|C_1,C_2\rangle$ satisfies the following constraint equation:\begin{footnotesize}
\bea \hat{\bf P}_{C}|C_1,C_2\rangle&=&\int \frac{d^3p}{(2\pi)^3}~{\bf p}~C^{\dagger}({\bf p})C({\bf p})|C_1,C_2\rangle=\prod_{{\bf k}}\int \frac{d^3p}{(2\pi)^3}~\frac{{\bf p}~C^{\dagger}({\bf p})C({\bf p})}{\sqrt{|C_1|}}\exp\left[\frac{C^*_2}{2C^*_1}~C^{\dagger}({\bf k})C^{\dagger}(-{\bf k})\right]|0\rangle\nonumber\\
                                    &=&\int \frac{d^3p}{(2\pi)^3}~\frac{{\bf p}~C^{\dagger}({\bf p})C({\bf p})}{\sqrt{|C_1|}}\exp\left[\frac{C^*_2}{2C^*_1}~\sum_{{\bf k}}C^{\dagger}({\bf k})C^{\dagger}(-{\bf k})\right]|0\rangle\\
                                    &=&\int \frac{d^3p}{(2\pi)^3}~\frac{{\bf p}~C^{\dagger}({\bf p})C({\bf p})}{\sqrt{|C_1|}}\exp\left[\frac{C^*_2}{2C^*_1}~\int \frac{d^3k}{(2\pi)^3}C^{\dagger}({\bf k})C^{\dagger}(-{\bf k})\right]|0\rangle= 0,\nonumber \eea\end{footnotesize}
\noindent{\hspace{-3pt}which is true for the quantum vacuum state $|D_1,D_2\rangle$ for tensor modes also. Since scalar modes are exactly similar to tensor modes we will not speak about tensor modes in the next part.}                                    

Additionally, here it is important to note that the~annihilation  and creation operators for Bunch–Davies vacuum ($a({\bf k}), a^{\dagger}({\bf k})$) and the arbitrary quantum vacuum $|C_{1},C_{2}\rangle$ state  ($C({\bf k}), C^{\dagger}({\bf k})$) are connected via the following sets of Bogoliubov transformations:
                                    \bea C({\bf k})&=&C^{*}_{1}~a({\bf k})-C^{*}_{2}~a^{\dagger}(-{\bf k}),\\
                                  a({\bf k})&=&C_{1}~C({\bf k})+C^{*}_{2}~C^{\dagger}(-{\bf k}).\eea        
                                  
      A very well-known feature of QFT is that it makes itself particularly manifest
in the context of curved space-time backgrounds, physically represents particle excitation. Consequently, a~thermal quantum state depends sensitively on the proper choice of quantum mechanical vacuum state. But~we know that for a generalized curved
space-time, no canonical or even preferred quantum vacuum state exists. In~the context of QFT in curved space-time, there is a huge class of quantum mechanical states over a background de Sitter space which are invariant under all the ${\bf SO(1,4)}$ isometries\footnote{Due to isometries, a time-like Killing vector field provides a natural physical explanation of partitioning the frequency modes into positive and negative categories, which is similar to the standard procedure performed in Minkowski flat space-time. Furthermore, one can associate these positive and negative frequency modes with annihilation and creation operators in the present context. Then a quantum mechanical vacuum state can be described by imposing the constraint condition that the state be annihilated by all the annihilation operators. In~the absence of time-like Killing vector, there exists no natural choice of quantum mechanical vacuum state. In~such a situation, one can apply different conditions to choose a particular quantum vacuum state. In~this specific situation, a~natural simplest possible choice is to consider a physical region of space-time for which a time-like Killing vector does exist, which can be further used to construct the corresponding quantum mechanical vacuum state. Similarly, if~the space-time asymptotically matches with the Minkowski flat case, then in that specific situation there exists another possibility to use the most generalized {\it Poincar'e} quantum mechanical vacuum state. In~an alternative prescription, one can consider a physical situation where the quantum mechanical vacuum state be annihilated by the physical generators of some specific symmetry group. On~the other hand, a~quantum mechanical vacuum state can also be treated as an un-physical if it fails to satisfy certain necessary physical constraints. To~demonstrate this explicitly one can consider an example, where the expectation value of the stress energy momentum tensor diverges at a non-singular point in space-time, such as at a horizon. In~that situation one can easily discard the possibility of the corresponding quantum mechanical vacuum state existing. In~our description we need a particular universal form for the cosmological correlation function and the associated spectrum which is actually dictated by the conservation of the stress energy.}\footnote{Here it is important to note that under~the application of an arbitrary  
de Sitter transformation, which is commonly identified as isometries, each positive frequency modes of de Sitter 
mix among themselves and each negative frequency
modes mix among themselves, but~precisely they do not mix among each other. This physically implies that the {\it Bunch–Davies vacuum}
state is invariant under the de Sitter isometry {\bf SO(1,4)} group. On~the other hand, for~massive scalar fields (or may be the scalar field have very tiny non-zero mass), if~we set the parameter $\beta=0$ then we get a one-parameter family of de Sitter invariant vacuums, which is commonly known as $\alpha$-vacuum, which physically represents the squeezed states. It is important to note that for our discussion we always consider non-zero mass of the scalar fields because of the fact that for massless scalar degrees of freedom quantum mechanical vacuum states are not invariant under the de Sitter isometry {\bf SO(1,4)} group.}, which is commonly known as the $\alpha,\beta$-vacuum and represent the excited quantum states. By~fixing the parameter $\beta=0$ one can explicitly show that $\alpha$-vacuum is CPT-invariant, where $\alpha$ plays the role of super-selection real parameter. Furthermore, fixing the real parameter $\alpha=0$ one can get back the well-known {\it Bunch–Davies} quantum vacuum state for de Sitter space where the Cosmological Constant, $\Lambda>0$. In~the limit, $\Lambda\rightarrow 0$ one can further show that with $\alpha=0$ we can get back the unique Minkowski quantum vacuum state. One can also choose a quantum mechanical vacuum $\alpha$ state as an initial condition, which at late time scale will give rise to long-range (Hubble scale) quantum correlations. In~this context the long-range quantum correlations are  manifestation of entanglement associated with the quantum mechanical vacuum state which is here identified as the initial~state. 

Among these classes of quantum mechanical vacuum, there is a specific type of vacuum state whose associated Green's functions verify the well-known {\it Hadamard condition} behaving on the light-cone as in flat Minkowski space. This quantum mechanical state is usually known as the {\it Bunch–Davies vacuum} or {\it Euclidean vacuum}. The~{\it Bunch–Davies vacuum} can also be described as being generated by an infinite time-trace operation from the condition that the associated scale of quantum fluctuations is much smaller than the cosmological Hubble~scale.

The {\it Bunch–Davies} vacuum state is treated as the zero-particle ground state in the context of QFT of curved space-time which is actually observed by a geodesic observer. This~quantum mechanical vacuum state is very useful which explains the origin of quantum mechanical fluctuations in the context of inflationary~models.                                             
    \begin{enumerate}
    \item \underline{{Bunch–Davies vacuum}:}\\
    {Bunch–Davies vacuum} is specified by fixing the coefficients to, 
   $C_{1}= 1=D_{1},~~~
         C_{2}=0=D_{2},$
         in the solution of the scalar and tensor mode fluctuation as derived earlier.
         In this case, the quantum vacuum state $|0\rangle$ is defined as the state that gets annihilated
         by the annihilation operator, as~given by, 
        $a({\bf k})|0\rangle= 0~\forall~{\bf k}.$ 
         Here the creation and annihilation operators $a({\bf k})$ and $a^{\dagger}({\bf k})$ satisfy the following canonical commutation relations:
         \bea \left[a({\bf k}),a({\bf k}^{'})\right]=0,~~
       \left[a^{\dagger}({\bf k}),a^{\dagger}({\bf k}^{'})\right]=0,~~ 
       \left[a({\bf k}),a^{\dagger}({\bf k}^{'})\right]=(2\pi)^3\delta^{(3)}({\bf k}+{\bf k}^{'}).
       \eea 
    \item \underline{{$\alpha,\beta$ vacuum}:}\\
    {$\alpha,\beta$ vacuum}  is specified by  fixing the coefficients to, $C_{1}= \cosh\alpha=D_{1},~~~
             C_{2}= e^{i\beta}\sinh\alpha=D_{2},$
                     in~the solution of the scalar and tensor mode fluctuation as derived earlier. In~this case the quantum vacuum state $|\alpha,\beta\rangle$ is defined as the state that gets annihilated
                              by the annihilation operator, as~given by,
                              $ b({\bf k})|\alpha,\beta\rangle= 0~\forall~{\bf k}.$
                              Here the creation and annihilation operators $b({\bf k})$ and $b^{\dagger}({\bf k})$ satisfy the following canonical commutation relations:
                              \bea \left[b({\bf k}),b({\bf k}^{'})\right]=0,~~
                            \left[b^{\dagger}({\bf k}),b^{\dagger}({\bf k}^{'})\right]=0,~~ 
                            \left[b({\bf k}),b^{\dagger}({\bf k}^{'})\right]=(2\pi)^3\delta^{(3)}({\bf k}+{\bf k}^{'}).
                            \eea
                            Here one can write the Bunch–Davies vacuum state $|0\rangle$ as a special class of $|\alpha,\beta\rangle$ vacuum state. Also using Bogoliubov transformation one can write down $|\alpha,\beta\rangle$ vacuum state in terms of the Bunch–Davies vacuum state $|0\rangle$, as~given by:
                            \bea  |\alpha,\beta\rangle&=&\prod_{\bf k}\frac{1}{\sqrt{|\cosh\alpha|}}\exp\left[-\frac{i}{2}e^{-i\beta}\tanh\alpha~a^{\dagger}({\bf k})a^{\dagger}(-{\bf k})\right]|0\rangle\nonumber\\
                            &=&\frac{1}{{\cal N}}~\exp\left[-\frac{i}{2}e^{-i\beta}\tanh\alpha\sum_{{\bf k}}~a^{\dagger}({\bf k})a^{\dagger}(-{\bf k})\right]|0\rangle\nonumber\\
                            &=&\frac{1}{{\cal N}}~\exp\left[-\frac{i}{2}e^{-i\beta}\tanh\alpha\int\frac{d^3k}{(2\pi)^3}~a^{\dagger}({\bf k})a^{\dagger}(-{\bf k})\right]|0\rangle,\eea
                            where ${\cal N}=\sqrt{|\cosh\alpha|}$ is the overall normalization constant.
                            Here it is important to mention that the 
                            $|\alpha,\beta\rangle$ vacuum state satisfies the following constraint equation:
                            \bea \hat{\bf P}|\alpha,\beta\rangle&=&\int \frac{d^3p}{(2\pi)^3}~{\bf p}~a^{\dagger}({\bf p})a({\bf p})|\alpha,\beta\rangle\nonumber\\
                            &=& \prod_{{\bf k}}\int \frac{d^3p}{(2\pi)^3}~\frac{{\bf p}~a^{\dagger}({\bf p})a({\bf p})}{\sqrt{|\cosh\alpha|}}\exp\left[-\frac{i}{2}e^{-i\beta}\tanh\alpha~a^{\dagger}({\bf k})a^{\dagger}(-{\bf k})\right]|0\rangle\nonumber\\
                                                                &=&\int \frac{d^3p}{(2\pi)^3}~\frac{{\bf p}~a^{\dagger}({\bf p})a({\bf p})}{\sqrt{|\cosh\alpha|}}\exp\left[-\frac{i}{2}e^{-i\beta}\tanh\alpha~\sum_{{\bf k}}a^{\dagger}({\bf k})a^{\dagger}(-{\bf k})\right]|0\rangle\\
                                                                &=&\int \frac{d^3p}{(2\pi)^3}~\frac{{\bf p}~a^{\dagger}({\bf p})a({\bf p})}{\sqrt{|\cosh\alpha|}}\exp\left[-\frac{i}{2}e^{-i\beta}\tanh\alpha~\int \frac{d^3k}{(2\pi)^3}a^{\dagger}({\bf k})a^{\dagger}(-{\bf k})\right]|0\rangle= 0,\nonumber\eea
                            Additionally, here it is important to note that the~creation and annihilation operators for Bunch–Davies vacuum and $|\alpha,\beta\rangle$ vacuum state are connected via the following sets of Bogoliubov~transformations:
                            \bea b({\bf k})&=&\cosh\alpha~a({\bf k})+i~e^{-i\beta}\sinh\alpha~a^{\dagger}(-{\bf k}),\\
                          a({\bf k})&=&\cosh\alpha~b({\bf k})-i~e^{-i\beta}\sinh\alpha~b^{\dagger}(-{\bf k}).\eea
    \end{enumerate}
    \section{{Useful Integrals as Appearing in Scalar Three-Point~Function}}
    \label{q4}
   	  
   		 All the useful integrals appearing in the scalar three-point function are appended bellow:
   		 \begin{footnotesize}
   		   	   		   	   \bea &1.&\int^{0}_{-\infty}d\eta~\eta^2~e^{\pm iK\tilde{c}_{S}\eta}=\mp\frac{2}{iK^3\tilde{c}^3_{S}},\\
   		   	   		   	  &2.&\int^{0}_{-\infty}d\eta~\eta^2~e^{\mp i(2k_{a}-K)\tilde{c}_{S}\eta}=\pm \frac{2}{i(2k_{a}-K)^3\tilde{c}^3_{S}},
   		   	   		   	   \\ &3.&\int^{\eta_{f}=0}_{\eta_{i}=-\infty}d\eta~(1\mp ik_{b}\tilde{c}_{S}\eta)(1\mp ik_{c}\tilde{c}_{S}\eta)~e^{\pm iK\tilde{c}_{S}\eta} 
   =\frac{1}{iK^3\tilde{c}_{S}}\left[K^2+2k_{b}k_{c}+K(K-k_{a})\right],~~~~~~~~~~~~\\
   		   	   		   	   &4.&\int^{0}_{-\infty}d\eta~(1- ik_{b}\tilde{c}_{S}\eta)(1- ik_{c}\tilde{c}_{S}\eta)~e^{ i(K-2k_{a})\tilde{c}_{S}\eta}=- \int^{0}_{-\infty}d\eta~(1+ik_{b}\tilde{c}_{S}\eta)(1+ ik_{c}\tilde{c}_{S}\eta)~e^{- i(K-2k_{a})\tilde{c}_{S}\eta}\nonumber\\
   		   	   		   	    		   	   		   	   &=&-\frac{1}{i(2k_{a}-K)^3\tilde{c}_{S}}\left[K^2+2k_{b}k_{c}+K(K-5k_{a})-2(K-k_{a})k_{a}+4k^2_a\right],~~~~~\\
   		   	   		   	    &5.&\int^{0}_{-\infty}d\eta~(1-ik_{b}\tilde{c}_{S}\eta)(1\mp ik_{c}\tilde{c}_{S}\eta)~e^{i(K-2k_{b})\tilde{c}_{S}\eta}=-\int^{0}_{-\infty}d\eta~(1\mp ik_{b}\tilde{c}_{S}\eta)(1\pm ik_{c}\tilde{c}_{S}\eta)~e^{\mp i(K-2k_{b})\tilde{c}_{S}\eta}\nonumber\\
   		   	   		   	    		   	   		   	    		   	   		   	   &=&-\frac{1}{i(2k_{b}-K)^3\tilde{c}_{S}}\left[K^2-4k_{b}k_{c}+K(k_{c}-5k_{b})+6k^2_b\right],~~~~~\\ &6.&\int^{0}_{-\infty}d\eta~(1\mp ik_{a}\tilde{c}_{S}\eta)~e^{\pm iK\tilde{c}_{S}\eta} 
   =\pm \frac{1}{iK^2\tilde{c}_{S}}(K+k_{a}),~~~~~~~~~~~~\\
 &7.&\int^{0}_{-\infty}d\eta~(1\pm ik_{a}\tilde{c}_{S}\eta)~e^{\pm i(K-2k_{a})\tilde{c}_{S}\eta}=\pm
   		   	   		   	    \frac{(K-3k_a)}{i(2k_{a}-K)^2\tilde{c}_{S}},~~~~~\\
   		   	   		   	   &8.&\int^{0}_{-\infty}d\eta~(1\mp ik_{a}\tilde{c}_{S}\eta)~e^{\pm i(K-2k_{b})\tilde{c}_{S}\eta}=\pm\frac{(K+k_a-2k_b)}{i(2k_{b}-K)^2\tilde{c}_{S}},~~~~~ \\
   				   	   		   	    &9.&\int^{0}_{-\infty}d\eta~(1-ik_{a}\tilde{c}_{S}\eta)(1-ik_{c}\tilde{c}_{S}\eta)~e^{i(K-2k_{b})\tilde{c}_{S}\eta}\nonumber\\
   				   	   		   	    		   	   		   	    		   	   		   	    		   	   		   	   &=&-\frac{1}{i(2k_{b}-K)^3\tilde{c}_{S}}\left[(K-2k_b)(K+k_a-2k_b)+(K+2k_a-2k_b)k_c\right].~~~~~ \eea   			   	   		   	    		   	   		   	    		   	   		   	    		   	   		   	   \end{footnotesize}
  \vspace{-6pt}


\end{document}